\pdfoutput=1
\documentclass[11pt,a4paper,english,amsmath,amssymb]{scrbook}
\usepackage{mathptmx}
\usepackage[T1]{fontenc}
\usepackage[latin9]{inputenc}
\usepackage{pifont}
\usepackage{float}
\usepackage{calc}
\usepackage{amsthm}
\usepackage{amsmath}
\usepackage{graphicx}
\usepackage{setspace}
\usepackage{amssymb}
\makeatletter

\providecommand{\tabularnewline}{\\}

 \theoremstyle{definition}
 \newtheorem*{defn*}{Definition}
  \theoremstyle{plain}
  \newtheorem*{thm*}{Theorem}
\theoremstyle{plain}
\newtheorem{thm}{Theorem}
  \theoremstyle{plain}
  \newtheorem{conjecture}[thm]{Conjecture}
\newenvironment{lyxcode}
{\par\begin{list}{}{
\setlength{\rightmargin}{\leftmargin}
\setlength{\listparindent}{0pt}
\raggedright
\setlength{\itemsep}{0pt}
\setlength{\parsep}{0pt}
\normalfont\ttfamily}%
 \item[]}
{\end{list}}

\usepackage{myupthesis}

\usepackage[numbers,sort&compress]{natbib}

\usepackage{multicol}

\usepackage{acronym}

\usepackage{bm}

\usepackage{bbm}

\usepackage{epsfig}

\usepackage{dcolumn}

\usepackage{wasysym}

\usepackage{marvosym}

\usepackage{color}

\usepackage[colorlinks=false]{hyperref}

\newcommand{\mychaptermark}[1]{\markboth{#1}{}}

\AtBeginDocument{

}

\makeatother

\usepackage{babel}

\begin{document}
\thispagestyle{empty}
\pagenumbering{gooble}~

\begin{center}

\par\end{center}

\begin{center}
\textbf{\Large Aires F. Ferreira}{\large{} }\\
 \vspace{1.5cm}
\par\end{center}

\begin{doublespace}
\begin{center}
\textbf{\Huge The Quantum-Classical Boundary: }\\
\textbf{\Huge from Opto-Mechanics to Solid-State}
\par\end{center}{\Huge \par}
\end{doublespace}

\vspace{3cm}

\begin{center}
\includegraphics[clip,width=0.25\paperwidth]{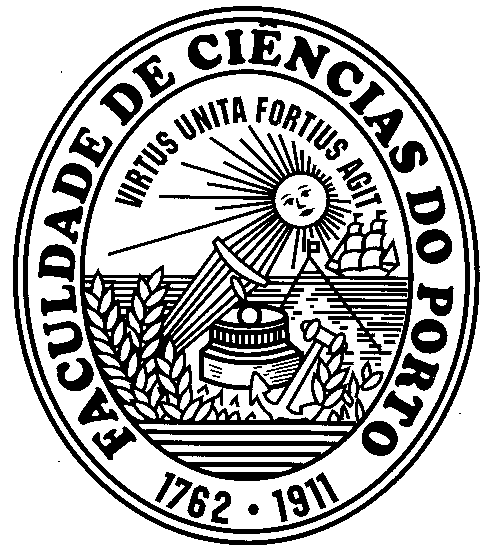}
\par\end{center}

\vspace{3cm}

\begin{center}
\textbf{\Large Departamento de Física}\\
\textbf{\Large Faculdade de Ciências da Universidade do Porto }\\
\textbf{\Large September / 2009}
\par\end{center}{\Large \par}

\newpage{}\thispagestyle{empty}~

\newpage{}\thispagestyle{empty}
\pagenumbering{arabic}~

\begin{center}
\textbf{\Large Aires F. Ferreira}{\large{} }\\
 \vspace{1.5cm}
\par\end{center}

\begin{doublespace}
\begin{center}
\textbf{\Huge The Quantum-Classical Boundary: }\\
\textbf{\Huge from Opto-Mechanics to Solid-State}
\par\end{center}{\Huge \par}
\end{doublespace}

\vspace{1.5cm}

\begin{center}
\includegraphics[width=0.25\paperwidth]{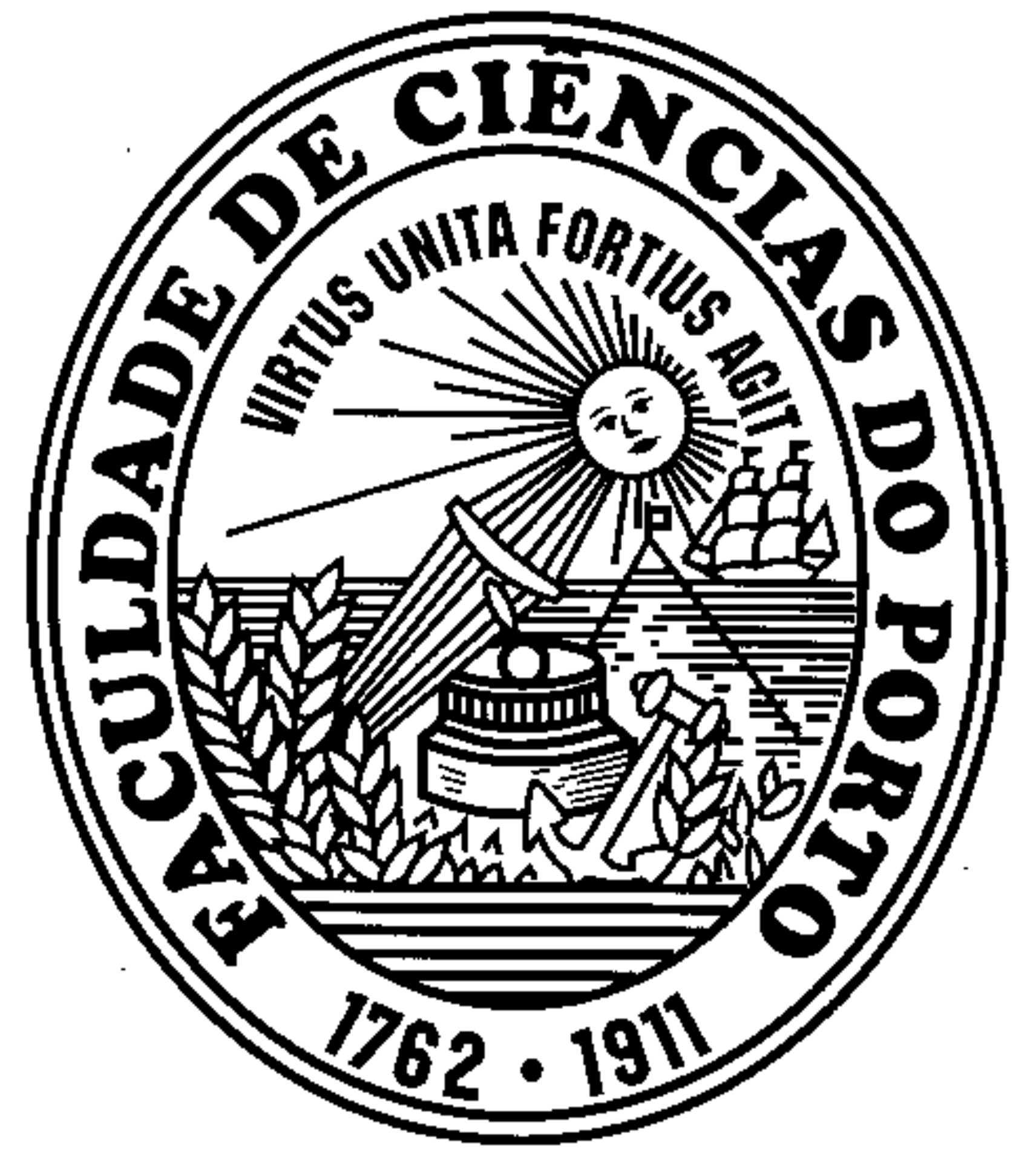}
\par\end{center}

\begin{center}
\vspace{1cm}
\par\end{center}

\begin{center}
\emph{PhD Thesis supervised by Prof. João M. B. Lope}s dos Santos
\par\end{center}

\vspace{1cm}

\begin{center}
{\Large Departamento de Física}\\
{\Large Faculdade de Ciências da Universidade do Porto }\\
{\Large September / 2009}
\par\end{center}{\Large \par}

\newpage{}~

\begin{center}
\newpage{}~\vspace{5cm}
\par\end{center}

\begin{center}
\textbf{\emph{To my adorable godfather}}\\
\textbf{\emph{Joaquim Francisco Gonçalves Ferreira de Sousa}}\\
\textbf{\emph{with love.}}
\par\end{center}

\newpage{}~

\newpage{}

\begin{onehalfspace}

\chapter*{Acknowledgments}
\end{onehalfspace}

First I would like to thank my parents for being there for me throughout
my studies, since primary school to university. I am also indebted
to my godfather, who had been my inspiration since I was a little
kid, not only for me but also to many people who had the chance to
meet him. Unfortunately, he is not among us anymore, but he would
have enjoyed sharing with me this enthusiastic journey through science.
This thesis is dedicated to him.

I offer my sincerest gratitude to my PhD advisors, Ariel Guerreiro
and João Lopes dos Santos. Without them this thesis would not have
been written. I would like to thank them for their encouragement and
effort throughout these years.

I was an undergraduate student when I had the first contact with research.
I thank Carlos Herdeiro for the time he had spent with me doing calculations
and discussing the physics of relativity --- this experience was most
important to my scientific culture. In 2005, at Vienna, I had the
chance to make part of the Anton Zeilinger\textquoteright{}s research
group. I would like to thank Markus Aspelmeyer for his hospitality
and good humour --- he made me feel at home. Thanks also to Sylvain
Gigan and Hannes Böhm for sharing my enthusiasm towards the \textquotedbl{}mirror
experiment\textquotedbl{}. Physics in Zeilinger\textquoteright{}s
group is taken seriously and most passionately. David Vitali, from
Camerino University, was also essential to augment the scope of our
second work. I hope the Vienna group is the first to be successful
in demonstrating macroscopic quantum entanglement.

At the beginning of this journey, Ariel Guerreiro, my Portuguese supervisor
at the time, was very understanding about my scientific options. In
a brief visit to Porto, we set forth the method that had led to the
main conjecture of the Vienna period. Still from this time, I am grateful
to \v{C}aslav Brukner, Marcin Wie\'{s}niak and J.Kofler for the fruitful
discussions, friendship and encouragement. Thanks too to Vlakto Vedral
for his determined spirit --- one felt fired with enthusiasm when
he was around. I am also indebted to Simon Gröblacher, Susana Anjos,
Mariana Meireles and my brother, Telmo Ricardo, for their friendship:
you all helped me to go beyond the occasionally loneliness of this
Austrian winter. 

The second half of my PhD took place in Portugal, at the CFP, under
the supervision of João Lopes dos Santos. I appreciate his courage
of embracing a project outside his lines of research at the moment.
His sui generis intuition in Physics was essential for the prediction
of results; I truly believe he has intuition for all fields of research.
Besides, one could not wish for a better and friendlier supervisor. 

At the CFP, a small group of Theoretical Physics, there was a dynamics
between senior and younger scientists that became very gainful, of
which I was lucky to make part. CFP was like a second home to me as
well. In regard to this, I thank to João Viana Lopes, a former CFP
PhD student, whose effort led to improve the conditions of work in
the centre. I also welcome his friendship and won\textquoteright{}t
forget the well-spent talks about science and politics that have helped
me to expand my perspectives about the world. During this period,
I also had the chance to embrace several challenges, namely, sessions
with high-school and undergraduate students, The Summer Physics School,
and collaborations with the financial consulting world. These experiences
helped me to become a more adaptable person in the face of these different
situations that turned out to be successful as a result of the group\textquoteright{}s
unity.

They are, Prof. Eduardo Lage, Miguel Sousa Costa, João Penedones,
João Viana Lopes, Vítor Pereira, Eduardo Castro, Pedro Gil, Miguel
Dias Costa, Filipe Paccetti, Miguel Zilhão, Jozinaldo Menezes, Joana
Espaín, Roberto Menezes, Carmen Rebelo, Teresa Martins, Filipe Zola,
Jaime Santos and Luís Bessa. Carla Rosa, Helder Crespo and Murilo
Baptista, you also made part of this journey. Thanks to Pedro Bernardino,
Filipe Paccetti and Mariana Marques for their helpful suggestions
during the writing of this thesis. I cannot forget Florbela Teixeira
and Fátima Pinheiro for their good humours and pragmatism in the solving
of any issue related to the faculty. I acknowledge the Algorithms
and Libraries for Physics Simulations project and the library \textquotedblleft{}looper\textquotedblright{}.
Also thanks to EU and FCT for its financial support. 

Last, but not least, I am grateful to Mariana for her love and care. 

\newpage{}~

\newpage{}

\chapter*{Resumo}
\begin{quotation}
{\small Recentes avanços experimentais em vários campos da Física,
desde a Óptica Quântica até ao Estado Sólido, trouxeram velhas questões
da Mecânica Quântica para a linha da frente do debate científico.
Até que ponto efeitos quânticos podem ser observados em sistemas grandes
é ainda uma questão em aberto. Apesar disto, no laboratório, a fronteira
quântica-clássica tem-se rapidamente movido em direcção ao mundo macroscópico:
foi observada interferência em moléculas tão grandes quanto flurenos
(1999, Viena), assim como sobreposição de currentes macroscópicas
em dispositivos supercondutores de interferência quântica (2000, Nova
Iorque).}{\small \par}

{\small Comum a todas estas experiências, é o fenómeno de perda de
coerência; a irreversibilidade inevitável dos sistemas abertos, dita
as escalas em que assinaturas quânticas podem ser observadas. A coerência
quântica pode ser extrememamente robusta, a saber, nos }\emph{\small spins}{\small{}
de electrões em defeitos de diamante devido ao grande hiato espectral
destes materiais, mas pode ser muito frágil em sistemas massivos,
dado os ínumeros canais de perda de coerência. Onde se encontra exactamente
a fronteira quântica-clássica vai depender das peculiariedades do
sistema físico em questão, em particular, da magnitude do acoplamento
ao ambiente e a sua temperatura. Esta última grandeza, crê-se ser
o principal obstáculo na obtenção de estados não-clássicos nas mais
leves nano-estruturas mecânicas, onde nenhum efeito quântico foi ainda
observado.}{\small \par}

{\small Motivados por estes desafios, e pela diferença fundamental
entre correlações clássicas e }\emph{\small entanglement}{\small ,
investigamos a fronteira quântico-clássica através das seguintes questões:
podem correlações quânticas macroscópicas (e portanto, o comportamento
quântico) persistir acima do limite de baixas temperaturas, e como
podem estes efeitos ser usados, por exemplo, para correlacionar quanticamente
outros sistemas?} 
\end{quotation}
A primeira parte é dedicada a um sistema paradigmático de opto-mecânica:
mostra-se que as quadraturas de um estado coerente da luz e o movimento
dum oscilador mecânico, acoplados através de pressão de radiação numa
geometria de cavidade, podem ficar substancialmente correlacionados
de forma genuinamente quântica. Inicialmente, consideramos um cenário
ideal (isto é, sem perda de coerência) e, através dum procedimento
de renormalização de subspaços do operador de estado, conjecturamos
que o \emph{entanglement} bipartido do sistema é robusto relativamente
à temperatura. De seguida, num breve capítulo, discutimos um cenário
realista onde uma fonte luminosa intensa é usada para popular uma
cavidade com espelhos parcialmente reflectivos. É mostrado que um
acoplamento efectivo, proporcional à amplitude interna da cavidade,
surge no estado estacionário, o que comprova as conclusões tiradas
acerca do sistema ideal: encontra-se \emph{entanglement} macroscópico
opto-mecânico persistente até temperatures muito acima da energia
de ponto zero do oscilador mecânico. 

Na segunda parte, focamo-nos num outro cenário promissor: a geração
de \emph{entanglement }robusto entre \emph{spins }distantes, através
de sistemas de Estado Sólido fortemente correlacionados. Com efeito,
através de uma teoria de perturbações adequada, é estudado o problema
de extracção de \emph{entanglement }a partir destes sistemas. Inicialmente,
consideram-se dois modelos de anti-ferromagnetismo a uma dimensão:
a cadeia finita de Heisenberg e a cadeia infinita de Affleck-Kennedy-Lieb-Tasaki.
Demonstramos que a geração de \emph{entanglement} entre os dois \emph{spins},
devido à interacção local com o sistema mediador, é extremamente eficiente:
os spins adquirem \emph{entanglement} quase completo mesmo a distâncias
grandes. O cálculo do Hamiltoniano efectivo de interacção entre os
spins, prova-se adequado para a investigação deste fenómeno em sistemas
de muitos corpos de hiato finito e quando o acoplamento \emph{spin}-sistema
é fraco. No último capítulo, estes resultados são generalizados para
possibilitar o cálculo das correlações \emph{spin-spin} no regime
não-perturbativo a temperatura finita. Comparando resultados analíticos
com dados de Monte Carlo Quântico, é provada a existência de correlações
quânticas nos spins a temperaturas muito mais elevadas do que anteriormente
se julgava ser possível. Isto acontece no cenário bi-dimensional,
devido ao surgimento de hiatos de energia consideráveis, mesmo em
sistemas com um hiato intrínseco muito pequeno (desaparecendo no limite
termodinâmico), devido apenas à presença dos \emph{spins }externos.

A presente tese, mostra que as ferramentas de Informação Quântica
podem ser usadas no sentido de melhor compreendermos a fronteira que
separa o mundo quântico do mundo clássico em sistemas macroscópicos.
As nossas descobertas sugerem uma forma de alargar esta fronteira
em direcção ao mundo macroscópico, através do acoplamento dum espelho
movível a um campo electromagnético confinado, e abrem novas possibilidades
relativamente à computação e processamento de informação quântica
em sistemas fortemente correlacionados a temperaturas realistas.

\chapter*{Summary}
\begin{quotation}
{\small Recent experimental breakthroughs in miscellaneous fields
of physics, from Quantum Optics to Solid State, have brought the old
questions of Quantum Mechanics into the front line of scientific debate.
To what extent }\emph{\small bona fide }{\small quantum effects are
observable in large systems is still an open question. Notwithstanding,
in the laboratory, the quantum-classical boundary has been moving
towards the macroscopic world very quickly: interference was observed
in molecules as large as fullerenes (1999, Vienna) and superposition
of macroscopic currents was achieved in superconducting quantum interference
devices (2000, New York).}{\small \par}

{\small Common to all these experiments is the decoherence phenomenon;
the unavoidable irreversibility of open systems, ultimately setting
the scales where a quantum signature is hoped to be observed. The
quantum coherence can be extremely robust, for instance, in single
electron spins in diamond defects, due to its natural large gap to
excited states, but can be extremely feeble in massive systems for
many decoherence channels are available. Where the quantum-classical
boundary exactly lies depends on the peculiarities of a given physical
system and especially on the strength of coupling to the environment
and on its temperature. In fact, it is believed that temperature is
the main obstacle in achieving non-classical effects in light mechanical
nano-structures where no quantum behaviour has been observed so far. }{\small \par}

{\small Motivated by these developments and the fundamental difference
between classical correlations and entanglement, in this thesis we
investigate the quantum-classical boundary by asking the following
questions; can macroscopic quantum correlations, an thus quantum behavior,
persist above the low-temperature threshold, and how such effects
can be used, for instance, to entangle other systems? }{\small \par}
\end{quotation}
In the first part, focusing on a paradigmatic opto-mechanical system,
it is shown that the quadratures of a coherent state of light and
the motion of a mechanical oscillator, coupled via radiation pressure
in a cavity geometry, can be substantially quantum correlated. Initially,
we consider an ideal scenario (\emph{i.e.}~no decoherence) and, by
employing a renormalization procedure to finite dimensional subspaces
of the complete density matrix, we conjecture that bipartite entanglement
is very robust against temperature. The entropy of the subsystems
discloses a macroscopic amount of quantum correlations and suggests
that the mirror-light entanglement can be enhanced by adding more
photons to the cavity. Afterwards, in a short chapter, we discuss
a realistic scenario where a pumping bright source is used to populate
a cavity with partially reflective mirrors. We show that an effective
coupling emerges in the stationary regime that is proportional to
the intra-cavity field amplitude, thus settling on solid grounds previous
conclusions about the ideal system. We find opto-mechanical entanglement
surviving at temperatures much above the mechanical oscillator\textquoteright{}s
ground state energy, therefore overcoming the conventional criterion
on temperature. 

In the second part, we focus on yet another encouraging scenario,
namely that of generating robust entanglement between distant spins
by exploiting the highly correlated ground states of solid state systems.
Indeed, by developing an adequate perturbation theory, we study the
problem of entanglement extraction from non-critical many-body systems
to probes endowed with a two-dimensional Hilbert space. Initially,
we focus on two models of one-dimensional anti-ferromagnetism, namely
the Heisenberg and the Affleck-Kennedy-Lieb-Tasaki spin chains, and
show that entanglement generation between initially uncorrelated probes,
weakling interacting with the many body bus, is extremely efficient,
as they can share quasi-perfect entanglement even at large distances.
The computation of the effective Hamiltonian of interaction of the
probes defines a suitable framework to investigate the phenomenon
in generic gapped quantum lattice systems. In the last chapter these
results are generalized as to allow the computation of probe correlations
in situations where adiabatic continuity between the eigenstates of
the full many-body Hamiltonian and the unperturbed system holds. This
encompasses the effect of temperature and the non-perturbative regime.
By comparing analytic results against Quantum Monte Carlo data, we
go far away perturbation theory limits and unveil probe-probe quantum
correlations at temperatures much higher than previously thought possible.
The latter happens in the two-dimensional scenario, where robust gaps
are shown to emerge even in lattices with a very small gap (vanishing
in the thermodynamic limit) solely due to the presence of the probes.

The present thesis shows that Quantum Information tools can be used
to better understand the quantum-to-classical boundary in mesoscopic
and macroscopic systems. Our findings suggest a way to push this boundary
towards the macroscopic world by coupling a moveable mirror to a confined
quasi-classical electromagnetic field, and opens new possibilities
towards quantum computation and information processing with strongly-correlated
systems at realistic temperatures, \emph{i.e.}~much above their natural
ground states. 

\tableofcontents{}

\listoffigures

\chapter*{List of Acronyms\label{cha:Acronyms}}

\mychaptermark{List of Acronyms}

\begin{acronym}

\acro{1D}{one-dimensional}
\acro{2D}{two-dimensional}
\acro{3D}{three-dimensional}
\acro{AF}{Anti-Ferromagnetic}
\acro{AKLT}{Affleck-Kennedy-Lieb-Tasaki}
\acro{CV}{Continuous Variable}
\acro{DMRG}{Density Matrix Renormalization Group}
\acro{EM}{Electromagnetic}
\acro{EPR}{Einstein-Podolsky-Rosen}
\acro{FM}{Ferromagnetic}
\acro{GS}{Ground State}
\acro{LASER}{Light Amplification by Stimulated Emission of Radiation}
\acro{LDE}{Long-Distance Entanglement}
\acro{LOCC}{Local Operations and Classical Communication}
\acro{PPT}{Positive Partial Transposition}
\acro{QI}{Quantum Information}
\acro{QMC}{Quantum Monte Carlo}
\acro{QND}{Quantum Non-Demolition}
\acro{SQL}{Standard Quantum Limit}
\acro{ZPF}{Zero-Point Fluctuations}

\end{acronym}

\begin{onehalfspace}

\chapter{Introduction\label{cha:Introduction}}
\end{onehalfspace}

This thesis is devoted to the study of the quantum-classical boundary
at finite temperatures, in two distinct physical scenarios, and to
its implications for fundamental physics.

This monograph consists of two large blocks as follows:

\begin{minipage}[t]{1\columnwidth}%
\begin{itemize}
\item in Chapter~\ref{cha:MacroscopicEntang}, we focus our attention on
a very popular subject in current state-of-the-art opto-mechanics:
the \emph{bona fide} quantum behavior of a macroscopic mechanical
oscillator driven by electromagnetic radiation pressure. We will argue
that this system accomplishes the possibility of \emph{macroscopic
entanglement}. In Chapter~\ref{cha:Stationary-optomechanical-entangl},
we will show that the same opto-mechanical setup, provided with an
appropriate measurement apparatus, turns out to be an excellent candidate
for an experiment aiming to test the quantum-to-classical transition
at realistic temperatures;
\item in Chapter~\ref{cha:LDE_Via_GS_GappedSpinChains}, we show that spin-$1/2$
probes develop quantum correlations when they are locally coupled
to gapped many-body systems; this is a different facet of the quantum-to-classical
transition when the bulk system is perceived as a model for an environmental
bath. The possibilities for quantum communication and computation
entirely based on solid-state devices at finite temperatures will
be analyzed in Chapter~\ref{cha:LDE_Finite_Temperature}, where the
emergence of gaps in various spin systems due to additional spin particles
will be shown to accommodate bipartite spin-spin entanglement at temperatures
much higher than previously considered possible.
\end{itemize}
\end{minipage}

The present chapter attempts to shed some light on the context and
relevance of the topics covered in this thesis (Sec.~\ref{sec:Context})
and also to make a comprehensive review of the basic results for the
characterization of quantum correlations (Sec.~\ref{sec:C-versus-Q}
and \ref{sec:Entang-th}): the theory of \emph{quantum entanglement}.
While experts may consider skipping the introduction, the reader new
to Quantum Information concepts should find this section particularly
helpful to learn the crucial difference between the classical correlations
and their non-classical counterparts. To ensure the readability of
the text, useful background related to major standard technical subjects
will be given in special appendices at the proper time, and informal
and easy to follow derivations will be favoured when possible over
more rigorous (but often less illuminating!) mathematical approaches.

\section{Context\label{sec:Context}}

Correlations between different systems have always been an active
subject of study in various branches of theoretical physics. Ranging
from criticality in classical statistical mechanics to many body effects
in electronic systems, correlations appear as a fundamental property
characterizing interacting systems. More recently, a new area of physics
has emerged mainly from the Quantum Optics community: \ac{QI} science.
Originally motivated by a close examination of the foundations of
Quantum Mechanics, the \ac{QI} community envisaged quantum communication
protocols and a new paradigm of computation based on the laws of Quantum
Mechanics.

In the past few years, the interests of this community have broadened
by extending the methods initially developed to characterize the quantum
correlations in small Hilbert spaces (such as the polarization degree
of freedom of two photons) to encompass the solid-state and condensed-matter
systems from a new perspective: the so-called \textquotedbl{}entanglement\textquotedbl{}
approach.

The entanglement theory provides a suitable framework to think about
non-classical quantum information processing tasks (\emph{e.g.}~teleportation),
and also paves the way for the resolution of the old question of the
quantum-to-classical transition: when does a physical system looses
every quantum signature and behaves classically? 

Throughout this monograph, we will show that applying the entanglement
approach to the study of interacting systems unveils how far we can
hope to go on pushing the genuine quantum behavior of the microscopic
world towards the macroscopic domain. The \textquotedbl{}boundary\textquotedbl{}
is not the same in distinct physical scenarios, where many different
kind of interactions may play a role, and it will show to be very
sensitive to the initial conditions. Notwithstanding, the entanglement
approach will prove to capture important subtleties of correlations
in an unified picture.

\section{Classical versus quantum correlations\label{sec:C-versus-Q}}

\subsection{Preliminaries }

In general, an experimentalist has no way to prepare with perfect
control a definite quantum state, \begin{eqnarray}
|\psi\rangle & = & \sum_{n}\langle n|\psi\rangle|n\rangle,\label{eq:Chap1-QuantumState}\end{eqnarray}
where $\left\{ |n\rangle\right\} \in\mathcal{H}$ denotes a basis
of the Hilbert space and $\langle n|\psi\rangle$ are arbitrary complex
amplitudes. Instead, he/she prepares a probabilistic ensemble of p\emph{ure
states} $\left\{ |\psi_{n}\rangle\right\} $ with associated probabilities
$\rho_{n}$. For the moment, and without loss of generality, this
ensemble will be taken to consist entirely of \emph{orthogonal} (and
\emph{normalized}) individual states. At the end of the day, one is
interested in averages of physical quantities, and thus it is instructive
to see how physical averages look like when an ensemble of quantum
states is assumed. 

Bearing in mind these considerations, the prediction $\langle A\rangle$
of a generic physical observable $\hat{A}$ must be the weighted average
of the expectation values $\langle\psi_{n}|\hat{A}|\psi_{n}\rangle$
for the pure states $|\psi_{n}\rangle$, \begin{eqnarray}
\langle A\rangle & = & \sum_{n}\rho_{n}\langle\psi_{n}|\hat{A}|\psi_{n}\rangle=\text{Tr}\left[\hat{\rho}\hat{A}\right],\label{eq:Chap1-QuantumAverage}\end{eqnarray}
where we have conveniently introduced the \emph{density matrix} operator,\begin{equation}
\hat{\rho}:=\rho=\sum_{n}\rho_{n}|\psi_{n}\rangle\langle\psi_{n}|.\label{eq:Chap1_DensityMatrix}\end{equation}
The spectrum of this operator may be thought as the quantum-mechanical
analogue of the familiar Boltzmann weights in classical statistical-mechanics,
$\rho_{n}\sim e^{-\beta E_{n}},$ describing the probability of finding
a canonical system in a configuration with definite energy $E_{n}$
at temperature $T\equiv1/k_{B}\beta$. The only difference being that
in Quantum Mechanics there are not many different microscopic configurations
contributing to the same macroscopic configuration $n$, but rather
an unique%
\footnote{Here it is assumed that the energy spectrum is non-degenerate. It
can happen (and many times it does) that a given state is degenerate.
This degeneracy is usually broken in realistic scenarios by the environment.%
} quantum state $|\psi_{n}\rangle$ with energy given by the Schr\"{o}dinger
equation,\begin{equation}
\hat{H}|\psi_{n}\rangle=E_{n}|\psi_{n}\rangle,\label{eq:Chap1-SchroedingerEq}\end{equation}
with $\hat{H}$ denoting the system's Hamiltonian. There will more
to say about this analogy soon. Let us now look into the properties
of the density matrix in more detail. From Eq.~(\ref{eq:Chap1_DensityMatrix})
we have,\begin{equation}
\text{Tr}[\rho]=\sum_{n}\rho_{n}\text{Tr}[|\psi_{n}\rangle\langle\psi_{n}|]=\sum_{n}\rho_{n}\langle\psi_{n}|\psi_{n}\rangle=1,\label{eq:Chap1-NormalizationCondition}\end{equation}
as both the $\left\{ \rho_{n}\right\} $ and the individual states
$\left\{ |\psi\rangle\right\} $ are assumed to be normalized, that
is $\sum_{n}\rho_{n}=1$ and $\langle\psi_{n}|\psi_{n}\rangle=1$.
The density matrix is Hermitian as can be seen by direct inspection
(hence assuring that $\rho_{n}\in\mathbb{R}$). Moreover, since $\rho_{n}$
must represent a probability, the density matrix $\rho$ is \emph{semi-positive
definite}. The latter is a strong restriction in the class of matrices
in $\mathcal{H}$ allowed as acceptable density matrices:\begin{equation}
\langle\psi|\rho|\psi\rangle=\sum_{n}|\langle\psi|n\rangle|^{2}\rho_{n}\geqslant0.\label{eq:Chap1_Semi-Positivity}\end{equation}
Eq.~(\ref{eq:Chap1_Semi-Positivity}) is, for calculus purposes,
an efficient way to check if a given matrix actually represents a
physical state. Before moving forward, a comment is in order: the
decomposition in Eq.~(\ref{eq:Chap1_DensityMatrix}) is not unique
for mixed states of two or more quantum systems. This makes no difference
at the time of evaluating a given physical quantity by means of Eq.~(\ref{eq:Chap1-QuantumAverage}),
but will have important consequences for the discussion of entanglement
in mixed states.

\subsection{The randomness of quantum states}

We have introduced the notation of density matrix defining ensembles
of quantum states and its most elementary features. Now we prepare
the grounds for the understanding of the difference between ordinary
correlations of every-day life and quantum correlations, by reviewing
the concept of entropy. 

The notion of statistical ensembles appeared a long time ago in the
description of systems surrounded by a (very large) heat bath, where
the microscopic details of the system-bath interaction are discarded
in favor of concrete answers for quantities of physical interest (such
as the specific heat, \emph{etc}.). The statistical description is
also the appropriate framework to study classical communication problems
where noise (as a source of uncertainty) cannot be neglected. 

In general, only probabilities for the outcomes of physical observables
can be predicted. This lack of knowledge about physical systems, lead
us to the concept of entropy: let us imagine that we own $N$ copies
of identical prepared systems (\emph{e.g.}~an atom in its ground
state) and that $p_{1},$$...,$$p_{d}$ are the \emph{a priori }known
probabilities of the different outcomes we may get ($d$ standing
for the number of such outcomes; typically the dimension of the system's
Hilbert space). How much information do we gain about a single system
by performing measurements on a large number of identically prepared
systems? The answer is related to the number of possible ways to arrange
the measurement's results: assuming that $N$ is large, we expect
that each outcome, labeled by $i$, materializes in average $\sim n_{i}=Np_{i}$
times, with $\sum_{i}n_{i}=N$. The number of arrangements is $k=N!/(n_{1}!...n_{d}!)$.
In the limit $N\rightarrow\infty$, we get $\log k=\log\left(N!/(n_{1}!...n_{d}!)\right)\simeq-N\sum_{i}p_{i}\log p_{i}$
, naturally suggesting the quantity%
\footnote{Other functions of $k$ could be considered as measures of entropy,
but this particular choice has suitable properties. For instance,
it is \emph{additive}: for two statistical independent systems it
yields the sum of individual entropies. %
},\begin{equation}
H:=-\sum_{i=1}^{d}p_{i}\log p_{i},\label{eq:Chap1_Shannon Entropy}\end{equation}
as a measure of the uncertainty. This is the famous Shannon entropy
for the classical distribution $\left\{ p_{1},...,p_{d}\right\} $.
It measures the ignorance we have about a physical system (prior to
measurement) and achieves its maximal value if all the $p_{i}$ are
equal. 

The classical information theory was founded by Shannon \cite{1948Shannon}
whose pioneering work is of paramount importance nowadays in technology
and science. In a classical information problem (\emph{e.g.}~the
transmission of a message from $A$ to $B$) $H$ quantifies the information
that has been transmitted, which may seem a bit awkward for the physicist
who usually thinks about entropy as ignorance rather than knowledge.
This false impression, however, becomes clear by noting that $B$
cannot learn anything if the outcome is known to him/her prior to
transmission, and does learn a good amount of information if he/she
cannot predict any of the outcomes, \emph{i.e.}~if $p_{i}=1/d$,
$\forall_{i}$.

In classical statistical mechanics, the entropy is the logarithm of
the number of microstates ($\Omega$) of the system, $S=\ln\Omega$
(defined up to a constant due to the arbitrary dense volume of the
phase space cells), where the Boltzmann's constant was set to unit.
In quantum statistical mechanics the Heisenberg principle imposes
a limit on how small the phase space cells can be; there is no arbitrariness
in counting the number of microstates. Let us consider the elementary
example of a \emph{totally random mixture} (also known as \emph{maximally
mixed state} in \ac{QI}) of eigenstates occurring with the same probability,
with $\dim{\cal H}=d$, \begin{eqnarray}
\rho_{d} & = & \frac{1}{d}\sum_{n=1}^{d}|\psi_{n}\rangle\langle\psi_{n}|=\frac{{\mathbbm{1}}_{d}}{d}.\label{eq:Chap1_random_mixture}\end{eqnarray}
The number of \textquotedbl{}microstates\textquotedbl{} is the number
of pure states $d$ and, therefore, the entropy reads $S=\ln d$ according
to Boltzmann's formula. In quantum mechanics, however, we have to
compute the entropy by the von Neumann formula,\begin{equation}
S(\rho):=-\text{Tr}[\rho\ln\rho].\label{eq:Chap1_VonNeumann}\end{equation}
This expression yields zero for every pure state, $S(|\psi\rangle)=0,$
and reaches its maximum value for a maximally mixed state, $S(\rho_{d})=\ln d$.
Although there are many possible entropies (for a complete review
on the subject see \cite{1978Wehrl}), von Neumann's formula has suitable
physical properties (Appendix~\ref{sec:App-Entropy}) and plays a
crucial role for the theory of entanglement as it will become apparent
in the next section. For the moment, it is enough to realize that
the familiar result for the Boltzmann entropy of a totally random
mixture, $S(\rho_{d})=\ln d$, is straightforwardly obtained using
Eq.~(\ref{eq:Chap1_VonNeumann}), and that it is the quantum extension
of the classical Shannon entropy. The latter can be seen by considering
a density matrix in the general form given by Eq.~(\ref{eq:Chap1_DensityMatrix})
and computing its von Neumann entropy. This yields,\begin{equation}
S=-\sum_{n}\rho_{n}\ln\rho_{n},\label{eq:-5}\end{equation}
which resembles the classical Shannon entropy {[}Eq.~(\ref{eq:Chap1_Shannon Entropy}){]}
for a random variable. 

The von Neumann entropy {[}Eq.~(\ref{eq:Chap1_VonNeumann}){]} is
the best measure of how \textquotedbl{}mixed\textquotedbl{} a state
is, although it has a major disadvantage compared to other types of
entropy; excluding rather special situations it is very difficult
to compute as it presupposes diagonalization of $\rho$. For practical
ends, one often adopts the so-called \emph{linear entropy} instead.
For a density matrix $\rho$ living in a Hilbert space of dimension
$d$ it reads\emph{ $S_{\mathcal{L}}(\rho):=d/(d-1)\left(1-\text{Tr}\left[\rho^{2}\right]\right).$}
It has the nice virtue of being easy to compute in all situations
and it is directly related to the \emph{purity }of a state ${\cal P}(\rho):=\text{Tr}[\rho^{2}]$,
\begin{equation}
S_{{\cal L}}(\rho)=\frac{d}{d-1}(1-{\cal P}(\rho)).\label{eq:Chap1_LinearEntropy}\end{equation}
The purity\emph{ ${\cal P}\in[0,1]$} is one for pure states and decreases
with the degree of mixture, \emph{i.e.} as soon as more non-zero eigenvalues
$\rho_{n}$ appear in decomposition {[}Eq.~(\ref{eq:Chap1_DensityMatrix}){]}.
Depending on the context it may be more convenient to employ the purity
rather than the linear entropy; indeed, both concepts will be employed
in this monograph and, unless stated otherwise, the word entropy will
be used to denote the von Neumann entropy {[}Eq.~(\ref{eq:Chap1_VonNeumann}){]}.

\subsection{The EPR \textquotedbl{}super-correlations\textquotedbl{}}

Very often, density matrices appear coupled to some sort of statistical
description of a physical system; this could be due to the inability
to control the states in a realistic experimental scenario (\emph{e.g.}~external
noise or an intrinsic random process), or even as a mathematical tool
to compute an average from an adequate statistical ensemble. This
is not, however, the full story. For a system with several components
(\emph{e.g.} particles, degrees of freedom) the density matrix is
an useful concept even when the full system is a pure one. In order
to see this, let us consider the Bohm version of the famous \ac{EPR}
paradox \cite{1935EPR}. A spin-1 particle in the $S=0$ state decays
into two spin-1/2 particles ($A$ and $B$) in a singlet state:\begin{equation}
|\psi_{AB}^{-}\rangle=\frac{1}{\sqrt{2}}\left(|\uparrow_{A},\downarrow_{B}\rangle-|\downarrow_{A},\uparrow_{B}\rangle\right).\label{eq:-6}\end{equation}
This is a very special state for reasons, which will become clear
in the course of this chapter. On one hand, it is rotationally invariant:
$\uparrow_{A}$ means that spin $A$ points in the positive direction
of the $z$-axis or any other direction in \ac{3D} space. This suggests
an alternative, and perhaps more revealing, way of writing the state
of the two particles:\begin{equation}
|\psi_{AB}^{-}\rangle=\frac{1}{\sqrt{2}}\left(|+_{A},-_{B}\rangle-|-_{A},+_{B}\rangle\right),\label{eq:Chap1_SingletState}\end{equation}
where now the meaning of $+_{A}$ is that the spin $A$ is pointing
in the positive direction of a given (arbitrary) axis. From (\ref{eq:Chap1_SingletState})
we immediately see that whenever $A$ is detected in the positive
direction, $B$ will be detected pointing out in the negative direction
(provided the measurement is made along the same axis). This is already
a curious feature of (\ref{eq:Chap1_SingletState}), but more is yet
to come; if an observer decides to measure $B$ along a different
direction, he/she will measure $+$ or $-$ with the same probability.
The \ac{EPR} paradox comes about when we imagine these particles
to be separated over a large distance%
\footnote{Whatever direction $A$ is detected being spinning then $B$ will
be found spinning in the opposite direction. This might suggest faster-than-light
signaling, but this is of course not the case since two distant observers
must agree on the measurement axis, which force them sending (perhaps
by a phone call) a \textquotedbl{}classical\textquotedbl{} information
which speed is bounded from above by the speed of light. %
}, in such a way that a measurement performed in one of the particles
cannot influence the result obtained for the other. If we measure
$A$ along $x$, we can infer the corresponding outcome for $B$ {[}Eq.~(\ref{eq:Chap1_SingletState}){]}.
If we now measure $B$ along $y$, we will have determined the spin
of $B$ along two orthogonal directions! On the other hand, Quantum
Mechanics does not allow for the simultaneous knowledge of the value
of two non-commuting variables (such as the spin operators along $x$
and $y$). Hence, one can conclude that either Quantum Mechanics is
incomplete or these two quantities cannot have simultaneous reality.
This is the essence of the \ac{EPR} paradox in a few lines. In order
to see where the density matrix shows up, we define the partial state
of two systems $A$ and $B$,\begin{eqnarray}
\rho_{A} & := & \text{Tr}_{B}\rho_{AB}\nonumber \\
\rho_{B} & := & \text{Tr}_{A}\rho_{AB},\label{eq:Chap1_PartialStates}\end{eqnarray}
where $\rho_{AB}=|\psi_{AB}\rangle\langle\psi_{AB}|$ for a generic
state of two spin-1 particles $|\psi_{AB}\rangle$ and $\text{Tr}_{A(B)}$
denotes the partial trace with respect to $A(B)$. For the singlet
{[}Eq.~(\ref{eq:Chap1_SingletState}){]}, this yields,\begin{equation}
\rho_{A}=\rho_{B}=\frac{1}{2}\left(|+\rangle\langle+|+|-\rangle\langle-|\right)=\left(\begin{array}{cc}
\frac{1}{2} & 0\\
0 & \frac{1}{2}\end{array}\right).\label{eq:Chap1_partial_sates_singlet}\end{equation}
The partial states are proportional to the identity matrix making
the entropy maximal ($S(\rho_{A})=S(\rho_{B})=\ln2$). These density
operators do not correspond to any state vector, as can be concluded
for lack of \emph{idempotence} (note that for a pure state one has
$\rho^{2}=\left(|\psi\rangle\langle\psi|\right)^{2}=\rho$, whereas
in our example $\rho_{A}^{2}=\rho_{B}^{2}=\frac{1}{4}{\mathbbm{1}}$).
We see from (\ref{eq:Chap1_partial_sates_singlet}) that there is
more entropy in considering the subsystems $A$ and $B$, individually,
than in considering the compound system '$A+B$', $S(\rho_{A})>S(\rho_{AB})=0$:
that is, we have lost information about correlations while performing
the trace! This is at odds with classical intuition; for two classical
random variables ($A$ and $B$), the Shannon entropy obeys $H(A,B)\ge\max\left\{ H(A),H(B)\right\} $,
as $A$ (or $B$) cannot have more entropy than the overall system. 

The special form of (\ref{eq:Chap1_partial_sates_singlet}) implies
that the average of any observable defined in the Hilbert space of
spin-1/2 will not change when a change of basis is performed --- this
is just what the rotational invariance of the singlet means --- and
hence the spins will be perfectly anti-correlated in every direction,
yielding subsystems totally mixed. Let us compare the singlet state
with the following \textquotedbl{}classical mixture\textquotedbl{}
of two spins:\begin{equation}
\sigma_{AB}=\frac{1}{2}\left(|\uparrow_{A}^{z},\downarrow_{B}^{z}\rangle\langle\uparrow_{A}^{z},\downarrow_{B}^{z}|+|\downarrow_{A}^{z},\uparrow_{B}^{z}\rangle\langle\downarrow_{A}^{z},\uparrow_{B}^{z}|\right).\label{eq:Chap1_classicalmixture}\end{equation}
The mixed state $\sigma_{AB}$ yields the same partial states of the
singlet {[}Eq.~(\ref{eq:Chap1_partial_sates_singlet}){]} and also
displays a perfect anti-correlation along the $z$ axis. However,
the situation for $\sigma_{AB}$ is distinct in two ways: when measured
along a different direction, say $x$, the particles appear uncorrelated,
and, this time, the partial entropies do not surpass the total entropy.
While one could attribute the violation of a classical bound on entropies
to the particular case of the \ac{EPR} pair in a singlet state, this
is a general feature of non-classical bipartite states.

\subsection{Beyond classical correlations}

On general grounds, the averages of operators defined in the Hilbert
space of one subsystem is completely determined by its partial state.
Hence, for every operator $O_{A}$ defined in the Hilbert space of
$A$, one has \begin{equation}
\langle O_{A}\rangle=\langle\psi_{AB}|O_{A}\otimes{\mathbbm{1}}_{B}|\psi_{AB}\rangle=\text{Tr}_{A}\left[O_{A}\rho_{A}\right].\label{eq:-7}\end{equation}
The last equality can be checked by inserting the definition of partial
state {[}Eq.~(\ref{eq:Chap1_PartialStates}){]}. The analogous holds
true for operators defined in the Hilbert space of $B$. This confirms
the idea expressed in the previous section: the knowledge of partial
states is sufficient, as long as we just care about the local properties
of $A$ and $B$, but as soon as we ask about non-local properties,
we need the information of the full density matrix $\rho_{AB}$ (the
state vector $|\psi_{AB}^{-}\rangle$, in the \ac{EPR} example). 

The \ac{AF} correlations in the singlet state {[}Eq.~(\ref{eq:Chap1_SingletState}){]}
are much stronger than what they could ever be classically, being
maximal in every spatial direction. Two particles in such states are
said to display the so-called \emph{entanglement} (\emph{i.e.}~genuine
quantum correlations). This is a unique feature of compound systems
in quantum mechanics, and will be addressed in detail in Sec.~\ref{sec:Entang-th}.
Here we make a glimpse of entanglement in two-level systems (also
known as \emph{qubits}).

In a classical framework, we would say that correlations among spins-$1/2$
do exist both for \ac{AF} and \ac{FM} states, since the \emph{connected
correlation,\begin{equation}
\langle O_{A}O_{B}\rangle_{c}:=\langle O_{A}O_{B}\rangle-\langle O_{A}\rangle\langle O_{B}\rangle\label{eq:Chap1_ConnectedCorrelation}\end{equation}
}can be non-zero in both cases. In order to see this, we take the
primary matrices of the Clifford algebra (the famous Pauli matrices)
--- which, together with the identity matrix, form an orthogonal basis
of the complex Hilbert space of all $2\times2$ matrices, and therefore
can be employed to write any physical state of the Hilbert space $\mathbb{C}^{2}\otimes\mathbb{C}^{2}$
of two spin-$1/2$ particles, \begin{equation}
\sigma^{x}=\left(\begin{array}{cc}
0 & 1\\
1 & 0\end{array}\right),\sigma^{y}=\left(\begin{array}{cc}
0 & -i\\
i & 0\end{array}\right),\sigma^{z}=\left(\begin{array}{cc}
1 & 0\\
0 & -1\end{array}\right),\label{eq:Chap1_PauliMatrices}\end{equation}
and consider the special set of $SU(2)$ rotational invariant states:\begin{equation}
\rho_{AB}(f)=\frac{1}{4}\left({\mathbbm{1}}_{A}\otimes{\mathbbm{1}}_{B}+f\sum_{\alpha=x,y,z}\sigma_{A}^{\alpha}\otimes\sigma_{B}^{\alpha}\right).\label{eq:Chap1_Rotational_inv_state}\end{equation}
The partial states are maximally mixed, $\rho_{A}(f)=\rho_{B}(f)=\frac{1}{2}{\mathbbm{1}}_{2}$,
from our symmetry requirement, and the correlations can be readily
obtained using the properties of the Pauli algebra: \begin{eqnarray}
\text{Tr}\left[\sigma^{i}\right] & = & 0\label{eq:Chap1_Pauli_Properties}\\
\sigma^{i}\sigma^{j} & = & \epsilon_{k}^{ij}\sigma^{k}+\delta^{ij}{\mathbbm{1}},\label{eq:Chap1_Pauli_Properties2}\end{eqnarray}
where $\epsilon_{c}^{ab}$ is the Levi-Cevitta symbol and the summation
over repeated indexes is implicit. Since $\langle\sigma_{A,B}^{\alpha}\rangle=0$,
we get,\emph{ }\begin{equation}
\langle\sigma_{A}^{z}\sigma_{B}^{z}\rangle_{c}=\text{Tr}\left[\sigma_{A}^{z}\sigma_{B}^{z}\rho_{AB}(f)\right]=f.\label{eq:-8}\end{equation}
There is a constraint on $f$ resulting from the the density matrix
being semi-positive definite, $\rho_{AB}(f)\ge0$. The eigenvalues
of Eq.~(\ref{eq:Chap1_Rotational_inv_state}) read:\begin{eqnarray}
\rho_{1} & = & \frac{1-3f}{4},\label{eq:Chap1_eigen1}\\
\rho_{2}=\rho_{3}=\rho_{4} & = & \frac{1+f}{4},\label{eq:Chap1_eigen234}\end{eqnarray}
from which we readly obtain, \begin{equation}
f\in[-1,\frac{1}{3}].\label{eq:-9}\end{equation}
This constraint is physically natural since the correlations are bounded
from definition (\ref{eq:Chap1_ConnectedCorrelation}). When $f=-1$
the singlet state (\ref{eq:Chap1_SingletState}) is recovered and,
in this case, we say that the spins are fully anti-correlated. The
reverse case happens when $f=1/3$ and the correlations are as much
\ac{FM} as a rotational invariant state can afford; to achieve higher
\ac{FM} correlations ($\max_{\rho}\langle\sigma_{A}^{z}\sigma_{B}^{z}\rangle=1)$
we need to consider different states (and thus break rotational invariance). 

It is no coincidence that \ac{AF} correlations can be \textquotedbl{}stronger\textquotedbl{}
than \ac{FM} correlations for rotational invariant states\emph{:
quantum correlations} not only are maximal for the singlet state $f=-1$,
but also cease to exist when $f\ge-1/3$ (Sec.~\ref{sub:Entanglement_MixedStates})!
Indeed, whereas for classical spins \ac{AF} and \ac{FM} correlations
are on equal footing, for quantum spins there are non-classical correlations
which only emerge for sufficiently robust \ac{AF} interactions. Hence,
for rotational invariant states, besides the usual distinction between
ferromagnetic correlations ($f\in]0,1/3]$) and antiferromagnetic
correlations ($f\in[-1,0[$), we have a more symmetrical classification
of correlations:\begin{eqnarray*}
f & \in & [-1,-1/3[\longrightarrow\text{quantum correlations;}\\
f & \in & ]-1/3,1/3]\longrightarrow\text{classical correlations.}\end{eqnarray*}
What is so special about states with \ac{AF} correlations in the
range $[-1,-1/3[$ besides the 'spooky action at distance' as conceived
by \ac{EPR} ? The answer is that the states with $f<-1/3$ are necessarily
written as an entangled superposition of different branches\emph{:}
they are entangled in any local basis! In the next section, we will
see that this forces the correlations to be highly non-classical in
a very precise sense. Fig.~\ref{fig:Chap1_Entang2qubitsDensity}
displays, in a density plot, the amount of quantum correlations for
two spins-1/2 interacting via an \ac{AF} Heisenberg Hamiltonian as
function of (isotropic) magnetic field and temperature.

\section{Entanglement theory\label{sec:Entang-th}}

Bipartite physical states displaying \ac{EPR} correlations cannot
be prepared by two observers by separate local operations (unitary
transformations, measurements, etc.) and classical communication (exchange
of information by classical means). As a matter of fact, the correlation
in the singlet state {[}Eq.~(\ref{eq:Chap1_SingletState}){]} is
so peculiar that leads to interesting \ac{QI} possibilities with
no correspondence in the classical information theory. Perhaps, the
most genuine \ac{QI} task is quantum teleportation \cite{1993Bennett}.
It has been experimentally implemented with photons \cite{1997Bouwmeester,1998Boschi,2003Marcikic,2004Ursin}
and also with atoms \cite{2004Riebe,2004Barrett}. In it, quantum
states of fields are \textquotedbl{}teleported\textquotedbl{} through
the space among observers sharing a invaluable kind of correlation:
\textbf{\emph{entanglement}}\emph{.}

\subsection{Local operations, separable states and the fundamental law of QI\label{sub:Sec1.3.1-Local op and sep states}}

Let us start by making precise the meaning of \ac{LOCC} before defining
entanglement in a rigorous way. There are several ways one can formalize
the concept of \ac{LOCC}, but here we will adopt the formalism of
separable actions --- the reader is referred to \cite{Thesis-2001-Eisert},
and references therein for a complete survey on the subject. Before
introducing the set of separable actions, it is useful to recall the
set of elementary quantum operations one can perform in a given system
which are allowable operations in the context of \ac{QI}:
\begin{enumerate}
\item unitary transformations (time evolution) --- a system evolves under
some Hamiltonian and its density matrix changes according to\begin{equation}
\rho\longmapsto\rho(t)=U\rho U^{\dagger},\label{eq:Chap1-ElemOperations1}\end{equation}
with $U=U(t)$ being the corresponding (unitary) evolution operator;
\item measurements --- a physical state can change via the process of measurement.
One can label the possible measurement outcomes by an index $i=1,...,K$,
where $K$ stands for the number of such possible outcomes. Associated
with each of these outcomes is a projector $P_{i}$ that obeys\begin{equation}
P_{i}P_{j}=\delta_{ij}P_{i},\qquad\sum_{i=1}^{K}P_{i}={\mathbbm{1}}_{K}.\label{eq:Chap1-ElemOperations2a}\end{equation}
A given outcome $i$ will happen with probability $p_{i}=\text{Tr}\left[P_{i}\rho P_{i}\right]$
and the state will change according to\begin{equation}
\rho\longmapsto\frac{P_{i}\rho P_{i}}{\text{Tr}\left[P_{i}\rho P_{i}\right]}.\label{eq:Chap1-ElemOperations2b}\end{equation}
The latter is called a \emph{selective projective measurement} and
it is said complete if all projectors $P_{i}$ are one-dimensional.
A \emph{non-selective projective measurement} will not discriminate
between the different subspaces spanned by $P_{i}$ entailing the
following map:\begin{equation}
\rho\longmapsto\sum_{i=1}^{K}P_{i}\rho P_{i}.\label{eq:Chap1-ElemOperations2c}\end{equation}

\item enlargement of the Hilbert space --- one can attach to a quantum system,
described by the Hilbert space ${\cal H}$, an \emph{ancilla }with
support in an auxiliary Hilbert space ${\cal V}$. Let $\rho$ and
$\sigma$ denote the state of the system and ancilla, respectively.
The appending of the ancilla to the original system is described by
\begin{equation}
\rho\longmapsto\rho\otimes\sigma.\label{eq:Chap1-ElemOperations2d}\end{equation}
The map above is a natural quantum operation, since any system can
be thought as being part of a larger system. 
\item partial trace --- finally one may be interested in discarding the
ancilla (with support in ${\cal V}$) and perform a quantum operation
in the final system $S$:\begin{equation}
\rho\longmapsto\rho_{S}=\text{Tr}_{{\cal V}}\rho.\label{eq:Chap1-ElemOperations2e}\end{equation}

\end{enumerate}
The classes $1$---$4$ of allowable quantum operations can be combined
together. The resulting action will be described by a positive linear
map from the whole state space onto itself. \ac{LOCC} is the class
of quantum operations in a bipartite scenario where two parties, $A$
and $B$, perform any combination of $1$---$4$ in their own Hilbert
space, $\mathcal{H}_{A}$ and $\mathcal{H}_{B}$, respectively, and
also in additional ancillas they might have at their disposal. In
the \ac{LOCC} setup, $A$ and $B$ are even allowed to exchange \textquotedbl{}classical
information\textquotedbl{} containing results of local measurements
they get; if $A$ communicates the result of his/her measurement to
the distant party $B$, then\textbf{ $B$} may apply a specific quantum
operation which can depend on the result obtained by $A$. Recall,
for instance, the teleportation protocol: $A$ performs a joint measurement
of two particles (one in an unknown state and another particle belonging
to an entangled pair that $A$ shares with $B$). The result of the
measurement is transmitted to $B$, who, depending on the outcome
$A$ obtained, performs an unitary transformation in his/her particle
\cite{1993Bennett}.

On quite general grounds, we can describe any combination of allowable
quantum operations by \emph{superoperators} acting in the compound
system . The superoperators are trace-preserving\emph{ completely
positive maps}%
\footnote{A map $\mathcal{S}$ is said completely positive if $\mathcal{S}\otimes{\mathbbm{1}}_{N}$
is also positive with $N\in\mathbb{N}$. This property is essential
since many actions will leave unchanged the state of auxiliary particles
that might exist; in their Hilbert space, the map acts as the identity
operator. Interestingly, this condition is more general than simple
positivity of $\mathcal{S}$ \cite{Thesis-2001-Eisert}: there are
positive maps that are not completely positive (in Sec.~\ref{sub:Entanglement_MixedStates}
we will see their implication to entanglement theory). %
}and, at the operator level, can always be written as \cite{1999Rains-2,1996Bennett}:\begin{equation}
\rho\longmapsto\rho^{\prime}=\mathcal{S}(\rho)=\sum_{i}S_{i}\rho S_{i}^{\dagger},\label{eq:Chap1-KrausOpAction}\end{equation}
where $\rho$ and $\rho^{\prime}$ are any two density matrices in
the state space. The $\{S_{i}\}$ are known as Kraus operators in
\ac{QI} and describe the action of superoperators. These states may
have different dimensions (if for instance, the quantum operation
above is meant to denote an enlargement of the Hilbert space by an
ancilla) and thus the Kraus operators are not operators in the usual
sense in Quantum Mechanics \cite{Book-Kraus-1983}. In a simple unitary
evolution, this map would just contain one such operator, $\mathcal{S}(\rho)=U\rho U^{\dagger}$,
but more involved situations can be considered. The trace-preserving
character of the superoperator manifests as $\sum_{i}S_{i}^{\dagger}S_{i}={\mathbbm{1}}$
for it implies $\text{Tr}\:\mathcal{S}(\rho)=1$. Curiously, any trace-preserving
quantum operation can be mapped to a situation where an ancilla $a$
(with density matrix $\sigma)$ is added to the system (initially
in the state $\rho$ uncorrelated with $a$), which is left to evolve
unitarily and then traced out \cite{1998Barnum}; for every quantum
operation $\mathcal{S}$ there exists a density matrix $\sigma$ and
an unitary operator $U$ such that,\begin{equation}
\mathcal{S}(\rho)=\text{Tr}_{a}\left[U\left(\rho\otimes\sigma\right)U^{\dagger}\right].\label{eq:Chap1-ElemOperationsA2}\end{equation}
The map (\ref{eq:Chap1-KrausOpAction}) being trace-preserving, it
does not encompass the class of measurements \textquotedbl{}reducing\textquotedbl{}
the wave-function, but can be easily generalized as to do so by relaxing
its trace-preserving character, that is allowing quantum operations
$\mathcal{S}_{m}$ which, although completely positive, are not trace-preserving:
$\text{Tr}\,\mathcal{S}_{m}(\rho)\le1$. To this end we let $\mathcal{S}$
to be decomposable in a sum, $\sum_{m}\mathcal{S}_{m}$, in which,
each $\mathcal{S}_{m}$ is not trace-preserving. The map now acquires
the following form: \begin{equation}
\rho\longmapsto\mathcal{S}_{m}(\rho)=\sum_{i}S_{m,i}\rho S_{m,i}^{\dagger},\label{eq:Chap1-ElemOperationsB}\end{equation}
and $\sum_{m}\sum_{i}S_{m,i}^{\dagger}S_{m,i}={\mathbbm{1}}$, such
that for each $m$ we can have $\sum_{i}S_{m,i}^{\dagger}S_{m,i}\le{\mathbbm{1}}$.
This decomposition of unity is called a Positive Operator Valued Measure
(POVM); the $S_{m,i}^{\dagger}S_{m,i}$ are the elements of the POVM
which is being measured and $S_{m,i}$ are Kraus operators. The \textquotedbl{}classical
information\textquotedbl{} associated with the outcome $m$ (\emph{e.g.}~a
spin projection $-1$ or $1$) indicates which superoperator $\mathcal{S}_{m}$
acted on the state space, and thus it is said that this kind of operation
is partially classical. This generalizes the concept of projective
measurements introduced above, and it is known as a \emph{generalized
measurement}. 

Now we are in position to define \ac{LOCC} more formally following
references \cite{1998Vedral,1999Rains-2}. For pedagogical purposes,
we first introduce the concept of Local Operations. Let $\mathcal{H}_{A}\otimes\mathcal{H}_{B}$
denote the Hilbert space as usual and $\rho_{AB}$ be the density
matrix of the compound system. Associated to the parties $A$ and
$B$ we have Kraus operators $A_{i}$ and $B_{j}$, respectively.
With these definitions, Local Operations is the class of operations
for which the corresponding superoperator is \textquotedbl{}nonmeasuring\textquotedbl{}
(\emph{i.e.}~trace-preserving), and it is a direct product of superoperators
acting alone in $A$ and $B$, that is $\mathcal{S}=\mathcal{S}_{A}\otimes\mathcal{S}_{B}$.
At the operator level this map reads,\begin{equation}
\rho_{AB}\underset{LO}{\longmapsto}\mathcal{S}(\rho_{AB})=\sum_{i,j}(A_{i}\otimes B_{j})\rho_{AB}(A_{i}^{\dagger}\otimes B_{j}^{\dagger}).\label{eq:Chap1-ElemOperationsC1}\end{equation}
Finally, we introduce the concept of separable actions; this is the
class in which each $\mathcal{S}_{m}$ (non trace-preserving in general)
is separable, in the sense that its Kraus operators are separable,
$S_{m,i}=A_{i}\otimes B_{i}$, resulting in the following map: \begin{equation}
\rho_{AB}\underset{LOCC}{\longmapsto}\mathcal{S}_{m}(\rho_{AB})=\sum_{i}(A_{i}\otimes B_{i})\rho(A_{i}^{\dagger}\otimes B_{i}^{\dagger}).\label{eq:Chap1-LOCC}\end{equation}
Note that in the map above the individual actions of $A$ and $B$
may be correlated in a way not necessarily decomposable into direct
products each acting in the respective subsystem, and therefore it
accomplishes the possibility for classical communications. The separable
map of Eq.~(\ref{eq:Chap1-LOCC}) can be implemented using local
operations and classical communication only \cite{2001Cirac}. Although
every \ac{LOCC} operation can be written in the above form, the contrary
is not true as shown in \cite{1999Bennett}. Here, the map (\ref{eq:Chap1-LOCC})
will be assumed to represent the change of a generic bipartite state
$\rho_{AB}$ under generic \ac{LOCC}. 

\begin{figure}[tb]
\noindent \centering{}\includegraphics[width=0.5\textwidth]{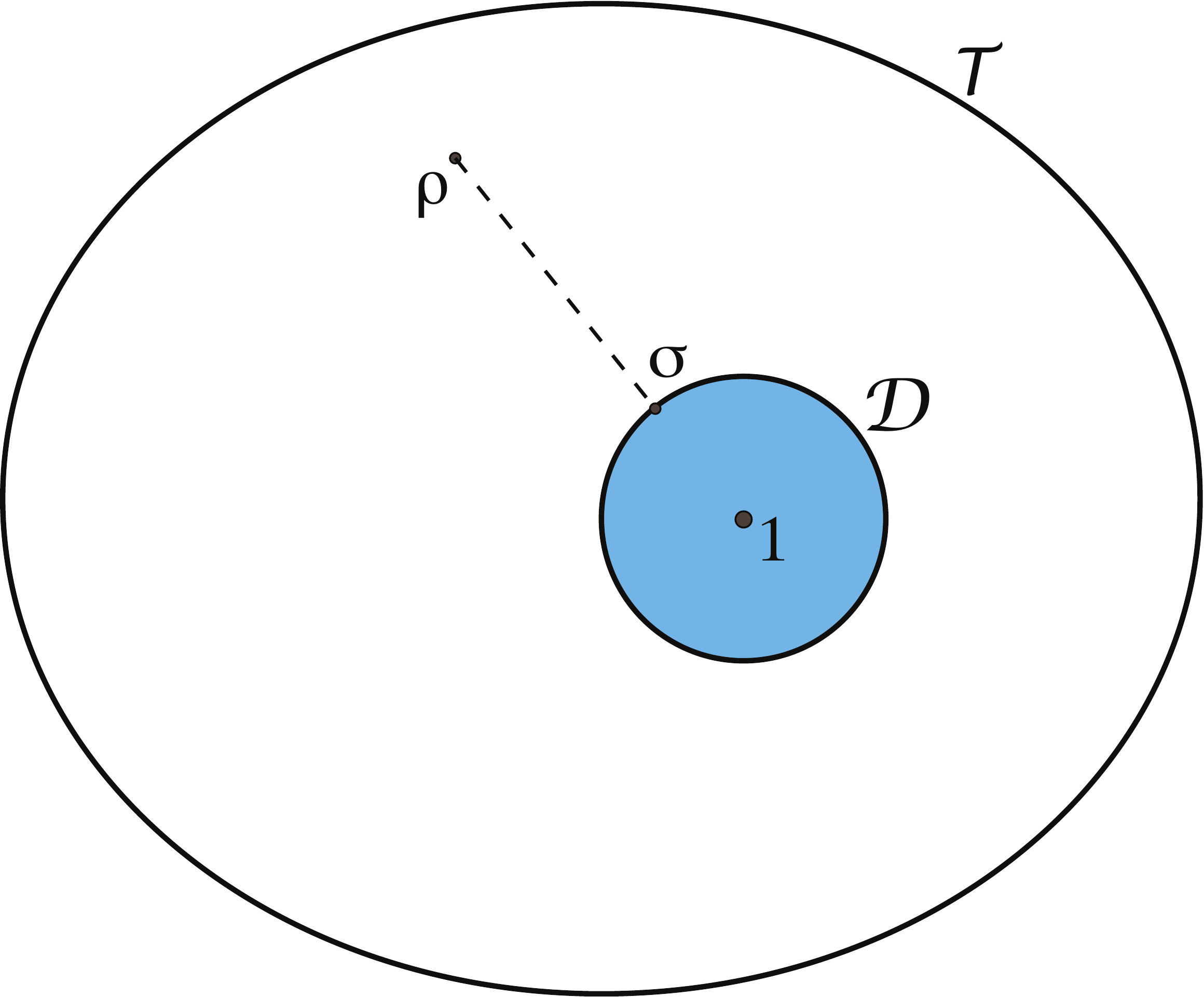}\caption[The set of all bipartite density matrices]{\label{fig:Chap1_set_bipartites_states}The amount of non-separability
of a state $\rho\equiv\rho_{AB}$ can be thought as the distance of
$\rho$ to the closest separable state $\sigma\equiv\sigma_{A}\otimes\sigma_{B}$
\cite{1997Vedral}. The set of all bipartite density matrices is represented
by ${\cal T}$ and the set of separable states by ${\cal D}.$ At
the center of the inner circle lies the most separable state for a
given $d$-dimensional Hilbert space: the identity matrix.}

\end{figure}

Having reviewed the concept of quantum operations on states and the
\ac{LOCC} scenario, we are know able to introduce formally the concept
of entanglement. Let us denote the set of all density matrices by
$\mathbf{{\cal T}}$ and the set of the \emph{separable} density matrices
by $\mathcal{D}$. The latter being the subset of bipartite states
which can be prepared by \ac{LOCC} alone.
\begin{defn*}
A bipartite state $\sigma_{AB}$ is said to be separable if it can
be written as a convex mixture of product states $\rho_{A}\otimes\rho_{B}$,
\begin{equation}
\sigma_{AB}:=\sum_{i}p_{i}\rho_{A}^{i}\otimes\rho_{B}^{i},\label{eq:Chap1_separable_states}\end{equation}
with $p_{i}$ representing probabilities. Any state $\rho_{AB}$ ,
which cannot be written in this form, is said to be entangled or to
display entanglement. 
\end{defn*}
Werner \cite{1989Werner} showed that, contrary to \ac{EPR}-like
states such as the spin singlet {[}Eq.~(\ref{eq:Chap1_SingletState}){]},
separable states {[}Eq.~(\ref{eq:Chap1_separable_states})){]} trivially
satisfy a local hidden variable model \cite{1964Bell}, hence not
violating Bell inequalities. The set (\ref{eq:Chap1_separable_states})
is clearly \emph{convex }since if $\sigma_{1}\in{\cal D}$ and $\sigma_{2}\in{\cal D}$,
then $\sigma=p\sigma_{1}+(1-p)\sigma_{2}$ with $p\in[0,1]$ is also
a separable state. Recalling that the definition of density matrix
implies that the convex mixture of any two density matrices is also
a valid state, and following \cite{1997Vedral}, we introduce a convenient
picture where the set of bipartite states is divided into two regions:
the separable set ${\cal D}$ and the entangled set ${\cal T}\setminus{\cal D}$
--- see Fig.~\ref{fig:Chap1_set_bipartites_states}.

The way to prepare a separable state (\ref{eq:Chap1_separable_states})
is straightforward: $A$ samples from the distribution $p_{i}$ and
communicates the result of the outcome to $B$, which in turn creates
$\rho_{B}^{i}.$ We remark that the correlations in these states are
said to be classical not because the sates $\rho_{A(B)}^{i}$ are
classical at all (they are arbitrary quantum states), but rather because
they do not permit non-classical tasks as teleportation or \emph{dense-coding}
(by which two classical bits of information are sent at the expense
of just one quantum-bit) \cite{1998Plenio,1992Bennett}. Moreover,
these states do not violate classical upper bounds in entropic inequalities
(see Appendix \ref{sec:App-Entropy}). 

The most important \ac{QI} contention is the postulate stating that
two parties, by means of \ac{LOCC}, cannot increase the amount of
quantum correlations they share. This obviously implies that one has
some well-defined entanglement measure, for instance, capable of giving
the distance of a given (potential) entangled state $\rho$ to the
closest separable state $\sigma$ in the sense of Fig.~\ref{fig:Chap1_set_bipartites_states}.
We have not introduced this measure yet, but we shall not be concerned
with it for the moment. Let us rather think of the particular case
where two parties share no entanglement at all. In this case, all
they can do by local operations and classical communication is to
prepare states which are a mixture of product states of the form \begin{equation}
|\Psi_{AB}\rangle=|\psi_{A}\rangle\otimes|\psi_{B}\rangle.\label{eq:Chap1_pure_separable_state}\end{equation}
This can be seen by taking a generic separable state {[}Eq.~(\ref{eq:Chap1_separable_states}){]}
and check that it remains separable under the action of a \ac{LOCC}
map {[}Eq.~(\ref{eq:Chap1-LOCC}){]}. The strong assertion that one
cannot create entanglement by means of \ac{LOCC}, in cases where
the initial state is not of the form given above, is not yet demonstrated
for generic states $\rho_{AB}$ (partially because the difficulty
of computing known entanglement measures for mixed states) and constitutes
what is known as the fundamental law of \ac{QI}. It can be formulated
in two different ways \cite{1998Plenio}:

\framebox{\begin{minipage}[t]{1\columnwidth}%
\begin{enumerate}
\item {[}\emph{restricted form}{]} The parties $A$ and $B$ cannot, with
no matter how small probability, by \ac{LOCC} transform a \emph{separable}
state into an \emph{entangled} state;
\item {[}\emph{general form}{]} The parties $A$ and $B$ cannot increase
the amount of entanglement they share solely by \ac{LOCC}.
\end{enumerate}
\end{minipage}}

\smallskip{}
Clearly the singlet state {[}Eq.~(\ref{eq:Chap1_SingletState}){]}
cannot be recast in the form (\ref{eq:Chap1_pure_separable_state})
by means of local unitary transformations $U_{A}\otimes U_{B}$ %
\footnote{If this were the case the entropy of the partial states would be zero
and not maximal, since $\text{Tr}_{A}\left[U_{A}\otimes U_{B}|\Psi_{AB}\rangle\langle\Psi_{AB}|U_{A}^{\dagger}\otimes U_{B}^{\dagger}\right]=|\psi_{B}\rangle\langle\psi_{B}|$
which again is a pure state yielding zero entropy.%
}, but according to the restricted form of the postulate, this will
be the case even if parties $A$ and $B$ communicate and make generalized
measurements in their particles. In what follows, we make further
insight into the fundamental postulate and its consequences for \ac{QI}
by introducing the entanglement measure for pure states.

\subsection{The entanglement measure for pure states\label{sub:Sec1.3.2-The entang meas for pure states}}

The problem of computing the entanglement for arbitrary bipartite
states (\emph{i.e.}~general mixed states) is an extremely hard one,
and has only been solved satisfactorily for Hilbert spaces of the
form $\mathbb{C}^{2}\otimes\mathbb{C}^{2}$ \cite{1997Wootters} and
for Gaussian states \cite{2000Simon} (for a survey of the subject
see \cite{1997Vedral,1998Vedral,Book-Nielsen-2000,2000Parker,2002Plenio}).
Remarkably, however, there exists an unique entanglement measure for
\emph{bipartite pure states} of arbitrary dimension resulting from
a link between entanglement and thermodynamics. In this section, we
outline the main ingredients leading to this conclusion. 
\begin{thm*}
\emph{<Schmidt}, \cite{Book-Peres-1993}\emph{> }Every pure state
$|\psi_{AB}\rangle$ of a composite system of two parties in $\mathbb{C}^{d_{A}}\otimes\mathbb{C}^{d_{B}}$,
with $d_{A(B)}=\dim\:\mathcal{H}_{A(B)}$, can be written in the Schmidt
form: \begin{equation}
\sum_{i=1}^{d_{A}}\sum_{j=1}^{d_{B}}c_{ij}|\phi_{i}\rangle_{A}\otimes|\phi_{j}\rangle_{B}\rightarrow\sum_{i=1}^{r}\sqrt{p_{i}}|u_{i}\rangle_{A}\otimes|v_{i}\rangle_{B},\label{eq:Chap1_schmidt_decomp}\end{equation}
where $c_{ij}\in\mathbb{C}$, $\left\{ |u_{i}\rangle_{A}\right\} $
and $\left\{ |v_{i}\rangle_{B}\right\} $ are orthogonal complete
bases \textbf{(}the Schmidt bases\textbf{)} of the Hilbert spaces
of subsystem $A$ and \textbf{$B$ ,} respectively, $r\le\min\left\{ d_{A},d_{B}\right\} $,
and $p_{i}$ are positive real numbers called Schmidt coefficients
(phase factors can always be absorbed in the Schmidt basis). 
\end{thm*}
The decomposition (\ref{eq:Chap1_schmidt_decomp}) is unique when
the coefficients $\sqrt{p_{i}}$ are all different from one another
and has some attractive properties. The partial states are diagonal
in the Schmidt basis and have a common eigenvalue spectrum:\begin{eqnarray}
\rho_{A} & = & \sum_{i=1}^{r}p_{i}|u_{i}\rangle\langle u_{i}|\label{eq:Chap1_Schmidt_Weights0}\\
\rho_{B} & = & \sum_{i=1}^{r}p_{i}|v_{i}\rangle\langle v_{i}|.\label{eq:Chap1_Schmidt_Weights}\end{eqnarray}
The latter, provides an useful shortcut to the Schmidt weights $p_{i}$
by simply computing the partial states and picking up its eigenvalues.
The number of non-zero eigenvalues ($r$) is the Schmidt rank of the
decomposition (\ref{eq:Chap1_schmidt_decomp}). Entangled states of
bipartite pure states are those with Schmidt rank higher than one,
and the respective degree of entanglement $E_{AB}$ can be measured
by the Shannon entropy {[}Eq.~(\ref{eq:Chap1_Shannon Entropy}){]}
of the Schmidt weights (or equivalently by the entropy of partial
states):\begin{equation}
E_{AB}=-\sum_{i=1}^{r}p_{i}\ln p_{i}.\label{eq:Chap1-EntangEntro}\end{equation}
In particular, for any $2\otimes2$ system, the maximal entanglement
occurs for $r=2$ and $p_{1}=p_{2}=1/2$, whereas for $r=1$ the state
is separable. The entropy associated with the Schmidt coefficients
$\left\{ p_{i}\right\} $ will not change under local unitary transformations.
Moreover, according to the fundamental law of \ac{QI}, the entropy
$E_{AB}$ can never increase under \ac{LOCC}. 
\begin{defn*}
The \emph{maximally entangled state} of a bipartite $\mathbb{C}^{d}\otimes\mathbb{C}^{d}$
system reads\begin{equation}
|\Psi_{AB}^{max}\rangle:=\frac{1}{\sqrt{d}}\left(|u_{1},v_{1}\rangle+...+|u_{d},v_{d}\rangle\right);\label{eq:Chap1_maximally_entang_state}\end{equation}
where $|u_{i}(v_{i})\rangle$ are Schmidt basis (or any other locally
equivalent basis). 
\end{defn*}
\smallskip{}

\begin{defn*}
The \emph{entropy of entanglement} of a bipartite pure state $|\psi_{AB}\rangle$
reads\begin{equation}
E(|\psi_{AB}\rangle\langle\psi_{AB}|):=-\text{Tr}\left[\rho_{A}\ln\rho_{A}\right]=-\text{Tr}\left[\rho_{B}\ln\rho_{B}\right].\label{eq:Chap1_entropy_of_entanglement}\end{equation}

\end{defn*}
For generic $d\otimes d$ systems, the maximal entropy of entanglement\emph{
}occurs for maximum Schmidt rank when all weights are equal, $\max_{\rho}E=\ln d$
{[}see Eq.~(\ref{eq:Chap1-EntangEntro}){]}. These are the states
that are locally unitarily equivalent to $|\Psi_{AB}^{max}\rangle$.
Maximally entangled states allow to prepare any bipartite state $\rho_{AB}$
solely by \ac{LOCC} \cite{1999Nielsen,1999Vidal,2001Lo,1999Jonathan,1999Hardy},
as well as enhancing many of the non-classical tasks (hence their
name \textquotedbl{}maximally entangled states\textquotedbl{}). 

For \ac{QI} purposes (\emph{e.g.}~perfect teleportation), we might
be interested in distilling a number of maximally entangled pairs
of particles out from a certain number of partially entangled particles
just by \ac{LOCC} (imagine the situation where no maximally entangled
pair is available but one has access to two or more partially entangled
pairs). This procedure is known as \emph{entanglement concentration,}
and its study in the early nineties turned out to yield seminal conclusions
for entanglement theory, as we will briefly see. The initial partially
entangled state shared by $A$ and $B$ is denoted by $\Psi_{i}$,
\begin{eqnarray}
\Psi_{i} & = & |\Psi_{A_{1}B_{1}}\rangle\otimes|\Psi_{A_{2}B_{2}}\rangle\otimes...|\Psi_{A_{n}B_{n}}\rangle\label{eq:Chap1-EntangConcentration1}\\
\Psi_{f} & = & |\Psi_{A_{1}B_{1}}^{max}\rangle\otimes|\Psi_{A_{2}B_{2}}^{max}\rangle\otimes...|\Psi_{A_{k}B_{k}}^{max}\rangle,\label{eq:Chap1-EntangConcentration2}\end{eqnarray}
whereas $\Psi_{f}$ denotes the state of the final product of the
entanglement concentration procedure, \emph{i.e.}~maximally entangled
pairs (the state of the remaining, non entangled, particles is not
represented). With this notation, the first particle of each pair,
namely $A_{1}...\:\text{and }A_{k(n)}$, belongs to $A$ and the remaining
particles, namely $B_{1}...\:\text{and }B_{k(n)}$, are in possession
of $B$; each of the $\Psi_{A_{i}B_{i}}$ represents the same partial
entangled state and thus, from now on, it will be simply denoted by
$|\Psi_{AB}\rangle$. How many maximally entangled pairs ($k$) can
$A$ and\textbf{ $B$} extract by means of standard quantum operations? 

\begin{figure}[tb]
\noindent \begin{centering}
\includegraphics[width=0.5\textwidth]{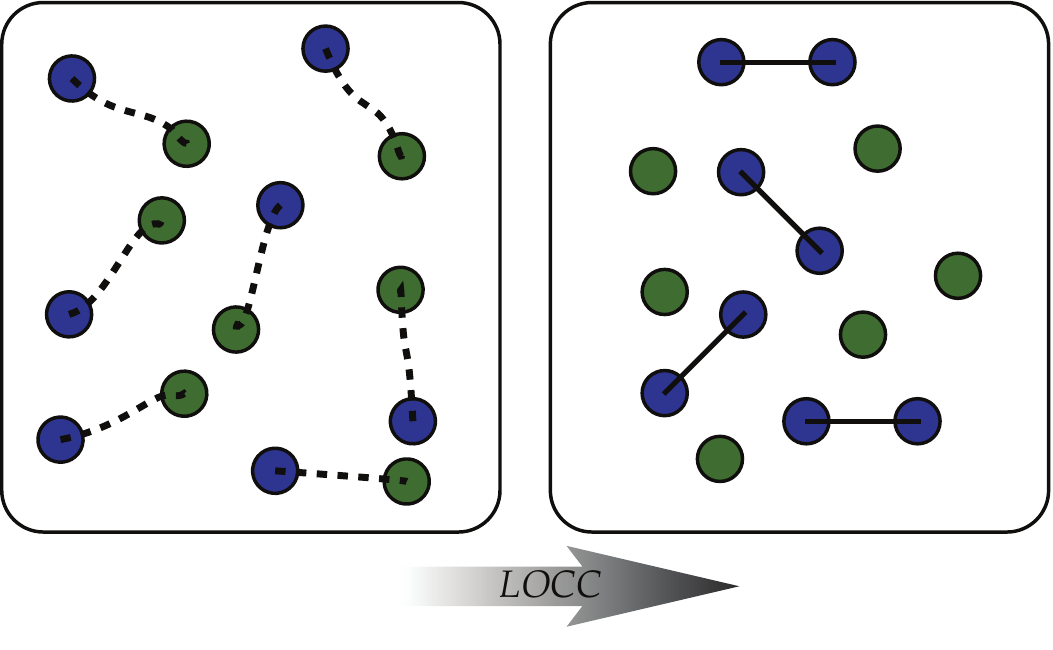}
\par\end{centering}

\caption[The entanglement concentration procedure]{\label{fig:Chap1_entang_concentration}The entanglement concentration
procedure (see \cite{1996Bennett}) takes $n$ partial entangled states
(represented in the left by particles connected with dashed lines)
and transform them into $k\leq n$ perfect singlets (blue particles
connected by lines) plus some separable pairs (in the right). In the
asymptotic limit, the conversion rate $k/n$ reads $E(\Psi_{AB})/E(\Psi_{AB}^{max})$,
where $\Psi_{AB}$ denotes the compound state of a pair in the left.}

\end{figure}

C. H. Bennett and co-authors showed that the entropy of entanglement
of the initial state $\Psi_{i}$ equals the number of maximally entangled
pairs (\emph{i.e.}~the number of pairs in $\Psi_{f}$) one can extract
asymptotically (\emph{i.e.}~$n,k\rightarrow\infty$ with $n/k$ kept
constant) by means of \ac{LOCC} \cite{1996Bennett} --- see Fig.~\ref{fig:Chap1_entang_concentration}:\[
n\text{ partially entangled pairs in the state }\Psi_{i}\underset{LOCC}{\longrightarrow}k=E(\Psi_{i})/E(\Psi_{AB}^{max})\text{ maximally entangled pairs.}\]
In other words, the initially amount of entanglement in $n$ pairs
of particles {[}which, according to Eq.~(\ref{eq:Chap1_entropy_of_entanglement}),
equals $E=nE(\Psi_{AB})$; see also Appendix~\ref{sec:App-Entropy}
for the additivity property of the von Neumann entropy{]} will determine
how many maximally entangled pairs we might get by standard quantum
operations. Let us outline some of the consequences of this conclusion;
suppose that two observers share some amount of entanglement in the
form of $n$ pairs of qubits, each one in a partial entangled state,\begin{equation}
|\Psi_{AB}\rangle=\cos\theta|\uparrow_{A}\uparrow_{B}\rangle+\sin\theta|\downarrow_{A}\downarrow_{B}\rangle.\label{eq:-11}\end{equation}
By \ac{LOCC} they can concentrate their amount of entanglement into
$k\le n$ pairs of particles, but never increase the amount of entanglement
(\emph{i.e.}~the number of singlets) they share. That nature does
not allow to create new entangled states from a previous entangled
state solely by local operations can be understood with a simple example.
Let us imagine that two parties share an entangled state of $k$ pairs
plus one extra pair in a separable state $|\Psi_{i}\rangle=|\Psi_{AB}\rangle^{\otimes k}\otimes|\Phi_{A}\otimes\Phi_{B}\rangle$,
and they wish to get a final entangled state of $k+1$ pairs, such
as $|\Psi_{f}\rangle=|\Psi_{AB}\rangle^{\otimes k+1}$. The final
state has an higher Schmidt rank%
\footnote{In the multipartite scenario of the present example ($2k+2$ qubits
in the state $|\Psi_{AB}\rangle^{\otimes k}\otimes|\Phi_{A}\otimes\Phi_{B}\rangle$)
the Schmidt rank is the minimal number of product terms in a decomposition
of $|\Psi_{i}\rangle$ in the form $\sum_{n=1}^{r}\sqrt{p_{n}}|u_{n}\rangle_{1}\otimes....\otimes|u_{n}\rangle_{2k+2}$
with $p_{n}\ge0$ and $|u_{n}\rangle_{i}\in\mathbb{C}^{2}$ --- compare
with Eq.~(\ref{eq:Chap1_schmidt_decomp}); see also reference \cite{2001Eisert}.
\smallskip{}
}. On the other hand, this can never happen with local operations and
classical communication%
\footnote{The impossibility of increasing the Schmidt rank under \ac{LOCC}
can be easily shown in the bipartite scenario \cite{2001Lo}; consider
a state $|\phi_{AB}\rangle$ with Schmidt decomposition given by $\sum_{i=1}^{r}\sqrt{p_{i}}|u_{i}\rangle_{A}\otimes|u_{i}\rangle_{B}$
with $|u_{i}\rangle_{A(B)}\in\mathbb{C}^{d_{A}(d_{B})}$. Unitary
local transformations will just re-define the vectors $|u_{i}\rangle_{A(B)}$
changing neither the number of terms, $\mbox{r},$ nor the probabilities
$p_{i}$. This, however, will no longer be the case if one of the
parties, say $A$, decides to make a projective measurement on his/her
particle; if the outcome $m$ is obtained then the state will be {[}Eq.~(\ref{eq:Chap1-ElemOperations2b}){]}
$P_{A}(m)|\phi_{AB}\rangle=\sum_{i=1}^{r}\sqrt{p_{i}}\left[P_{A}(m)|u_{i}\rangle_{A}\right]\otimes|u_{i}\rangle_{B}$,
where $P_{A}(m)$ is the projector corresponding to the outcome $m$.
The new state has at most $r$ non-vanishing terms in accordance with
the general statement that the Schmidt rank never increases under
\ac{LOCC}.%
} \cite{1999Nielsen,1999Vidal,1999Jonathan,2001Lo}, and hence making
this transformation impossible (in agreement with the the fundamental
law of \ac{QI}). 

The entanglement concentration procedure is \emph{reversible} in the
sense that the two parties can start with $k$ maximally entangled
pairs and distribute their entanglement among $n$ pairs:\begin{equation}
|\Psi_{AB}^{max}\rangle^{\otimes k}\otimes\left(|\Phi_{A}\otimes\Phi_{B}\rangle\right)^{n-k}\underset{LOCC}{\longleftrightarrow}|\Psi_{AB}\rangle^{\otimes n}.\label{eq:Chap1_entang_concentration}\end{equation}
Moreover, by local operations, the entanglement can be shifted from
one pair to another pair,\begin{equation}
|\Psi_{AB}^{max}\rangle\otimes\left(|\Phi_{A}\otimes\Phi_{B}\rangle\right)\underset{LOCC}{\rightarrow}\left(|\Phi_{A}\otimes\Phi_{B}\rangle\right)\otimes|\Psi_{AB}^{max}\rangle.\label{eq:-12}\end{equation}
None of these transformations violates the fundamental law, which
by using Eq.~(\ref{eq:Chap1_entropy_of_entanglement}) can now be
expressed as,\begin{equation}
E(\rho_{AB})\ge E(\Phi(\rho_{AB})),\label{eq:Chap1_Fund_Law}\end{equation}
where $\Phi(X)=\sum_{i}A_{i}\otimes B_{i}XA_{i}^{\dagger}\otimes B_{i}^{\dagger}$
is a \ac{LOCC} map {[}Eq.~(\ref{eq:Chap1-LOCC}){]} and $A_{i}(B_{i})$
the Kraus operators of $A$ and $B$, respectively. The considerations
made so far and specially the above equation suggest an analogy with
the second law of thermodynamics, which goes much beyond the common
definitions we have encountered (namely, the entanglement entropy
and the von Neumann entropy). 

The existence of a \emph{reversible} transformation gathering the
entanglement of $n$ systems into a smaller number of pairs $k$ {[}Eq.~(\ref{eq:Chap1_entang_concentration}){]}
when approaching the thermodynamic limit (with $n/k$ finite) is the
crucial result that lead S. Popescu and D. Rohrlich to the unique
measure of entanglement \cite{1997Popescu}. It is instructive to
review their argument; the key observation is that the good entanglement
measure should be the one yielding the same value for any of two reversibly
convertible entangled states {[}Eq.~(\ref{eq:Chap1_entang_concentration}){]}.
Indeed, the problem of finding the unique entanglement measure ${\cal E}$
for pure states is reduced to a much simpler one: finding the proper
measure for $k$ maximally entangled pairs. This measure must be proportional
to $k$ \cite{1997Popescu}, which comes about since the reversibility
of the entanglement concentration procedure is just strictly true
in the asymptotic limit ($n\rightarrow\infty$, see \cite{1996Bennett})
forcing to consider intensive quantities like the ratio of the total
entanglement to the entanglement of a pair, instead of the total entanglement.
The entanglement measure for a single pair in the initial state ${\cal E}(|\Psi_{AB}\rangle)$
is related to the entanglement of a maximally pair ${\cal E}(|\Psi_{AB}^{max}\rangle)$
by: \begin{equation}
{\cal E}(|\Psi_{AB}\rangle)=\lim_{n,k\rightarrow\infty}\left[\frac{k}{n}{\cal E}(|\Psi_{AB}^{max}\rangle)\right].\label{eq:-13}\end{equation}
On the other hand, for the entanglement concentration problem {[}Eq.~(\ref{eq:Chap1_entang_concentration}){]},
we have seen that this limit was computed to be the entropy of entanglement.
Thus, ${\cal E}(|\psi_{AB}\rangle)=E(|\psi_{AB}\rangle)$ for every
pure state $|\psi_{AB}\rangle$: thermodynamic arguments and the results
from entanglement concentration uniquely determine the measure of
entanglement for pure states obeying the fundamental law (\ref{eq:Chap1_Fund_Law}):
the entropy of entanglement {[}Eq.~(\ref{eq:Chap1_maximally_entang_state}){]}.

\subsection{The entanglement of mixed states\label{sub:Entanglement_MixedStates}}

In the \ac{QI} literature we find a considerable number of proposals
for entanglement measures. Some of these quantities have a well defined
operational meaning, as the entanglement entropy in the previous section.
Other measures, like the relative entropy of entanglement \cite{1997Vedral,1998Plenio},
lack direct physical significance, but still may be very useful in
multiple contexts (providing a simple interpretation of the amount
of entanglement in a given state, classifying correlations in quantum
many-body systems, \emph{etc.}). Before introducing the entanglement
for mixed states, we outline the main mathematical properties that
a \textquotedbl{}good\textquotedbl{} measure of entanglement should
satisfy:

\framebox{\begin{minipage}[t]{1\columnwidth}%
\begin{enumerate}
\item The entanglement in a \emph{bipartite} state $\rho_{AB}$ in the Hilbert
space ${\cal H}_{A}\otimes{\cal H}_{B}$ is a mapping $E(.)$ from
density matrices into positive real numbers:\[
\rho_{AB}\in{\cal {\cal T}}\rightarrow E(\rho_{AB})\in\mathbb{R}^{+},\]
and it is maximum for states locally equivalent to the maximally entangled
state:$U_{A}\otimes U_{B}|\Psi_{AB}^{max}\rangle$;
\item $E(\rho_{AB})=0$ if and only if $\rho_{AB}\in{\cal D}$;
\item $E(\rho_{AB})$ does not increase under \ac{LOCC} {[}Eq.~(\ref{eq:Chap1-LOCC}){]};
\item For pure states it reduces to the entropy of entanglement, $E(|\psi_{AB}\rangle\langle\psi_{AB}|)=S(\rho_{A})=S(\rho_{B}).$
\end{enumerate}
\end{minipage}}

\smallskip{}
When the mapping $E(.)$ satisfies the first three conditions, we
call it an \emph{entanglement monotone}, and if besides that it satisfies
the last condition we call it an \emph{entanglement measure}. Presently,
no entanglement measure $E(.)$ for mixed states is known, for the
constraints (3-4) are hard to realize together within the general
space state ${\cal T}$, and there is no guaranteed reversibility
in entanglement manipulations {[}Eq.~(\ref{eq:Chap1_entang_concentration}){]},
as in the pure state case, just to name a few reasons. However, for
the special case of two qubits (${\cal H}_{A}={\cal H}_{B}=\mathbb{C}^{2}$),
two important and widely-used entanglement monotones do exist: the
\emph{concurrence} \cite{1997Wootters} and the \emph{negativity}
\cite{2002Vidal}. The former is an explicit formula for $\rho_{AB}\in2\otimes2$
that equals the minimum over all the possible decompositions of $\rho_{AB}$
into pure states {[}Eq.~(\ref{eq:Chap1_DensityMatrix}){]} ,\begin{equation}
E(\rho_{AB})=\min\sum_{i}p_{i}E(|\psi_{i}\rangle\langle\psi_{i}|),\label{eq:Chap1-EntangFormation}\end{equation}
with $\rho_{AB}=\sum_{i}p_{i}|\psi_{i}\rangle\langle\psi_{i}|$. The
latter is based on the so-called \ac{PPT} criterion \cite{1996Peres}.
It is instructive to review the main argument leading to the concept
of \ac{PPT}. We consider a generic \emph{bipartite} separable state,
$\sigma_{AB}=\sum_{i}p_{i}\rho_{A}^{i}\otimes\rho_{B}^{i},$ and note
that partial transposition $\Lambda(.)$ with respect to one of the
subsystems,\emph{ }say $A$, still yields a valid density matrix,\begin{equation}
(\Lambda_{A}\otimes{\mathbbm{1}}{}_{B})\sigma_{AB}=\sum_{i}p_{i}\Lambda(\rho_{A}^{i})\otimes\rho_{B}^{i}\in{\cal T},\label{eq:Chap1-AuxPPT}\end{equation}
Hence, like any density matrix, the state after partial transposition
must remain \emph{positive-semidefinite}. This is the \ac{PPT} criterion,
\begin{equation}
(\Lambda_{A}\otimes{\mathbbm{1}}{}_{B})\sigma_{AB}\ge0.\label{eq:Chap1_PPT}\end{equation}
The transposition map $\Lambda_{A}$ is positive but not completely
positive {[}Eq.~(\ref{eq:Chap1-KrausOpAction}){]}; there will be
states $\rho_{AB}$ for which $(\Lambda_{A}\otimes{\mathbbm{1}}{}_{B})\rho_{AB}\ngeq0$
thus violating \ac{PPT}: these states are entangled. This extraordinary
simple separability condition by Asher Peres was of breakthrough importance
in the entanglement theory of mixed states, and alone is already a
stronger marker of non-separability than the usual violation of Bell
inequalities \cite{1996Peres}. In order to see how it works, we apply
partial transposition to a rotational invariant state of two qubits
{[}Eq.~(\ref{eq:Chap1_Rotational_inv_state}){]} and compute the
eigenvalues of the outcome,\begin{equation}
(\Lambda_{A}\otimes{\bf 1}_{B})\rho_{AB}(f)=\frac{1}{4}\left({\mathbbm{1}}_{A}\otimes{\mathbbm{1}}_{B}+f\Lambda_{A}(\vec{\sigma_{A})}\cdot\vec{\sigma_{B}}\right).\label{eq:Chap1-ExampleforPPT}\end{equation}
The transposition only affects the $y$-component of the Pauli matrices
{[}Eq.~(\ref{eq:Chap1_PauliMatrices}){]}, $\Lambda(\sigma^{y})=-\sigma^{y}$,
yielding the following set of eigenvalues {[}compare with Eqs.~(\ref{eq:Chap1_eigen1})-(\ref{eq:Chap1_eigen234}){]}:
$\rho_{1}=\rho_{2}=\rho_{3}=(1-f)/4$ and $\rho_{4}=(1+3f)/4$. From
direct inspection, we conclude that the rotational invariant state
is entangled for $f\in[-1,-1/3[$, whereas for $f\in[-1/3,1/3]$ is
separable (as mentioned without proof in Sec.~\ref{sec:C-versus-Q}). 

Remarkably, it has been shown that \ac{PPT} is sufficient and necessary
for separability of $2\otimes2$ and $2\otimes3$ cases \cite{1996Horodecki}.
In larger Hilbert spaces, however, there are states which are not
separable, but still remain positive after partial transposition.
The kind of entanglement present in those states is referred to as
\emph{bound entanglement} in opposition to the ordinary entanglement
(as the one shared among two qubits), also known as \emph{free entanglement}.
The reason for this distinction stems from the non-distillability
of the bound entangled states: the procedure of entanglement concentration
under \ac{LOCC} {[}Eq.~(\ref{eq:Chap1_entang_concentration}){]}
is only possible when the parties share free entanglement. The amount
of negativity of the state after partial transposition (\emph{i.e.}~the
sum of the negative eigenvalues) is related to the actual free entanglement
existing in the state, and it was shown to be a full entanglement
monotone \cite{2002Vidal},\begin{equation}
N(\rho_{AB}):=\frac{||(\Lambda\otimes{\mathbbm{1}})\rho_{AB}||-1}{2},\label{eq:Chap1_Negativity}\end{equation}
where $||A||:=Tr[\sqrt{A^{\dagger}A}]$ is the trace norm. For the
two qubits rotational invariant state {[}Eq.~(\ref{eq:Chap1_Rotational_inv_state}){]}
the negativity\emph{ }is a linear function of the correlation $\langle\sigma_{A}^{z}\sigma_{B}^{z}\rangle$
and it has a discontinuity in the first derivative for $f=-1/3$,\begin{equation}
N(\rho_{AB}(f))=-\theta\left(-f-\frac{1}{3}\right)\frac{1+3f}{4}.\label{eq:Chap1_Negativity_Inv_State}\end{equation}
\begin{figure}[tb]
\noindent \begin{centering}
\includegraphics[width=0.5\columnwidth]{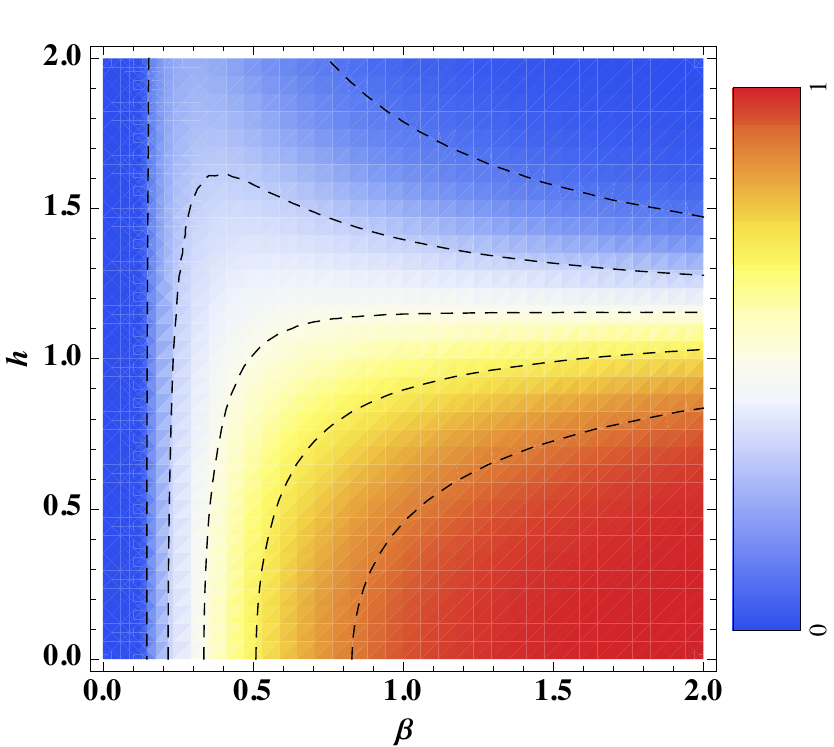}
\par\end{centering}

\caption[Density plot of entanglement in the Heisenberg magnet]{\label{fig:Chap1_Entang2qubitsDensity}The density plot shows the
degree of entanglement of the \ac{AF} Heisenberg magnet $\Xi$ {[}Eq.~(\ref{eq:Chap1_ThermalState_2spins}){]}
as measured by the concurrence. The dashed lines separate regions
where the entanglement differs by more than $20\%$. The first dashed
line (on the left hand-side) is a transition line ($\beta\simeq0.14$)
separating separable states (left) from entangled ones (right). For
small $\beta$ the entanglement completely vanishes, \emph{i.e.} the
temperature is so high that no quantum correlations survive. At the
top, $E(\Xi)$ disappears exponentially fast since large magnetic
fields tend to align the spins in the opposite direction of $\vec{B}$
producing a separable state. At bottom-right the system is practically
in its ground-state (a singlet state) and thus entanglement is nearly
maximal.}

\end{figure}
The maximal entanglement occurs for $f=-1$ (the singlet state) in
agreement with what we expect from an entanglement monotone. When
dealing with more than two particles, the negativity defined as (\ref{eq:Chap1_Negativity})
has a drawback however; it\emph{ }suffers from non-additivity, \emph{i.e.}~$N(\rho_{AB}\otimes\rho_{CD})\neq N(\rho_{AB})+N(\rho_{CD})$,
and, occasionally, it may be more convenient to use the \emph{logarithmic
negativity }instead\emph{,}\begin{equation}
E_{N}(\rho_{AB}):=\log_{2}||(\Lambda\otimes{\mathbbm{1}})\rho_{AB}||.\label{eq:Chap1_LogNegativity}\end{equation}
The \emph{logarithmic negativity} is an additive entanglement monotone
with two desirable properties: it has a clear operational meaning
and is an upper bound to the distillable entanglement \cite{2003Audenaert,2005Plenio}.

Before ending this section, let us apply the concepts we have learned
to a simple physical model: two quantum spins in a isotropic magnetic
field $\vec{B}=h(1,1,1)/\sqrt{3}$ interacting via an \ac{AF} Heisenberg
model. Adopting the standard summation convention for repeated indices
and dropping the subscripts identifying the particles and denoting
the partition function by $\boldsymbol{\mathcal{Z}}$, the canonical
thermal state $\Xi$ of the system becomes,\begin{eqnarray}
\Xi & = & \mathit{\boldsymbol{\mathcal{Z}}}^{-1}e^{-\beta\left(\sigma^{\mu}\otimes\sigma_{\mu}+B^{\mu}\sigma_{\mu}\otimes1+B^{\mu}1\otimes\sigma_{\mu}\right)}\label{eq:Chap1_ThermalState_2spins}\\
 & = & \frac{1}{4}\left({\mathbbm{1}}\otimes{\mathbbm{1}}+C(\beta,h)\sigma^{\mu}\otimes\sigma_{\mu}+m(\beta,h)\sum_{\mu}(\sigma^{\mu}\otimes{\mathbbm{1}}+{\mathbbm{1}}\otimes\sigma^{\mu})\right),\label{eq:-14}\end{eqnarray}
with $\sigma_{\mu}:=\sigma^{\mu}$ and $\mu=x,y,z$. The last line
follows from the and the properties of Pauli matrices {[}Eqs.~(\ref{eq:Chap1_Pauli_Properties})
and (\ref{eq:Chap1_Pauli_Properties2}){]}. The two-body correlation
$C(\beta,h)$ and the local magnetization $m(\beta,h)$ then univocally
determine the entanglement. When applying partial transposition,\begin{equation}
(\Lambda\otimes{\mathbbm{1}})\Xi\Rightarrow\sigma^{y}\rightarrow-\sigma^{y},\label{eq:-15}\end{equation}
according to \ac{PPT} the eigenvalues will remain positive (and thus
defining a physical density matrix) only if $\Xi$ is separable (\emph{i.e.}~if
it can be written as a convex sum of product states) --- see Fig.~\ref{fig:Chap1_Entang2qubitsDensity}
for a detailed discussion about entanglement in this model.

We have seen that entanglement in bipartite pure states is a well
understood problem and that a single measure singles out from adequate
\ac{QI} definitions and thermodynamics considerations. On the other
hand, other types of entanglement (\emph{e.g.}~\emph{tripartite}
entanglement), or bipartite entanglement of mixed states is not so
well understood (\emph{e.g.}~the open problem of \ac{LOCC} interchangeability
for mixed states), although some important conclusions can be drawn:
i) for two qubits systems the entanglement is a monotone function
of the concurrence or negativity, and therefore can be properly quantified,
and ii) for particles living in higher-dimensional Hilbert spaces
the negativity yields the amount of free entanglement detected by
the \ac{PPT} criterion.

\section{Continuous variable entanglement\label{sec:Sec1.4-Continuous_variab}}

When the Hilbert state is no longer finite, the pure states are usually
described by wavefunctions defined in the continuous phase-space.
The study of \ac{CV} entanglement encounters many difficulties in
infinite-dimensional systems (see, for instance, \cite{2000Parker,2002Plenio}),
but an enormous simplification is found if we consider the special
set of \emph{Gaussian} quantum states. This set encompasses the most
important states of the quantum harmonic oscillator, such as thermal
states and the coherent states describing the \ac{EM} field of a
coherent light source (\emph{e.g.}~a \ac{LASER}), just to name a
few. Gaussian states of \ac{CV} systems are fundamental in experimental
Quantum Optics/\ac{QI} \cite{Book-Mandel-1995,1963Glauber,Book-Leonhardt-1998,Book-Schleich-2001}.
They are known to be invariant under the action of linear optical
devices (beam splitters, phase shifts, \emph{etc.}), and can be used
to securely send/receive information, carry quantum error correction
and teleport \cite{1998Furusawa,2003Grosshans,2003Zhang}. In the
following, we review the main results of the theory of bipartite entanglement
of Gaussian states (for a detailed survey of the subject the reader
is referred to \cite{2005Braunstein,2005Ferraro}).

\subsection{Preliminaries}

In order to set up the basic definitions and introduce the sympletic
group, let us focus onto systems made of $n$ bosons (these could
represent $n$ modes of the \ac{EM} field, the positional degrees
of freedom of $n$ atoms in a lattice, \emph{etc}.). The Hilbert space
is $\mathcal{H}=\bigotimes_{k=1}^{n}{\cal F}_{k}$, where ${\cal F}_{k}$
is the infinite dimensional Fock space spanned by the number basis
$\left\{ |m\rangle_{k}\right\} _{m\in\mathbb{N}}$ (\emph{i.e.}~the
eigenstates of the number operator $n_{k}:=a_{k}^{\dagger}a_{k}$,
with $\left[a_{k},a_{k^{\prime}}^{\dagger}\right]=\delta_{kk^{\prime}}$).
Considering, for the moment, all the modes having the same frequency
and $\omega_{k}\hbar=1$, the free Hamiltonian reads,\begin{equation}
H_{0}=\sum_{k=1}^{n}\left(a_{k}^{\dagger}a_{k}+\frac{1}{2}\right).\label{eq:Chap1_FreeBosonsHamiltonian}\end{equation}
The position- and momentum- like operators of each mode are defined
through the canonical \emph{Cartesian decomposition,} \begin{eqnarray}
a_{k} & = & 1/\sqrt{2}\left(X_{k}+iP_{k}\right)\nonumber \\
a_{k}^{\dagger} & = & 1/\sqrt{2}\left(X_{k}-iP_{k}\right).\label{eq:Chap1_Quadratures}\end{eqnarray}
Introducing the vector of operators $\boldsymbol{O}=\left(X_{1},P_{1},...,X_{n},P_{n}\right)^{T}$
living in the phase-space $\boldsymbol{\Gamma}=\mathbb{R}^{2n}$,
the canonical commutation relations $[X_{j},P_{k}]=\imath\delta_{jk}$
assume the compact form $[O_{i},O_{j}]=i\sigma_{ij}$ , where we have
introduced the \emph{sympletic matrix}, \begin{equation}
\sigma:=\bigoplus_{k=1}^{n}\left(\begin{array}{rr}
0 & 1\\
-1 & 0\end{array}\right).\label{eq:Chap1_SympleticMatrix}\end{equation}
The correlations between the modes are captured by the \emph{covariance
matrix, $V_{ij}:=\frac{1}{2}\langle\left\{ O_{i},O_{j}\right\} \rangle-\langle O_{i}\rangle\langle O_{j}\rangle.$}
Having all possible correlations in its entries, we shall see that
the covariance matrix is the adequate object for the characterization
and quantification of the entanglement of Gaussian states. To acquaint
the reader with the notation, we explicitly write the covariance matrix
of a canonical thermal state of $n$ modes with $\bar{n}_{k}=1/(\exp\beta_{k}-1)$
bosons, in average, in each mode:\begin{equation}
V_{thermal}=\bigoplus_{k=1}^{n}\left(\begin{array}{rr}
\bar{n}_{k}+\frac{1}{2} & 0\\
0 & \bar{n}_{k}+\frac{1}{2}\end{array}\right).\label{eq:Chap1_Cov_ThermalState}\end{equation}
The uncertainty principle, stemming from the non-commutativity of
quantum observables, $[O_{i},O_{j}]=i\sigma_{ij}$ , also has a compact
form in phase space,\begin{equation}
V+i\frac{\sigma}{2}\ge0.\label{eq:Chap1_UncertaintyRelations}\end{equation}
This inequality is derived from the positivity of $\rho$ and the
uncertainty relations of the operators $\boldsymbol{O}$, and is the
only condition a $2n\times2n$ symmetric matrix has to satisfy in
order to be a \emph{bona fide} covariance matrix of a physical state
\cite{1987Simon,1994Simon,2000Simon}. 

\begin{figure}[tb]
\noindent \centering{}\begin{tabular}{cc}
\includegraphics[width=0.45\textwidth]{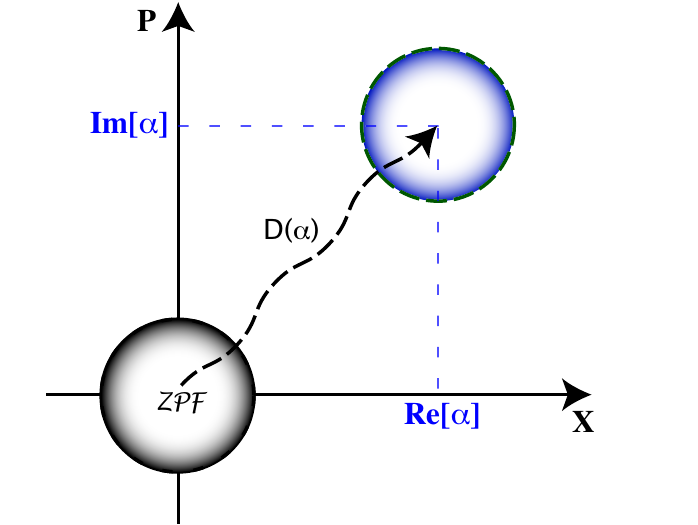} & \includegraphics[width=0.45\textwidth]{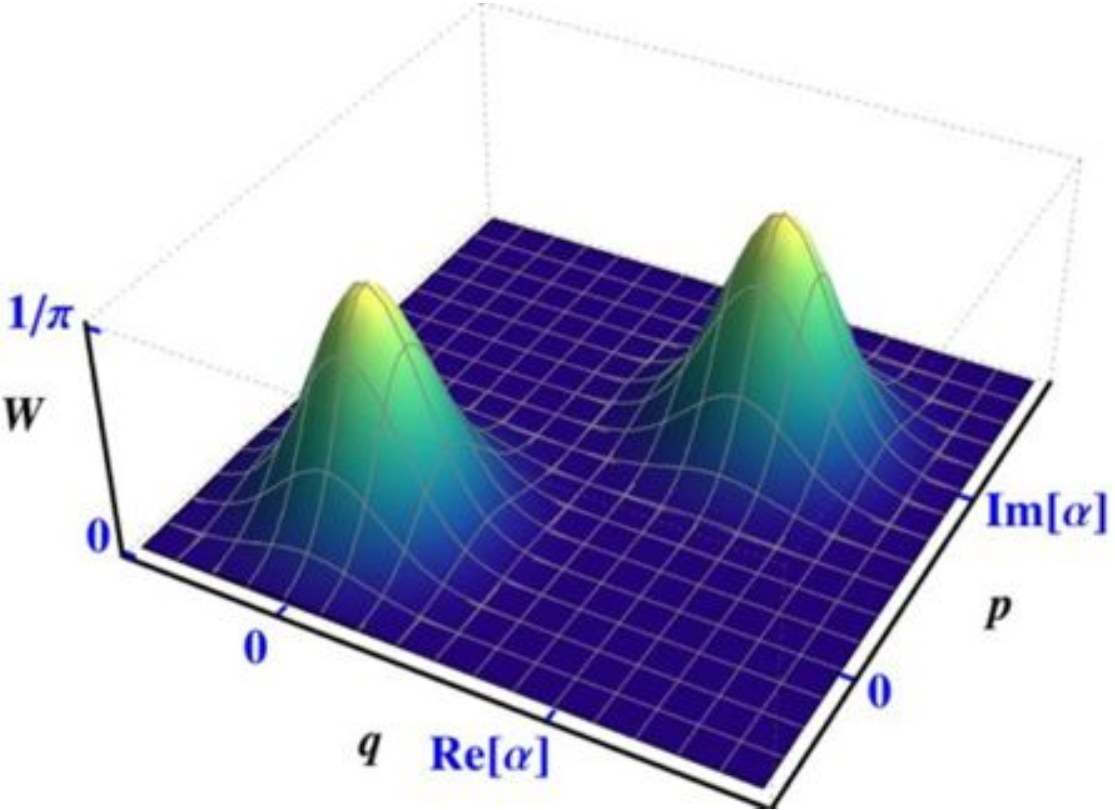}\tabularnewline
\end{tabular}\caption[The Wigner function of the displaced vacuum]{\label{fig:Chap1_Displaced Vacuum} Left - The zero-point fluctuations
of the \ac{EM} are coherently displaced $|0\rangle\rightarrow D(\boldsymbol{\alpha})|0\rangle$.
The field acquires $n=\langle a^{\dagger}a\rangle=|\alpha|^{2}$ photons
and non-zero quadratures: $X=\text{Re}[\alpha]$ and $P=\text{Im}[\alpha]$.
Right - The Wigner function $W(q,p)$ is plotted (before and after
displacement of vacuum). }

\end{figure}

In Quantum Optics, where one is often interested in photon statistics,
it is useful to describe the physics via the \emph{characteristic
function} {[}or its Fourier transform (the \emph{quasiprobability
function}){]} from which all statistical quantities can be predicted
\cite{1932Wigner,Book-Mandel-1995,Book-Leonhardt-1998,Book-Schleich-2001}.
The most interesting states of canonical systems of many modes are
fully determined by their covariance matrix; these are the so-called
Gaussian states. We would like to define a Gaussian state more carefully,
thus we introduce the \emph{displacement operator} and the characteristic
function of bosonic fields. The displacement operator for $n$ bosons
is defined as \cite{1963Glauber,1969Glauber},\begin{equation}
\mathit{D}\left(\boldsymbol{\lambda}\right)=\bigotimes_{k=1}^{n}\exp\left(\lambda_{k}a_{k}^{\dagger}-\lambda_{k}^{*}a_{k}\right),\label{eq:Chap1_DisplacementOperator}\end{equation}
with $\boldsymbol{\lambda}\in\mathbb{C}^{n}$. The quasi-classical
(coherent) states of the \ac{EM} field are obtained from the vacuum
through the action of the displacement operator $|\boldsymbol{\alpha}\rangle=\mathit{D}(\boldsymbol{\alpha})|0\rangle$.
This operator displaces the vacuum to another point of the phase-space
(\emph{i.e.}~populates it with bosons) preserving the uncertainty
of the canonical operators (see Fig.~\ref{fig:Chap1_Displaced Vacuum}).
Hence, the covariance matrix is the same as for the vacuum: $V_{ij}=\frac{1}{2}\delta_{ij}$.
The set of displacement operators is complete in the sense that every
operator $O$ on $\mathcal{H}$ can be written as,\begin{equation}
O=\int_{\mathbb{C}^{n}}\frac{d^{2n}\boldsymbol{\lambda}}{\pi^{n}}\text{Tr}\left[O\mathit{D}(\boldsymbol{\lambda})\right]\mathit{D}^{\dagger}\left(\boldsymbol{\lambda}\right).\label{eq:-16}\end{equation}
The previous formula is due to Glauber and \begin{equation}
\chi[O](\boldsymbol{\lambda}):=\text{Tr}\left[O\mathit{D}(\boldsymbol{\lambda})\right].\label{eq:Chap1_Characteristic_func}\end{equation}
is the characteristic function of the operator $O$ \cite{1969Glauber}.
\begin{defn*}
A state $\rho$ of a \ac{CV} canonical system with $2n$ degrees
of freedom is Gaussian if its characteristic function (or equivalently,
its quasiprobability distribution) is Gaussian,~\emph{i.e}.\begin{equation}
\chi[\rho](\boldsymbol{\lambda})=\chi[\rho](0)e^{-\boldsymbol{\lambda}^{T}\Sigma\boldsymbol{\lambda}+\boldsymbol{d}^{T}\boldsymbol{\lambda}},\label{eq:Chap1_GaussianState}\end{equation}
where $\Sigma$ is a $2n\times2n$ matrix and $\boldsymbol{d}\in\mathbb{R}^{2n}$. 
\end{defn*}
From its definition, a Gaussian state is completely characterized
via its first and second moments (higher moments can be obtained by
taking partial derivatives of $\chi$, see \cite{2005Ferraro} for
instance). The formal link to the vector of operators of the modes
$\boldsymbol{O}$ and the covariance matrix $V$ is,\begin{eqnarray}
\boldsymbol{d} & = & \sigma\text{Tr}[\boldsymbol{O}\rho]\label{eq:-17}\\
\Sigma & = & \sigma^{T}V\sigma.\label{eq:-18}\end{eqnarray}
The quantum correlations of a Gaussian state are encoded in the second
moments only, for $\boldsymbol{d}$ can be trivially set to zero with
unitary transformations of the individual modes. In alternative, we
can also consider the quasiprobability distribution, $W[\rho](\boldsymbol{O})$,
in phase-space, $\boldsymbol{\Gamma}=\mathbb{R}^{2n}$, associated
with the state $\rho$. However, we must have some care when interpreting
it as a classical probability distribution, for in Quantum Mechanics
the expected value of observables have an intrinsic uncertainty (ultimately
due to vacuum fluctuations), and the notion of phase-space cannot
be pushed too far. Nevertheless, we can always postulate the properties
of such distribution and use it solely as a tool to compute statistics
of observables. The price to pay is that a quantum distribution must
have some defect as a phase-space in a classical fashion does not
exist; in particular, it may become negative or ill-behaved (hence
the name \textquotedbl{}quasiprobability\textquotedbl{}). Bertrand
and Bertrand showed that one postulate is enough to define the quasiprobability
function \cite{1987Bertrand}: $W(\boldsymbol{O})$ is a joint probability
distribution for the operators  $\boldsymbol{O}=(X_{1},P_{1},...)$
(by bearing in mind that simultaneous determination of non-commuting
observables, such as $X_{1}$ and $P_{1}$, is not possible). The
\emph{marginal distributions} yield the positions or the momenta distributions,\begin{eqnarray}
P(X_{1},...,X_{n}) & := & \int_{\mathbb{\mathbb{R^{\mathrm{n}}}}}\prod_{k}dP_{k}W[\rho](X_{1},P_{1},...,X_{n},P_{n})\label{eq:-19}\\
P(P_{1},...,P_{n}) & := & \int_{\mathbb{R^{\mathrm{\textrm{n}}}}}\prod_{k}dX_{k}W[\rho](X_{1},P_{1},...,X_{n},P_{n}).\label{eq:-20}\end{eqnarray}
In the literature, one can find several functions $W[\rho]$ obeying
the above equations (see \cite{Book-Schleich-2001} for an introduction
to the phase-space methods). The most famous is the Wigner function,
which for the single-mode case $n=1$ reads \cite{1932Wigner}, \begin{equation}
W(q,p)=\frac{1}{2\pi}\int_{\mathbb{R}}dxe^{ipx}\langle q-\frac{x}{2}|\rho|q+\frac{x}{2}\rangle.\label{eq:Chap1_WignerDist}\end{equation}
The Wigner function of the vacuum is displayed in Fig.~\ref{fig:Chap1_Displaced Vacuum}.
In it, we see the action of the displacement operator {[}Eq.~(\ref{eq:Chap1_DisplacementOperator}){]}
for the one-mode \ac{EM}: $D(\boldsymbol{\alpha}=\langle q\rangle+i\langle p\rangle)$.
The fluctuations are kept to its minimum value --- the so-called \ac{ZPF}
--- but the quadratures of the \ac{EM} field no longer have a zero
mean-value, \emph{i.e.}~the vacuum is coherently populated with photons. 

Gaussian states --- like the coherent states --- play a notable role,
both in view of their conceptual importance and their relevance in
experimental Quantum Optics. They are the states more easily prepared
and controlled in the laboratory, and have been successfully employed
in quantum cryptography \cite{2003Grosshans} and quantum teleportation
protocols \cite{1998Furusawa,2003Zhang}. Moreover, it turns out that,
when properly endowed with \emph{sympletic transformations}, the complex
problem of finding entanglement monotones for generic mixed states
becomes much easier for Gaussian states.

\subsection{Sympletic local invariants of Gaussian states}

With the basic definitions established, we now review a fundamental
class of transformations\emph{ }known for a long time in classical
mechanics that makes the connection between unitary transformations
$U$ and its counterpart at the phase-space level; the sympletic transformations.
The Hamilton equations of motion for the canonical variables $\boldsymbol{R}=\left(q_{1},p_{1},...,q_{n},p_{n}\right)^{T}$
have a compact form in the sympletic formalism, \begin{equation}
\dot{\boldsymbol{R}_{i}}=\sigma_{ij}\partial_{j}H.\label{eq:Chap1_HAmilton}\end{equation}
A transformation of coordinates ($\boldsymbol{R}\rightarrow S\boldsymbol{R}$)
leaves the form of the previous equation invariant \emph{iff $S\sigma S^{T}=\sigma$.}
\begin{defn*}
The \emph{sympletic group }$S_{p}(2n,\mathbb{R})$ is the set of $2n\times2n$
real matrices $S$ satisfying\begin{equation}
S\sigma S^{T}=\sigma.\label{eq:Chap1_sympletic_group}\end{equation}
This group has dimension $n(2n+1)$, and its elements are called canonical
or sympletic transformations. 
\end{defn*}
In Quantum Mechanics the elements of $S_{p}(2n,\mathbb{R})$ preserve
the commutation relations, and all mode transformations generated
by linear and bilinear interactions are sympletic. Interestingly,
the opposite is also true: every sympletic transformation is generated
by a bilinear Hamiltonian. This relation between unitary operations
and sympletic transformations is a consequence of the Stone- von Neumann
theorem: every sympletic transformation ($S$) in the $2n$-dimensional
phase space $\Gamma=\mathbb{R}^{2n}$ has its counterpart at the Hilbert
space level ${\cal H}$ via an unitary transformation ($U$) \cite{1994Simon}.
The table \ref{tab:Chap1_HilbertVSPhaseSpace} summarizes the main
differences between the description of physical states within the
realm of the Hilbert space and that of the phase-space.

\begin{table}[h]
\noindent \begin{centering}
\begin{tabular}{c||c|c}
 & Hilbert space ${\cal H}$ & Phase-space $\Gamma$\tabularnewline
\hline
\hline 
dimension & $\infty$ & $2n$\tabularnewline
\hline 
structure & $\bigotimes$ & $\bigoplus$\tabularnewline
\hline 
description & $\rho$ & $V$\tabularnewline
\hline 
physical states & $\rho\ge0$ & $V+\frac{i}{2}\sigma\ge0$\tabularnewline
\hline 
spectrum & $0\le p_{i}\le1$ & $1/2\le\lambda_{i}<\infty$\tabularnewline
\hline
\end{tabular}
\par\end{centering}

\caption{The Hilbert space versus phase-space description of physical states.\label{tab:Chap1_HilbertVSPhaseSpace}}

\end{table}

The sympletic theory turns out to be the proper playground for the
study of entanglement in \ac{CV} states. This originates from the
fact that the covariance matrix can always be brought to a suitable
form by applying local unitary operations $U_{A}\otimes U_{B}\Leftrightarrow S_{A}\oplus S_{B}$
(which, as we have seen in the Sec.~\ref{sec:Entang-th}, do not
affect the amount of entanglement shared by two parties), and that
a necessary and sufficient condition for separability of Gaussian
states exists that can be easily expressed within the sympletic framework.
In order to see this, we recall an important theorem due to Williamson
and explore its implications for Gaussian states. 
\begin{thm*}
\emph{<Williamson, \cite{2005Ferraro}>} Given $V\in M(2n,\mathbb{R})$
satisfying $V^{T}=V$ and $V>0$ there exists a sympletic transformation
$S\in S_{p}(2n,\mathbb{R})$ and a diagonal matrix $D\in M(n,\mathbb{R})$
positive definite such that,\begin{equation}
V=S^{T}\left(\begin{array}{rr}
D & 0\\
0 & D\end{array}\right)S.\label{eq:Chap1_Williamson_theorem}\end{equation}
The matrices $S$ and $D$ are unique, up to a permutation of the
elements of $D$. The eigenvalues of $D$ are called sympletic eigenvalues\emph{.} 
\end{thm*}
We conclude that every Gaussian state $\rho$ can be obtained from
a thermal state $\Xi$, with covariance matrix given by Eq.~(\ref{eq:Chap1_Cov_ThermalState}),
through an adequate unitary transformation $U_{S}$ associated with
the sympletic matrix $S$: \begin{equation}
\rho=U_{S}\Xi U_{S}^{\dagger}.\label{eq:Chap1_ThermalState_decomp}\end{equation}
This is a direct consequence of the Williamson theorem {[}Eq.~(\ref{eq:Chap1_Williamson_theorem}){]},
which will be very useful to discuss separability of bipartite states.
Due to its relevance in what will follow, we focus on the particular
case of two-mode Gaussian states, $\boldsymbol{O_{AB}}=\left(X_{A},P_{A},X_{B},P_{B}\right)^{T}$.
With all generality, we can write the covariance matrix in the $2\times2$
block form,\begin{equation}
V=\left(\begin{array}{cc}
A & C\\
C^{T} & B\end{array}\right).\label{eq:Chap1_CovarianceMatrix2modes}\end{equation}
In it, $A$ ($B$) and $C$ are $2\times2$ matrices containing the
information about the reduced state of $A$($B$) and the correlations
between the two subsystems, respectively. Local invariants with respect
to $S_{p}(2,\mathbb{R})\otimes S_{p}(2,\mathbb{R})$ can be straightforwardly
derived by considering the action of the generic local sympletic transformation
$S_{A}\oplus S_{B}$, with $S_{A},S_{B}\in S_{p}(2,\mathbb{R})$,
on $V$,\begin{eqnarray}
A & \rightarrow & S_{A}AS_{A}^{T}\label{eq:-21}\\
B & \rightarrow & S_{B}BS_{B}^{T}\label{eq:-22}\\
C & \rightarrow & S_{A}CS_{B}^{T}.\label{eq:-23}\end{eqnarray}
The determinant of every block will not change under the action of
sympletic transformations $S_{p}(2,\mathbb{R})\otimes S_{p}(2,\mathbb{R})\subset S_{p}(4,\mathbb{R})$:
$\det A$, $\det B$, $\det C$ and $\det V$ are sympletic invariants.
Theorem~(\ref{eq:Chap1_Williamson_theorem}) allows to perform a
sympletic diagonalization of matrices $A$ and $B$ by a proper choice
of $S_{A}$ and $S_{B}$, respectively,\begin{eqnarray}
S_{A}AS_{A}^{T} & = & D_{A}:=a{\mathbbm{1}}_{2},\label{eq:-24}\\
S_{B}BS_{B}^{T} & = & D_{B}:=b{\mathbbm{1}}_{2}.\label{eq:-25}\end{eqnarray}
We make a final simplification of Eq.~(\ref{eq:Chap1_CovarianceMatrix2modes})
by noting that matrices $C$ and $C^{T}$, being $2\times2$ real
matrices, admit diagonalization by a proper orthogonal matrix $O_{AB}$
(naturally not affecting $D_{A}$ and $D_{B}$ being proportional
to the identity matrix):\begin{equation}
V=\left(\begin{array}{cc}
\begin{array}{cc}
a & 0\\
0 & a\end{array} & \begin{array}{cc}
c_{1} & 0\\
0 & c_{2}\end{array}\\
\begin{array}{cc}
c_{1} & 0\\
0 & c_{2}\end{array} & \begin{array}{cc}
b & 0\\
0 & b\end{array}\end{array}\right).\label{eq:Chap1_CovMatrixNormalForm}\end{equation}
The covariance matrix (\ref{eq:Chap1_CovMatrixNormalForm}) is said
to be in its normal form, and the three independent sympletic invariants
now read: $\det A=a^{2}$, $\det B=b^{2}$ and $\det C=c1c2$. These
invariants provide us an handy way to get the sympletic eigenvalues
of the covariance matrix \cite{1994Simon,2004Serafini}:\begin{equation}
\sqrt{2}d_{\pm}=\sqrt{\Sigma[V]\pm\sqrt{\left(\Sigma[V]\right)^{2}-4\det V}},\label{eq:Chap1_SympleticEig2modes}\end{equation}
with $\Sigma[V]:=\det A+\det B+2\det C$. Note that $d_{\pm}$ are
the eigenvalues of $D$ {[}Eq.~(\ref{eq:Chap1_Williamson_theorem}){]}
and that the Williamson form of Eq.~(\ref{eq:Chap1_CovMatrixNormalForm})
is simply:\begin{equation}
V=S^{T}\left(\begin{array}{cc}
d_{+} & 0\\
0 & d_{-}\end{array}\right)\oplus\left(\begin{array}{cc}
d_{+} & 0\\
0 & d_{-}\end{array}\right)S.\label{eq:Eq:Chap1_WilliamsonDecomposition2mode}\end{equation}

\subsection{The separability of Gaussian states\label{sub:The-separability-of}}

We would like to use the phase-space picture to say about the degree
of non-separability of quantum states in \ac{CV}. Like in \ac{PPT}
for density matrices {[}Eq.~(\ref{eq:Chap1_PPT}){]}, we should start
with a well-defined criterion for separability. Simon's approach to
this problem is based on the observation that transposition (of a
density matrix) is equivalent to a mirror reflection in the \ac{CV}
scenario: $\boldsymbol{O}\rightarrow\Lambda\boldsymbol{O}$ with $\Lambda=\text{Diag}\left(1,-1,...,1,-1\right)$,
as the transposition of a Hermitian matrix corresponds to complex
conjugation, and this, in its turn, amounts to time reversal in the
Schr\"{o}dinger picture \cite{2000Simon}. Another way of seeing
this is to take the Wigner distribution {[}Eq.~(\ref{eq:Chap1_WignerDist}){]}
and make the transposition of the density matrix elements. For a bipartite
system, ${\cal H}={\cal H}_{A}\otimes{\cal H}_{B}$, partial transposition
with respect to system $A$ will be rendered on the phase space through
the action of the matrix,\begin{equation}
\Lambda_{A}=\Lambda\oplus{\mathbbm{1}}.\label{eq:-26}\end{equation}
According to what we have learned in Sec.~\ref{sec:Entang-th}, a
necessary condition for separability is then that partial transposition
still yields a semi-positive defined operator (\emph{i.e.}~a physical
state) {[}Eq.~(\ref{eq:Chap1_UncertaintyRelations}){]},\begin{equation}
\Lambda_{A}V\Lambda_{A}+\frac{i}{2}\sigma\ge0,\label{eq:Chap1_SimonCriterion}\end{equation}
This is the Simon's criterion for \ac{CV} separability. It is instructive
to recast the above inequality in an intrinsically $S_{p}(2,\mathbb{R})\otimes S_{p}(2,\mathbb{R})$
invariant form. To this end, we take advantage of the Williamson decomposition
for two-mode states {[}Eq.~(\ref{eq:Eq:Chap1_WilliamsonDecomposition2mode}){]}
and write the positivity condition for physical states {[}Eq.~(\ref{eq:Chap1_UncertaintyRelations}){]}
as function of the sympletic eigenvalues: $d_{-}\ge1/2.$ This expression
has a straightforward physical meaning: the product of the variances
of canonical conjugate operators cannot be below the \ac{ZPF} . Hence,
when performing partial transposition of a separable state, the vacuum
still yields the absolute lower bound for the uncertainties (recall
that $V_{vacuum}=\frac{1}{2}{\mathbbm{1}}$),\begin{equation}
\tilde{d_{-}}:=d_{-}\left(\Lambda_{A}V\Lambda_{A}\right)\ge\frac{1}{2}.\label{eq:-27}\end{equation}
The partial transposition ($\Lambda_{A}V\Lambda_{A}$) affects only
the off-diagonal blocks ($c_{2}\rightarrow-c_{2}$), and hence a single
sympletic invariant: $\det C\rightarrow-\det C$. Using the explicit
sympletic invariant formula for $d_{-}$ {[}Eq.~(\ref{eq:Chap1_SympleticEig2modes}){]}
the following criterion is obtained:\begin{equation}
\det A+\det B+2|\det C|\le\frac{1}{4}+4\det V.\label{eq:Chap1_SimonCriterionInvariantForm}\end{equation}
Before partial transposition, the covariance matrix already obeys
the above inequality with $|\det C|\rightarrow\det C$. The inclusion
of the absolute value operation above, leads then to a more restrictive
separability condition. It should be noted that the above criterion
is valid for any \ac{CV} (Gaussian or not). The bottom line for Gaussian
states {[}Eq.~(\ref{eq:Chap1_GaussianState}){]} separability is
the remarkable conclusion that Simon's criterion (\ref{eq:Chap1_SimonCriterion}-\ref{eq:Chap1_SimonCriterionInvariantForm})
is also necessary:
\begin{thm*}
\emph{<Simon, \cite{2000Simon}>} The \ac{PPT} is a necessary and
sufficient condition for separability, for all bipartite Gaussian
states.
\end{thm*}
In the same spirit as the $2\otimes2$ and $2\otimes3$ cases where
\ac{PPT} is sufficient and necessary, the quantification of entanglement
of Gaussian states is conveniently given by the logarithmic negativity
{[}Eq.~(\ref{eq:Chap1_LogNegativity}){]}. For the bipartite scenario,
it is a decreasing function of the smallest sympletic eigenvalue $\tilde{d_{-}}$
\cite{2002Vidal},\begin{eqnarray}
E_{N}(\rho_{AB}) & = & \max\left[0,-\ln\left(2\tilde{d_{-}}\right)\right].\label{eq:Chap1_LogNegGaussianStates}\end{eqnarray}
Here, for pedagogical reasons, we apply Simon's criterion to the \emph{two-mode
squeezed thermal state}. This state arises when a pair of bosonic
modes, $a=(X_{A}+iP_{A})/\sqrt{2}$ and $b=(X_{B}+iP_{B})/\sqrt{2}$,
interact via an Hamiltonian of the form $H=gab+g^{*}a^{\dagger}b^{\dagger}$.
The production and detection of squeezed states represent one of the
major topics of Quantum Optics \cite{1987Loudon,1992Fabre}; the name
\textquotedbl{}squeezed\textquotedbl{} was appropriately adopted since,
under evolution through the squeezing operator, $S_{1}(z)=\exp\left(za^{2}-z^{*}a^{\dagger2}\right)$,
with $z=re^{i\theta}$, the single-mode \ac{EM} radiation sees one
of its quadratures, say $X_{1}$, going below the \ac{ZPF} level:\begin{eqnarray}
S_{1}^{\dagger}(z)aS_{1}(z) & = & \cosh(r)a+e^{i\theta}\sinh(r)a^{\dagger}\label{eq:-28}\\
S_{1}^{\dagger}(z)a^{\dagger}S_{1}(z) & = & \cosh(r)a^{\dagger}+e^{-i\theta}\sinh(r)a.\label{eq:-29}\end{eqnarray}
This is a manifest quantum phenomenon, and it can be easily checked
that the above equations effectively squeezes the variances of the
quadratures {[}Eq.~(\ref{eq:Chap1_Quadratures}){]} maintaining the
product of the variances $\Delta X_{A}\Delta P_{A}$ unchanged. 

Likewise, we define the two-mode squeezed vacuum as $S_{2}^{\dagger}(z)|0,0\rangle_{A,B}$,
where \begin{equation}
S_{2}(z)=\exp\left(zab-z^{*}a^{\dagger}b^{\dagger}\right).\label{eq:Chap1-2modeSqueezingOperator}\end{equation}
If instead of vacuum we had a thermal state of the modes, we would
get the two-mode squeezed thermal state, \begin{equation}
\Xi_{AB}(\bar{n},r):=S_{2}^{\dagger}(r)\Xi_{A}(\bar{n})\otimes\Xi_{B}(\bar{n})S_{2}(r).\label{eq:-30}\end{equation}
Once more the squeezing is nicely captured in the Heisenberg picture,\begin{eqnarray}
S_{2}^{\dagger}(z)aS_{2}(z) & = & \cosh(r)a+e^{i\theta}\sinh(r)b^{\dagger}\label{eq:-31}\\
S_{2}^{\dagger}(z)b^{\dagger}S_{2}(z) & = & \cosh(r)b^{\dagger}+e^{-i\theta}\sinh(r)a,\label{eq:-32}\end{eqnarray}
yielding the following change quadrature's transformations: $\boldsymbol{O_{AB}}\rightarrow\Omega(z)\boldsymbol{O_{AB}}$.
The correlations of the thermal state $\Xi_{A}\otimes\Xi_{B}$ will
change in agreement, \begin{equation}
V\rightarrow\Omega(z)V(\bar{n})\Omega^{T}(z),\label{eq:Chap1_CovarianceSqueezedState}\end{equation}
where $V(\bar{n})$ is the covariance matrix of the two-mode thermal
state $\Xi_{A}(\bar{n})\otimes\Xi_{B}(\bar{n})$ without squeezing,
that is {[}see Eq.~(\ref{eq:Chap1_Cov_ThermalState}){]}: \begin{equation}
V(\bar{n})=\bigoplus_{k=A,B}\left(\begin{array}{rr}
\bar{n}+\frac{1}{2} & 0\\
0 & \bar{n}+\frac{1}{2}\end{array}\right).\label{eq:Chap1-eqaux_V_thermal}\end{equation}
The $4\times4$ sympletic matrix $\Omega$ encodes the two-mode squeezing
,

\noindent \begin{center}
\begin{tabular}{cc}
$\Omega(z)=\left(\begin{array}{cc}
\cosh r{\mathbbm{1}}_{2} & \sinh r\boldsymbol{R}(\theta)\\
\sinh r\boldsymbol{R}(\theta) & \cosh r{\mathbbm{1}}_{2}\end{array}\right),$ & $\boldsymbol{R}(\theta)=\left(\begin{array}{cc}
\cos\theta & \sin\theta\\
\sin\theta & -\cos\theta\end{array}\right)$.\tabularnewline
\end{tabular}
\par\end{center}

The possibility for entanglement can be investigated via inequality
Eq.~(\ref{eq:Chap1_SimonCriterionInvariantForm}). This is accomplished
by computing the determinant $V$ {[}Eq.~(\ref{eq:Chap1_CovarianceSqueezedState}){]}
and the determinant of its $2\times2$ blocks (namely, $A$, \textbf{$B$}
and $C$). We find, \begin{eqnarray}
\det V & = & \frac{1}{16}\left(1+2\bar{n}\right)^{4}\label{eq:Chap1-SqueezeAux1}\\
\det A=\det B & = & \frac{1}{4}\left(1+2\bar{n}\right)^{2}\cosh\left(2r\right)^{2}\label{eq:Chap1-SqueezeAux2}\\
\det C & = & -\left(1+2\bar{n}\right)^{2}\cosh\left(r\right)^{2}\sinh\left(r\right)^{2}\label{eq:Chap1-SqueezeAux3}\end{eqnarray}
The important case of the two-mode squeezed vacuum ($\bar{n}=0$)
yields a violation of Eq.~(\ref{eq:Chap1_SimonCriterionInvariantForm})
for any $r>0$. This means that any finite squeezing will generate
entanglement from the vacuum. This makes all sense as $S_{2}^{\dagger}(z)|0,0\rangle_{A,B}$
is highly non-local in the Fock basis \cite{Book-Leonhardt-1998},\begin{equation}
S_{2}^{\dagger}(z)|0,0\rangle_{A,B}=\frac{1}{\sqrt{\cosh r}}\sum_{n=0}^{\infty}\left(e^{i\theta}\tanh r\right)^{n}|n,n\rangle_{A,B}.\label{eq:-33}\end{equation}
The case $\tanh r=1$ corresponds to infinite squeezing (like in the
original \ac{EPR} pair) giving rise to a maximally entangled state
$\sim\sum_{n}|n,n\rangle$ with diverging entanglement entropy {[}Eq.~(\ref{eq:Chap1_maximally_entang_state}){]}.
The log-negativity {[}Eq.~(\ref{eq:Chap1_LogNegGaussianStates}){]}
for the two-mode squeezed thermal state can be computed using Eqs.~(\ref{eq:Chap1-SqueezeAux1})-(\ref{eq:Chap1-SqueezeAux3})
together with the expression for the sympletic eigenvalue $\tilde{d_{-}}$
{[}Eq.~(\ref{eq:-27}); see also Eq.~(\ref{eq:Chap1_SympleticEig2modes}){]}
and yields\begin{equation}
E_{N}(\Xi_{AB})=\max\left[0,2r-\ln\left(2\bar{n}+1\right)\right].\label{eq:-34}\end{equation}
When the vacuum is populated with photons, \emph{i.e.}~$\bar{n}>0$,
the violation of the separability criterion may still happen, but
the amount of entanglement decreases according to the formula above.
This should be no surprise, as \textquotedbl{}mixing\textquotedbl{}
usually destroys entanglement (Sec.~\ref{sec:Entang-th}). Hence,
like in the Heisenberg magnet (see Fig.~\ref{fig:Chap1_Entang2qubitsDensity}),
there is a critical temperature $\bar{n}_{c}$ defined by: $\exp\left(-2r\right)(2\bar{n}_{c}+1)=1$. 

In the next chapter, we will show the first evidence of an ideal macroscopic
system capable of sustaining bipartite entanglement at high temperature
(compared to typical energy scales) by allowing one part of the system
to be initialized in a pure state, rather than a thermal state. A
similar phenomenon had been reported before for an \ac{EM} mode interacting
with a two-level atom \cite{2001Bose}, and we will see it remains
true in the macroscopic domain by exploiting the vastness of the Hilbert
space.

%

\chapter{Macroscopic entanglement at finite temperatures: an ideal scenario\label{cha:MacroscopicEntang}}

%

\emph{This chapter is based on the following publication by the author:}
\begin{itemize}
\begin{onehalfspace}
\item \emph{Macroscopic thermal entanglement due to radiation pressure},
AIRES FERREIRA, A. Guerreiro, and V. Vedral, Phys. Rev. Lett. \textbf{96},
060407 (2006).\end{onehalfspace}

\end{itemize}

\section{Overview\label{sec:Sec2.1-Overview}}

The mind-puzzling question of the quantum-to-classical transition
is not yet fully understood despite all the effort made in that direction,
since the very birth of Quantum Mechanics \cite{Review-Zureck-2003,Book-Zurek-1996,Book-Penrose-1986}.
Does the center-of-mass motion of macroscopic bodies obey the Schr\"{o}dinger
equation? Many condensed matter physicists would answer positively
to this question; it is known that superconducting quantum interference
devices allow for the superposition of a clockwise and anti-clockwise
current consisting of billions of electrons \cite{1999Mooij,2000Friedman}:
a genuine superposition of macroscopic states! 

Nothing in the principles of standard Quantum Mechanics says it would
be different with, let us say, the center-of-mass motion of an apple
provided that the degrees of freedom of the macroscopic object are
sufficiently decoupled from the environment (as first noted by Caldeira
and Leggett \cite{1981Caldeira}). Although this might seem clear
for some, it is still a question of great debate and, despite the
recent technological and experimental advances in such direction,
no experiment capable of testing such limits has been performed so
far.

Within the Quantum Optics community, a few opto-mechanical experiments
have already been proposed endeavoring to reach the quantum-classical
boundary from top to bottom. For instance, in \cite{2003Marshall}
(like in the original Schr\"{o}dinger-cat \emph{gedanken} experiment
\cite{1935Schroedinger}) an entangled photon state induces quantum
superpositions of a mirror, and in \cite{1997Bose_new} multi-component
cats of the \ac{EM} field are created in the interaction of a cavity
mode with a moveable tiny mirror. All these proposals have the common
feature of considering special states of light (\emph{e.g.}~Fock
states, squeezed states, etc.) to create superpositions involving
macroscopic subsystems. If there is no \textquotedbl{}collapse\textquotedbl{}
of the wave function of the macroscopic object, then one would be
able to measure its interference with a photon in the spirit of the
original thought experiment by Penrose \cite{Book-Penrose-1986,Book-Penrose-1986b}. 

Opto-mechanical systems represent a natural candidate to test the
quantum-to-classical transition for several reasons. To begin with,
mechanical oscillators resemble a prototype of \textquotedbl{}classical\textquotedbl{}
systems, and thus any genuine quantum signature would constitute a
major progress. From the experimental side, a fine control of the
\ac{EM} field is possible in the laboratory, and high quality mechanical
oscillators can be manufactured with state-of-the-art microfabrication
techniques.

The interaction of light with mechanical oscillators was well studied
throughout the last century, mainly because of detection of gravitational
waves \cite{Book-Braginsky-1992}. As the waves travel, their energy
impinges a very weak force onto the mechanical oscillator, thus requiring
an unprecedented level of precision in monitoring the oscillator's
position: the \ac{SQL}. For a mechanical oscillator with mass $m$
and natural frequency $\omega_{m}$, the \ac{SQL} equals the uncertainty
in the position due to vacuum fluctuations,\emph{ i.e.}~$\Delta x_{SQL}=\sqrt{\hbar/2m\omega_{m}}$. 

The seminal work on single quantum measurements by Braginsky \cite{Book-Braginsky-1992}
has taught us that the actual conditions to observe quantum signatures
in a mechanical oscillator depends on the details of the specific
experiment due to the quantum back action of the measurement device.
As a consequence, the common cited criterion for an oscillator to
behave quantum mechanically,\begin{equation}
k_{B}T\le\frac{\hbar\omega_{m}}{2},\label{eq:Chap2-Criterion_QuantumBehavior}\end{equation}
has to be corrected depending on the measurement time and the oscillator's
relaxation time (we refer the reader to Appendix~\ref{sec:App-Standard_Quantum_Limit}
for more details on the \ac{SQL}). While the quantum-classical transition
has been observed in microscopic systems and even in mesoscopic systems
\cite{1999Arndt,2004Arndt}, for a macroscopic mechanical oscillator
(with $\omega_{m}$ typically in the $\text{Mhz}$ range) achieving
the quantum realm requires extremely low temperatures (a rough estimate
is provided by the above criterion: $T\sim1\mu\text{K}$). 

The temperature is indeed the main obstacle in most of the proposals,
but not the only one, though. For instance, in the Penrose experiment,
the back-action of the environment (by friction) rapidly destroys
the coherence of the macroscopic superpositions --- a particular case
of a phenomenon traditionally referred to as \emph{decoherence} in
the context of open quantum systems (see \cite{2005Bassi,2005Adler,2006Bernard}
for a quantitative description of the difficulties of such experiments).
\begin{figure}[h]
\noindent \begin{centering}
\includegraphics[width=0.4\textwidth]{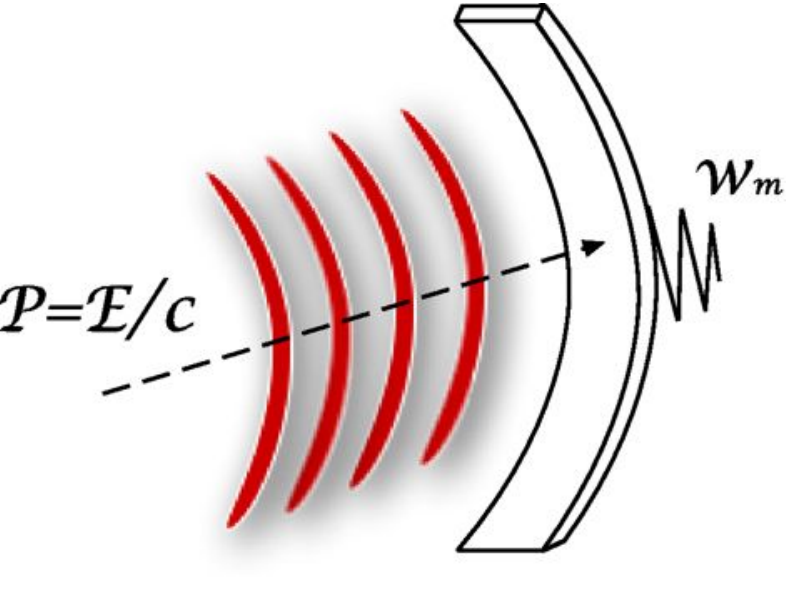}
\par\end{centering}

\caption[The radiation-pressure mechanism]{\label{fig:Chap2-Photon_meets_mirror}A photon carrying momentum
$\boldsymbol{P}=\boldsymbol{E}/c$ meets a totally reflective mirror
attached to a vibrating spring with natural frequency $\omega_{m}$.
A very feeble momentum ($=2\boldsymbol{P}$) is transferred to the
mirror which starts to oscillate. A correlation between the \ac{EM}
field quadratures and the mirror canonical observables develops in
time. Can we speak about genuine quantum entanglement in this case?}

\end{figure}

Inspired by the fundamental difference between classical and quantum
correlations discussed in Chap.~\ref{cha:Introduction}, we will
tackle the aforementioned problem by asking the question of the quantum-to-classical
transition in a slightly different way; can a macroscopic system be
entangled with another system? We know that teleportation of photons
\cite{1997Bouwmeester,1998Boschi,2003Marcikic} is possible due to
maximal polarization entanglement created by a standard Quantum Optics
phenomenon: the so-called parametric-down conversion \cite{1970Burnham}.
In this process, a photon interacts with a non-linear medium producing
two entangled photons with half of the frequency of the original photon.
In general, however, there are no generic processes leading to entanglement
between other degrees of freedom, but interactions can always be \textquotedbl{}designed\textquotedbl{}
so that the resulting wave function is entangled. Photons are known
to carry a momentum given by its frequency, $\boldsymbol{P}=\hbar\omega/c$,
and we may feel tempted to think that, due to the ponderomotive interaction,
the quadratures of a \ac{EM} field consisting of many photons could
actually become entangled with the positional degree of freedom of
a massive oscillator. If the latter is true, then we could speak about
\emph{macroscopic entanglement} (see Fig.~\ref{fig:Chap2-Photon_meets_mirror}).

In what follows, we present a very simple heuristic calculation in
favor of the idea expressed in the previous paragraph. Indeed, let
us consider two mirrors mounted in a Fabry-Perot geometry, and allow
the end mirror to move under radiation pressure (see the details in
Fig.~\ref{fig:Chap2_FabryPerot}). A single mode of the \ac{EM}
field is prepared in a superposition of Fock states, $|\psi(0)\rangle=|0\rangle+|1\rangle$,
and we let the system evolve unitarily. If the temperature is negligible,
$T\approx0$, one can approximate the initial state of the mirror
by the vacuum state, $|0\rangle_{m}$. By confining the \ac{EM} field
into a sufficiently small cavity, we expect an instantaneous displacement
of the mirror due to radiation pressure. We can then describe the
coherent evolution of the coupled system by,\begin{equation}
(|0\rangle+|1\rangle)\otimes|0\rangle_{m}\rightarrow|0\rangle\otimes|0\rangle_{m}+f(k,t)|1\rangle\otimes|\phi(t)\rangle_{m},\label{eq:Chap2-Eq_Exemplo1}\end{equation}
where $k$ is the strength of the radiation pressure interaction in
units of the initial wave packet size, $f(k,t)$ is an unknown function
and $|\phi(t)\rangle_{m}$ describes the state of the mirror after
the momentum of the \ac{EM} field has been transferred. We require
the wave function of the mirror to preserve its Gaussian character
and, hence describe it as a coherent state. We take the displacement
of the mirror to be proportional to the radiation pressure coupling,
$\phi(t)\sim k$, and, by noticing that the amplitude of a coherent
state is proportional to $\text{Re}[\phi(t)]$, define, \begin{equation}
\phi(t)\equiv k\eta(t).\label{eq:Chap2-Eq2_Exemplo1}\end{equation}
The unitary evolution will make the mirror to oscillate coherently.
This implies a few restrictions on the properties of the unknown functions:
$\eta(0)=\eta(2\pi/\omega_{m})=0$ and $f(k,t)=e^{i\theta(t)}$ with
$\theta(t)\in\mathbb{R}$. Hence, \begin{equation}
|\psi(t)\rangle=\frac{1}{\sqrt{2}}\left(|0\rangle\otimes|0\rangle_{m}+e^{i\theta(t)}|1\rangle\otimes|k\eta(t)\rangle_{m}\right).\label{eq:Chap2-Eq3_Exemplo1}\end{equation}
In the above formula, $|k\eta(t)\rangle_{m}$ describes a coherent
state obtained from the mirror's vacuum through the action of the
displacement operator $D(z)=\exp\left(za^{\dagger}-z^{*}a\right)$
{[}with $a$ and $a^{\dagger}$ being standard bosonic operators satisfying
$\left[a,a^{\dagger}\right]=1$ and $z\in\mathbb{C}$ --- see also
Eqs.~(\ref{eq:Chap1_DisplacementOperator}) and (\ref{eq:App2-Coh_States-DisplacementOperator}){]}
according to,\begin{equation}
|k\eta(t)\rangle_{m}=D_{m}(k\eta(t))|0\rangle=e^{-|k\eta(t)|^{2}/2}\sum_{n=0}^{\infty}\frac{1}{\sqrt{n!}}\left(k\eta(t)\right)^{n}|n\rangle.\label{eq:Chap2-Eq4_Exemplo1}\end{equation}
The last equality is straightforwardly obtained by factorizing the
displacement operator $D_{m}(k\eta(t))$ with the Baker-Campbell-Hausdorff
formula \cite{Book-Schleich-2001,Book-Mandel-1995}. The entanglement
present in such bipartite system, being pure, is determined by the
partial states. The Schmidt decomposition (\ref{eq:Chap1_schmidt_decomp})
guarantees that the eigenvalues spectrum of the mirror and cavity
are the same {[}Eq.~(\ref{eq:Chap1_Schmidt_Weights}){]}, and hence
we conveniently choose to compute the partial state of the cavity
field due to its low dimensionality. Arranging the result in a matrix
in the Fock basis, $\left\{ |0\rangle,|1\rangle\right\} $, we have,
\begin{equation}
\rho_{cav}(t)=\text{Tr}_{m}\left[|\psi(t)\rangle\langle\psi(t)|\right]=\frac{1}{2}\left(\begin{array}{cc}
1 & e^{i\theta(t)}e^{-|k\eta(t)|^{2}/2}\\
e^{-i\theta(t)}e^{-|k\eta(t)|^{2}/2} & 1\end{array}\right).\label{eq:Chap2-Eq5_Exemplo1}\end{equation}
The density matrix $\rho_{cav}(t)$ determines the full dynamics of
the \ac{EM} field quadratures. The entropy of entanglement {[}Eq.~(\ref{eq:Chap1_entropy_of_entanglement}{]}
is directly obtained from its eigenvalues. \begin{eqnarray}
E(|\psi(t)\rangle) & = & -x_{+}(t)\ln x_{+}(t)-x_{-}(t)\ln x_{-}(t)\label{eq:-35}\\
x_{\pm}(t) & = & \frac{1\pm e^{-|k\eta(t)|^{2}}}{2}.\label{eq:Chap2-Eq6_Exemplo1}\end{eqnarray}
The entanglement shared by two parties is maximal when the reduced
states have no information, \emph{i.e.}~the partial state is maximally
mixed {[}Eq.~(\ref{eq:Chap1_partial_sates_singlet}){]} and \emph{$E=\ln2$}.
The latter happens for large coupling and $t\in]0,2\pi/\omega_{m}[$
as $k\eta(t)\gg1$ implies $x_{\pm}(t)\simeq1/2$. For $t=0,2\pi/\omega_{m},4\pi/\omega_{m},...$
the system returns to its initial state and the wave function displays
no entanglement.

\section{Towards high-temperature macroscopic entanglement\label{sec:2.2.Towards-high-temp}}

In this section we will study the interaction of a tiny mirror with
a coherent state of the \ac{EM} field, and show for the first time
that \emph{macroscopic bipartite entanglement} can persist at finite
temperatures. When referring to \emph{macroscopic entanglement}, we
mean that at least one of the subsystems has many internal degrees
of freedom or a macroscopic mass%
\footnote{A criterion to decide whether or not a given mass is macroscopic could
be that its mass is above the Planck scale, $m>m_{\text{Planck}}\simeq4\mu\text{g}$.%
}. In either case, the systems are allowed to exchange large amounts
of energy due to an intrinsic large Hilbert space, in opposition to
single photons interacting with two-level atoms, for instance (\emph{i.e.}
microscopic systems). 

Our results will lead us to the conclusion that the standard criterion
for a macroscopic oscillator to behave quantum mechanically (\ref{eq:Chap2-Criterion_QuantumBehavior})
can be surpassed, and that in fact the quantum behavior of two coupled
systems can survive at moderately high temperatures. This clearly
paves the way to a realistic experiment aiming to test the limits
of quantum mechanics, and we shall discuss such possibility with more
detail in Chap.~\ref{cha:Stationary-optomechanical-entangl} by computing
the critical temperature above which no measurable entanglement is
expected in a realistic scenario.

Our motivation is the naive optimistic result expressed in Eq.~(\ref{eq:Chap2-Eq6_Exemplo1}):
at zero temperature, the entanglement between the mirror and the cavity
field, $|0\rangle+|1\rangle$, is large when the coupling between
the photon and the mechanical oscillator is high enough, $k\gtrsim1$.
Generally, the effect of temperature unavoidably destroys entanglement,
for mixing together pure states corresponds to entanglement dilution
(compare with the entanglement reduction by temperature in the Heisenberg
magnet, Fig.~\ref{fig:Chap1_Entang2qubitsDensity}). Nevertheless,
it can happen that the radiation-pressure mechanism is robust enough
to attain entanglement even at high-temperatures in some physical
regime exploiting the vastness of the Hilbert space of a macroscopic
system. In order to see whether this turns out to be true, a more
sophisticated calculation is needed, which is able to take into account
the important issue of temperature and the possibility of preparing
the cavity with many photons. The drawback of this approach is that
one will not be able to compute the exact entanglement as for the
simple case of Eq.~(\ref{eq:Chap2-Eq3_Exemplo1}); but still, the
most important question addressing the possibility of macroscopic
entanglement will be answered. %
\begin{figure}[h]
\noindent \begin{centering}
\includegraphics[width=0.5\textwidth]{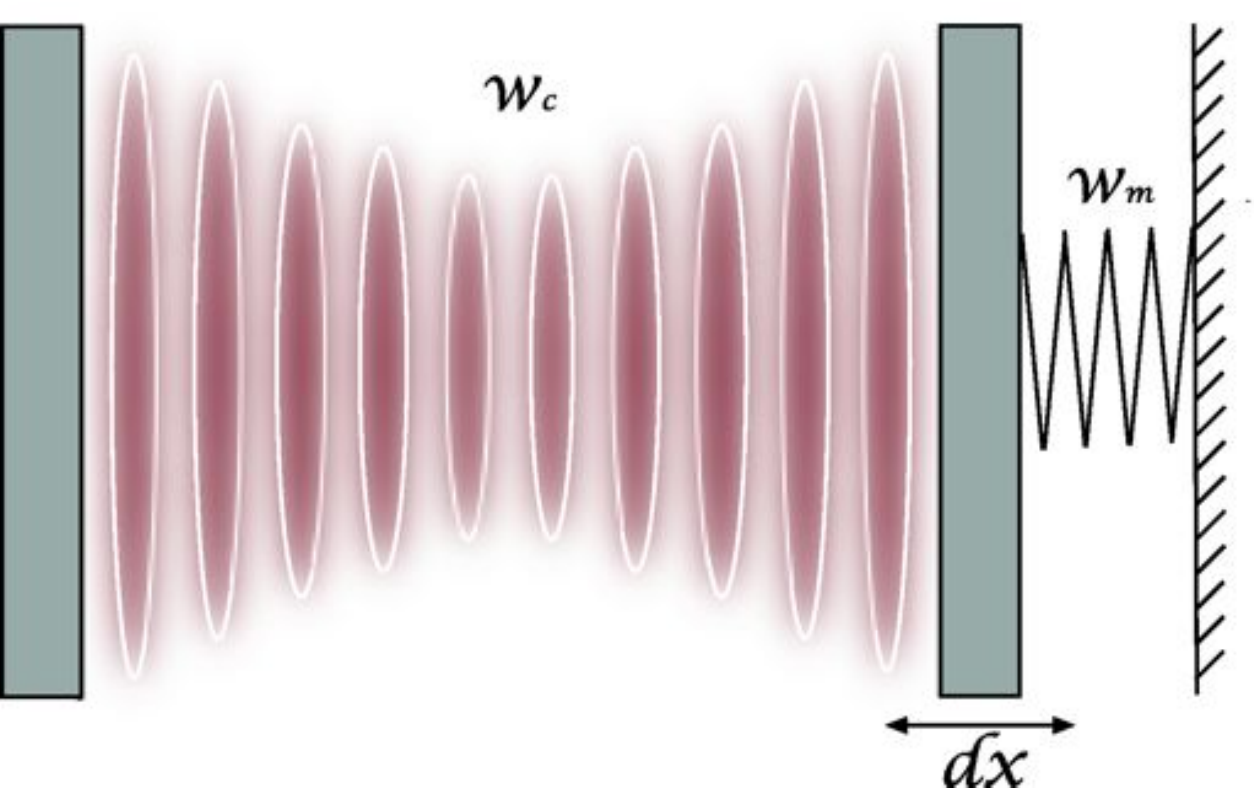}
\par\end{centering}

\caption[Schematic picture of the Fabry-Perot cavity with a moveable mirror]{\label{fig:Chap2_FabryPerot}At equilibrium the Fabry-Perot cavity
consists of two mirrors separated by a distance $L$. Hence, the cavity
has resonance frequencies given by $\omega_{n}=n\omega_{c}$ (with
$n\in\mathbb{N}$ and $\omega_{c}=\pi c/L$) and the mirrors are assumed
to be perfect reflective. The end mirror is able to move under radiation
pressure (experimentally this can be done by placing the mirror on
a oscillator cantilever), and as soon as a large coherent field is
prepared inside the cavity (for instance, by a pumping \ac{LASER})
the mirror starts to oscillate with natural frequency $\omega_{m}$
and amplitude larger than the \ac{ZPF}. }

\end{figure}

We start by studying more carefully the interaction of the cavity
\ac{EM} field with a movable mirror (see Figure~\ref{fig:Chap2_FabryPerot}).
In a quantum treatment, the mirror is modeled by an harmonic oscillator
with operators $b$ and $b^{\dagger}$ acting in the Fock space of
phononic occupation. The full opto-mechanical Hamiltonian includes
all the modes of the cavity as they can be excited by the motion of
the mirror. A rigorous derivation of the interaction non-relativistic
Hamiltonian was given in \cite{1995Law}. Here, instead, we will derive
it heuristically for the case of interest: the adiabatically driven
mechanical oscillator, \emph{i.e}.~$\omega_{m}\ll\omega_{c}$. Typically,
the resonant frequencies of an optical cavity and mechanical oscillator
are of the order of $10^{12}-10^{15}\mathrm{Hz}$ and $10^{6}\mathrm{Hz}$,
respectively, which is well inside the mentioned limit. This simplifies
very much the treatment, as photons do not get scattered to higher
modes. The free Hamiltonian of the system simply reads,\begin{equation}
H_{0}=\hbar\omega_{c}a^{\dagger}a+\hbar\omega_{m}b^{\dagger}b,\label{eq:Chap2_FreeHamiltonian}\end{equation}
where $[a,a^{\dagger}]=1$ and $[b,b^{\dagger}]=1$ are the standard
commutation rules of bosonic operators (see the table~\ref{tab:Chap2_GeneralizedCoord}
for the relation between these operators and the quadratures of the
fields).

\begin{table}[H]
\noindent \begin{centering}
\begin{tabular}{c||c|c|}
 & Position ($Q$) & Momentum ($P$)\tabularnewline
\hline
\hline 
mechanical mode & $q=\sqrt{\frac{\hbar}{2m\omega_{m}}}(b+b^{\dagger})$ & $p=\frac{1}{i}\sqrt{\frac{\hbar m\omega_{m}}{2}}(b-b^{\dagger})$\tabularnewline
\hline 
cavity field & $X=\frac{1}{\sqrt{2}}(a+a^{\dagger})$ & $Y=\frac{1}{i\sqrt{2}}(a-a^{\dagger})$\tabularnewline
\hline
\end{tabular}
\par\end{centering}

\caption{This table summarizes the relation between the canonical bosonic operators
and the generalized coordinates (or quadratures) of the mechanical
mode and the cavity field, respectively.\label{tab:Chap2_GeneralizedCoord}}

\end{table}

At equilibrium, the lowest frequency of the cavity reads $\omega_{c}(L)=\pi c/L$
where $L$ is the length of the cavity. For small displacements of
the mirror, $q\ll L$, the frequency $\omega_{c}$ can be Taylor expanded
around the equilibrium position, $\omega_{c}(q)\simeq\omega_{c}(1-q/L)$,
and the adiabatic interaction Hamiltonian can be immediately written,\begin{eqnarray}
H_{int} & = & -\frac{\hbar\omega_{c}}{L}a^{\dagger}aq\label{eq:}\\
 & = & -\frac{\hbar\omega_{c}}{L}x_{ZPF}a^{\dagger}a(b+b^{\dagger})\\
 & = & -\hbar ga^{\dagger}a(b+b^{\dagger}).\label{eq:Chap2-Interaction_Hamilt}\end{eqnarray}
The radiation pressure coupling can be written as function of the
\ac{ZPF} {[}$x_{ZPF}=\sqrt{\hbar/2m\omega_{m}}${]} of the mechanical
oscillator. The intuitive idea expressed in the previous section that
a sufficiently small cavity leads to large couplings is then confirmed,\begin{equation}
g=\frac{\omega_{c}}{L}\sqrt{\frac{\hbar}{2m\omega_{m}}}.\label{eq:Chap2-InteractionStrength}\end{equation}
Each resonant photon transfers momentum to the mirror in each of the
reflection it undergoes causing the enhancement of the mechanical
effects of light. The full Hamiltonian reads,\begin{equation}
H=\hbar\omega_{c}a^{\dagger}a+\hbar\omega_{m}b^{\dagger}b-\hbar ga^{\dagger}a(b+b^{\dagger}).\label{eq:Chap2-Hamiltonian}\end{equation}
Some comments on the validity of an unitary evolution under Hamiltonian
(\ref{eq:Chap2-Hamiltonian}) are in order. As mentioned before, the
derivation is strictly valid in the adiabatic limit, where the resonant
frequency of the mirror is much smaller than the free-spectral range
of the cavity, $\omega_{m}\ll\pi c/L$ . In this limit, the coupling
between different cavity field modes (leading to the Casimir effect,
\emph{etc.}) can be neglected as the single mode picture captures
all the relevant physics. 

In practice, perfect cavities do not exist, and the photons have some
probability to leak out destroying unitarity. However, for cavities
with very high quality factor $Q$ (defined as the number of average
photon round trips inside the cavity), the damping is negligible,
as it occurs on a time scale much longer than it takes for the photons
to perform several round trips. As long as photon leakage is the most
relevant source of decoherence, the unitary evolution under Hamiltonian
(\ref{eq:Chap2-Hamiltonian}) is expected to be a good description
of the problem for times $t\ll QL/c$. The effect of the cavity damping
and a finite viscosity (the main sources of decoherence in a realistic
scenario) will be taken in account in Chap.~\ref{cha:Stationary-optomechanical-entangl}.
In what follows, we discuss the unitary evolution of a thermalized
mirror and its entanglement dynamics with the cavity mode. A summary
of the main features of optical cavities is given in Appendix~\ref{sec:App-Cavities}.

The evolution operator associated with the Hamiltonian (\ref{eq:Chap2-Hamiltonian})
has a closed formula and it was derived in \cite{1997Mancini}, using
the Campbell-Baker-Hausdorff formula for the Lie algebra, and in \cite{1997Bose_new},
using operator algebra methods: \begin{equation}
U(t)=e^{-i\omega_{c}a^{\dagger}at}e^{ik^{2}(a^{\dagger}a)^{2}\Lambda(t)}D_{m}(\eta(t)ka^{\dagger}a)e^{-i\omega_{m}b^{\dagger}bt}.\label{eq:Chap2-EvolutionOperator}\end{equation}
In the above formula, $\Lambda(t)=\omega_{m}t-sin(\omega_{m}t)$,
$\eta(t)=1-e^{-i\omega_{m}t}$, $k=g/\omega_{m}$ and $D_{m}(\eta(t)ka^{\dagger}a)=e^{ka^{\dagger}a(\eta(t)b^{\dagger}-\eta(t)^{*}b)}$
is the displacement operator of the mirror, $D_{m}(\gamma)|0\rangle=|\gamma\rangle$.
It is pedagogical to apply the evolution operator to the simple example
of the Cat-like state considered above, $|\psi(0)\rangle=(|0\rangle+|1\rangle)\otimes|0\rangle_{m}.$
The evolution operator consists of two free evolution phases, a Kerr-like
non-linear phase and a displacement operator whose amplitude depends
upon the photon pressure $\sim a^{\dagger}a$. They yield for the
\textquotedbl{}Cat-like\textquotedbl{} state,\begin{equation}
(|0\rangle+|1\rangle)\otimes|0\rangle_{m}\rightarrow|0\rangle\otimes|0\rangle_{m}+e^{if(t)}|1\rangle\otimes|k\eta(t)\rangle_{m}.\label{eq:Chap2-HeuristicEvConfirma}\end{equation}
Since $f(t)=k^{2}\Lambda(t)-\omega_{c}t$ is a phase, we have exactly
recovered the heuristic result derived earlier (\ref{eq:Chap2-Eq3_Exemplo1}):
the interaction term of the Hamiltonian has the potential to entangle
the cavity field modes with the vibrational modes of the mirror for
intermediate times {[}Eq.~(\ref{eq:Chap2-Eq6_Exemplo1}){]}. The
entanglement results from the evolution of the term $|1\rangle\otimes|0\rangle_{m}$,
which can be interpreted as the transference of momentum from the
photon $|1\rangle$ to the mirror $|0\rangle_{m}$, as the photon
kicks the mirror. Though the Cat-like state of light, $|0\rangle+|1\rangle$,
is easy to produce experimentally, the radiation pressure in this
case is so small that it is virtually impossible to detect any entanglement
using present day technology. However, we will see in this chapter,
that a detectable amount of thermal entanglement is expected when
the cavity is initially in a coherent state with sufficiently high
amplitude, \emph{i.e}.~when many photons are considered instead of
just one. 

The \ac{EM} field is prepared in a coherent state of the light, $|\alpha\rangle=D(\alpha)|0\rangle$,
using a driving \ac{LASER} tuned to resonance with the cavity mode,
whereas the mirror is considered to be initially in a Gibbs state
with temperature $T$. Expressing the thermal state of the mirror
in the coherent state basis (see Appendix~\ref{sec:App-CoherentStates}
), the composite state of the system reads,\begin{equation}
\rho(t_{0})=\frac{1}{\bar{n}}\int_{\mathbb{C}}\frac{d^{2}z}{\pi}e^{-|z|^{2}/\bar{n}}|\alpha\rangle\langle\alpha|\otimes|z\rangle\langle z|,\label{eq:Chap2-InitialState}\end{equation}
where $\bar{n}=1/(e^{\hbar\omega_{m}/K_{B}T}-1)$ is the mean number
of phononic excitations and $z\in\mathbb{C}$ represents all the possible
coherent states of the mirror. The density matrix $\rho(t_{0})$ evolves
according to $\rho(t)=U(t)\rho(0)U^{\dagger}(t)$ and it can be readily
obtained as soon as the evolution of a pure coherent state $|z\rangle_{m}$,
for any population of the cavity $|n\rangle$, is written in a suitable
basis. From Eq.~(\ref{eq:Chap2-EvolutionOperator}) we have,\begin{equation}
U(t)|n\rangle\otimes|z\rangle_{m}=e^{-i\phi_{n}(t)}|n\rangle\otimes|ze^{-i\omega_{m}t}+kn\eta(t)\rangle_{m},\label{eq:Chap2-InitialCoherentStateEv}\end{equation}
with $\phi_{n}(t)=n\omega_{c}t-k^{2}n^{2}\Lambda(t)$. The amplitude
$z$ in (\ref{eq:Chap2-InitialCoherentStateEv}) is displaced by the
evolution operator depending on radiation pressure ($\sim a^{\dagger}a$)
exactly as we found before {[}Eq.~(\ref{eq:Chap2-HeuristicEvConfirma}){]}.
A coherent state of the mirror will evolve according to,\begin{equation}
U(t)|\alpha\rangle\otimes|z\rangle_{m}=e^{-|\alpha|^{2}/2}\sum_{n\in\mathbb{N}_{0}}e^{-i\phi_{n}(t)}\frac{\alpha^{n}}{\sqrt{n!}}|n\rangle\otimes|ze^{-i\omega_{m}t}+kn\eta(t)\rangle_{m}.\label{eq:Chap2-Aux}\end{equation}
Finally, the evolution of the density matrix is obtained by averaging
the latter expression with the corresponding Boltzmann's weights {[}see
Eq.~(\ref{eq:Chap2-InitialState}){]}. Here we express $\rho(t)$
in the Fock basis, \begin{equation}
\rho(t)=\sum_{\mu,\nu,n,m=0}^{\infty}\rho_{\mu\nu nm}(t)|n\rangle\langle m|\otimes|\mu\rangle\langle\nu|,\label{eq:Chap2-DensityDecompFock}\end{equation}
where the Latin indexes refer to the radiation and the Greek indexes
refer to the mirror. It is useful to define the following functions,\begin{eqnarray}
\Phi_{\mu\nu nm}(t) & := & \alpha^{n}\alpha^{\ast m}e^{i\Lambda(t)\left(n^{2}-m^{2}\right)-i\omega_{c}\left(n-m\right)-|\alpha|^{2}}/(n!m!\mu!\nu!)^{1/2}\label{eq:-1}\\
K_{nm}(z) & := & |F_{n}(z)|^{2}/2+|F_{m}(z)|^{2}/2+|z|^{2}/\overline{n},\label{eq:Chap2_DensityAuxFunc}\end{eqnarray}
where $F_{n}(z)=z+kn\eta(t)$. With these definitions the elements
of the density matrix read (see Appendix~\ref{sec:App-DensityMatrix}
for a detailed derivation): \begin{equation}
\rho_{\mu\nu nm}(t)=\Phi_{\mu\nu nm}(t)\int_{\mathbb{C}}\frac{d^{2}z}{\bar{n}\pi}F_{n}(z)^{\mu}F_{m}^{*}(z)^{\nu}e^{-K_{nm}(z)}.\label{eq:Chap2-DensityMatrix_Int_form}\end{equation}
This integral can be analytically solved (see Appendices~\ref{sec:App-CoherentStates}
and \ref{sec:App-DensityMatrix} for the integration techniques for
bosons). Up to a normalization constant it yields {[}Eq.~(\ref{eq:App2-DensityM-21}){]},\begin{eqnarray}
\rho_{\mu\nu nm}(t) & = & \Phi_{\mu\nu nm}(t)e^{-k^{2}|\eta(t)|^{2}\frac{n^{2}+m^{2}}{2}}\left[\partial_{a}^{\mu}\partial_{b}^{\nu}G_{nm}(a,b,\bar{n},k,t)\right]_{(a=0,b=0)}\label{eq:Chap2-DensityMatrixTOTAL}\\
G_{nm}(a,b,\bar{n},k,t) & := & \exp\left[\frac{\bar{n}}{\bar{n}+1}G1_{nm}(a,-t)G1(b,t)+G2_{nm}(a,b,t)\right],\label{eq:Chap2-DensityMatrix}\end{eqnarray}

where $G1_{nm}(X,t):=X-\eta(t)k(n+m)/2$ and $G2_{nm}(a,b,t):=akn\eta(t)+bkm\eta(-t)$.
These equations contain all the physics of the cavity-mirror problem
for the initial condition {[}Eq. (\ref{eq:Chap2-InitialState}){]}
and will be the basis of our discussion for the rest of the present
chapter.

For infinite dimensional density matrices, the separability problem
is solved for pure states {[}through the entropy of entanglement Eq.~(\ref{eq:Chap1_entropy_of_entanglement}){]}
and for Gaussian states (see Sec.~\ref{sec:Sec1.4-Continuous_variab}).
However, the state (\ref{eq:Chap2-DensityMatrix}) is neither pure
(except for $\bar{n}=0$) nor Gaussian, and thus we have to study
entanglement by less standard means --- note that even for the pure
state ($\bar{n}=0$) it is non-trivial to get the eigenvalues of the
matrix (\ref{eq:Chap2-DensityDecompFock}). Quantifying entanglement
in mixed states is generally a difficult problem, unless the Hilbert
dimension is sufficiently small%
\footnote{Recall that for $2\otimes2$ and $2\otimes3$ systems, \ac{PPT} and
separability are equivalent {[}Eq.~(\ref{eq:Chap1_PPT}){]}.%
}. Encouraged by the study of the entanglement between a two-level
atom and the \ac{EM} field by Bose \emph{et al.} \cite{2001Bose},
in this thesis we develop a method inspired by Boses's approach that
will allow us to discuss entanglement for arbitrary temperatures ---
we will refer to it as (discrete variable) projection method. The
projection method consists of two steps;

\framebox{\begin{minipage}[t]{1\columnwidth}%
\begin{enumerate}
\item projecting the original density matrix (\ref{eq:Chap2_DensityAuxFunc})
into subspaces of low dimensionality,\[
\rho(t)\rightarrow P\rho(t)P;\]

\item computing entanglement markers and monotones for the projected subspaces,\[
E(P\rho(t)P).\]

\end{enumerate}
\end{minipage}}

\smallskip{}
The projection (1) into a subspace of lower dimension corresponds
to a local action {[}Eq.~(\ref{eq:Chap1-LOCC}){]}, thus not increasing
the global amount of entanglement, $E(\rho(t))\ge E(P\rho(t)P)$,
as guaranteed by the fundamental law of \ac{QI} as it applies to
\ac{LOCC} actions {[}Eq.~(\ref{eq:Chap1_Fund_Law}){]}. Thus, if
succeed in showing the existence of entanglement in the smaller subspaces,
we will have proven the existence of genuine quantum correlations
among the \ac{EM} field and the mechanical oscillator. In other words,
the non-separability within the projected subspaces implies non-separability
of the full density matrix (\ref{eq:Chap2-DensityMatrix}), and thus
an evidence of quantumness of the coupled system. Finally, we remark
that, by using this simple method, lower bounds for the entanglement
can be obtained from the study of the smaller subspaces.

\subsection{Zero-Temperature analysis}

For the sake of simplicity, we begin our analysis by projecting the
density matrix into the smallest possible subspace capable of attaining
bipartite entanglement, that is a $2\otimes2$ subspace. Let $P_{nm\mu\nu}=P_{nm}\otimes P_{\mu\nu}$
be the projection operator onto the subspace spanned by $n$, $m$
excitations of the cavity field and $\mu$,$\nu$ excitations of the
mirror, with: \begin{equation}
P_{ab}:=|a\rangle\langle a|+|b\rangle\langle b|.\label{eq:Chap2-ProjectorAB}\end{equation}
The subspace spanned by the Fock states $\left\{ |n\rangle,|m\rangle\right\} \otimes\left\{ |\mu\rangle,|\upsilon\rangle\right\} $
will be denoted by $[n,m;\mu,\upsilon]$ and the expression \textquotedbl{}lowest
subspace\textquotedbl{} will be used to mean that the Fock states
we are looking at are close to the ground state\emph{ }of the system
--- \emph{i.e.}~the vacuum ($|0\rangle$) and a single excitation
of the fields ($|1\rangle$). In the system's ground state, the density
matrix (\ref{eq:Chap2-DensityMatrix}) is pure ($\rho^{2}=\rho$)
and it is advantageous to use a specific entanglement monotone ---
the so-called \emph{tangle --- }rather than the negativity {[}Eq.~(\ref{eq:Chap1_Negativity}){]}.
The tangle is a full entanglement monotone for bipartite pure states
$\rho_{AB}$ in $\mathcal{H}=\mathbb{C}^{2}\otimes\mathbb{C}^{2}$
defined as \cite{1997Wootters},\begin{eqnarray}
\tau(\rho_{AB}) & := & 4\det\rho_{A(B)},\label{eq:Chap2-Tangle}\end{eqnarray}
where $\rho_{A(B)}$ denotes the (normalized) partial state of subsystem
$A(B)$ --- recall that Schmidt spectrum of a pure state is common
to both partial states {[}Eqs.~(\ref{eq:Chap1_Schmidt_Weights0})-(\ref{eq:Chap1_Schmidt_Weights}){]}. 

Like the entanglement entropy, the tangle is valid for bipartite pure
states, but with the advantage of being much easier to calculate.
Let us focus momentarily on the lowest subspace, that is $[0,1;0,1]$;
the $4\times4$ elements of the density matrix projected onto this
subspace, $\rho_{AB}$, are computed from Eqs.~(\ref{eq:Chap2-DensityMatrixTOTAL})-(\ref{eq:Chap2-DensityMatrix})
by choosing the adequate values of $n$,$m$, $\mu$ and $\nu$. With
this notation $A$ refers to the cavity field and $B$ to the mirror.
Arranging the result in a matrix in the Fock basis: $\left\{ |0\rangle\otimes|0\rangle,|1\rangle\otimes|0\rangle,|0\rangle\otimes|1\rangle,|1\rangle\otimes|1\rangle\right\} $,
we have, \begin{equation}
\rho_{AB}=\frac{1}{1+\left[|\alpha|K(t)H(t)\right]^{2}}\left(\begin{array}{cccc}
1 & \alpha^{*}K(t) & 0 & \eta(-t)k\alpha^{*}K(t)\\
\alpha K(t) & |\alpha|^{2}K(t)^{2} & 0 & \eta(-t)k|\alpha|^{2}K(t)^{2}\\
0 & 0 & 0 & 0\\
\eta(t)k\alpha K(t) & \eta(t)k|\alpha|^{2}K(t)^{2} & 0 & 4k^{2}|\alpha|^{2}\sin(\omega_{m}t/2)^{2}K(t)^{2}\end{array}\right),\label{eq:Chap2-ProjectedDensityMatrix}\end{equation}
with $K(t):=\exp\left\{ -2k^{2}\sin(\omega_{m}t/2)^{2}\right\} $
and $H(t)=1+4k^{2}\sin(\omega_{m}t/2)^{2}$. The tangle is calculated
by tracing one of the subsystems in $\rho_{AB}$ and computing the
determinant of the remaining $2\times2$ matrix. When the mirror is
maximally displaced from its equilibrium position ($t=\pi/\omega_{m}$),
it reads, \begin{equation}
\tau(\pi/\omega_{m})=\frac{16k^{2}|\alpha|^{2}e^{4k^{2}}}{\left(e^{4k^{2}}+\left(1+4k^{2}\right)|\alpha|^{2}\right)^{2}}.\label{eq:Chap2-Tangle_for_subspace01}\end{equation}
It is apparent that a large mean number of photons ($=|\alpha|^{2}$)
does not favor entanglement; there exists an optimal value of $\alpha$
in every $2\otimes2$ subspace. We will briefly explain why this is
so. From this behaviour, however, it cannot be inferred that preparing
a cavity with a quasi-classical state, $\alpha\gg1$, in pursuit of
genuine quantum phenomena, such as entanglement, is inadequate (note
that no conclusions can be drawn about the overall entanglement contained
in $\rho(t)$ by peculiar phenomena occurring in low dimensional subspaces).
The crucial role played by $\alpha$ in the entanglement of the overall
density matrix will be discussed later, in Sec.~\ref{sub:Sec2.2.3-Mac_therm_entang}. 

\begin{figure}[tb]
\noindent \begin{centering}
\includegraphics[width=0.5\textwidth]{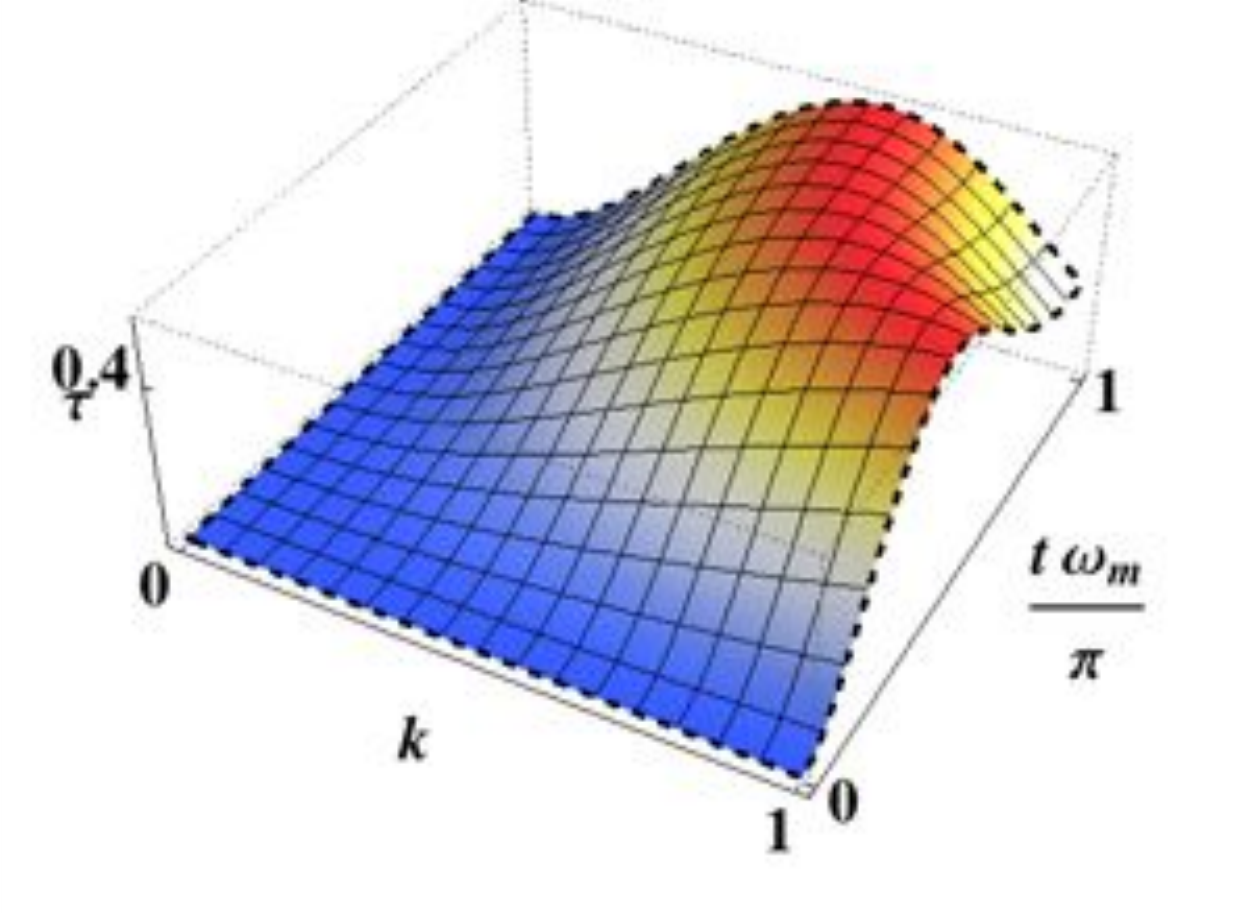}\includegraphics[width=0.5\textwidth]{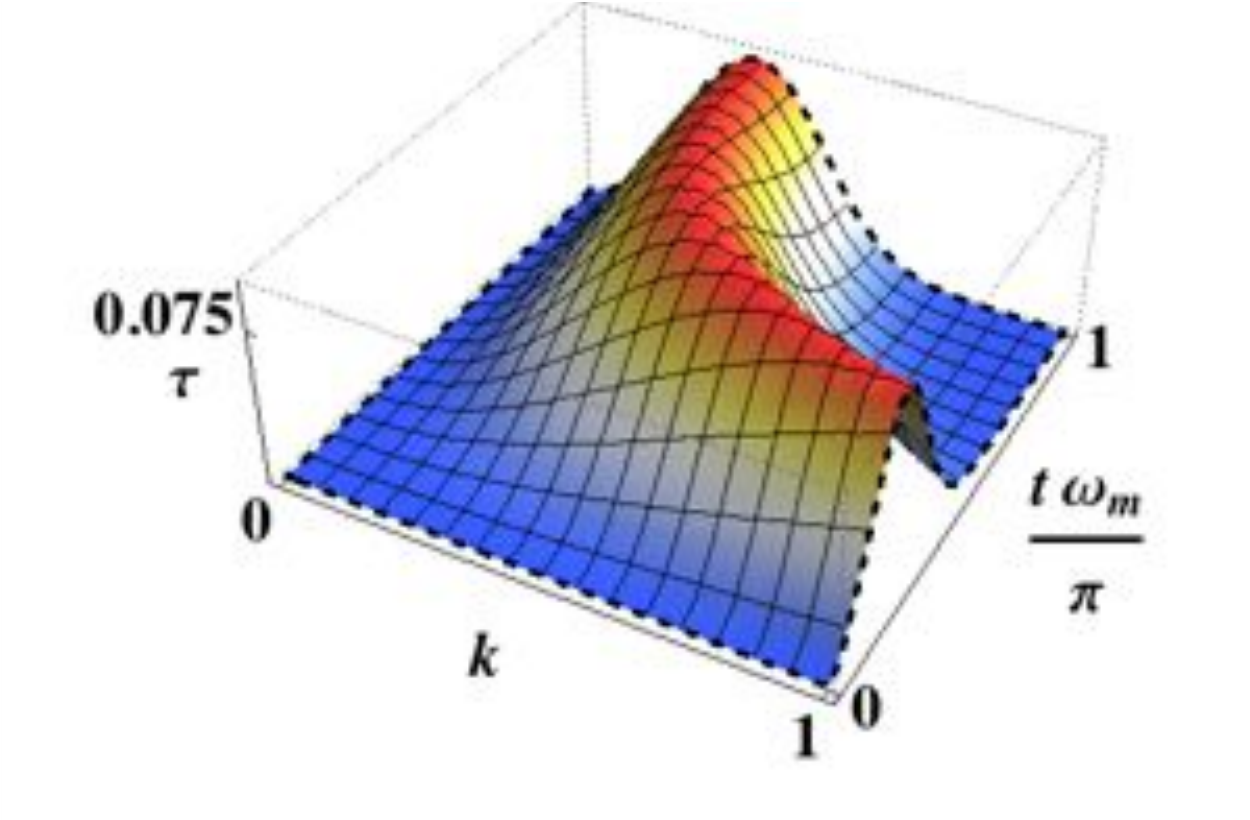}
\par\end{centering}

\caption[The tangle at zero temperature in two different subspaces]{\label{fig:Chap2-Tangle_T=00003D0}Tangle $\tau$ as a function of
the scaled coupling $k=g/\omega_{c}$ and the scaled time for $\bar{n}=0$
and $\alpha=1$. The projection subspace is $[0,1;0,1]$ (Left) and
$[1,2;1,2]$ (Right). Only half of the evolution is plotted $t\in[0,\pi/\omega_{m}]$
as the system possesses reflection symmetry around $t=\pi/\omega_{m}$.}

\end{figure}

The projected subspaces give us important hints about the entanglement
performance of this simple opto-mechanical system: Fig.~\ref{fig:Chap2-Tangle_T=00003D0}
shows the tangle as function of time $t$ for two different subspaces. 

The tangle reaches higher values in the lowest subspace as a null
temperature will favor the low occupation numbers. Thus, is not surprising
that by moving upwards from subspace spanned by $\left\{ |0\rangle,|1\rangle\right\} \otimes\left\{ |0\rangle,|1\rangle\right\} $
to the subspace spanned by $\left\{ |1\rangle,|2\rangle\right\} \otimes\left\{ |1\rangle,|2\rangle\right\} $
we loose most of the entanglement. The figure also shows a curious
dynamical transition from one regime where the maximal entanglement
is achieved for maximum displacement of the mirror, to a regime where
the maximum of the tangle is achieved faster. 

For small $k$ the system reaches the maximum of entanglement at $t=\pi/\omega_{m}$,
simultaneously with the maximum displacement of the mirror (Fig.~\ref{fig:Chap2-Tangle_T=00003D0}).
For $k$ above a critical value, say $k_{c}$, the maximum of entanglement
is achieved before $t=\pi/\omega_{m}$. Clearly, the time of maximum
entanglement depends on the balance between the interaction time $t_{int}\sim1/g$,
\emph{i.e.}~the time scale of the interaction term in the Hamiltonian,
and the time of oscillation of the mirror, $t_{m}\sim1/\omega_{m}$. 

It is worth understanding the importance of the amplitude of the coherent
state, $\alpha$, in establishing the value of $k_{c}$. Naively,
we would expect that increasing $\alpha$ would decrease $k_{c}$
because more photons interact with the mirror, for larger $\alpha$,
resulting in a larger effective coupling ($\sim g\langle a^{\dagger}a\rangle$).
Curiously, this is not the case: the value of $k_{c}$ increases with
$\alpha$! This can be understood as follows; the ratio between the
weight of the $|n+1\rangle$ number state and the weight of $|n\rangle$
number state in the expansion of the coherent state {[}Eq.~(\ref{eq:App2-Coh_States-FockExp}){]},
being given by $\alpha/\sqrt{n+1}$, increases with $\alpha$, weakening
the entanglement generated after interaction with the mirror {[}hence
explaining the tangle dependence on $\alpha$, see Eq.~(\ref{eq:Chap2-Tangle_for_subspace01}){]}.
Regarding entanglement, the best situation occurs when the weights
of the states are the most equally distributed {[}Eq.~(\ref{eq:Chap1_entropy_of_entanglement}){]}.
Hence, a higher coupling helps the entanglement generation to have
the same efficiency when $\alpha$ is increased.

For completeness, we give the explicit formula of $k_{c}$ for the
subspace $[1,2;1,2]$:\begin{equation}
-1+14k_{c}^{2}+24k_{c}^{4}=2\alpha^{2}(1+4k_{c}^{2}+96k_{c}^{4})e^{-12k_{c}^{2}}\geq0.\label{eq:Chap2-Critical_coupling_sub12}\end{equation}
The right-hand side of equation (\ref{eq:Chap2-Critical_coupling_sub12})
is non-negative resulting in a restriction for $k_{c}$, \emph{i.e.}~$k_{c}$
is lower bounded. Also, it can be deduced from equation (\ref{eq:Chap2-Critical_coupling_sub12})
that $k_{c}$ increases with $\alpha$. This confirms, at least for
this subspace, that a higher coupling is necessary for reaching the
maximum of entanglement before $t=\pi/\omega_{m}$ if the amplitude
of the cavity field is increased. Although the actual value of entanglement
differs from subspace to subspace (Fig.~\ref{fig:Chap2-Tangle_T=00003D0}),
there are quite universal characteristics; for example, the asymptotic
behavior of the tangle at $t=\pi$ as function of $\alpha$ is always\begin{equation}
\tau(\pi/\omega_{m})=\begin{cases}
\sim|\alpha|^{2} & ,\:\alpha\ll1\\
\sim|\alpha|^{-2} & ,\:\alpha\gg1\end{cases}\label{eq:-2}\end{equation}

\subsection{Finite temperature entanglement\label{sub:Chap2-Sec-Finite-temperature-entanglement}}

In practice, unless very low temperatures are considered, any mirror
will be populated with thermal phonons (even if few), and the previous
results should be seen as the limiting case $\bar{n}\simeq0$. At
$T>0$ the system is in a mixed state and the entanglement must be
investigated by other means than the tangle. It can be inferred from
the plots of the negativity {[}Eq.~(\ref{eq:Chap1_Negativity}){]}
that increasing the temperature transfers the correlations to higher
subspaces as higher excitations get populated by thermal phonons,
while the peak of negativity is reduced compared to the $T=0$ case;
also, it can be deduced that $k_{c}$ increases slowly with the temperature
(at least for the subspaces of Fig.~\ref{fig:Chap2-Tangle_T=00003D0}),
and hence thermal fluctuations make it more difficult to achieve maximum
entanglement before the mirror being maximally displaced from its
equilibrium position ($t=\pi/\omega_{m}$). 

We introduce a convenient marker of entanglement based on \ac{PPT}
which is valid for projected subspaces with arbitrary dimensionality.
In going to larger subspaces we will be able to strength our conclusions
about the nature of the entanglement generated by radiation pressure
and, especially, conclude about its robustness against thermal fluctuations.
First, we introduce the marker,\begin{equation}
\Upsilon(\rho):=-\det\left[(\Lambda_{A}\otimes{\bf 1}_{B})\rho\right].\label{eq:Chap2-MarkerEntang}\end{equation}
\begin{figure}[H]
\noindent \begin{centering}
\includegraphics[width=0.45\textwidth]{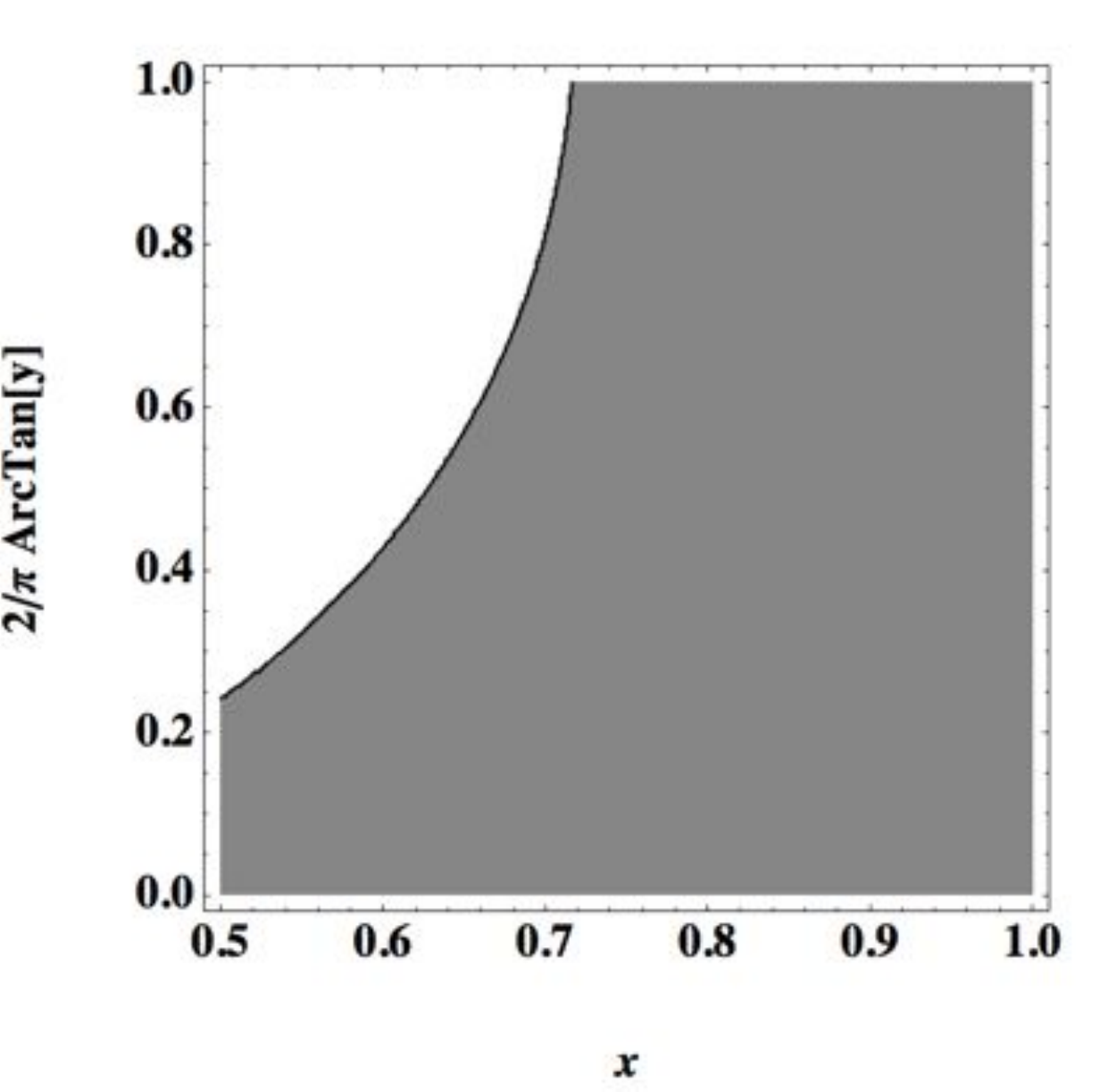}\includegraphics[width=0.45\textwidth]{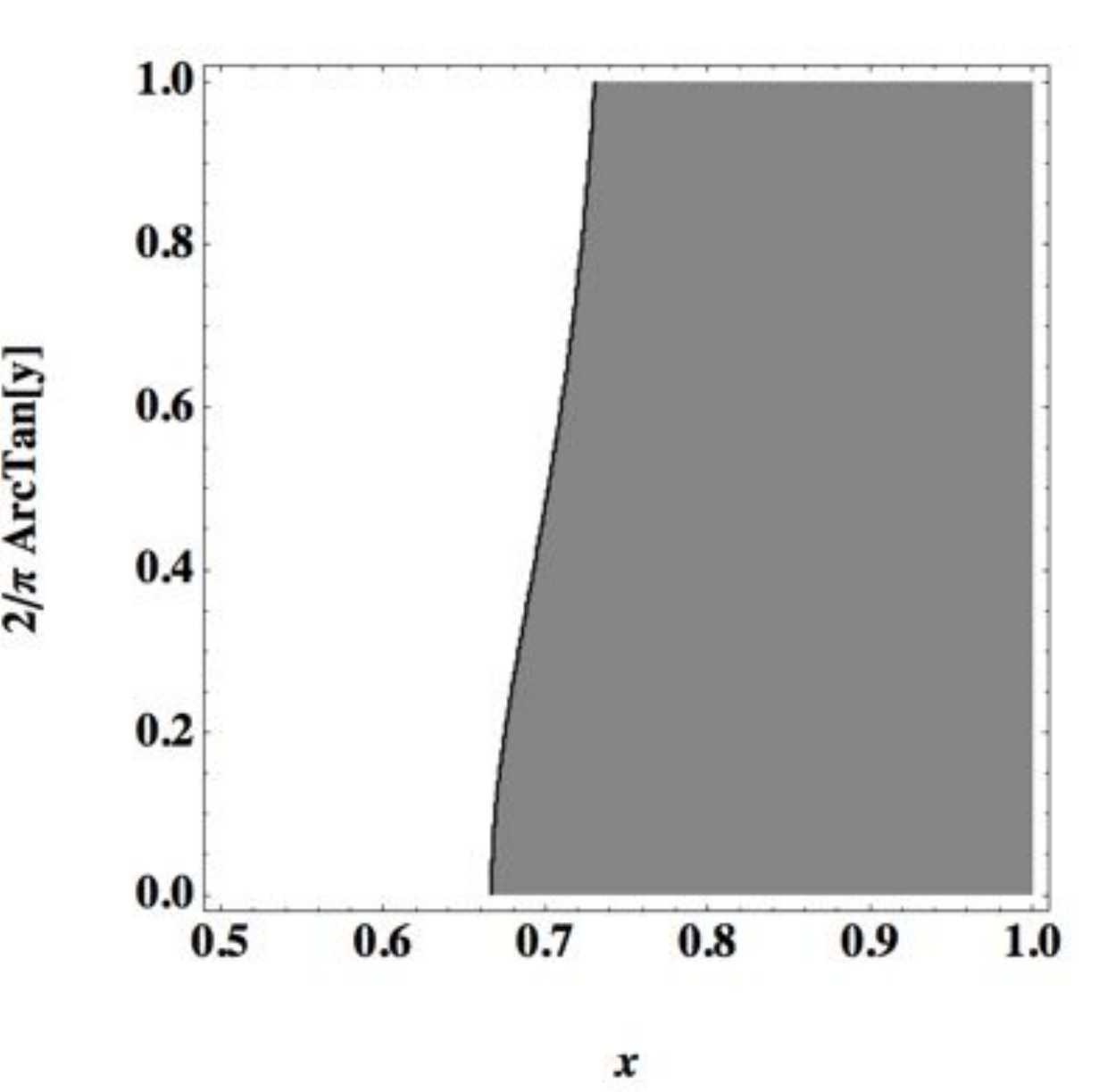}
\par\end{centering}

\noindent \begin{centering}
\includegraphics[width=0.45\textwidth]{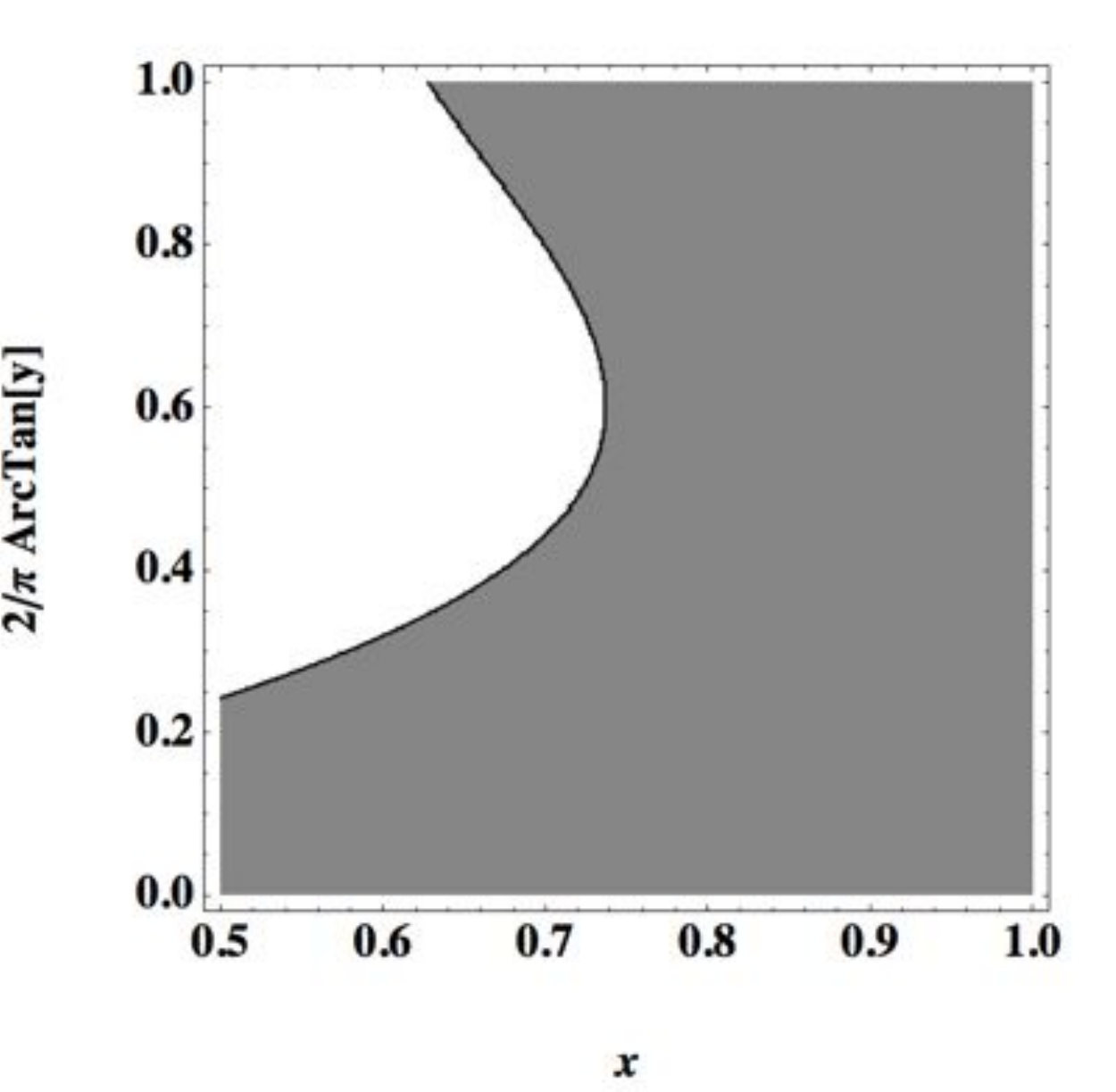}\includegraphics[width=0.45\textwidth]{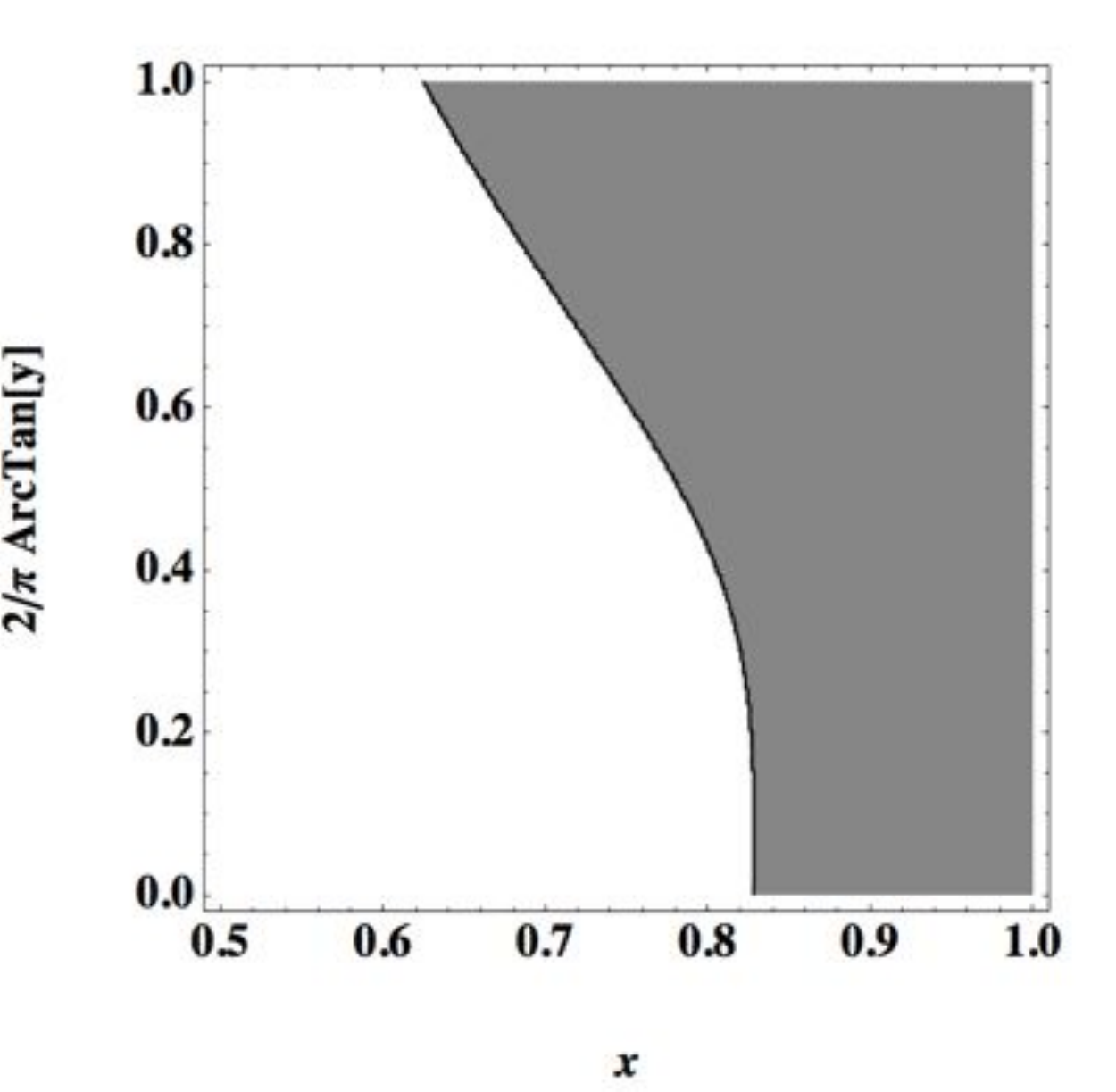}
\par\end{centering}

\caption[The marker of entanglement as function of coupling and temperature]{\label{fig:Chap2-Marker_2times2_sub}The marker of entanglement,
$\Upsilon$, for $\alpha=1$ as function of $(2/\pi)\arctan(y)$ and
$x$. The subspaces considered are: $[0,2;0,2]$ (Top-Left), $[1,2;2,3]$
(Top-Right), $[1,4;1,4]$ (Bottom-Left) and $[1,5;0,1]$ (Bottom-Right).
In these plots, the temperature varies such that $x$ lies in the
range $[.5,1[\Leftrightarrow\bar{n}\in[1,\infty[$. The coupling and
time assume any value as $y\in[0,\infty[$. For a given temperature
and coupling ($x$,$k$), by going downwards in the $y$ axis we can
imagine that we are decreasing time while fixing $k$ {[}$y=2k|\sin(\omega_{m}t/2)|${]}
in the interval $t\in[0,\pi/\omega_{m}]$ . The white regions witness
the presence of entanglement in the respective projected subspaces,
whereas the grey regions correspond to separable projected states.
The most important feature common to all the plots is the existence
of a critical temperature above which, no matter the value of $k$,
entanglement completely vanishes in the projected subspaces. }

\end{figure}

From \ac{PPT} {[}Eq.~(\ref{eq:Chap1_PPT}){]} it is clear that $\Upsilon(\rho)>0$
implies non-separability, since at least one eigenvalue of the partial
transposed matrix $(\Lambda_{A}\otimes{\bf 1}_{B})\rho$ is negative.
A careful inspection shows that, for the cases under study, this marker
is equivalent to the existence of entanglement as the partial transposed
matrix has at most one negative eigenvalue (it never happens that
two negative eigenvalues exist, in which case, of course, the marker
would not detect entanglement). Defining $x:=\bar{n}/(\bar{n}+1)$,
the marker for the lowest subspace reads,\begin{equation}
\Upsilon[0,1;0,1]=F(\alpha,t,k,x)\left(e^{x|k\eta(t)|^{2}/2}(x-2)^{4}|k\eta(t)|^{4}-16\left(-1+e^{x|k\eta(t)|^{2}/2}\right)^{2}x^{2}\right),\label{eq:Chap2-MarkerSub0101}\end{equation}

where $F(\alpha,t,k,x)$ is a positive function, and thus not relevant
to our discussion. By setting $\bar{n}=0$ ($\Leftrightarrow x=0)$
in the above expression, we get $\Upsilon[0,1;0,1]=16|k\eta(t)|^{4}F(\alpha,t,k,x)$
which is positive for every $t$ (except for $t=2p\pi/\omega_{m}$
with $p\in\mathbb{N}_{0}$), hence detecting entanglement; in accordance
with the result derived earlier {[}see Eq.~(\ref{eq:Chap2-Tangle_for_subspace01}){]}.
For other subspaces, the marker gets cumbersome but its sign is a
function of just $x$ and \begin{equation}
y(t):=|k\eta(t)|=2k|\sin\left(\omega_{m}t/2\right)|.\label{eq:-3}\end{equation}
In Fig.~\ref{fig:Chap2-Marker_2times2_sub} the marker is plotted
for various subspaces as function of $x$ and $y$. A curious feature
of $\Upsilon$ in $2\otimes2$ subspaces is the fact that its sign
does not depend on $\alpha$, meaning that the role played by the
amplitude of the coherent state will be in determining the exact amount
of entanglement of the complete density matrix (this will be confirmed
in Sec.~\ref{sub:Sec2.2.3-Mac_therm_entang} by an explicit calculation).

So far we have discussed entanglement in subspaces equivalent to two
spin-1/2 particles (\emph{qubits}) and neglected all the entanglement
shared between branches corresponding to different number occupations.
In doing so, we have found (see Fig.~\ref{fig:Chap2-Marker_2times2_sub})
that sometimes a thermal occupation as low as $\bar{n}=2$ is surprisingly
enough to destroy the quantum correlations within the low dimensional
Fock subspaces. A more realistic lower bound to entanglement, and
consequently stronger conclusions, however, can be obtained by enlarging
just a bit the projected subspace. Indeed, let us investigate the
entanglement dynamics for subspaces of dimension $d=6$, concretely
in Hilbert spaces of the form $\mathcal{H}=\mathbb{C}^{2}\otimes\mathbb{C}^{3}$. 

In the line with the projection method, we define the projector operator
to be now, \begin{equation}
P=P_{nm}\otimes P_{\mu\nu\epsilon},\label{eq:-4}\end{equation}
with $P_{\mu\nu\varepsilon}=|\mu\rangle\langle\mu|+|\nu\rangle\langle\nu|+|\epsilon\rangle\langle\epsilon|$
and denote the respective projected subspace by $[n,m;\mu,\nu,\epsilon]$.
The Fig.~\ref{fig:Chap2-SubPurity_as_Enforcer} shows two plots ---
at the left-hand side the negativity for the subspace $[0,1;3,4,5]$
and at the right-hand side the respective marker. Is is apparent from
these results that for some couplings the entanglement persists at
arbitrary high-temperatures (even if in a very small quantity). This
is at odds with what we have found in subspaces with $d=4$ and it
may seem an anomalous result {[}we have learned in the Introduction
that increasing the temperature too much unavoidably destroys entanglement
(see, for instance, the comments in Fig.~\ref{fig:Chap1_Entang2qubitsDensity}
or the explicit calculation of Sec.~\ref{sub:The-separability-of}){]}.
This is explained by the fact that the cavity is prepared in a pure
state; the purity of the subsystem works as an \emph{enforcer of entanglement}.
This was first discovered by Bose \emph{et al. }in \cite{2001Bose}
for one \ac{EM} field mode interacting with a two-level atom in a
thermal state and here we find a similar phenomenon. An even more
curious peculiarity of this system will be disclosed by studying the
implications of the photon conservation in the entanglement structure.

\begin{figure}[tb]
\noindent \begin{centering}
\includegraphics[width=0.5\columnwidth]{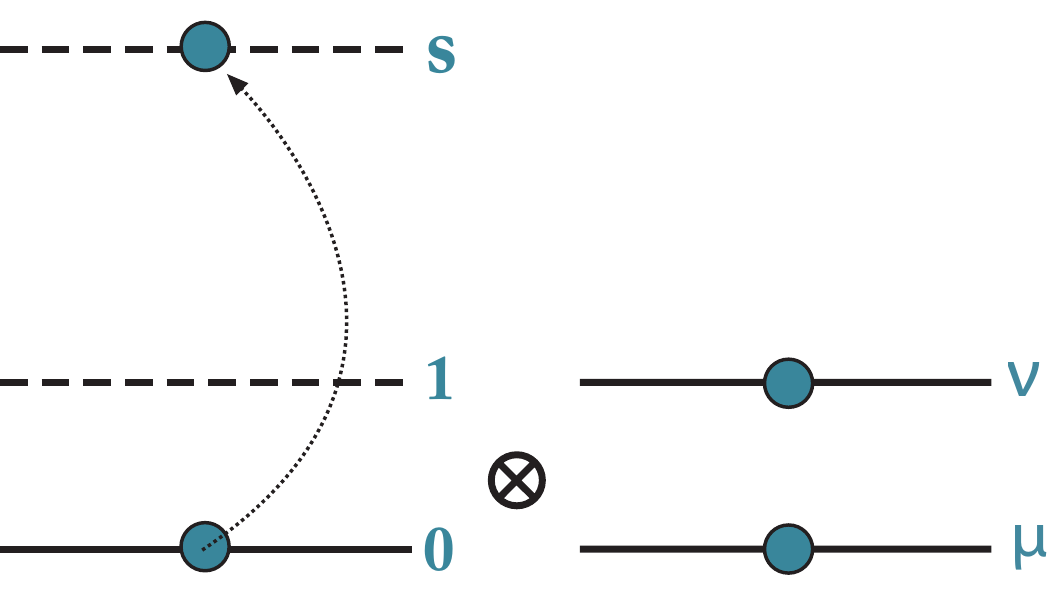}
\par\end{centering}

\caption[The subspace renormalization procedure]{\label{fig:Chap2-SubspaceRenorma}This figure illustrates the \textquotedbl{}subspace
renormalization\textquotedbl{} procedure discussed in the text; the
opto-mechanical state is projected onto the state having $0$ and
$1$ excitations of the cavity \ac{EM} field and $\mu$, $\nu$,
$...$ excitations of the mirror. If for a given temperature and $k$
the subspace $[0,1;\mu,\nu,...]$ is able to support entanglement,
then the subspace $[0,s;\mu,\nu,...]$ will attain entanglement for
the smaller coupling defined by $k_{s}:=k/s$ {[}see Eq.~(\ref{eq:Chap2-MarkerRelation}){]}. }

\end{figure}

The density matrix {[}Eq.~(\ref{eq:Chap2-DensityMatrix_Int_form}){]}
has a pattern regarding the matrix elements of the cavity; $k$ always
appears multiplied by $n$ or $m$. This follows directly from photon
conservation (the number operator $\hat{n}=a^{\dagger}a$ commutes
with the Hamiltonian {[}Eq.~(\ref{eq:Chap2-Hamiltonian}){]}). Bearing
this in mind, and denoting by $\Upsilon_{[0,s]}$ the marker for the
subspaces spanned by the Fock states $\left\{ |0\rangle,|s\rangle\right\} $
of the cavity (whereas the specific projection on the side of the
mirror is arbitrary), then it is straightforward to verify that $\Upsilon_{[0,s]}$
is proportional to $\Upsilon_{[0,1]}$ if, in the latter, the coupling
$k_{s}=k/s$ is chosen. This has an immediate important consequence:
if we choose a large $k$ capable of producing entanglement in the
cavity subspace $[0,1]$, then there must be entanglement in the subspace
$[0,s]$ for coupling constant $k_{s}:=k/s$, even though $k_{s}$
might not lead to entanglement in the subspace $[0,1]$. For instance,
in the right-hand side plot of Fig.~\ref{fig:Chap2-SubPurity_as_Enforcer},
entanglement actually survives for the parameter region concerning
the bottom-left corner, provided a proper choice of the projection
subspace $[0,s]$ is made. The grey region at the top-right is not
as important, since we can always, with fixed $k$, move downwards
in the $y$-axis by decreasing time (recall that $y=2k|\sin(\omega_{m}t/2)|$).
Mathematically, this can be expressed as follows,\begin{equation}
|\alpha|^{-2s}s!\Upsilon_{[0,s]}(k_{s})=|\alpha|^{-2}\Upsilon_{[0,1]}(k).\label{eq:Chap2-MarkerRelation}\end{equation}
This result implies that for a given Fabry-Perot geometry and mirror
--- characterized by its own natural frequency and mass --- the existence
of entanglement in a suitable $2\otimes3$ subspace of (\ref{eq:Chap2-DensityDecompFock})
is guaranteed for any temperature, even for small coupling $k=g/\omega_{m}$,
provided we choose a sufficiently high subspace (Fig.~\ref{fig:Chap2-SubspaceRenorma}).
Inspired by this result, we put forward the following conjecture,
\begin{conjecture}
The opto-mechanical system consisting of a thermalized mechanical
oscillator interacting via radiation pressure with a cavity \ac{EM}
field, initially prepared in a large coherent-state, supports macroscopic
entanglement.\label{con:Chap2-MacroscopicEntang}
\end{conjecture}
\begin{figure}[H]
\noindent \begin{centering}
\includegraphics[width=0.5\textwidth]{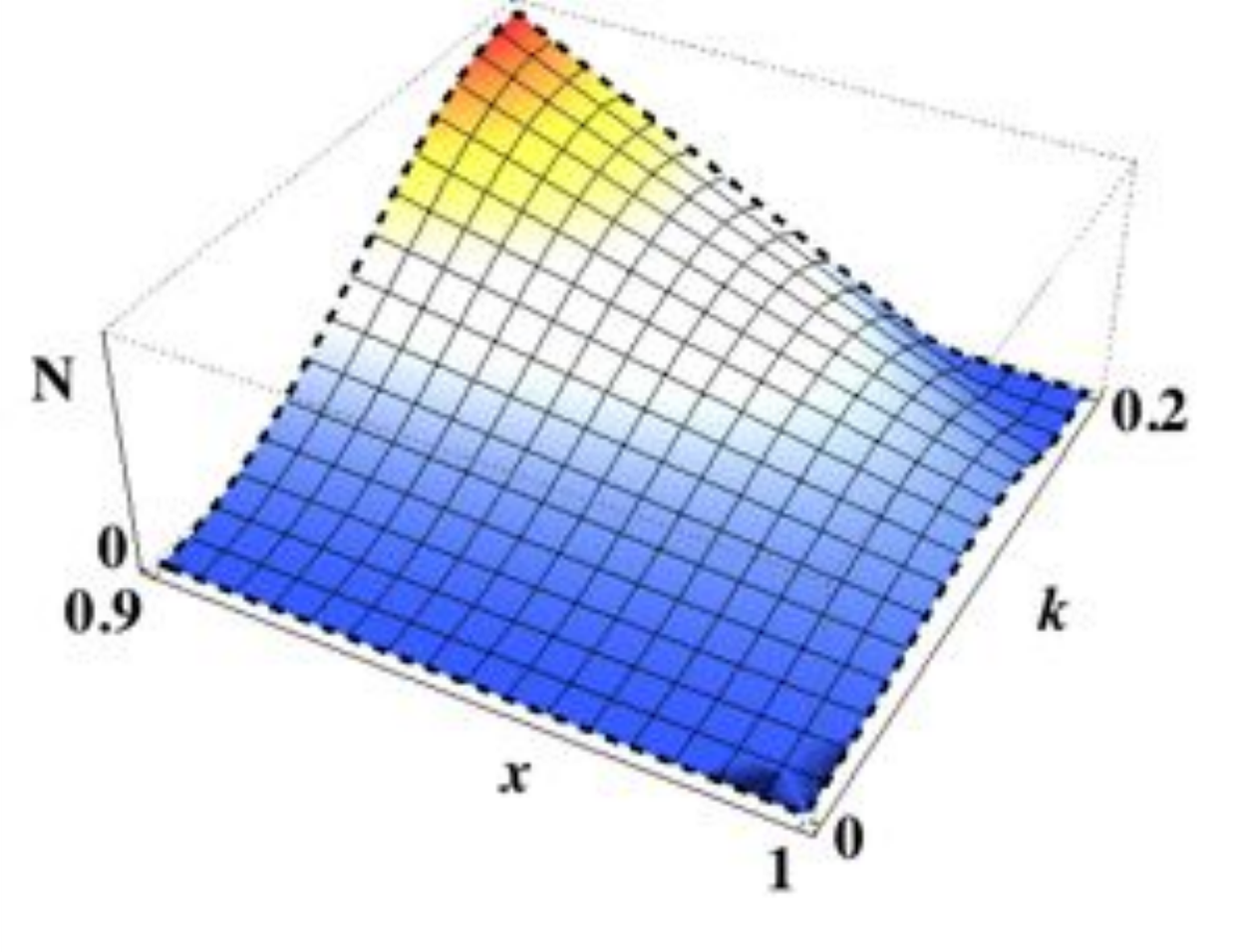}\includegraphics[scale=0.5]{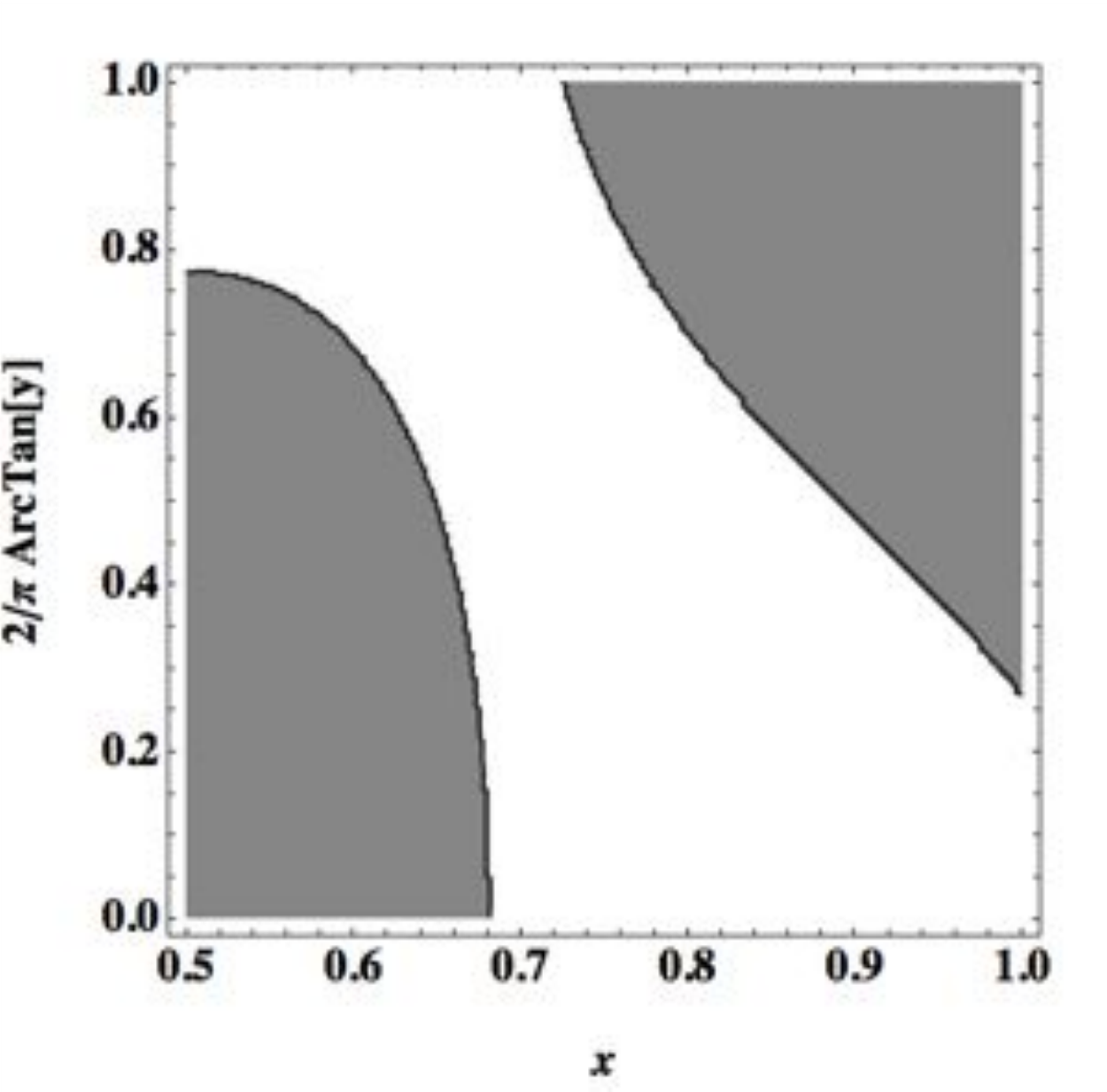}
\par\end{centering}

\caption[The robustness of thermal entanglement in $2\otimes3$ subspaces]{\label{fig:Chap2-SubPurity_as_Enforcer}The entanglement in the $2\otimes3$
subspaces is more robust against temperature. In fact, it persists
at arbitrary high-temperatures for some couplings (right-hand plot).
The condition in $k$ can even be relaxed if we choose sufficiently
high cavity excitations (see text and Fig.~\ref{fig:Chap2-SubspaceRenorma}).
Left - The negativity is plotted as function of coupling $k$ and
$x=\bar{n}/(1+\bar{n})$ for the subspace $[0,1;3,4,5]$ with $\alpha=1$
and $t=\pi/\omega_{m}$. In the range considered, the peak of the
negativity happens for $x=0.9\wedge k=0.2$, yielding $N\simeq3*10^{-4}$.
This value can be improved by a proper choice of parameters. Right
- The marker, $\Upsilon$, for the same subspace, as a function of
$(2/\pi)\arctan(y)$ and $x$. The white regions witness the presence
of entanglement, $\Upsilon>0$.}

\end{figure}

\subsection{Macroscopic thermal entanglement\label{sub:Sec2.2.3-Mac_therm_entang}}

At this point, fair criticism to the projection method is in order.
In Sec.~\ref{sub:Sec1.3.2-The entang meas for pure states}, we have
learned that the maximal amount of entanglement in a $d\otimes d$
dimensional density matrix is precisely $\ln d$. In \ac{QI}, by
comparison with the entanglement entropy of the singlet, one speaks
about $\ln d/(\ln2)$ \emph{e-bits} (\emph{i.e.}~entanglement bits)
in honour of Schumacher\textquoteright{}s seminal work on quantum
communication with qubits. In the macroscopic limit, we have a diverging
number of \emph{e-bits} in a maximally entangled state {[}Eq.~(\ref{eq:Chap1_maximally_entang_state}){]}.
The presence of entanglement in the $2\otimes3$ subspaces is not
a guarantee of macroscopic entanglement as these subspaces support
a maximum of just $\simeq2.6$ e-bits. The relevant question that
should be asked is then; does a finite amount of thermal entanglement
in the low dimensional subspaces imply a macroscopic amount of quantum
correlations in the complete density matrix? We have conjecture that,
indeed, the answer is positive, since the particular entanglement
structure shows that entanglement survives in many subspaces%
\footnote{The marker will actually detect entanglement in infinitely many subspaces,
at least for small temperatures (despite only a small subset will
yield a detectable amount of negativity). This can be seen as follows:
for a given set of parameters ($\bar{n}$, $t$, and $\alpha)$ consider
a subspace of the form $[0,1;....]$ leading to non-zero negativity
for couplings $k$ and $\bar{k}:=k*p$,\emph{ }with $p$\emph{ }an
integer. That is, the subspace of the mirror fulfills $\Upsilon_{[0,1]}(k)>0\wedge\Upsilon_{[0,1]}(\bar{k})>0$
--- an example is provided in Fig.~\ref{fig:Chap2-SubPurity_as_Enforcer}
for $x\lesssim0.7$. Then, by virtue of the renormalization procedure
{[}Eq.~(\ref{eq:Chap2-MarkerRelation}){]}, the subspace $[0,p;....]$
entails: $\Upsilon_{[0,p]}(k_{p}=\bar{k}/p)=\Upsilon_{[0,p]}(k)\sim\Upsilon_{[0,1]}(\bar{k})$.
Hence, $\Upsilon_{[0,p]}(k)>0$, for every $p$ since $\Upsilon_{[0,1]}(\bar{k})>0$. %
}. Thus, at least in principle, one could distill a detectable amount
of \emph{e-bits} when adding up the contribution of each subspace.

Further insight can be made by studying the mutual information {[}Eq.~(\ref{eq:App1-Mutual_Info}){]}
between the cavity field and the mirror. This quantity, with roots
in classical statistical theory, has two clear operational meanings
in the context of quantum systems: it equals the minimum amount of
randomness required to erase all the correlations shared by two subsystems
(in a many copy scenario like the one discussed in Sec.~\ref{sec:Entang-th})
\cite{Groisman2005}, and is equivalent to the relative entropy between
$\rho_{AB}$ and the separable state from the partial states: $\rho_{A}\otimes\rho_{B}$.
That is,\begin{equation}
I(\rho_{AB})=S(\rho_{AB}|\rho_{A}\otimes\rho_{B}).\label{eq:Chap2-RelativeEntropy-MutualInfo}\end{equation}
Indeed, $I(\rho_{AB})$ measures how much information the compound
system $'A+B'$ has more than the respective partial states. The connection
of $I(\rho_{AB})$ to entanglement is clear for the amount of non-separability
of a state $\rho_{AB}$ can be thought as the distance of $\rho_{AB}$
to the closest separable state $\sigma\equiv\sigma_{A}\otimes\sigma_{B}$
(Fig.~\ref{fig:Chap1_set_bipartites_states}). For pure states it
equals twice the von Neumann entropy of the reduced state, and therefore
is an entanglement measure. For mixed states it quantifies entanglement
no more, but can still be used to speak about the total amount of
correlations shared between two parties. 

Here, we consider a related quantity (which we will refer to as the
\emph{normalized mutual information}; see Appendix~\ref{sec:App-Entropy})
detecting non-classical correlations directly. For practical reasons,
we adopt the linear entropy {[}Eq.~(\ref{eq:Chap1_LinearEntropy}){]}
as our measure of mixdness. Defining $\phi_{nm}(t):=\Phi_{00nm}(t)$
{[}see~(\ref{eq:-1}){]}, the partial states of the density matrix
read (see Appendix~\ref{sec:App-CoherentStates}),\begin{eqnarray}
\rho_{cav}(t) & = & \sum_{nm}\phi_{nm}(t)e^{-|k\eta(t)|^{2}(n-m)^{2}(2+\bar{n})/4}|n\rangle\langle m|,\label{eq:Chap2-PartialStates1}\\
\rho_{m}(t) & = & \sum_{n}\phi_{nn}(t)\int_{\mathcal{\mathbb{C}}}\frac{d^{2}z}{\bar{n}\pi}e^{-|z|^{2}/\bar{n}}|z+kn\eta(t)\rangle\langle z+kn\eta(t)|.\label{eq:Chap2-PartialStates2}\end{eqnarray}
The entropy is invariant under unitary evolution, and hence the linear
entropy of the compound system, $S_{\mathcal{L}}(t)=1-\text{Tr}\left[\rho(t)^{2}\right]$,
can be evaluated at $t=0$, taking advantage of the state being separable
there {[}Eq.~(\ref{eq:Chap2-InitialState}){]}: \begin{equation}
S_{\mathcal{L}}(t)=1-\frac{1}{2\bar{n}+1}.\label{eq:Chap2-LinearEntropyAB}\end{equation}
The entropy of the partial states can be obtained through the same
methods leading to the partial entropies,\begin{eqnarray}
S_{\mathcal{L},cav}(t) & = & 1-e^{-2|\alpha|^{2}}\sum_{p,q}\frac{|\alpha|^{2(p+q)}}{p!q!}e^{-|k\eta(t)|^{2}(p-q)^{2}(2+\bar{n})/2},\label{eq:Chap2-PartialEntropy1}\\
S_{\mathcal{L},m}(t) & = & 1-\frac{1}{2\bar{n}+1}e^{-2|\alpha|^{2}}\sum_{p,q}\frac{|\alpha|^{2(p+q)}}{p!q!}e^{-|k\eta(t)|^{2}(p-q)^{2}g(\bar{n})},\label{eq:Chap2-PartialEntropy2}\end{eqnarray}
with $g(\bar{n})=x^{2}/(1+x)$ and $x=\bar{n}/(1+\bar{n})$ as before.
The normalized mutual information (\ref{eq:App1-Normalized_mutual_information})
guarantees the presence of quantum correlations whenever $\mathcal{I}>1/2$.
It reads,\begin{equation}
\mathcal{I}(t)=1-\frac{S_{\mathcal{L}}(t)}{S_{\mathcal{L},cav}(t)+S_{\mathcal{L},m}(t)}.\label{eq:Chap2-NormalizedLinearMutualInfo}\end{equation}
\begin{figure}[tb]
\noindent \centering{}\includegraphics[width=1\textwidth]{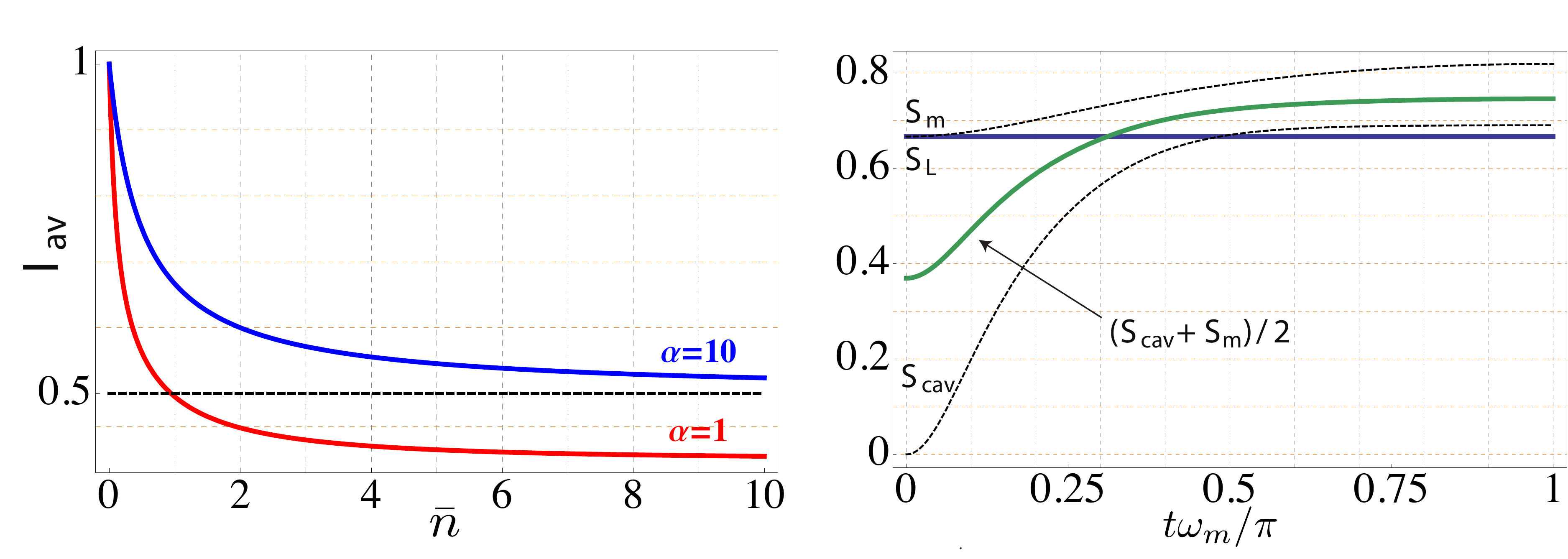}\caption[The linear mutual information of the full density matrix]{\label{fig:Chap2_MutualInfo}Left - The normalized mutual information
averaged through one period of the mechanical oscillator {[}Eq.~(\ref{eq:Chap2-AverageMutual}){]}
as function of the average number of phonons, $\bar{n}$, for a cavity
prepared with a small coherent state (one photon in average, red curve)
and a relative large amplitude (one hundred photons in average, blue
curve). The classical upper bound on correlations {[}Eq.~(\ref{eq:App1-Upper_bound_classical}){]}
is represented by the dashed curve. We see a clear signature of macroscopic
quantum correlations for the larger coherent state. In both plots
the coupling was chosen to be one (in units of the mirror's natural
frequency). Right - Various linear entropies are plotted for a mirror
at very low-temperature ($\bar{n}=1$) and $\alpha=1$ as function
of the scaled time $t\omega/\pi$ (only half of the evolution is plotted):
$S_{\mathcal{L}}(t)$ (thick, blue), $S_{m}(t)$ (dashed, black),
$S_{cav}(t)$ (dashed, black) and the average linear entropy of the
subsystems (green line). The latter is a signature of entanglement
every time it exceeds the global entropy $S_{\mathcal{L}}(t)$, which
happens in the interval $1.69\gtrsim\omega_{m}t/\pi\gtrsim0.31$. }

\end{figure}

From the above equations, we can infer that $\mathcal{I}$ increases
with $\alpha$ initially and that then it stabilizes; the quantum
correlations emerging from the radiation-pressure mechanism should
therefore increase with $\alpha$, at least in the range where $\mathcal{I}$
increases monotonously above the classical upper bound {[}Eq.~(\ref{eq:App1-Upper_bound_classical}){]}.
Within this range, a detectable amount of entanglement is expected. 

We finish the present chapter giving a numerical value for the averaged
normalized mutual information over a period of oscillation of the
mirror, \begin{equation}
\mathcal{I}_{av}:=\frac{1}{2\pi}\int_{0}^{2\pi}I(\tau)d\tau.\label{eq:Chap2-AverageMutual}\end{equation}
We choose a strong mirror-light coupling $k\simeq1$, a low (but not
too low) thermal occupancy $\bar{n}=10$, and a relative small cavity
amplitude $\alpha\simeq10$. With these values, the averaged normalized
mutual information yields $\mathcal{I}_{av}\simeq0.52>\mathcal{I}_{c}$,
which should be interpreted as an indicator of quantum behaviour. 

How does the temperature affect entanglement in the total density
matrix? According to Fig.~\ref{fig:Chap2_MutualInfo}, the correlations
become totally classical after a few thermal photons ($\bar{n}\sim10$)
have populated the mirror --- a careful study, though, shows that
this behavior is little sensitive to the radiation pressure coupling
and even $\alpha$; an indicator that the linear entropy does not
capture all the subtleness of the correlations from the radiation
pressure interaction.

\section{Concluding remarks}

We have seen that entanglement between the motion degree of freedom
of a mechanical oscillator and the \ac{EM} quadratures of light does
exist at any temperature in low dimensional subspaces of the density
matrix, although it may be very feeble for very high temperatures.
The findings of Sec.~\ref{sec:2.2.Towards-high-temp} show the coexistence
of entanglement in many subspaces; the radiation-pressure mechanism
is robust, in the sense that small opto-mechanical couplings, leading
to little or no entanglement in a given subspace, attain entanglement
in suitable subspaces, as implied by the subspace renormalization
condition {[}Eq.~(\ref{eq:Chap2-MarkerRelation}){]}. 

Our results provide the first plausible evidence for macroscopic entanglement
in opto-mechanical systems in the experimental relevant regime, $k_{B}T\gtrsim\hbar\omega_{m}$.
To our knowledge this was the first time that entanglement (and therefore
the quantum behavior) was shown to survive at finite temperatures
in bipartite macroscopic systems. This has confirmed Bose's expectation
about the capability of macroscopic compound systems to attain entanglement
when one of the subsystems is initialized in a pure state \cite{1999Bose-b,2001Bose}.
A complementary argument validating this conclusion will be given
in the following chapter, where the important issues of mirror's friction,
damping of the cavity and especially decoherence will be added to
Hamiltonian (\ref{eq:Chap2-Hamiltonian}). 

Our findings hence suggest a way to explore the quantum-classical
boundary in mesoscopic and macroscopic mechanical oscillators by coupling
them to a confined \ac{EM} field. The reason why opto-mechanical
systems are a laboratory for quantum effects stems from the radiation-pressure
mechanism; the thermal phonons of the mirror are coherently displaced
in phase space by virtue of the interaction term (\ref{eq:Chap2-EvolutionOperator})
in a manner that is only sensitive to the number of photons,\[
\langle b(t)\rangle=k\eta(t)|\alpha|^{2}.\]
By monitoring the position of the mirror, we actually perform a \ac{QND}
measurement of the resonator's energy as first envisaged by Braginsky
\cite{1980Braginsky} (as can be seen by the latter equation); a measurement
of $\langle b(t)\rangle$ gives information about the cavity energy
$E_{cav}=\hbar\omega_{c}|\alpha|^{2}$ up to any desired accuracy
(recall that $[a^{\dagger}a,H_{int}]=0$), when the adiabatic limit
is considered. In the line of Penrose's proposal, an equivalent setup
has been considered recently in the pursuit of macroscopic superpositions
of a tiny mirror \cite{2003Marshall}. There, they consider the effect
of a single photon, and hence the entanglement generation is too feeble
to survive at any realistic low-temperature \cite{2006Bernard}. Here
we have showed that one increases substantially the chances to probe
quantum macroscopic behavior by considering a coherent \ac{EM} field,
instead of a single photon%
\footnote{This could seem ironic since coherent states are \textquotedbl{}quasi-classical\textquotedbl{}
states of light (they resemble a classical \ac{EM} wave due to their
small uncertainty around the classical values). Nevertheless, photons
still are quantum in nature and their interaction with a free-standing
mirror, for instance, will be characterized by Hamiltonian {[}Eq.
(\ref{eq:Chap2-Hamiltonian}){]} in the adiabatic limit. Surprisingly,
as we have seen in the present chapter, these photons get strongly
entangled with the mirror as each one of them contributes to a sort
of macroscopic net effect.%
}. The applications of the simple radiation-pressure mechanism are
numerous; they comprise detection of gravitational waves and the study
of the quantum-to-classical transition.

The present chapter has shed some light on the celebrated radiation-pressure
mechanism in a cavity quantum electrodynamics setup. This was essentially
done by solving for the system density matrix in a exact way and taking
advantage from the already well-established tools of the young field
of \ac{QI} science. Many questions are still open, though. The exact
dependence of the entanglement with the radiation pressure coupling
(an thus the mirror's mass, for instance) was not addressed satisfactorily
--- only qualitative conclusions can be drawn by the study of the
linear entropy of the subsystems as argued in Sec.~\ref{sub:Sec2.2.3-Mac_therm_entang}.
Also, our method does not provide an entanglement measure (see Sec.
\ref{sub:Entanglement_MixedStates}) for the total density matrix.

These questions will be partially answered in the following chapter;
others we leave as possible research topics: can the spectrum of the
density-matrix {[}Eq.~(\ref{eq:Chap2-DensityMatrix_Int_form}){]}
be obtained, and, therefore, the mutual information (based on the
von Neumann entropy rather than the linear entropy)? Even more importantly,
perhaps, would be, not to assume the somewhat artificial initial state
{[}Eq.~(\ref{eq:Chap2-InitialState}){]},\[
\rho_{ab}(t_{0})=|\alpha\rangle\langle\alpha|\otimes\Xi(\beta),\]
(where $\Xi(\beta)$ is a thermal state of the mirror), but a more
realistic initial condition. This can be accomplished by considering
that an external coherent source populates a cavity initially with
no photons (which is a good approximation even at room temperature
given the high energy of optical photons),\[
\rho_{abc}(t_{0})=\left(|0\rangle\langle0|\otimes\Xi(\beta)\right)\otimes|\alpha\rangle\langle\alpha|,\]
which then would evolve via the total Hamiltonian,\[
H=\hbar\omega_{c}a^{\dagger}a+\hbar\omega_{m}b^{\dagger}b+\hbar\omega_{0}c^{\dagger}c-\hbar ga^{\dagger}a(b+b^{\dagger})+\hbar\left(Gac^{\dagger}+G^{*}a^{\dagger}c\right),\]
where $c$ and $c^{\dagger}$ are the bosonic operators of the driving
source photons with frequency $\omega_{0}$ and $G$ their coupling
to the intra-cavity field. The density operator would be computed
by performing the partial trace of the driving photons,\[
\rho_{ab}(t)=\text{Tr}_{c}\left[U(t)\rho(t_{0})U^{\dagger}(t)\right].\]
The main challenge is to solve for the evolution operator in a closed
fashion. If we were able to do so, then the same method leading to
the analytical solution of equations (\ref{eq:Chap2-PartialStates1})
and (\ref{eq:Chap2-PartialStates2}) could be used to compute $\rho_{ab}(t)$.
This would strength the validity of our conclusions since a more realistic
initial condition, breaking down the renormalization procedure, would
be considered. Nevertheless, this unitary approach, although introducing
mixdness in the reduced state, does not take into account decoherence
as a real active surrounding bath unavoidably leads to. We study the
effect of such bath in the next chapter when proper assumptions are
made about the dynamics of the system.

\chapter{Stationary opto-mechanical entanglement at moderately high temperatures\label{cha:Stationary-optomechanical-entangl} }

\emph{This chapter is ba}~\emph{sed on the following publications
by the author:}
\begin{itemize}
\begin{onehalfspace}
\item \emph{Macroscopic thermal entanglement due to radiation pressure},
AIRES FERREIRA, A. Guerreiro, and V. Vedral, Phys. Rev. Lett. \textbf{96},
060407 (2006).\end{onehalfspace}

\item \emph{Optomechanical entanglement between a movable mirror and a cavity
field, }D. Vitali, S. Gigan, AIRES FERREIRA, \emph{et al., }Phys.
Rev. Lett. \textbf{98}, 030405 (2007).
\end{itemize}

\section{Overview\label{sec:Overview}}

In the previous chapter, we have learned that the radiation pressure
mechanism is able to entangle the center-of-mass motion degree of
an object consisting of many particles and the \ac{EM} quadratures
of light itself. The ideal scenario --- with no dissipation and no
active environment (\emph{i.e}.~no thermal and quantum noise) ---
was considered through a global unitary evolution of the opto-mechanical
system. Based on solid arguments, we conjectured that a macroscopic
mirror, at one end of an optical cavity, shares genuine quantum correlations,
at finite temperature, if a sufficiently large coherent state of the
light is prepared inside the cavity. However, in the laboratory, the
unavoidable sources of noise, diffraction and imperfections in the
mirror break down the simple unitary description. Moreover, we expect
that only very moderate temperatures will accommodate a finite amount
of entanglement as the interaction with an active bath of quantum
oscillators (a real environment) will destroy entanglement above some
critical temperature. This phenomenon is referred to as \emph{decoherence},
and many people believe that eventually it is the responsible for
the quantum-to-classical transition observed many times in nature.
Nevertheless, whether decoherence is the actual mechanism explaining
classicality or not, it gives an appropriate description of the results
of experiments and can be analyzed within several frameworks, namely,
\begin{enumerate}
\item the master equation (Schr\"{o}dinger picture);
\item the Fokker-Planck equation (phase space);
\item the quantum Langevin equation (Heisenberg picture). 
\end{enumerate}
Decoherence is enhanced with temperature and the system size and,
in most of the cases, is so fast that quantum interference is never
observable; every bipartite system (or any system for that matter)
gets entangled with the environment causing entanglement dilution
within the degrees of freedom of the system. Very generally it manifests
as a suppression of the off-diagonal entries of the reduced density
matrix and it is present even at zero-temperature.

Indeed, opto-mechanical entanglement between a massive mirror and
the \ac{EM} field will only be observable if two conditions are met,
namely that the temperature is not high enough as to suppress quantum
coherence, and that the radiation pressure coupling, $g$, is sufficiently
robust compared to energy scales associated with noise. We already
gave arguments in favor of the last point, but our unitary approach
gives little clues about the former.

The thermal robustness of entanglement in this system is paramount
in order to achieve an experimental demonstration, since all types
of ground-state cooling techniques encounter many difficulties making
the regime $k_{B}T\lesssim\hbar\omega_{m}$ prohibitive even with
state-of-the-art technology. Some promising experiments have been
performed in this direction but no real quantum behavior of the mechanical
oscillator was probed so far \cite{2006Arcizet,2006Gigan,2008Thompson,2009Park,2009Schliesser,2009Groblacher,2006Naik}.
Recent advances in theory of opto-mechanical cooling \cite{1998Mancini,1999Cohadon,2007Marquardt,2007Wilson-Rae}
have raised the interest in opto-mechanical systems, either as a way
to test the quantum-classical boundary, or due to their potential
to enhance the sensitivity of displacement measurements (which is
crucial for many applications being the detection of gravitation waves
the most famed). Interestingly, the radiation-pressure mechanism by
itself is able to cool the motion of the mirror when the cavity field
is pumped by \ac{LASER} radiation \cite{2007Wilson-Rae,2008Dantan}
--- the so-called dynamical back-action cooling or simply self-cooling
--- in a spirit very close to what we have already seen in atomic
and molecular physics (\emph{e.g.}~with cold atoms, ion traps, etc.).
The opto-mechanical cooling is a very interesting and fast-developing
field of theoretical and experimental physics.

An experimental demonstration of entanglement in these systems may
be reachable after a sufficient experimental improvement is achieved
in the physics of opto-mechanical cooling. In what follows, we review
the dynamics of driven optical cavities and compute the stationary
entanglement taking into account the most relevant sources of dissipation
and decoherence. The reader unfamiliar with Langevin equations and
the input-output formalism of Quantum Optics will find some guidance
in short appendices. The results of this chapter will settle on quite
solid grounds the preceding conclusions about macroscopic entanglement
and, together with the recent effort in understanding and reaching
the quantum regime in the laboratory \cite{2004Metzger,2006Kleckner,2006Naik,2006Arcizet,2006Gigan,2006Schliesser,2007Corbitt,2008Thompson,2009Park,2009Schliesser,2009Groblacher},
open a very promising direction in the endeavour to bring quantum
effects to the macroscopic domain.

\section{The dynamics of a driven cavity\label{sec:The-dynamics}}

In the past two decades, the study of small systems interacting with
the \ac{EM} field of an optical cavity had been a fertile ground
producing many outstanding results, especially in the so-called cavity
quantum electrodynamics\emph{ }experiments\emph{. }Among many experimental
demonstrations, we find\emph{ }the energy quantum jumps between Fock
states of a cavity field \cite{2008Brune}, and the quantum coherent
control of atomic collisions inside a cavity, just to name a few (see
\cite{Review-Walther-2006} for a review). The reason why cavities
are so special is because they intrinsically impose boundary conditions
on the \ac{EM} field leading to a discrete energy spectrum. This
not only amplifies the radiation pressure coupling as we have seen
in Sec.~\ref{sec:2.2.Towards-high-temp}, but it also introduces
a myriad of new physics (\emph{e.g.}~the modification of the spontaneous
emission rate of atoms \cite{1985Hulet}).

According to the results of Chap.~\ref{cha:MacroscopicEntang}, a
large coherent-state of the \ac{EM} field must be prepared inside
the optical resonator if one hopes to reach a detectable amount of
entanglement. In fact, this represents no problem at all, for it became
a standard enterprise in Quantum Optics the preparation of coherent-states
of light by means of a pumping coherent-source, \emph{i.e.}~\ac{LASER}
light. Harder experimental constraints, however, come about when we
make the requirement that the life-time of photons is large enough
as to enhance the momentum-transfer to the mirror. Discarding decoherence
effects on the mechanical oscillator, and if no photons leak out of
the cavity (perfect reflecting mirrors), the \ac{EM} excited mode
of the cavity will essentially follow an unitary evolution under the
Hamiltonian of Eq.~(\ref{eq:Chap2-Hamiltonian}). Real cavities,
though, have many modes beside the resonant ones, and hence ans unitary
description will not apply; they are characterized by a finite photon
life-time, $\tau$. A poor \emph{optical finesse }{[}Eq.~(\ref{eq:AppB1-OpticalFinesse}){]}
will blur the effectiveness of the radiation pressure-mechanism and
thus one must guarantee that a high-quality cavity is used in a real
experiment. 

\begin{figure}[tb]
\noindent \begin{centering}
\includegraphics[width=0.6\textwidth]{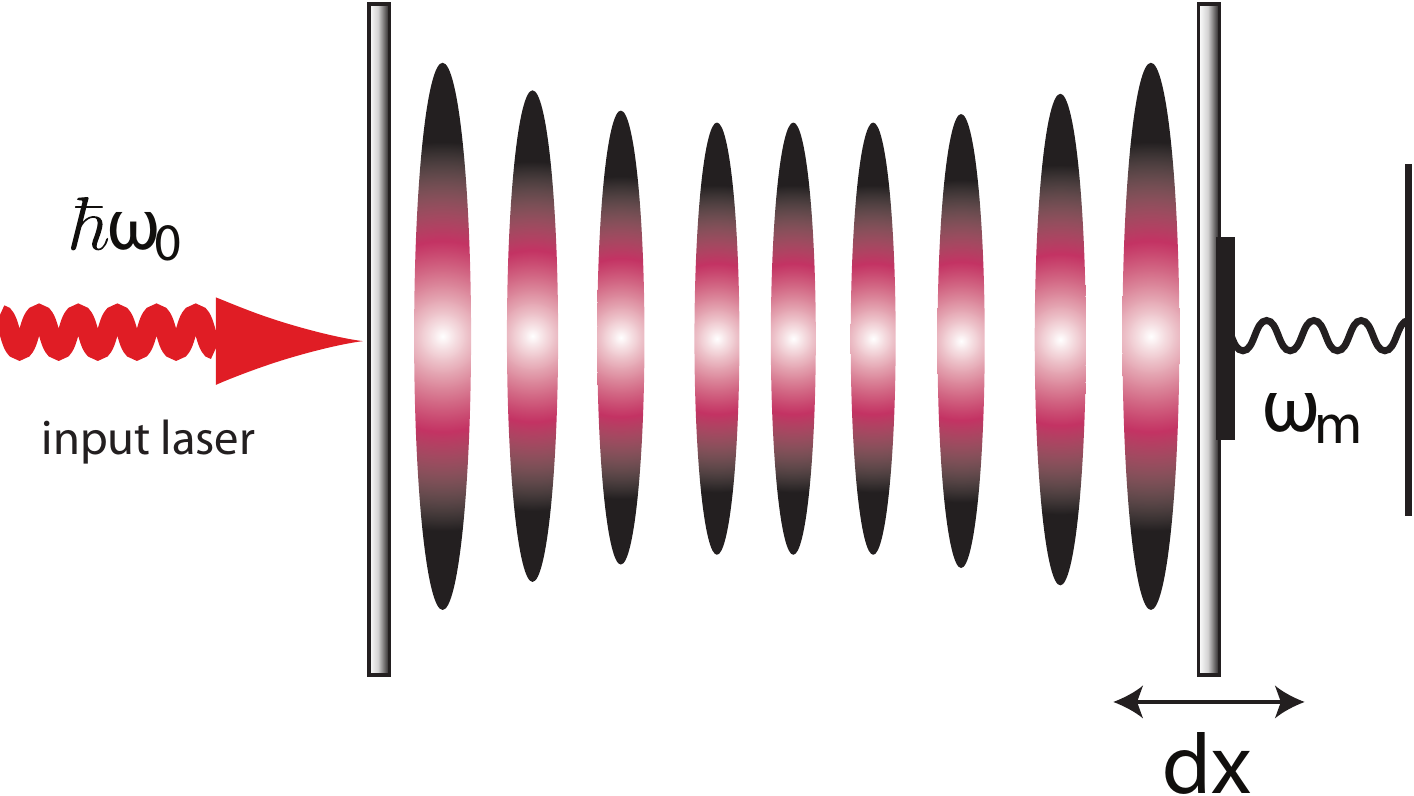}
\par\end{centering}

\caption[Schematic picture of the opto-mechanical interaction]{\label{fig:Chap3_Schematic_optomechanical}Schematic of the opto-mechanical
interaction of a cavity field and a moveable mirror. The input \ac{LASER}
field feds the cavity with high-energetic optical photons, with almost
null thermal occupancy at room temperature, lowering the effective
temperature of the mirror and opening the possibility to the detection
of genuine macroscopic entanglement.}

\end{figure}

In order to study how a finite finesse and a damped mirror will affect
the entanglement (and hence, genuine quantum effects) of the opto-mechanical
system, we must study the non-unitary evolution of the system due
to dissipation and decoherence, even at zero temperature. This is
easier to accomplish with the Langevin equations of motion, which
are often used in Quantum Optics when quantum and thermal noise must
be taken into account. A master equation approach for the full density-matrix,
in the same spirit as Leggett and Caldeira \cite{1981Caldeira,1983Caldeira},
would raise many difficulties (for instance, in finding the exact
low-temperature limit of the master equation of quantum friction \cite{2005Bassi,2005Adler,2006Bernard}).
Following tradition, we adopt the Heisenberg equations of motions
for the fields as the starting point to the study of the dynamics
(and later the entanglement). Hence, a study of quantum correlations
in the spirit of Chap.~\ref{cha:MacroscopicEntang} will not be possible.
This is, however, not a problem at all, since the \ac{CV} approach
of the present section will yield all the relevant information in
order to reconstruct the covariance matrix in the relevant regimes. 

We consider a single mechanical mode of the mirror only, which can
be modeled as an harmonic oscillator with natural frequency $\omega_{m}$.
The notation is essentially the same as of Chap.~\ref{cha:MacroscopicEntang},
but now with dimensionless position and momentum operators satisfying
$[q,p]=i$. This is obtained from the canonical conjugate operators
{[}see table~\ref{tab:Chap2_GeneralizedCoord}{]} through the transformation,\begin{eqnarray}
q & \rightarrow & \sqrt{\frac{m\omega_{m}}{\hbar}}q,\label{eq:Chap3_renormaliz1}\\
p & \rightarrow & \sqrt{\frac{1}{m\hbar\omega_{m}}}p.\label{eq:Chap3_renormaliz}\end{eqnarray}
The Heisenberg equations of motion for the operators will therefore
contain parameters with dimensions of frequency. This will simplify
the study of the covariance matrix as all its entries become dimensionless
and allow an easier investigation of entanglement through the parameter
region. We model the driving coherent-source as a classical field
with strength $E$ and frequency $\omega_{0}$, and hence the Hamiltonian
reads \cite{2001Giovannetti,2007Vitali}, \begin{equation}
H=\underset{\text{radiation-pressure}}{\underbrace{\hbar\omega_{c}a^{\dagger}a+\frac{\hbar\omega_{m}}{2}(p^{2}+q^{2})-\hbar ga^{\dagger}aq}}+\underset{\text{driving source}}{\underbrace{i\hbar E(e^{-i\omega_{0}t}a^{\dagger}-e^{i\omega_{0}t}a)}}.\label{eq:Chap3-FullHamiltonian}\end{equation}
The extra term describes the effect of adding photons by a driving
coherent source. The amplitude $E$ is related to the input \ac{LASER}
power $P$ by $|E|=\sqrt{2P\kappa/\hbar\omega_{0}}$, where $\kappa$
is the decay rate of the cavity. The radiation-pressure coupling $g$
{[}Eq.~(\ref{eq:Chap2-InteractionStrength}){]} gets renormalized
by a factor of $\sqrt{2}$ due to the transformation above and reads,\begin{equation}
g=\frac{\omega_{c}}{L}\sqrt{\frac{\hbar}{m\omega_{m}}.}\label{eq:Chap3-RadiationPressureCoupling}\end{equation}
Although several degrees of freedom, which have different resonant
frequencies, will be excited by the motion of the mirror, the single-mode
description of Eq.~(\ref{eq:Chap3-FullHamiltonian}) will capture
the physics as long as mode-mode coupling is negligible. This happens
in the so-called adiabatic regime (Sec.~\ref{sec:2.2.Towards-high-temp}),
where the frequency of the mirror is much smaller than the free-spectral
range of the cavity, \begin{equation}
\omega_{m}\ll\Delta\omega_{c}=\pi c/L.\label{eq:Chap3_AdiabaticCondition}\end{equation}
We are well inside this limit as typically $\Delta\omega_{c}\sim10^{12}-10^{15}\:\mathrm{Hz}$
for optical cavities, and $\omega_{m}\sim10^{5}-10^{7}\:\mathrm{Hz}$
for macroscopic mirrors --- please refer to Appendix~\ref{sec:App-Cavities}
for an outline of the physical parameters controlling the operation
of an optical resonator. From the detection side, a single frequency
mode can also be addressed via a bandpass filter in the detection
scheme \cite{1999Pinard}. This, together with the discussion of Sec.~\ref{sec:2.2.Towards-high-temp}
on the derivation of the radiation-pressure interaction, give the
grounds for the use of Hamiltonian (\ref{eq:Chap3-FullHamiltonian}).

The dynamics of the cavity field will be strongly influenced by the
motion of the mirror as any slightly change in the mirror's position
changes the cavity length and thus the mode spectrum (Fig.~\ref{fig:AppB_ModeSpectrum}).
The Heisenberg equation of motion for the operators assume a more
elegant form when written in the frame rotating at frequency $\omega_{0}$.
This corresponds to write $a$ and $a^{\dagger}$ in the interaction
picture with respect to $\hbar\omega_{0}a^{\dagger}a$:\begin{equation}
a_{I}(t):=e^{i\hbar\omega_{0}a^{\dagger}at}ae^{-i\hbar\omega_{0}a^{\dagger}at}=ae^{i\omega_{0}t}.\label{eq:Chap3-InteractionPicture}\end{equation}
The equation of motion for $a_{I}$ is $\dot{a}_{I}=\partial_{t}a_{I}+(i/\hbar)\left[H,a_{I}\right]$
and the Hamiltonian has no time dependence in the rotating frame (\emph{i.e.}~when
written in terms of $a_{I}$ and $a_{I}^{\dagger}$). The transformation
$a\rightarrow a_{I}$ corresponds to a local unitary action, thus
not changing the correlations properties whatsoever. In what follows,
we drop the subscript $I$ by bearing in mind that we are in the rotating
frame. The equations of motion read \begin{eqnarray}
\dot{q} & = & \omega_{m}p,\\
\dot{p} & = & -\omega_{m}q+ga^{\dagger}a,\\
\dot{a} & = & -i\Delta_{0}a+igaq+E,\end{eqnarray}
where $\Delta_{0}=\omega_{c}-\omega_{0}$ is the source-cavity detuning.
The coupled dynamics is entailed by the non-linear term proportional
to the radiation-pressure coupling, $g$ and its classical orbits
display a multitude of rich phenomena (\emph{e.g.}~static bistability
\cite{1983Dorsel} and dynamical multistability leading to self-induced
oscillations \cite{2006Marquardt}). 

The mirror and the cavity are not isolated from the rest of the world,
and thus the above equations are not the full story yet. The two main
sources of noise must be taken into account. They are,
\begin{enumerate}
\item the dissipative effects affecting the mirror, which in the absence
of radiation-pressure would essentially follow a Brownian quantum
motion (even at zero temperature);
\item the damping of the cavity dynamics due to photon leakage (no perfect
reflective mirrors do exist).
\end{enumerate}
The interaction of the cavity \ac{EM} field with its environment
is correctly described by the input-output theory due to Gardiner
and Collett \cite{1984Collett,1985Gardiner}. The Appendix~\ref{sec:App-The-input-output-theory}
contains the essential of this theory for the present study. Finally,
the derivation of the Langevin equations for the mechanical oscillator
is outlined in Appendix~\ref{sec:App-QuantumBrownianMotion}. 

We denote the mechanical and cavity damping rates by $\gamma_{m}$
and $\kappa$, respectively. Indeed, the Langevin equations for the
opto-mechanical system read,\begin{eqnarray}
\dot{q} & = & \omega_{m}p,\label{eq:Chap3_Langevin1}\\
\dot{p} & = & -\omega_{m}q-\gamma_{m}p+ga^{\dagger}a+\xi,\label{eq:Chap3_Langevin2}\\
\dot{a} & = & -(\kappa+i\Delta_{0})a+igaq+E+\sqrt{2\kappa}a_{in}.\label{eq:Chap3_Langevin3}\end{eqnarray}
Consistently with Eq.~(\ref{eq:AppB1-LangevinEquation}), we have
introduced the vacuum radiation input noise, $a_{in}$, whose only
nonzero correlation function is {[}Eq.~(\ref{eq:AppB1-Averages_a_in}){]}
\begin{equation}
\langle a_{in}(t)a_{in}^{\dagger}(t')\rangle=\left(\bar{n}(\omega_{c})+1\right)\delta(t-t'),\label{eq:Chap3_CorrelationAi}\end{equation}
and the Hermitian Brownian noise operator $\xi$ {[}with correlation
function given by Eq.~(\ref{eq:AppB3-Correlation}){]}. The quantum
Langevin equations {[}Eqs.~(\ref{eq:Chap3_Langevin1})-(\ref{eq:Chap3_Langevin3}){]}
very much resemble the long-established Langevin equation from classical
physics. Remarkably, however, the quantum Brownian motion is not a
Markovian process in general {[}Eq.~(\ref{eq:AppB3-LangevinCommutation}){]}.
Quantum effects, on the other hand, are only achievable by using oscillators
with a large mechanical quality factor $Q:=\omega_{m}/\gamma_{m}\gg1$.
In this limit, $\xi(t)$ becomes delta-correlated {[}see Eq.~(\ref{eq:AppB3-CorrelationNoiseHighTHighQ})
and comments therein{]}: \begin{equation}
\left\langle \xi(t)\xi(t')\right\rangle \simeq\gamma_{m}\left(2\bar{n}+1\right)\delta(t-t'),\label{eq:Chap3_correlation_noise_appr}\end{equation}
where $\bar{n}=\left(\exp\{\hbar\omega_{m}/k_{B}T\}-1\right)^{-1}$
is the mean thermal excitation number of the mirror, and one recovers
a Markovian process. 

The noise auto-correlation function (\ref{eq:Chap3_CorrelationAi})
can be significantly simplified by noting that optical photons are
very energetic ($\sim1\text{eV}$) and thus thermal occupation is
insignificant even at room temperature: $\bar{n}(\omega_{c})\simeq0$.
Although these simplifications constitute reasonable progress, the
Langevin equations {[}Eqs.~(\ref{eq:Chap3_Langevin2}) and (\ref{eq:Chap3_Langevin3}){]}
are a non-linear dynamical system and, hence, do not admit a simple
general solution. We are interested in the steady-state regime, however.
In this case an analytical solution can be bound following the tradition
in Quantum Optics \cite{1994Fabre,1994Mancini,1998Mancini} of considering
small fluctuations around the steady-state. 

Indeed, we rewrite each Heisenberg operator as a $c$-number steady
state value plus an additional fluctuation operator with zero mean
value, $a=\alpha_{s}+\delta a$, $q=q_{s}+\delta q$, $p=p_{s}+\delta p$.
By inserting these expressions into the Langevin equations {[}Eqs.~(\ref{eq:Chap3_Langevin1}),
(\ref{eq:Chap3_Langevin2}) and (\ref{eq:Chap3_Langevin3}){]}, these
latter decouple into a set of nonlinear algebraic equations for the
steady state values and a set of quantum Langevin equations for the
fluctuation operators~\cite{1994Fabre,1994Mancini}. The steady state
values are given by,\begin{eqnarray}
p_{s} & = & 0,\label{eq:Chap3-steady_state1}\\
q_{s} & = & \frac{g}{\omega_{m}}|\alpha_{s}|^{2},\label{eq:Chap3-steady_state2}\\
\alpha_{s} & = & \frac{E}{\kappa+i\left(\Delta_{0}-gq_{s}\right)}.\label{eq:Chap3-steady_state3}\end{eqnarray}
The latter equation is, in fact, a nonlinear equation determining
the stationary intra-cavity field amplitude, $\alpha_{s}$, since
the effective cavity detuning, $\Delta\equiv\Delta_{0}-gq_{s}$, including
radiation pressure effects, is given by $\Delta=\Delta_{0}-g^{2}|\alpha_{s}|^{2}/\omega_{m}$,
and thus depends on $\alpha_{s}$. Given our conjecture (Sec.~\ref{sub:Chap2-Sec-Finite-temperature-entanglement}),
we expect the parameter regime relevant for generating opto-mechanical
entanglement is that of a very large input power $P$, \emph{i.e.}~$|\alpha_{s}|\gg1$.
In this case, one can safely neglect the nonlinear terms $\delta a^{\dagger}\delta a$
and $\delta a\delta q$, and one gets the linearized Langevin equations,
\begin{eqnarray}
\delta\dot{q} & = & \omega_{m}\delta p,\label{eq:Chap3-LinearizedLangevin1}\\
\delta\dot{p} & = & -\omega_{m}\delta q-\gamma_{m}\delta p+G\delta X+\xi,\label{eq:Chap3-LinearizedLangevin2}\\
\delta\dot{X} & = & -\kappa\delta X+\Delta\delta Y+\sqrt{2\kappa}X_{in},\label{eq:Chap3-LinearizedLangevin3}\\
\delta\dot{Y} & = & -\kappa\delta Y-\Delta\delta X+G\delta q+\sqrt{2\kappa}Y_{in},\label{eq:Chap3-LinearizedLangevin4}\end{eqnarray}
where we have re-written the cavity operators as function of the cavity
field quadratures (see table~\ref{tab:Chap2_GeneralizedCoord}):
$\delta X\equiv(\delta a+\delta a^{\dagger})/\sqrt{2}$ and $\delta Y\equiv(\delta a-\delta a^{\dagger})/i\sqrt{2}$.
The corresponding Hermitian input noise operators read $X_{in}\equiv(a_{in}+a_{in}^{\dagger})/\sqrt{2}$
and $Y_{in}\equiv(a_{in}-a_{in}^{\dagger})/i\sqrt{2}$. The most important
aspect of the linearized Langevin equations is that the quantum fluctuations
of the field and the oscillator are now coupled by the much larger
\emph{effective} opto-mechanical coupling, \begin{equation}
G\equiv g\alpha_{s}\sqrt{2}.\label{eq:Chap3-EffectiveOptoMechanicalCoupling}\end{equation}
The latter can be very large by increasing the intra-cavity field
amplitude and a significant amount of entanglement will be possible
as we shall see in a moment. For the sake of simplicity, in what follows
we choose the phase reference of the cavity field so that $\alpha_{s}$
is real. Its amplitude reads, \begin{equation}
\alpha_{s}=\sqrt{\frac{P\kappa}{2\hbar\omega_{0}(\kappa^{2}+\Delta^{2})}}.\label{eq:Chap3_IntraFieldAmplitude}\end{equation}
When the system is stable it reaches a unique steady state, independently
of the initial condition. Since the quantum noises $\xi$ and $a_{in}$
are zero-mean quantum Gaussian noises and the dynamics is linearized,
the quantum steady state for the fluctuations is a zero-mean bipartite
Gaussian state, fully characterized by its $4\times4$ correlation
matrix {[}Eq.~(\ref{sec:Sec1.4-Continuous_variab}){]}, \begin{equation}
V_{ij}=\left(\langle u_{i}(\infty)u_{j}(\infty)+u_{j}(\infty)u_{i}(\infty)\rangle\right)/2,\label{eq:Chap3_CovarianceMatrix}\end{equation}
where $u^{T}(\infty)=(\delta q(\infty),\delta p(\infty),\delta X(\infty),\delta Y(\infty))$
is the vector of \ac{CV} fluctuation operators at the steady state
($t\to\infty$). Defining the vector of noises $n^{T}(t)=(0,\xi(t),\sqrt{2\kappa}X_{in}(t),\sqrt{2\kappa}Y_{in}(t))$
and the matrix, \begin{equation}
A=\left(\begin{array}{cccc}
0 & \omega_{m} & 0 & 0\\
-\omega_{m} & -\gamma_{m} & G & 0\\
0 & 0 & -\kappa & \Delta\\
G & 0 & -\Delta & -\kappa\end{array}\right).\label{dynmat}\end{equation}
This matrix determines the dynamical stability of the physical system,
and it also provides a measure of the correlations between its two
subsystems, the intra-cavity field and the mirror. Eqs.~(\ref{eq:Chap3-LinearizedLangevin1})-(\ref{eq:Chap3-LinearizedLangevin4})
can be now written in compact form as \begin{equation}
\dot{u}(t)=Au(t)+n(t),\label{eq:Chap3_LangevinEqsCompactForm}\end{equation}
whose formal solution is \begin{equation}
u(t)=M(t)u(0)+\int_{0}^{t}dsM(s)n(t-s),\label{eq:Chap3_LangevinSolution}\end{equation}
with $M(t)=\exp\{At\}$. The system is stable and reaches its steady
state when all the eigenvalues of $A$ have negative real parts so
that $M(\infty)=0$. The stability conditions can be derived by applying
the Routh-Hurwitz criterion \cite{1987DeJesus}, yielding the following
two nontrivial conditions on the system parameters \begin{eqnarray}
\mathcal{S}_{1} & = & 2\gamma_{m}\kappa\left[\Delta^{4}+\Delta^{2}(\gamma_{m}^{2}+2\gamma_{m}\kappa+2\kappa^{2}-2\omega_{m}^{2})\right.\\
 &  & \left.+\left(\gamma_{m}\kappa+\kappa^{2}+\omega_{m}^{2}\right)^{2}\right]+\omega_{m}G^{2}\Delta(\gamma_{m}+2\kappa)^{2}>0,\label{eq:Chap3_StableCond1}\\
\mathcal{S}_{2} & = & \omega_{m}^{2}\left(\Delta^{2}+\kappa^{2}\right)-\omega_{m}G^{2}\Delta>0.\label{eq:Chap3_StableCond2}\end{eqnarray}
A careful stability analysis of this system is found in \cite{2009Genes}.
From now on we will consider the above conditions to be satisfied.
When the system is stable Eq.~(\ref{eq:Chap3_CovarianceMatrix})
becomes, \begin{equation}
V_{ij}=\sum_{k,l}\int_{0}^{\infty}ds\int_{0}^{\infty}ds'M_{ik}(s)M_{jl}(s')\Phi_{kl}(s-s'),\label{eq:Chap3_cm2}\end{equation}
where $\Phi_{kl}(s-s')=\left(\langle n_{k}(s)n_{l}(s')+n_{l}(s')n_{k}(s)\rangle\right)/2$
is the matrix of the stationary noise correlation functions. Using
the fact that the components of $n(t)$ are uncorrelated, we get $\Phi_{kl}(s-s')=D_{kl}\delta(s-s')$,
where $D=\mathrm{Diag}[0,\gamma_{m}(2\bar{n}+1),\kappa,\kappa]$ is
a diagonal matrix, and Eq.~(\ref{eq:Chap3_cm2}) becomes $V=\int_{0}^{\infty}dsM(s)DM(s)^{T}$.
When the stability conditions are satisfied, $M(\infty)=0$, we get
the following equation for the steady-state covariance-matrix (see
Appendix~\ref{sec:CovarianceMatrix} for the derivation), \begin{equation}
AV+VA^{T}=-D.\label{eq:Chap3_Lyaponov}\end{equation}
The latter is a linear equation for $V$ containing the linearized
dynamics of the full system. The linearized dynamics of the system
can be studied in all the parameter region obeying inequalities (\ref{eq:Chap3_StableCond1})
and (\ref{eq:Chap3_StableCond2}). The analytic expression for $V$
can be straightforwardly derived either from Eq.~(\ref{eq:Chap3_Lyaponov})
or from the Fourier transform $\mathcal{F}$ of the Langevin equations,
\begin{equation}
-i\omega\left(\mathcal{F}u\right)[\omega]=A\left(\mathcal{F}u\right)[\omega]+n[\omega],\label{eq:Chap3-Fourier}\end{equation}
by expressing all the correlation functions in the frequency domain.
For instance, \begin{equation}
V_{11}=\langle\delta q(\infty)\delta q(\infty)\rangle=\frac{1}{2\pi}\int_{-\infty}^{\infty}d\omega\left|\mathcal{F}(\delta q)[\omega]\right|^{2}.\label{eq:Chap3-V11-FT}\end{equation}
The exact form of $V$ in the whole parameter region is complicated
and not very enlightening. Indeed, we just present the analytic solution
for the mirror's reduced covariance matrix (containing the mirror's
correlations) at zero-detuning ($\Delta=0$), where the system is
stable everywhere {[}see equations~(\ref{eq:Chap3_StableCond1})
and (\ref{eq:Chap3_StableCond2}){]}. We find $V_{12}=V_{21}=0$,
and \begin{eqnarray}
V_{11} & = & \frac{1}{2}+\bar{n}+\frac{G^{2}\left(\kappa+\gamma_{m}\right)}{2\gamma_{m}\left(\kappa^{2}+\kappa\gamma_{m}+\omega_{m}\right)},\label{eq:Chap3_SolutionV11Delta0}\\
V_{22} & = & \frac{1}{2}+\bar{n}+\frac{G^{2}\kappa}{2\gamma_{m}\left(\kappa^{2}+\kappa\gamma_{m}+\omega_{m}\right)}.\label{eq:Chap3_SolutionV22Delta0}\end{eqnarray}
Recalling that the effective temperature of the mirror (and thus the
effective phononic occupation number, $n_{\text{eff}}$) is determined
by the average of the fluctuations undergoing Brownian motion, \emph{i.e.}
\begin{equation}
\hbar\omega_{m}\frac{\langle\delta p(\infty)^{2}\rangle+\langle\delta q(\infty)^{2}\rangle}{2}\equiv\hbar\omega_{m}\left(n_{\text{eff}}+\frac{1}{2}\right),\label{eq:Chap3_EffectiveOccupationNumber}\end{equation}
we conclude that, in the zero-detuning case, the mirror is in a squeezed
thermal state with an effective thermal occupation number given by\begin{equation}
n_{\text{eff}}=\bar{n}+\frac{G^{2}\left(2\kappa+\gamma_{m}\right)}{4\gamma_{m}\left(\kappa^{2}+\kappa\gamma_{m}+\omega_{m}^{2}\right)}.\label{eq:Chap3-EffectiveTemperature}\end{equation}
This shows the heating of the mirror above its environmental temperature
($n_{\text{eff}}>\bar{n}$) for $\Delta=0$. Whether this results
in detectable entanglement or not, will depend on the particular statistics
of the cavity field%
\footnote{Recall that in the unitary evolution scenario the heating of the mirror
was a signature of quantum correlations only when the partial sate
of the cavity field was sufficiently mixed, see Fig. \ref{fig:Chap2_MutualInfo}:
after some time has passed the partial entropies, $S_{cav}$ and $S_{m}$,
are sufficiently high producing a violation of the classical bound
{[}Eq. (\ref{eq:App1-Upper_bound_classical}){]}, $S_{cav}+S_{m}>2S$,
where $S$ is the total entropy of the system. In this case, also,
the effective temperature of the mirror increases, in agreement with
the increase of the linear entropy. In fact the mirror's effective
phononic occupancy can be derived quite easily from the evolution
operator {[}Eq.~(\ref{eq:Chap2-EvolutionOperator}){]}:\begin{eqnarray*}
\langle(b^{\dagger}b)(t)\rangle & = & \langle U^{\dagger}(t)b^{\dagger}bU(t)\rangle\\
 & = & \langle U^{\dagger}(t)b^{\dagger}U(t)U^{\dagger}(t)bU(t)\rangle\\
 & = & \langle\left(b^{\dagger}+k\eta(-t)a^{\dagger}a\right)\left(b+k\eta(t)a^{\dagger}a\right)\rangle\\
 & = & \bar{n}+k^{2}|\eta(t)|^{2}|\alpha|^{2}\left(1+|\alpha|^{2}\right),\end{eqnarray*}
from which we confirm the heating of the mirror for $t\neq2n\pi/\omega$
(with $n\in\mathbb{N}_{0}$). %
}. 

In the following section, we will see that, in fact, stationary entanglement
is strictly null for $\Delta=0$ and that the situation is radically
different for non-zero detuning $\Delta>0$ as the mirror is effectively
cooled down to its ground state.

\section{Approaching stationary entanglement\label{sec:Approaching-stationary-entanglement}}

\begin{figure}[tb]
\noindent \begin{centering}
\includegraphics[width=1\textwidth]{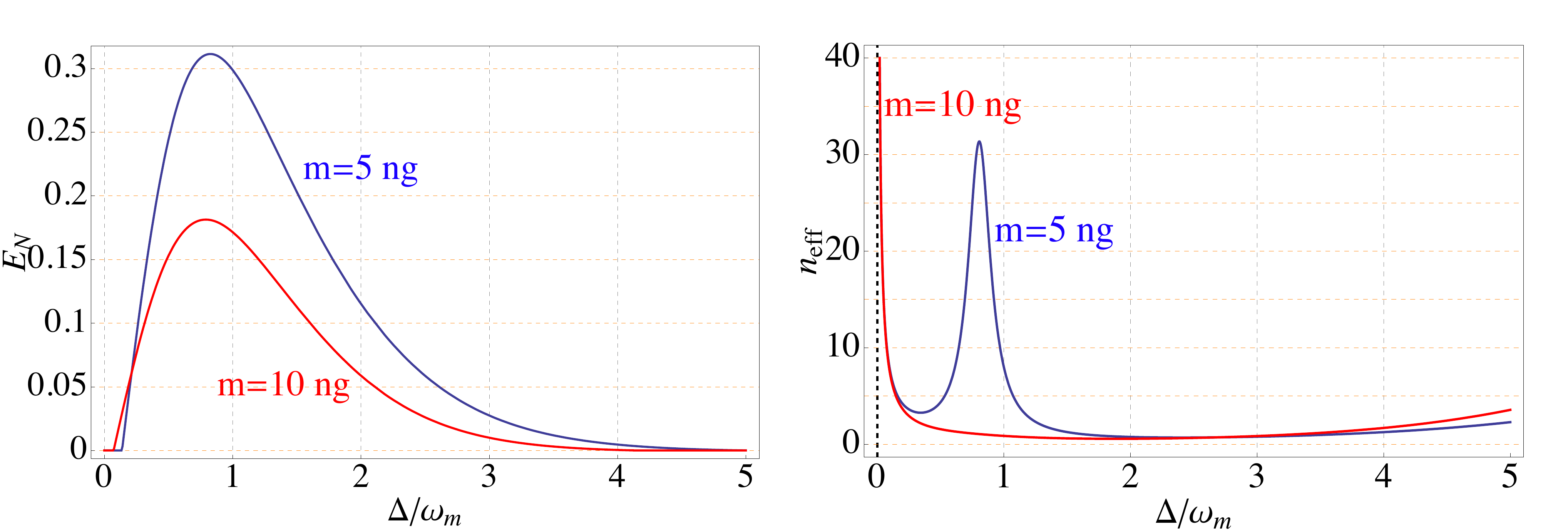}
\par\end{centering}

\caption[Stationary high-temperature entanglement for realistic parameters]{\label{fig:Plot1a-1b}Left - The entanglement {[}Eq.~(\ref{logneg}){]}
is plotted as function of the rescaled detuning, $\Delta/\omega_{m}$,
for a cavity with $L=1\:\mathrm{mm}$ and optical finesse {[}Eq.~(\ref{eq:AppB1-OpticalFinesse}){]}
$\mathcal{F}=1.07*10^{4}$ (corresponding to a damping rate of $\simeq88\:\mathrm{MHz}$)
and $T=0.4\:\mathrm{K}$ (corresponding to a thermal occupancy of
$\bar{n}\simeq832$). The cavity is driven by a \ac{LASER} with wavelength
$\lambda=810\:\mathrm{nm}$ and power $P=50\:\mathrm{mW}$. The mirror
has quality factor {[}Eq.~(\ref{eq:AppB3-Qfactor}){]} $Q=10^{5}$
and natural frequency $\omega_{m}/2\pi=10\:\mathrm{MHz}$. Right -
The effective phononic occupation number, $n_{\text{eff}}$, is plotted
for the same physical parameters. We clearly see a giant suppression
of the thermal fluctuations of the mirror as soon as $\Delta>0$;
an effective self-induced cooling has lowered the phononic occupation
of the mechanical oscillator and enhanced the quantum correlations
of the compound system.}

\end{figure}
In order to establish the conditions under which the optical mode
and the mirror vibrational mode are entangled, we consider the logarithmic
negativity, $E_{\mathcal{N}}$, a quantity that correctly quantifies
entanglement for Gaussian bipartite states. The nature of the correlations
described by $V$ depends only on the smallest sympletic eigenvalue
of the partial transposed covariance matrix, $\tilde{d}_{-}\equiv d_{-}(\Lambda_{A}V\Lambda_{A})$
{[}Eq.~(\ref{eq:Chap1_SympleticEig2modes}){]}, \begin{equation}
\tilde{d}_{-}\equiv2^{-1/2}\left[\tilde{\Sigma}(V)-\left[\tilde{\Sigma}(V)^{2}-4\det V\right]^{1/2}\right]^{1/2}\label{eq:Chap3-Sympletic}\end{equation}
with $\tilde{\Sigma}(V)\equiv\det A+\det B-2\det C$, and we have
used the $2\times2$ block form of the covariance matrix, \begin{equation}
V\equiv\left(\begin{array}{cc}
A & C\\
C^{T} & B\end{array}\right).\label{blocks}\end{equation}
\begin{figure}[H]
\noindent \begin{centering}
\includegraphics[width=0.5\textwidth]{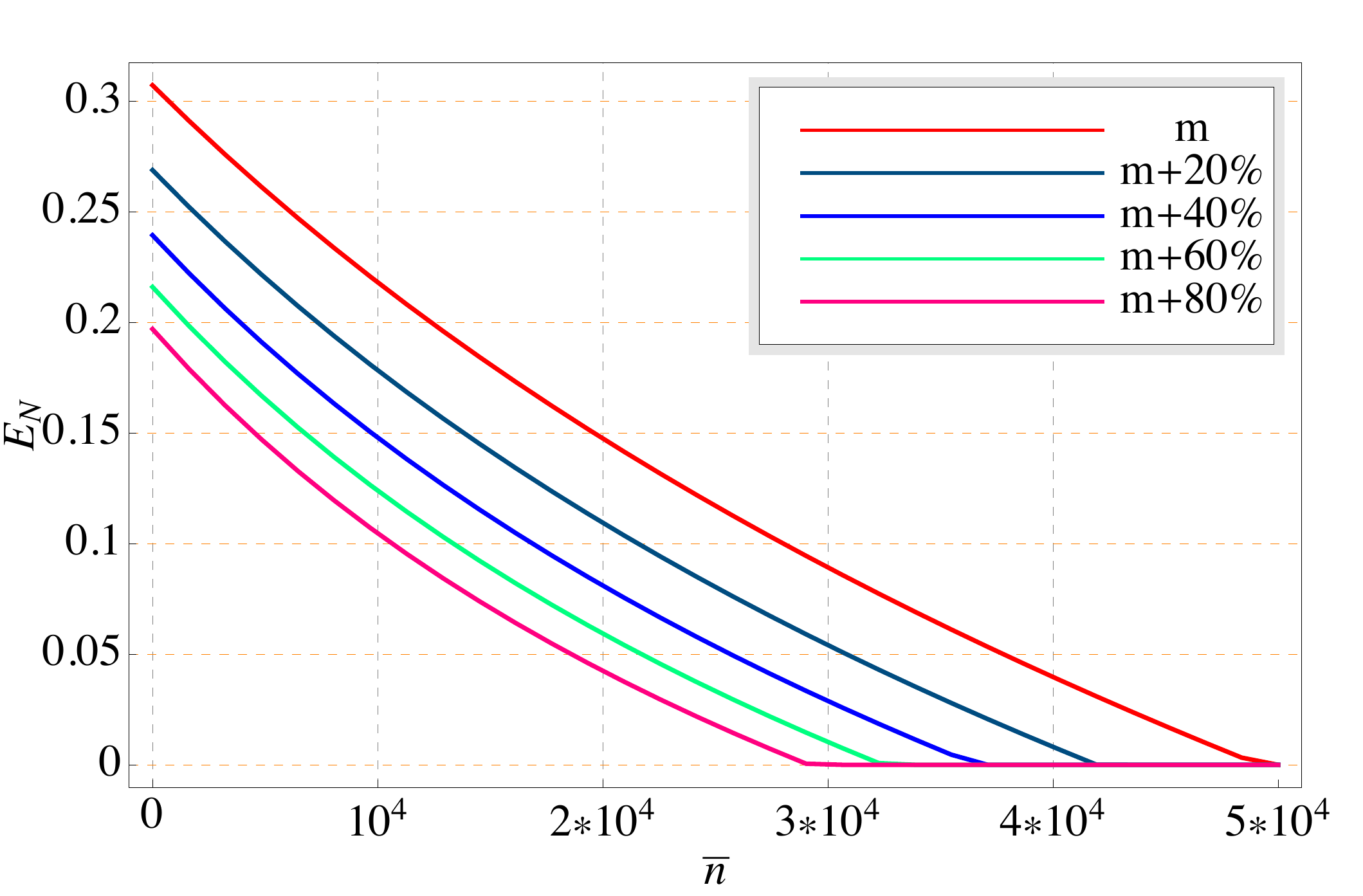}\includegraphics[width=0.5\textwidth]{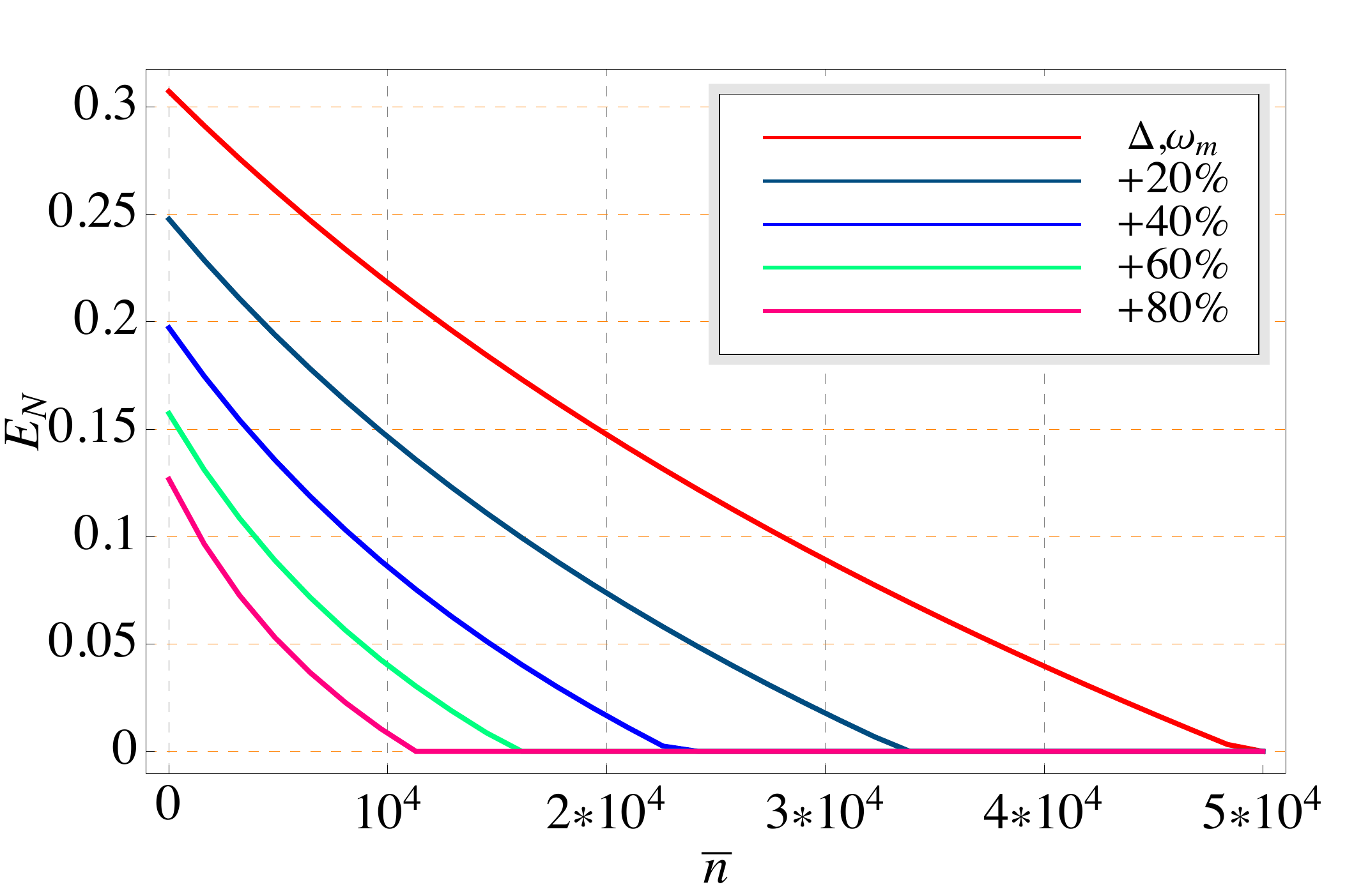}
\par\end{centering}

\noindent \begin{centering}
\includegraphics[width=0.5\textwidth]{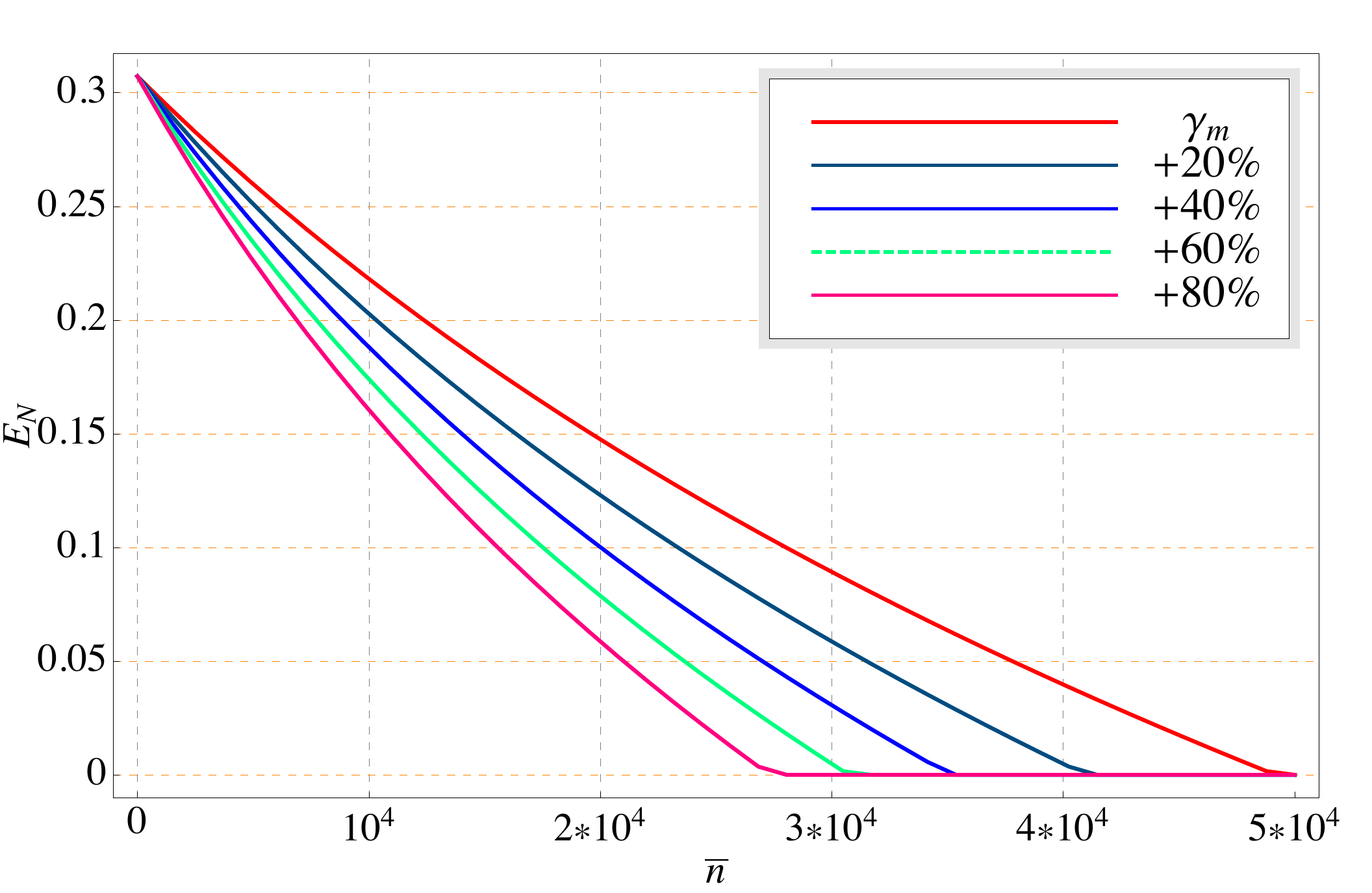}\includegraphics[width=0.5\textwidth]{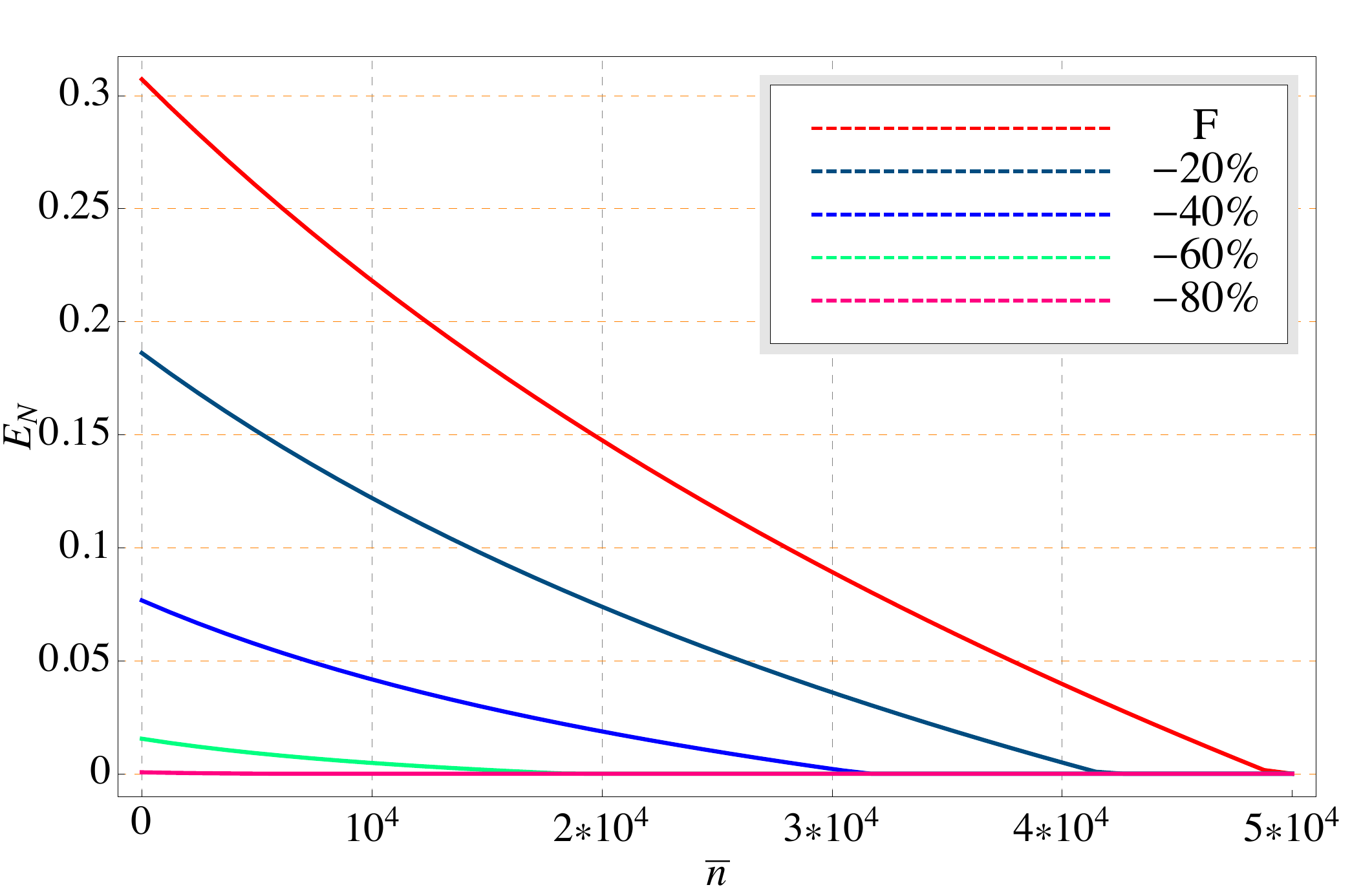}
\par\end{centering}

\caption[The entanglement dependence on the physical parameters.]{\label{fig:Chap3-Plot2}These figures show the entanglement dependence
on the physical parameters. The values of Fig.~\ref{fig:Plot1a-1b}
($m=5\:\mathrm{ng}$, $\omega_{m}=2\pi*10\:\mathrm{MHz}$, $\mathcal{F}=1.07*10^{4}$
and $Q=10^{5}$) are taken as reference. The different curves show
the change in $E_{\mathcal{N}}$ by increasing the reference values
(decreasing in the case of the finesse) by $20\%$, $40\%$, $60\%$
and $80\%$. The top-right plot shows $E_{\mathcal{N}}$ when both
$\Delta$ and $\omega_{m}$ are increased while keeping the ratio
$\Delta/\omega_{m}$ constant. The driving \ac{LASER} has $P=50\:\mathrm{mW}$
and $\lambda=810\:\mathrm{nm}$; the cavity size is $1\:\mathrm{mm}$. }

\end{figure}
A Gaussian state is entangled if and only if $\tilde{d}_{-}<1/2$,
which is equivalent to Simon's necessary and sufficient transpose
criterion for Gaussian states (Sec.~\ref{sec:Sec1.4-Continuous_variab}).
The entanglement of a Gaussian bipartite state is properly quantified
via the logarithmic negativity {[}Eq.~(\ref{eq:Chap1_LogNegGaussianStates}){]},\begin{equation}
E_{\mathcal{N}}=\max[0,-\ln2\tilde{d}_{-}].\label{logneg}\end{equation}
As mentioned in the previous section, we can show that $\tilde{d}_{-}\ge1/2$
for $\Delta=0$ , at any temperature, and thus no genuine quantum
correlations are shared by the photons and the phononic mode of the
mirror. On the contrary, by a proper choice of detuning, we can reach
a steady state with entanglement: Fig.~\ref{fig:Plot1a-1b} shows
that a detectable amount of steady-state entanglement is expected
for $\Delta\sim\omega_{m}$ at environmental temperatures much above
the mirror's ground state for state-of-the-art experimental parameters.
For the parameters values of Fig.~\ref{fig:Plot1a-1b} the effective
coupling, $G$, varies between $10^{7}$ and $10^{8}$ (in units of
frequency). This large effective coupling leads to the establishment
of \emph{bona fide} macroscopic entanglement at relatively high-temperature
($T=.4\:\mathrm{K}\Leftrightarrow\bar{n}\simeq832)$. Curiously, the
effective temperature of the mirror suffers an abrupt change as soon
as $\Delta>0$ (see the right-hand side plot in Fig.~\ref{fig:Plot1a-1b}).
We find a mirror's effective thermal occupancy as low as $n_{\text{eff}}\simeq0.75$
for a mirror of mass $m=5\:\mathrm{ng}$ and $\Delta=2\omega_{m}$,
which is in deep contrast to the situation found for $\Delta=0$ {[}Eq.~(\ref{eq:Chap3-EffectiveTemperature}){]}.
A careful study in the stable parameter region shows that entering
in the low coupling regime ($G\lesssim\omega_{m}$), completely destroys
the quantum correlations even at low temperatures (see Fig.~\ref{fig:Chap3-Plot2}). 

It is apparent from these results that the usual criterion on temperature
for quantum behavior {[}Eq.~(\ref{eq:App1-SQL-TempCriterion1}){]}
is misleading in the context of complex open quantum systems. Braginsky
found a similar situation for an oscillator measured for a short time
compared to its own relaxation time {[}see Appendix~\ref{sec:App-Standard_Quantum_Limit},
and Eq.~(\ref{eq:App1-SQL-TempCriterion2}){]}. Indeed, the rigid
condition $k_{B}T\lesssim\hbar\omega_{m}/2$ can be very much relaxed
when we look carefully to all the physical scales playing role in
a given setup. Ultimately, with optimal experimental conditions, our
results predict stationary entanglement up to $\bar{n}\simeq10^{4}\Leftrightarrow T\simeq20\:\mathrm{K}$
(for a mirror with $m=5\,\mathrm{ng}$ and $\omega_{m}/2\pi=10^{7}\,\mathrm{Hz}$).

In our case, the mechanism by which we beat the naive $k_{B}T\lesssim\hbar\omega_{m}/2$
criterion has no direct correspondence with the presence of a measuring
apparatus in the Braginsky calculation. But still we can relate the
quantum behaviour at high-temperature we found to the low entropy
continuously flowing into the cavity, which enhances the effective
coupling, thus making particularly efficient the radiation-pressure
mechanism {[}for $P\gg1$ we can have $\alpha_{s}\gg1$, and hence
$G\gg g$, see Eqs.~(\ref{eq:Chap3-EffectiveOptoMechanicalCoupling})~and~(\ref{eq:Chap3_IntraFieldAmplitude}){]}.
The functional dependence of entanglement with all parameters is rather
complicated but Fig.~\ref{fig:Chap3-Plot2} shows the general trends:
increasing the mass suppresses entanglement, as decreasing the cavity
optical finesse, for instance; the plots also indicate that steady-state
entanglement is much more sensitive to the mirror's frequency, the
detuning or the optical finesse rather than the mirror's mass or its
quality factor.

\section{Concluding remarks}

Opto-mechanical coupling via the radiation pressure mechanism, as
Braginsky conceived \cite{Review-Braginsky-1996,Book-Braginsky-1992},
is a promising approach to prepare and manipulate quantum states of
mesoscopic and macroscopic mechanical oscillators. We proposed an
experimentally achievable setup to create opto-mechanical entanglement
between a light field and a mechanical oscillator (see \cite{2007Vitali}
for the details of covariance matrix detection). This is accomplished
by using a bright \ac{LASER} field that resonates inside a cavity
and couples to the position and momentum of the moving (micro)mirror. 

Our proposal is based on feasible experimental parameters in accordance
with current state-of-the-art optics and microfabrication \cite{2006Gigan},
although its practical implementation is not yet within reach (the
main obstacle probably being combining in the same experiment all
the requirements leading to entanglement, see for instance \cite{2009Groblacher}).
The noticeable feature of this proposal is the fact that, in contrast
to previous proposals \cite{2003Zhangb,2005Pinard}, it neither requires
non-classical states of light, nor temperatures close to the oscillator's
ground state. 

Our calculation was based on two assumptions, namely,
\begin{enumerate}
\item the single-mode description justified by the adiabatic limit, $\omega_{m}\ll\Delta\omega_{c},$
and
\item the linearization of the Langevin equations, which is accurate for
large intra-field amplitudes, $\alpha_{s}\gg1$.
\end{enumerate}
The fluctuations of the cavity intra-field around its classical value
can be very small (for $\alpha_{s}\gg1$), thus making the linearization
procedure very accurate and commonly used when dealing with optical
cavities. These simplifications allowed to take into full account
the quantum Brownian motion of the mirror and the main source of the
cavity field decoherence (\emph{i.e.}~leakage through the mirrors).
In the linearized regime,  the bipartite state (mirror$+$cavity)
is well-described by a Gaussian state. Thus, in resemblance to the
cases $2\otimes2$ and $2\otimes3$, where \ac{PPT} is a sufficient
and necessary condition for separability, we were able to completely
characterize entanglement, albeit the Hilbert space being infinite
dimensional. This method is to be compared with perturbative approaches
based on the master equation for the reduced state of the mirror \cite{2007Wilson-Rae}
which are only valid in the weak-coupling regime $G\ll\omega_{m}$
of little interest for entanglement generation according to our calculation. 

Also, we have settled on quite solid grounds the main conclusions
of Chapter~\ref{cha:MacroscopicEntang}: the radiation-pressure mechanism
is extremely robust as it accommodates high-temperature macroscopic
entanglement --- a phenomenon which we believe is very rare (recall
the discussions of Chapters~\ref{cha:Introduction} and \ref{cha:MacroscopicEntang}
regarding the damage caused by thermal noise on quantum correlations).
Some interesting questions are open, \emph{e.g.}~how does the non-linear
regime affect the stability region of our setup and the amount of
stationary entanglement? Other questions were already answered at
the time of writing this thesis, for instance, the entanglement between
the \emph{output} field (instead of the intra-cavity field) and the
moveable mirror was computed recently in Ref.~\cite{Genes2008}.

\chapter{Entanglement mediated by the ground-state of gapped spin chains \label{cha:LDE_Via_GS_GappedSpinChains} }

\emph{This chapter is based on the following publication by the author:}
\begin{itemize}
\begin{onehalfspace}
\item \emph{Analytic results on long-distance entanglement mediated by gapped
spin chains}, AIRES FERREIRA, and J. M. B. Lopes dos Santos, Phys.
Rev. A\textbf{ 77}, 034301 (2008). \end{onehalfspace}

\end{itemize}

\section{Overview\label{sec:Chap4-Overview}}

When macroscopic degrees of freedom are addressed with high precision
and the decoherence effect of the environment is sufficiently suppressed,
the \ac{EM} field and a macroscopic mirror can get entangled at high-temperatures
--- this was the essential conclusion of the previous chapters, where
by means of a \ac{QI} approach the quantum-classical boundary of
opto-mechanical systems was studied.

The present and the following chapters are devoted to solid-state
systems, where a multitude of many-body quantum behavior is well known
for several decades (\emph{e.g.}~the superconductivity or the fractional
Hall effect) and new phenomena is discovered regularly (\emph{e.g.}~the
exotic physics of graphene). The study of entanglement in many-body
physics is a topic in its infancy but already lead to at least two
significant contributions, namely:
\begin{itemize}
\item strict bounds to the scaling of von-Neumann entropy; area laws have
been shown to emerge for the latter quantity in the ground state of
quantum lattice systems with short-range interactions (\emph{e.g.}~one-dimensional
spin chains, bosonic harmonic lattices, disordered systems, \emph{etc}.).
This says that quantum correlations (entanglement) between a region
$\boldsymbol{R}$ and the rest of the lattice $\mathcal{L}\setminus\boldsymbol{R}$
are encoded in the boundary $\partial\boldsymbol{R}$. This remarkable
result resembles the black hole entropy and it is at odds with the
volume law usually satisfied by regions sharing classical correlations
(such those arising in thermal states) --- for a complete review see
\cite{2009Eisert} and references therein;
\item a deeper understanding of numerical methods, in particular their ability
in simulating efficiently complex many-body systems. For instance,
it was understood that if little entanglement is present in the ground
state then a matrix-product state yields a good approximation to this
ground state \cite{2005Verstraete}, a area law would be observed
and algorithms such as the \ac{DMRG} would perform well. By looking
to the problem of simulating many-body physics with a \ac{QI} perspective,
the tensor product structure of the Hilbert space of a quantum lattice
system could be better manipulated, and powerful new numerical methods,
like the multiscale entanglement renormalization ansatz, were proposed
capable of reliably simulating \ac{1D} and \ac{2D} systems \cite{2007Vidal}.
\end{itemize}
\begin{figure}[tb]
\noindent \begin{centering}
\includegraphics[width=0.5\textwidth]{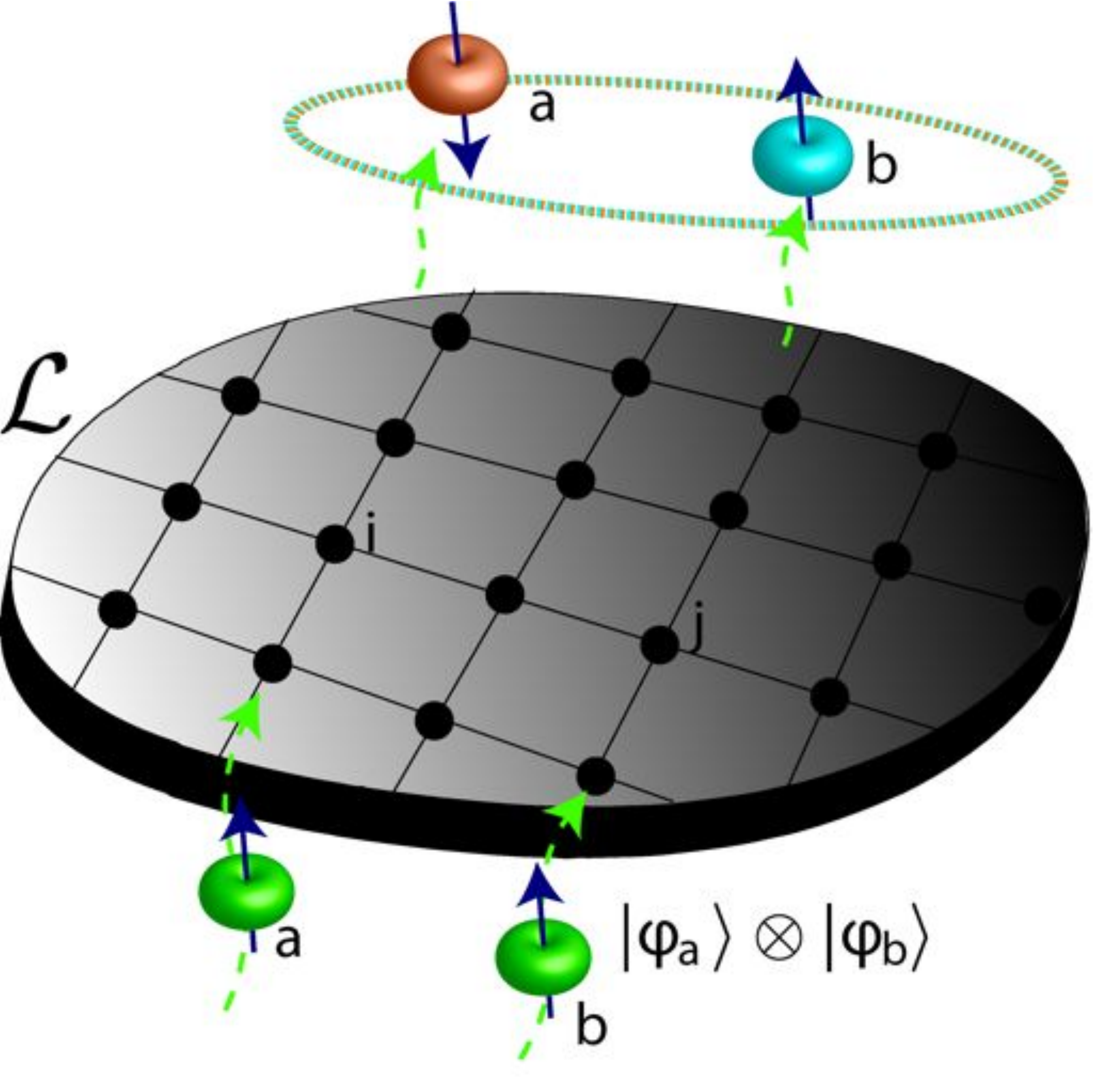}
\par\end{centering}

\caption[Schematic of two spin probes getting entangled by a many-body lattice]{\label{fig:SchematicLDE}Schematic of two probe particles, $a$ and
$b$, initially in a product state $|\varphi_{a}\rangle\otimes|\varphi_{b}\rangle$,
getting entangled after interacting with a quantum many-body lattice,
$\mathcal{L}$, through sites, $i$ and $j$, respectively. When the
probes are separated by distances of the order of the lattice's size,
we say the lattice has mediated long-distance entanglement between
the probes. }

\end{figure}
Other topics in this field are related to the characterization of
the entanglement properties of many-body systems at zero temperature,
particularly near quantum phase transitions and also at fi{}nite temperatures
(see \cite{Review-Amico-2008} for a review). Although up to our knowledge
this approach has not unveiled new properties regarding the phases
of matter, it already led to a more complete understanding of physics
known for a long time, such that of quantum phase transitions. 

The possibility of using quantum many-body systems as quantum channels
(systems capable of transporting quantum information), or even as
\textquotedbl{}all-in-one\textquotedbl{} devices for quantum computation
is also a promising subject where entanglement plays an important
role. Encouraging works showed that the collective dynamics of the
low excitations of the ferromagnetic \ac{1D} Heisenberg spin-$1/2$
model is able to transfer quantum states of a qubit with high fidelity
\cite{2003Bose}, and that finite spin-$1/2$ chains with \textquotedbl{}always
on\textquotedbl{} interactions encompass all the features of a processing
core model for quantum computation \cite{2006Yung}.

\begin{figure}[tb]
\noindent \begin{centering}
\includegraphics[width=0.5\columnwidth]{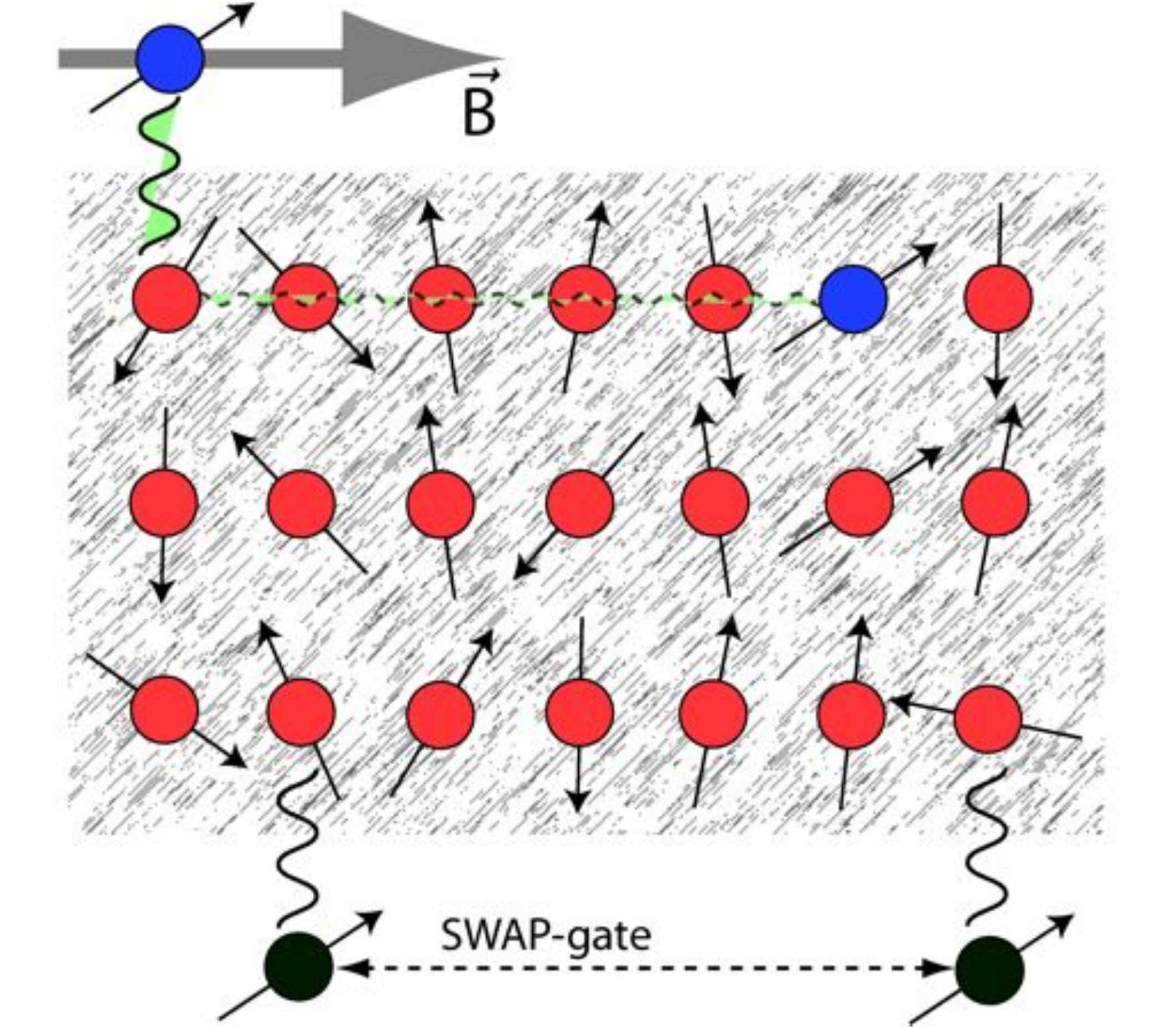}
\par\end{centering}

\caption[Schematic of a hypothetical solid-state quantum computer]{\label{fig:Chap4-QuantumComputer}Schematic of a hypothetical solid-state
quantum computer. A highly correlated spin system (inside box) serves
as a quantum bus. A magnetic field $\vec{B}$ prepares a qubit (blue
on top) in a given quantum state. This qubit is dynamically transferred
to a particle within the quantum bus (the blue spin). In the bottom
array, long-range order among two spins is used to prepare a SWAP
gate: the fundamental gate of quantum information (see \cite{Book-Nielsen-2000}
for a review). The latter simply swaps the $z$-th component of the
spin state of a bipartite system (blue spins on bottom), $U_{SWAP}|\chi_{a},\chi_{b}\rangle=|\chi_{b},\chi_{a}\rangle$.
This gate together with single qubit operations (rotations) encompasses
universal quantum computing. }

\end{figure}

The reason why spin-$1/2$ chains have been explored in \ac{QI} is
because their particles naturally embody the $SU(2)$ algebra of a
qubit, therefore allowing quantum information processing and manipulation
along the traditional lines of quantum computing, \emph{i.e.}~via
the establishment of quantum gates \cite{Book-Nielsen-2000} (see
Figure \ref{fig:Chap4-QuantumComputer}). Recent \ac{DMRG} results
by Campos Venuti and collaborators \cite{2006Venuti,2007Venuti} showed
that spin systems can also mediate entanglement between two spin probes
separated by large distances. This possibility had already been suggested
earlier, in the proposal for entanglement extraction from solids by
De Chiara \cite{2006Chiara}, \emph{i.e}.~that entanglement in a
many-body system could be ''swapped'' to neutrons interacting with
the bulk during a flight. 

The possibility of extracting entanglement from a large system might
seem a bit awkward, as generally the coupling to a system with many
degrees of freedom usually destroys entanglement very quickly due
to suppression of off-diagonal elements of the density matrix \cite{1982Zurek,Book-Zurek-1996,Review-Zureck-2003}.
However, a few notable exceptions do exist; if two qubits, not interacting
directly, are coupled in a symmetric way to a bath of harmonic oscillators,
their entanglement will partially survive (or even be created if initially
their state was separable) during their evolution when these qubits
have degenerate energy eigenstates \cite{2002Braun,2003Benatti},
or when the bath has a gap in its spectrum \cite{2006Oh}. A considerable
quenching of decoherence is also found in bosonic systems, such as
two harmonic oscillators interacting with a common bath \cite{2008Paz}.
A similar phenomenon, where the effect of decoherence is largely avoided,
is found, for instance, in quantum computing using the so-called decoherence-free
subspaces \cite{1998Lidar,2000Beige}.%
The present chapter is devoted to the physics of \ac{LDE} in the
ground-state of gapped quantum lattice systems. In contrast with the
references cited in the previous paragraph, our focus will be on situations
where all particles involved are spins. We will demonstrate the emergence
of quasi-perfect entanglement among the spin probes by tuning their
coupling to the \textquotedbl{}entangler\textquotedbl{} bulk to very
small values. 

Since we are mainly interested in the quantum-classical boundary,
the concern of our investigation will be on the conditions which make
favorable the emergence of \ac{LDE}, as well as its quantification
and its robustness against temperature. For qubits interacting with
bosonic baths under Markovian and non-Markovian dynamics, the problem
is already well understood (see \cite{2008Paz} and references therein).
For the case of spins systems, the many-body physics is quite complicated
and we cannot make exact computations; nevertheless, in the present
chapter we will derive some analytical results in perturbation theory
strictly valid at $T=0$. The generalization to stronger probe-bulk
coupling and finite temperature will be made in the following chapter
with the help of large numerical simulations. 

In what follows, we review the basic ideas behind using many-body
systems to accomplish \ac{QI} tasks and outline the main difficulties
of computing \ac{LDE} in these systems. Afterwards, we derive an
adequate perturbation theory for the \ac{LDE} problem. Our main result
will be to show that two (or more) uncorrelated qubit probes, separated
by large distances, form highly entangled states, when interacting
locally with gapped \ac{1D} antiferromagnetic systems; a very appealing
situation for \ac{QI}.

\section{Many-body quantum channels (main ideas and difficulties)\label{sec:Many-body-systems-as}}

Feasible mechanisms of entanglement extraction from real solid state
and their ability to transfer entanglement between distant parties
are of crucial importance for the implementation of \ac{QI} protocols,
such as teleportation, information transfer, quantum secure protocols,
or superdense coding \cite{Book-Nielsen-2000}. Regarding information
transfer, for instance, there are two ways of implementing it in many-body
systems: a) by unitary dynamical evolution, and b) via bulk ground
states (or another equilibrium state), either with or without the
need of special measurements (in one or more particles). In the former
scheme, one particle is prepared in a superposition of states, say,
\begin{equation}
|\psi_{i}\rangle=\frac{1}{\sqrt{2}}\left(a|0\rangle+b|1\rangle\right),\label{eq:Chap4-QubitSuperposition}\end{equation}
(with $a,b\in\mathbb{C}$ and $\left\{ |0\rangle,|1\rangle\right\} $
denoting an orthogonal basis). Afterwards, the compound state (particle
$+$ system) is let to evolve unitarily; in suitable systems, after
some time has passed, the particular superposition of Eq. (\ref{eq:Chap4-QubitSuperposition})
will be transferred with high fidelity to another particle in the
system. That is, the overlap between the partial state of a distant
particle $\rho_{j}$ (at time $t^{*}$) and the fiducial particle
$i$ (in the state $|\psi_{i}\rangle$ at time $t=0$) will be nearly
maximal,\begin{equation}
\langle\psi_{i}|\rho_{j}(t^{*})|\psi_{i}\rangle\simeq1.\label{eq:Chap4-Fidelity}\end{equation}
The time elapsed $\Delta t=t^{*}-t$ depends on the velocity of the
excitations; in some spin systems this could be the spin-wave velocity.
The study of capabilities and limitations of information transfer
via dynamical evolution is a rapidly developing field; the interested
reader is referred to the introduction to quantum communication via
spin chains by Bose \cite{2008Bose}; a study of entanglement and
state transfer via dynamical evolution of harmonic spin chains and
the $XY$ spin chain can be found in references \cite{2006Hartmann,2005Plenio-new}.
Regarding possibility b), here we just mention that bipartite entanglement
works as a figure of merit for the capabilities of a physical system
towards quantum information processing. Thus, all that is found on
mediation of entanglement in spin lattices (such as  \ac{LDE}) will
tell us about performance in quantum processing via a \textquotedbl{}quantum
bus\textquotedbl{} in equilibrium. For these reasons, we focus on
the fundamental issue of quantification of \ac{LDE}, rather presenting
particular consequences for \ac{QI} protocols (specific implementations
may be found in many texts, see for instance\emph{ }\cite{2007Venuti}).

Finally, systems of spins have also been suggested to integrate \ac{QI}
tasks and accomplish quantum computation in a single processing core,
although many questions regarding their robustness against temperature
and decoherence remain open. Among the DiVicenzo's requirements to
achieve quantum computation, the ability to generate rapid elementary
gates between well-characterized qubits is central \cite{2000DiVincenzo}.
Due to significant technical difficulties in switching on direct interactions
between qubits, various proposals have been put forward where a quantum
sub-system usually denominated as \emph{bus} is used to mediate the
fundamental universal gates. The physical embodiment of such bus could
be, for instance, the phononic mode of cold ions in the famous ion-trap
quantum computer \cite{1995Cirac}, or the magnetic degrees of freedom
of a quantum spin chain in \textquotedbl{}all-in-one\textquotedbl{}
solid state device (see \cite{2000DiVincenzo-b} and references therein). 

Central to the discussion of \ac{LDE} and nearly all \ac{QI} enterprise
is the understanding how bipartite entanglement is distributed in
typical solid-state systems and how it is rearranged in the presence
of two (or more) probes. Coffman \emph{et al.} were the first to grasp
the complexity of this issue by considering pure states of three-qubits:
entanglement cannot be distributed arbitrarily but obeys the so-called
\emph{monogamy} relations \cite{2000Coffman-new}; one consequence
is the impossibility of a spin-$1/2$ in a singlet state to form a
singlet with a third spin --- the \emph{frustration} mechanism well-known
in condensed matter physics. Interestingly, restrictions from entanglement
theory are even stronger implying, in the particular case of $3$
qubits, for instance, that a singlet (being a maximal entangled state)
does not even allow one of its particles to be partially entangled
with a third system.

We are then led to the conclusion that, unlike classical correlations,
entanglement cannot be freely distributed among the parties. How does
nature distribute entanglement among the tiny magnetic moments constituting
a solid-state system? It is known that in systems with short-range
interactions, entanglement between two particles usually decays quickly
with distance between them \cite{2002Osborne,2002Osterloh}, in opposition
to the usual classical correlations which can persist for large distances.
As a pedagogical example, think about the paradigmatic model of antiferromagnetism:
the \ac{AF} spin-$1/2$ Heisenberg chain, whose Hamiltonian reads\begin{equation}
H_{AF}=J\sum_{i=1}^{N-1}\vec{\boldsymbol{S}}_{i}\cdot\vec{\boldsymbol{S}}_{i+1},\label{eq:Chap4_HeisenbergChain}\end{equation}
where $\vec{\boldsymbol{S}}_{i}:=(1/2)\vec{\boldsymbol{\sigma}}_{i}$
are spin-$1/2$ operators {[}$\sigma_{i}^{\alpha}$ denotes Pauli
matrices, see Eq.~(\ref{eq:Chap1_PauliMatrices}){]}. The spin-spin
correlations in the thermodynamic limit read, $\langle\vec{\boldsymbol{S}}_{i}\cdot\vec{\boldsymbol{S}}_{i+r}\rangle\sim\left(-1\right)^{r}\ln r/r$
\cite{1998Affleck}. The form of the partial state of the $i$-th
and $j$-th spins, $\hat{\rho}_{ij}$, is trivially fixed by the global
$SU(2)$ symmetry of the Hamiltonian,\begin{equation}
\hat{\rho}_{ij}=\frac{{\mathbbm{1}}}{4}+C_{ij}\mathcal{\hat{O}}_{ij},\label{eq:Chap4_PartialState_HeisenbChain}\end{equation}
where $\mathcal{\hat{O}}_{i,i}=(4/3)\vec{\boldsymbol{S}}_{i}\cdot\vec{\boldsymbol{S}}_{j}$
is the only $SU(2)$ invariant (besides the identity) for the two
spins, and $C_{ij}:=\langle\vec{\boldsymbol{S}}_{i}\cdot\vec{\boldsymbol{S}}_{j}\rangle$
is the spin-spin correlation function. From the above formula, one
can show (for instance, via \ac{PPT}, see Sec.~\ref{sec:Entang-th})
that asymptotic correlations are entirely classical. As a matter of
fact, in the thermodynamic limit, only next-neighbor spins are entangled
as it can be concluded by the study of the exact correlations for
this model. Take for instance, the nearest-neighbor correlator,\[
C_{i,i+1}=\frac{\partial E_{0}}{\partial J}=-\frac{1}{4}+\ln2\simeq-0.44,\]
where $E_{0}$ is the energy per site (derived via Bethe ansatz by
Hulthén \cite{1938Hulthen}). The latter value is low enough that
correlations are quantum (recall that two spins-$1/2$, $i$ and $j$,
display entanglement \emph{iff} $C_{i,j}<-1/4$ --- see Sec.~\ref{sec:C-versus-Q}).
In this case, each spin is entangled with its nearest neighbors but
the specific form of the many-body wave-function entails a very short
entanglement correlation length, despite the chain being critical
and correlations decaying slowly. 

In other spin chains, entanglement can persist at larger distances
(for instance reaching next-nearest-neighbors, see \cite{2002Osborne}),
but generally decaying much faster than correlations. On the other
hand, entanglement between a block of spins and the rest of the chain
can be very large in critical chains, reflecting that spin-spin entanglement
is just one facet of quantum correlations in ground-state of many-body
systems \cite{2009Eisert}.

The fact that bipartite spin-spin entanglement is highly restrained
in spin chains make us questioning about the possibility of establishing
entanglement between distant particles. However, as mentioned in the
previous section, numerical studies show that certain spin chains
are able to establish \ac{LDE} between probes to which they couple,
without the need of an optimal measurement strategy onto the rest
of the spins \cite{2006Venuti}, raising the question: which classes
of strongly correlated systems are able to produce \ac{LDE} ?

\section{A perturbation theory for long-distance entanglement\label{sec:A-perturbation-theory}}

In this section we answer the latter question by considering the particular
and important case of weakly coupled probes. The other limit, namely
that of strongly interacting probes is of little interest regarding
\ac{LDE} --- this fact can be understood by realizing that if a probe
interacts strongly with a given site in the lattice it will develop
entanglement with it, avoiding entanglement with the other probe (the
frustration mechanism mentioned previously). 

Regarding the nature of the quantum many-body system, our choice will
be directed to those leading to strongly-correlated ground states
with antiferromagnetic correlations. We can anticipate that ground
states with strong classical order (ferromagnetism, for instance)
will not be able to entangle external particles. Think for instance
in a one-dimensional ferromagnet; weakly coupled spin probes would
perceive this system as a strong magnetic field and would align producing
a product state with well-defined spin directions. 

We start by defining \ac{LDE} at zero-temperature; denoting the ground-state
of the total system by $|\psi\rangle$ and the degrees of freedom
of the many-body lattice by $\mathcal{L}$, then if the partial state
of two probes $a$ and $b$ (not interacting directly),\begin{equation}
\rho_{ab}=\text{Tr}_{\mathcal{L}}\left[|\psi\rangle\langle\psi|\right],\label{eq:Chap4-LDE-Definition}\end{equation}
is entangled for distances of the order of the system size, $d_{AB}\sim O(L)$,
they are said to display long-distance entanglement. The trace operation
in Eq.~(\ref{eq:Chap4-LDE-Definition}) is most of the times impossible
to perform analytically as the dimension of $\mathcal{H}$ grows exponentially
with the system size. If the quantum systems living on the sites of
$\mathcal{L}$ are spin-$1/2$ particles, then an exact calculation
can be carried out for the $XY$ model where a remarkable mapping
to a theory of free fermions exists \cite{1928Jordan}. In other cases,
such that of the \ac{AF} Heisenberg model, we have to address the
physics of the probes by other means. Here we use degenerate second-order
perturbation theory to derive a quantitative description of the effective
Hamiltonian of the probes. This will be sufficient to prove the existence
of quasi-perfect \ac{LDE} in \ac{1D} gapped systems and in the next
chapter, by means of a different approach, we will quantify the amount
of \ac{LDE}. 

The Hamiltonian of the \ac{LDE} problem has the following form, \begin{equation}
H=H_{0}+V_{m,n},\label{eq:Chap4-HamiltonianGeneral}\end{equation}
where $H_{0}$ is the full many-body Hamiltonian of the bulk and $V_{m,n}=V_{a,m}+V_{b,n}$
describes the interaction between the probes, $a$ and $b$, and the
bulk through sites $m$ and $n$, respectively. Although the present
formalism can be employed to general gapped many-body systems, we
focus on one-dimensional spin chains and probes with the same Hilbert
space (\emph{i.e.}~$\mathcal{H}_{A}=\mathcal{H}_{B}$). 

The requirement of weakly coupled probes reads \begin{equation}
J_{p}\ll J,\label{eq:Chap4-Weakly coupled probes}\end{equation}
where $J_{p}$ is the interaction strength between the probes and
the spin chains, and $J$ is a typical energy scale for the spin system
(for instance, a nearest neighbor exchange interaction). When $J_{p}=0$
the state of the probes becomes totally uncorrelated and the \ac{GS}
of the entire system becomes $d\times d$ fold-degenerate, where $d$
is the dimension of the probe's Hilbert space. In this case we may
write $|\psi\rangle=|\psi_{0}\rangle\otimes|\chi_{a}\rangle\otimes|\chi_{b}\rangle$,
where $|\psi_{0}\rangle$ is assumed non-degenerate and $|\chi_{\gamma}\rangle$
stands for the the state of the probe $\gamma=a,b$. The role of the
interaction ($J_{p}>0$) is to lift this degeneracy causing the probes
to develop correlations. 

On quite general grounds, this interaction can be recast in the following
form, \begin{equation}
V=\sum_{\alpha=1}^{p}\left(\gamma_{\alpha}^{a}O_{m}^{\alpha}\otimes A^{\alpha}\otimes{\mathbbm{1}}_{b}+\gamma_{\beta}^{b}O_{n}^{\alpha}\otimes{\mathbbm{1}}_{a}\otimes B^{\alpha}\right),\label{eq:Chap4-Hamiltonian_Probes_System}\end{equation}
where $\mbox{\ensuremath{A}\ensuremath{(B)}}$ denotes an (vector)
operator with components $\alpha=x,y,z$ acting on the Hilbert space
of the probe $a(b)$ and ${\mathbbm{1}}{}_{a(b)}$ the corresponding
identity operators. The many-body system operators on sites $m$ are
represented by $O_{m}^{\alpha}$ and $\gamma_{\alpha}^{a(b)}$ stand
for coupling strengths for each of the terms in $V$. 

A projection on the spin chain ground state, integrating their degrees
of freedom, is the key feature of our method and thus it is useful
to define the projector onto the states with unperturbed energy $E_{0}\equiv\langle\psi_{0}|H_{0}|\psi_{0}\rangle$
: \begin{equation}
\mathcal{P}_{0}=|\psi_{0}\rangle\langle\psi_{0}|\otimes{\mathbbm{1}}{}_{a}\otimes{\mathbbm{1}}_{b}.\label{eq:Chap4-ProjectorOntoGS}\end{equation}
The projector onto the subspace of higher energy $E_{k}>E_{0}$ is
denoted by $\mathcal{P}_{k}$ (with $k>0$) and thus ${\mathbbm{1}}=\mathcal{P}_{0}+\sum_{k>0}\mathcal{P}_{k}$.
Using second order degenerate perturbation theory we can determine
the probes \ac{GS} by diagonalizing an effective Hamiltonian in the
subspace spanned by $\mathcal{P}_{0}$. This is a familiar concept
that finds many applications in condensed matter physics, such as,
for instance, in the derivation of the Ruderman-Kittel-Kasuya-Yosida
magnetic interaction between local moments in a metal \cite{1956Kasuya,1956Kasuya-b,1957Yosida}. 

In what follows, we set $\mathcal{P}_{0}H_{0}\mathcal{P}_{0}$ to
zero as it contributes with a constant, and thus not changing the
physics. The derivation of the effective Hamiltonian is made in Appendix~\ref{sec:App-Degenerate-perturbation-theory}.
It reads \begin{equation}
H^{(ab)}=-\sum_{k>0}\langle\bar{V}\mathcal{P}_{k}\bar{V}\rangle_{0}(E_{k}-E_{0})^{-1}+\text{"local terms"},\label{eq:Chap4-Aux}\end{equation}
where the average is taken with respect to the \ac{GS} of the spin
chain, $\langle\bar{V}P_{k}\bar{V}\rangle=\langle\psi_{0}|\bar{V}P_{k}\bar{V}|\psi_{0}\rangle$,
$\bar{V}:=V-\langle\psi_{0}|V|\psi_{0}\rangle$ and constants are
absorbed in the local terms whose form we do not make explicit yet. 

Entanglement between the probes arises from $H^{(ab)}$ since it contains
non-local terms such as $\bar{V}_{a,m}\mathcal{P}_{k}\bar{V}_{b,n}$
\cite{2005Li}. The probe Hamiltonian can be transformed by straightforward
manipulations into an explicit form involving time dependent correlation
functions of the spin chain. A similar procedure is used to express
cross sections of scattering by many-body systems in terms of its
correlation functions \cite{Book-Squires-1978}. We obtain (we set
$\hbar=1$, see Appendix~\ref{sec:App-Degenerate-perturbation-theory}),
\begin{equation}
H^{(ab)}=-\frac{1}{2\pi}\int_{-\infty}^{+\infty}\frac{dE}{E}\int_{-\infty}^{+\infty}dt\langle\bar{V}(t)\bar{V}\rangle_{0}e^{iEt}.\label{eq:Chap4-Aux2}\end{equation}
We now introduce the explicit form of $\bar{V}$ to arrive at the
desired result: $H^{(ab)}=H_{L}^{(a)}+H_{L}^{(b)}+H_{NL}^{(ab)}$.
Defining the two-body connected correlation in the usual form \begin{equation}
\langle O_{m}^{\alpha}(t)O_{n}^{\beta}\rangle_{c}=\langle\psi_{0}|O_{m}^{\alpha}(t)O_{n}^{\beta}|\psi_{0}\rangle-\langle\psi_{0}|O_{m}^{\alpha}(t)|\psi_{0}\rangle\langle\psi_{0}|O_{n}^{\beta}|\psi_{0}\rangle,\label{eq:Chap4-Connected-Correlation}\end{equation}
the term coupling the two probes reads, \begin{eqnarray}
H_{NL}^{(ab)} & = & \sum_{\alpha,\beta=1}^{p}\gamma_{\alpha}^{a}\gamma_{\beta}^{b}(C_{m\alpha;n\beta}+C_{n\beta;m\alpha})A^{\alpha}\otimes B^{\beta}\label{eq:Chap4-EffecHam_Gen1}\\
C_{m\alpha;n\beta} & = & \frac{1}{2i}\int_{-\infty}^{\infty}dte^{-0^{+}|t|}sign(t)\langle O_{m}^{\alpha}(t)O_{n}^{\beta}(0)\rangle_{c}.\label{eq:Chap4-EffecCouplings}\end{eqnarray}
The form of local terms $H_{L}^{(a)}+H_{L}^{(b)}$ will not be given
since they play no role in the systems of our interest, \emph{i.e.}~those
with full rotational symmetry%
\footnote{Systems with magnetic field, for instance, will produce local terms
in the effective Hamiltonian which reduces the amount of entanglement
(compare with the effect a local magnetic field on the Heisenberg
magnet considered in the introduction, see for instance Fig.~\ref{fig:Chap1_Entang2qubitsDensity}).%
}. The coupling between the probes can be expressed in terms of the
response function (or adiabatic susceptibility) $\chi_{m\alpha;n\beta}(t)=-i\langle[O_{m}^{\alpha}(t),O_{n}^{\beta}]\rangle\theta(t)$,
where $\theta(t)$ is the Heaviside step function. Using the Lehman
representation at $T=0$ one can show that \begin{eqnarray}
\tilde{\chi}{}_{m\alpha;n\beta}(0) & = & C_{m\alpha;n\beta}+C_{n\beta;m\alpha},\label{eq:Chap4-RelationBetweenEffectiveCouplingsandXhi}\end{eqnarray}
where $\tilde{\chi}_{m\alpha;n\beta}(\omega)$ is the time Fourier
transform of $\chi_{m\alpha;n\beta}(t)$ (Appendix~\ref{sec:Time-correlation-functions}).
This formula says that any interaction mediated by the spin chain
will be encoded in the response function. Usually, the study of external
perturbations require the knowledge of the susceptibility $\chi(t)$
for all times, but since we are interested in the equilibrium physics
an integration in the time domain emerges in our equations. The fact
that we are able to write the effective couplings in this language
will be central later on, when specializing for concrete spin chains. 

Some comments about the validity of perturbation theory are in order:
the effective Hamiltonian {[}Eq.~(\ref{eq:Chap4-EffecHam_Gen1}){]}
lifts the degeneracy of the \ac{GS} level of the uncoupled system
($J_{p}=0$) and, as long as the couplings appearing in $H_{NL}^{(ab)}$
are small, compared to typical energy scales of the spin chain, like
the gap to first excited state, $\Delta$, the low energy physics
of this system, $E\ll\Delta$, with no real excitations of the spin-chain,
will be well described by $H_{NL}^{(ab)}$. This condition limits
the strength of the chain-probe interaction, but is shown by numerical
results to be the appropriate limit to maximize \ac{LDE} \cite{2006Venuti}
(stronger couplings will be studied in Chapter~\ref{cha:LDE_Finite_Temperature}).

\subsection{Quasi-perfect LDE in the spin-$1/2$ Heisenberg chain\label{sub:LDE_HeisenbergAF}}

Neutron-scattering experiments reveal that the low energy physics
of many magnetic compounds is described by the Heisenberg model \cite{Review-Manousakis}.
The same model emerges as the effective low-energy Hamiltonian in
strongly correlated systems, such as the Hubbard model at half-filling
\cite{Book-Auerbach-1994}. For these reasons this model has been
the basis of many studies in condensed matter physics for many years.
On the other hand, the Heisenberg model is also very important in
\ac{QI}. The reason is twofold; spin chains are candidates for quantum
computers (Sec~\ref{sec:Many-body-systems-as}), and recent advances
in the field of atomic and molecular physics, namely in optical cooling,
made it possible to engineer (via the Coulomb coupling of neighboring
ions) effective short-range Hamiltonians acting on internal degrees
of freedom of trapped ions --- a laboratory to investigate typical
condensed matter phenomena in different physical scenarios (the interested
reader is referred to \cite{2007Lewenstein} for an extensive review
on cold atoms in optical lattices and their applications in \ac{QI}). 

For the aforementioned reasons, we start by addressing the Heisenberg
antiferromagnetic spin chain and its capacity towards \ac{LDE}. It
is useful to write Eq.~(\ref{eq:Chap4-EffecHam_Gen1}) in terms of
spin operators $\vec{\boldsymbol{S}}_{m}$ for the spin chain and
$\vec{\boldsymbol{\tau}}_{a(b)}$ for the probes. Considering that
the probes couple with the spin chain via an Heisenberg interaction,
the most common situation, \begin{equation}
V_{m,n}=J_{a}\vec{\boldsymbol{S}}_{m}\cdot\vec{\boldsymbol{\tau}}_{a}+J_{b}\vec{\boldsymbol{S}_{n}}\cdot\vec{\boldsymbol{\tau}}_{b},\label{eq:Chap4-Probe-Chain Interaction}\end{equation}
the connection with the previous notation becomes straightforward
(see table~\ref{tab:Chap4_Descriptions}). The effective Hamiltonian
becomes simply, \begin{equation}
H^{(ab)}=J_{\text{eff}}\vec{\boldsymbol{\tau}}_{a}\cdot\vec{\boldsymbol{\tau}}_{b},\label{eq:Chap4-H_eff_spins}\end{equation}
 where $J_{\text{eff}}=J_{a}J_{b}\tilde{\chi}_{mz;nz}(0)$. %
\begin{table}[h]
\noindent \begin{centering}
\begin{tabular}{c||c|c}
 & generic quantum lattice & spin chain\tabularnewline
\hline
\hline 
many-body operators & $O_{m}^{\alpha}$ & $S_{m}^{\alpha}$\tabularnewline
\hline 
probe operators & $A^{\alpha}(B^{\alpha})$ & $\tau_{a(b)}^{\alpha}$\tabularnewline
\hline 
couplings & $\gamma_{\alpha}^{a(b)}$ & $J_{a(b)}$(isotropic)\tabularnewline
\end{tabular}
\par\end{centering}

\caption{Connection between the generic quantum lattice notation and the spin
chain notation.\label{tab:Chap4_Descriptions}}

\end{table}
The effective Hamiltonian in the general form (\ref{eq:Chap4-H_eff_spins})
already involves the important conclusion that for a correct choice
of signs of $J_{a}$ and $J_{b}$ any spin chain is a potential entangler
for the probes as long as the susceptibility is finite and sufficiently
small, so that one remains well inside perturbation theory limits.
To see if this is the case in the \ac{AF} Heisenberg chain, we have
to compute explicitly the effective coupling. The Hamiltonian of an
\ac{AF} Heisenberg chain with $L$ spins reads \begin{equation}
H_{0}:=J\sum_{i=1}^{L-1}\vec{\boldsymbol{S}}_{i}\cdot\vec{\boldsymbol{S}}_{i+1},\label{eq:Chap4-Heisenbeg}\end{equation}
with $J$ standing for the exchange coupling (in this chapter, without
any loss of generality, we set $J=1$). Our formalism only applies
to the \emph{finite} chain which has gapped excitations although its
size can be arbitrarily large. The effective Hamiltonian Eq.~(\ref{eq:Chap4-H_eff_spins})
will preserve the full $SU(2)$ symmetry of the interaction Hamiltonian
$H_{0}+V_{m,n}$, \emph{i.e.}~no local terms will give additional
contribution to $H^{(ab)}$. From now on, we assume that the probes
couple to the spin chain with the same strength, $J_{a}=J_{b}\equiv J_{p}$.
Hence, the effective Hamiltonian, $H_{\text{eff}}$, takes the very
compact form \begin{equation}
H_{\text{eff}}=J_{\text{eff}}\vec{\boldsymbol{\tau}}_{a}\cdot\vec{\boldsymbol{\tau}}_{b}\underset{SU(2)}{=}J_{p}^{2}\tilde{\chi}_{mz;nz}(0)\vec{\boldsymbol{\tau}}_{a}\cdot\vec{\boldsymbol{\tau}}_{b}.\label{eq:Chap4-Heff_heisenberg}\end{equation}
The partial state, $\rho_{ab}$, will correspond to a quasi-perfect
singlet-state as long as $\tilde{\chi}_{mz;nz}(0)>0$ is bounded and
$J_{p}$ is chosen such that $J_{p}^{2}\tilde{\chi}_{mz;nz}(0)\ll\Delta$,
where $\Delta$ is the gap to the first excited state of $H_{0}$.
Given the above result, one could be tempted to conclude that $\rho_{ab}$
is in fact a perfect singlet state since the \ac{GS} of Hamiltonian
(\ref{eq:Chap4-Heff_heisenberg}) is precisely the single-state for
the probes. However, due to the perturbative nature of our formalism
with expansion parameter $J_{p}/\Delta$ (Appendix~\ref{sec:App-Degenerate-perturbation-theory}),
the inspection of the perturbed wave-function shows that the negativity
(or any other entanglement monotone) will deviate from the value of
maximal entanglement (the singlet state) with corrections of the order
of $(J_{p}/\Delta)^{2}$, which can be very small for weakly interacting
probes, \emph{i.e.} \begin{equation}
E(\rho_{ab})=1-O(J_{p}^{2}/\Delta^{2}).\label{eq:Chap4-QuasiPerfectLDE}\end{equation}
The impossibility of perfect \ac{LDE} is in agreement with the intuition
that tracing out the degrees of freedom in the bulk (in general) introduces
mixdness, which in its turn reduces the amount of entanglement even
at $T=0$ (Sec.~\ref{sec:Entang-th}); this represents no limitation
from the point of view of \ac{QI} applications, since entanglement
extraction procedures do exist that convert partial entangled states
in maximally entangled pairs (Fig.~\ref{fig:Chap1_entang_concentration}). 

In Chapter~\ref{cha:LDE_Finite_Temperature}, we will see that equilibrium
averages of operators $\vec{\boldsymbol{\tau}}_{a(b)}$ computed directly
from Eq.~(\ref{eq:Chap4-Heff_heisenberg}) must be renormalized,
explaining why $\rho_{ab}$ does not follow directly from $H_{\text{eff}}$.
As a consequence, Eq.~(\ref{eq:Chap4-QuasiPerfectLDE}) is in fact
a particular case of a more general situation encompassing gapped
systems with different characteristics. For the moment, however, we
do not attempt to quantify the amount of entanglement but rather focus
on the \ac{LDE} capability of one-dimensional \ac{AF} systems, where
corrections can be so small that quasi-perfect \ac{LDE} {[}Eq.~(\ref{eq:Chap4-QuasiPerfectLDE}){]}
is guaranteed.

\subsubsection{Computation of correlation functions from conformal theory}

To calculate the \ac{GS} time-dependent correlation functions\emph{,}
$\langle S_{m}^{\alpha}(t)S_{n}^{\beta}(0)\rangle_{c}$, necessary
for the computation of the adiabatic susceptibility, we will use the
conformal invariance of the critical \emph{infinite} chain ($L\rightarrow\infty$)
since its time-dependent correlations are enough to extract the effective
coupling $J_{ab}$ for the finite chain. The reason why we adopt this
method (rather than trying to solve for $\tilde{\chi}$ directly in
the finite size scenario) its because the conformal character of the
infinite chain can be used to relate the physics in distinct geometries,
as we will briefly see. 

General results for correlations of critical spin-$1/2$ chains are
known from bosonization theory (\emph{e.g.}~the asymptotic behavior
of time-dependent correlations \cite{Book-Giamarchi-2004}), but how
can the physics of these systems be mapped to the physics of a gapped
(non-critical) spin chain? The basic idea is to generalize the scale
invariance of a classical system at criticality to encompass a broader
class of transformations than the usual rotations, changes of scale,
\emph{etc}. 

We outline the basic notions of the so-called conformal mapping; the
reader is referred to the review by Cardy for more details \cite{Review-Cardy-2008}.
We introduce some notation; let $\phi_{i}$ denote the fields of a
given theory defined in a lattice (assumed critical) and $b$ denote
a scale transformation $r_{i}^{\prime}=b^{-1}r_{i}$ . The critical
behavior of the theory entails the following scaling transformation
law, \begin{equation}
\langle\phi_{1}(r_{1})...\phi_{n}(r_{n})\rangle_{\mathcal{B}}=b^{-\mu_{1}}...b^{-\mu_{n}}\langle\phi_{1}(r_{1}^{\prime})...\phi_{n}(r_{n}^{\prime})\rangle_{\mathcal{B}^{\prime}},\label{eq:Chap4-CriticalBeh}\end{equation}
where $\langle...\rangle_{\mathcal{B}(\mathcal{B}^{\prime})}$ is
an equilibrium average of the fields defined in lattice $\mathcal{B}(\mathcal{B}^{\prime})$
and $\left\{ \mu_{i}\right\} $ denote the scaling dimensions of the
operators $\phi_{i}$. A trivial example of such scaling transformation
is a classical critical spin model whose lattice spacing $a$ is shrunk
by a constant factor $\gamma$; its correlations, such as the two-body
correlator, \begin{equation}
\langle\phi_{1}(r_{1})\phi_{2}(r_{2})\rangle_{\mathcal{B}}\sim\left(\frac{1}{r_{1}-r_{2}}\right)^{2\mu},\label{eq:Chap4-CriticalTwoPoint}\end{equation}
will change according to \begin{equation}
\langle\phi_{1}(r_{1})\phi_{2}(r_{2})\rangle_{\mathcal{B}}\underset{r^{\prime}=r/\gamma}{\longrightarrow}\gamma^{2\mu}\left(\frac{1}{r_{1}-r_{2}}\right)^{2\mu}=\gamma^{2\mu}\langle\phi_{1}(r_{1})\phi_{2}(r_{2})\rangle_{\mathcal{B}},\label{eq:Chap4-AuxCriticality1}\end{equation}
which clearly agrees with the transformation law (\ref{eq:Chap4-CriticalBeh})
with $\mu_{1}=\mu_{2}=\mu$. This result is not surprising as we would
not expect that changing the lattice spacing $a$ would modify the
asymptotic physics if the underlying model is the same; after all,
critical models look the same in all scales, a fact encoded in the
power-law behavior of their correlators. 

What is surprising though is that this transformation law holds to
all conformal transformations (~\emph{i.e.}~the transformations
preserving locally the angles between a triplet of points). In \ac{3D}
these are just rotations, changes of scale and translations which
do not produce essentially different lattices (hence the correlators
having similar structures). However, in \ac{2D} the group of conformal
transformations is a much larger class! The latter stems from the
well-known fact in complex analysis that any analytic transformation
$f(z)$ of the plane ($z$ being a complex coordinate) is conformal.
Accordingly, we can generalize (\ref{eq:Chap4-CriticalBeh}) by considering
all transformations $f(z)$ preserving locally the metric of $\mathcal{B}$:
\begin{equation}
\langle\phi_{1}(z_{1},\bar{z}_{1})...\phi_{n}(z_{n},\bar{z}_{n})\rangle_{\mathcal{B}}=\left|f^{\prime}(z_{1})\right|^{\mu1}...\left|f^{\prime}(z_{n})\right|^{\mu_{n}}\langle\phi_{1}\left(z_{1}^{\prime}=f(z_{1})\right)...\phi_{n}\left(z_{n}^{\prime}=f(z_{n})\right)\rangle_{\mathcal{B}^{\prime}}.\label{eq:Chap4-ConformalTransformationLaw}\end{equation}
This method is very powerful for it relates the physics of lattices
with different geometries simply by finding the appropriate analytic
function $f(z)$ mapping the points in $\mathcal{B}$ to points in
$\mathcal{B}^{\prime}$. The physical motivation behind generalization
(\ref{eq:Chap4-ConformalTransformationLaw}) is the following; if
the Hamiltonian contains only local terms (such the Heisenberg model)
then, in principle, conformal transformations will not change the
asymptotic physics as they are (locally) just simple rotations, rescalings
or translations.

\subsubsection{Adiabatic susceptibility of the Heisenberg AF spin-1/2 chain}

\begin{figure}[tb]
\noindent \begin{centering}
\includegraphics[width=0.7\columnwidth]{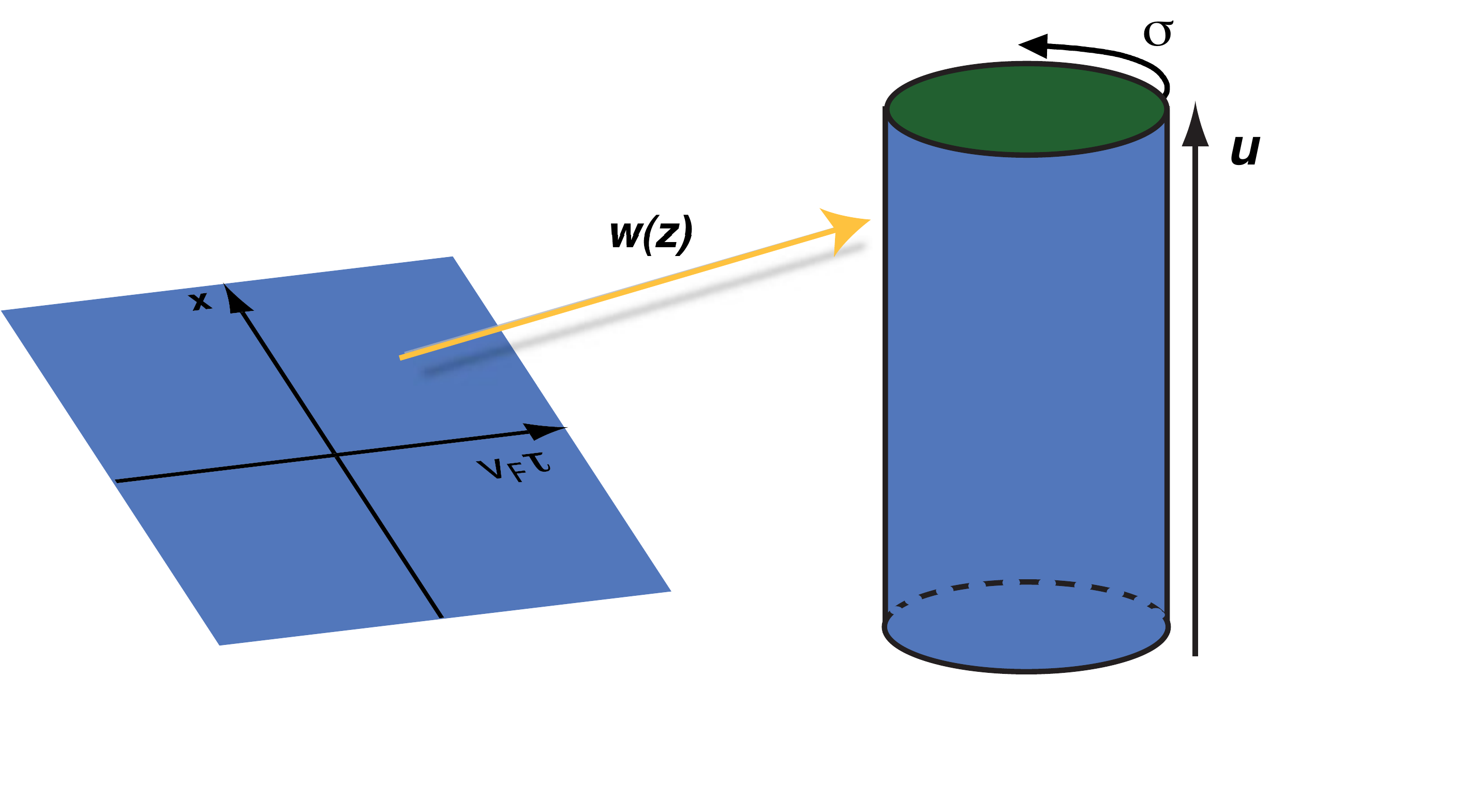}
\par\end{centering}

\caption[Conformal transformation between the plane and the cylinder]{\label{fig:Chap4-The-conformal-transformation}The conformal transformation
$w(z)=\sigma+ir$ maps every point $(v_{F}\tau,x)$ in the plane into
the strip geometry ($\sigma\in]-\infty,\infty[$, $r\in[-L/2,L/2]$)
with periodic boundary conditions along the $r$ direction (see Appendix~\ref{sec:Correlations-from-conformal}).}

\end{figure}

Let us apply conformal mapping methods to the critical Heisenberg
\ac{AF} chain so to extract $\tilde{\chi}_{mz;nz}(0)$ for the ring
geometry. It is convenient to express the correlations in terms of
the staggered magnetization, $M_{j}^{z}:=(-1)^{j}S_{j}^{z}$, exploiting
the fact that dominant long-distance correlations of the \ac{GS}
of Hamiltonian Eq.~(\ref{eq:Chap4-Heisenbeg}) oscillate with a $\pi$
phase change between neighbor spins \cite{Book-Auerbach-1994,Book-Giamarchi-2004}.
Indeed, we define the retarded Green's function for the staggered
magnetization, \begin{equation}
G_{mn}^{R}(t):=-i\langle[M_{m}^{z},M_{n}^{z}(t)]\rangle\theta(t),\label{eq:Chap4-GreenFunction_StaggeredMag}\end{equation}
from which the response function can be obtained simply by a sign
exchange {[}Eq.~(\ref{eq:App4-Relation-5}){]} and compensating the
$\pi$ phase change: $\chi{}_{mz;nz}(t)=(-1)^{|m-n|+1}G_{mn}^{R}(t)$.
In its turn, using conformal mapping, the retarded Green's function
can be obtained from the asymptotic Matsubara Green's function of
the critical chain. The latter is defined as $G(x,\tau):=\langle\hat{T}_{\tau}M_{m}^{z}(x_{m},\tau)M_{n}^{z}(x_{n},0)\rangle$
where $\hat{T}_{\tau}$ is the imaginary time-ordering operator {[}Eq.~(\ref{eq:App4-Analytic-4b}){]}
and $\tau\in[0,\beta]$ the imaginary time variable. In the limit
$|x_{m}-x_{n}|\gg1$, we may take $x=|x_{m}-x_{n}|$ as a continuum
variable, and the Matsubara function for gapless spin-$1/2$ chains
reads \cite{Book-Giamarchi-2004,Book-Sachdev-2004} \begin{equation}
G(x,\tau)=\frac{\mathcal{A}}{\left(v_{F}^{2}\tau^{2}+x^{2}\right)^{K/2}},\label{eq:Chap4-MatsubaraGreenFunc}\end{equation}
where $\mathcal{A}$ is an amplitude (from bosonization theory), $K=1$
for the \ac{AF} $SU(2)$ chain and $v_{F}$ stands for the Fermi
velocity of excitations. This result embodies a divergent $\tilde{\chi}_{mz;nz}(0)$;
a direct consequence of a zero gap and a signal of the critical nature
of the spin chain at $T=0$ --- external perturbations in such systems
are not described by linear response theory anymore. 

The mapping of the infinite chain to the finite chain is achieved
by the following analytic transformation (see Fig.~\ref{fig:Chap4-The-conformal-transformation}):
$w=\left(L/2\pi\right)\ln z\equiv\sigma+ir,$ where $z=v_{F}\tau+ix$.
Using the transformation law for conformal invariant theories {[}Eq.
(\ref{eq:Chap4-ConformalTransformationLaw}){]} the Matsubara Green
function for the finite \ac{AF} Heisenberg chain with periodic boundary
conditions in the spatial coordinate $r$ reads (Appendix~\ref{sec:Correlations-from-conformal},
\cite{Book-Giamarchi-2004}): \begin{equation}
G_{cyl}(r,\sigma)=\frac{2\mathcal{\pi A}}{L}(2\cosh(2\pi\sigma/L)-2\cos(2\pi r/L))^{-\frac{1}{2}}.\label{eq:Chap4-Matsubara_in_Cylinder}\end{equation}
The analytic continuation to real time is made by Wick rotation, $\sigma\rightarrow iv_{F}t+0^{+}sign(t)$,
yielding a time-ordered Green Function. Finally, using standard analytic
continuation methods (Appendix~\ref{sec:Analytic-continuation})
the latter function gives $G^{R}$ defined in the cylinder (and hence,
the susceptibility). 

Setting the branch cut of the logarithm in the negative real axis
(Fig.~\ref{fig:App4-PB_Log}), we get:

\begin{equation}
G_{cyl}^{R}(r,t)=-\frac{2\mathcal{\pi A}}{\sqrt{2}L}\frac{\theta\left(t\right)\theta\left[F(r,t)\right]sign\left[\sin(2\pi t/L)\right]}{\sqrt{F(r,t)}},\label{eq:Chap4-Retarded-Green-Function}\end{equation}
where $F(r,t)=2\cos(2\pi r/L)-2\cos(2\pi t/L)$ --- a detailed calculation
is given in Appendix~\ref{sec:Correlations-from-conformal} and an
outline of the main properties of spectral representation of Green's
functions can be found in Appendix~\ref{sec:Time-correlation-functions}. 

\begin{figure}[tb]
\noindent \begin{centering}
\includegraphics[width=0.5\columnwidth]{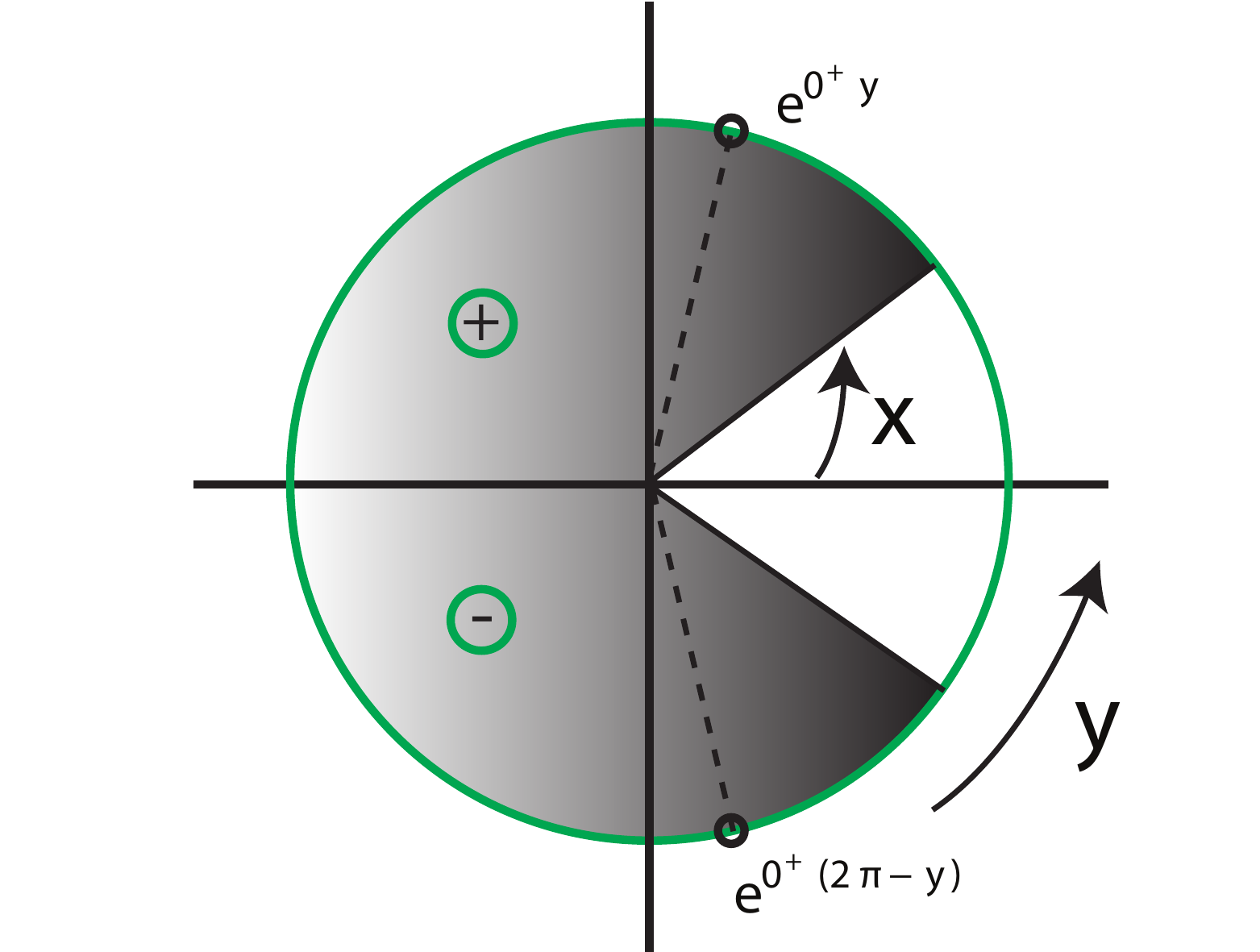}\includegraphics[width=0.5\columnwidth]{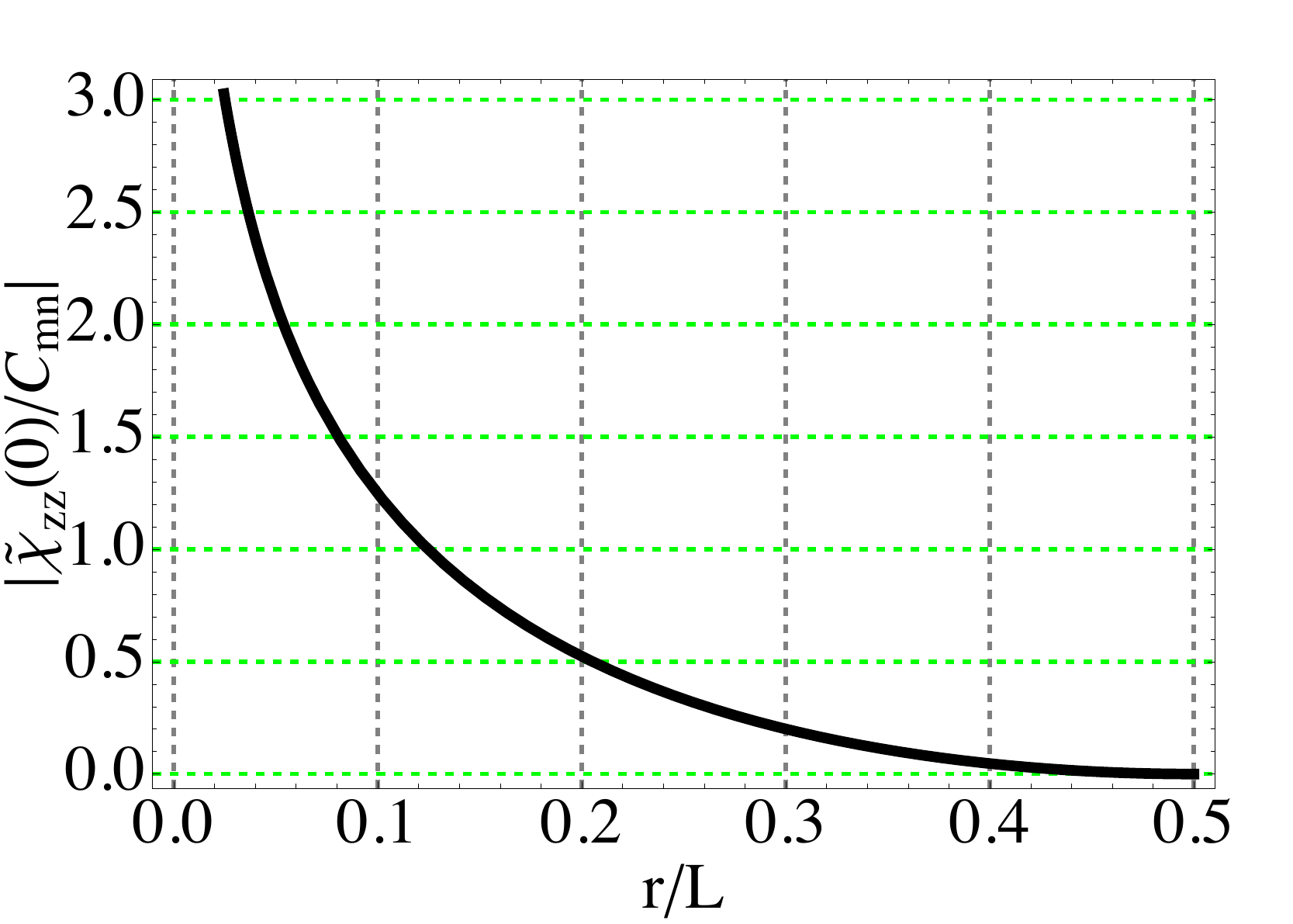}
\par\end{centering}

\caption[The adiabatic susceptibility for the Heisenberg ring as function of
distance]{\label{fig:Chap4-HeisenbergSusceptibility} Left - The integration
of the adiabatic susceptibility is done by thinking of the variable
$y$ {[}Eq.~(\ref{eq:Chap4-Aux_Retarded_3}){]} as living on the
circle. Then Eqs.~(\ref{eq:Chap4-Aux_Retarded_2})-(\ref{eq:Chap4-Aux_Retarded_5})
simply subtract the grey area on the top (with plus sign in the figure
and weight proportional to $e^{0^{+}y}$ ) from the grey area on the
bottom (with minus sign in the figure and weight proportional to $e^{0^{+}(2\pi-y)}$).
This procedure is repeated many times as $y$ crosses the same points
in the circle. At the end, the limit of $0^{+}\rightarrow0$ is done
to get Eq.~(\ref{eq:Chap4-Aux_Retarded_6}).  Right - The absolute
value of the response function at zero frequency for the finite antiferromagnetic
Heisenberg chain. We have assumed $r\gg1$, so that the results from
bosonization theory are accurate. Only $r/L\in]0,1/2]$ is represented,
since boundary conditions imply $\tilde{\chi}_{r/L}(0)=\tilde{\chi}_{1-r/L}(0)$.}

\end{figure}

We now compute the response function at zero frequency according to
Eq.~(\ref{eq:App4-Relation-24}), $\tilde{\chi}_{mz;nz}(0)=(-1)^{|m-n|+1}\int_{0}^{\infty}dtG^{R}(r,t)e^{-0^{+}t}.$
To this end, we define $\tau:=2\pi t/L$ and $x:=2\pi r/L$ and separate
the integral in many parts using $\sin(\tau)>0$ for $\tau\in]0,\pi[+2n\pi$
with $n\in\mathbb{N}_{0}$ (we omit, for the moment, multiplicative
constants and denote the integral by $\tilde{\chi}$),\begin{eqnarray}
\tilde{\chi} & \sim & \int_{0}^{\infty}d\tau e^{-0^{+}\tau}\frac{\theta\left[\cos(x)-\cos(\tau)\right]}{\sqrt{\cos(x)-\cos(\tau)}}sign\left(\sin(\tau)\right)\label{eq:Chap4-Aux_Retarded_1}\\
 & = & \sum_{n=0}^{\infty}\left(\int_{x+2n\pi}^{\pi+2n\pi}d\tau-\int_{\pi+2n\pi}^{2\pi-x+2n\pi}d\tau\right)\frac{e^{-0^{+}\tau}}{\sqrt{\cos(x)-\cos(\tau)}}\label{eq:Chap4-Aux_Retarded_2}\\
 & = & \sum_{n=0}^{\infty}\left(\int_{x}^{\pi}dy-\int_{\pi}^{2\pi-x}dy\right)\frac{e^{-0^{+}y}e^{-0^{+}2n\pi}}{\sqrt{\cos(x)-\cos(y)}},\label{eq:Chap4-Aux_Retarded_3}\end{eqnarray}
where we have defined $y:=\tau-2n\pi$. The next step is to perform
the summation over $n$ {[}the substitution $0^{+}\rightarrow0$ is
made only at the end to assure convergence (see Figure~\ref{fig:Chap4-HeisenbergSusceptibility}
- Left){]}:\begin{eqnarray}
\tilde{\chi} & = & \frac{1}{1-e^{-2\pi0^{+}}}\left(\int_{x}^{\pi}dy-\int_{\pi}^{2\pi-x}dy\right)\frac{e^{-0^{+}y}}{\sqrt{\cos(x)-\cos(y)}},\label{eq:Chap4-Aux_Retarded_4}\\
 & = & \frac{1}{1-e^{-2\pi0^{+}}}\int_{x}^{\pi}dy\frac{e^{-0^{+}y}-e^{-0^{+}\left(2\pi-y\right)}}{\sqrt{\cos(x)-\cos(y)}}\label{eq:Chap4-Aux_Retarded_5}\\
 & \underset{0^{+}\rightarrow0}{=} & \int_{x}^{\pi}dy\frac{1-y/\pi}{\sqrt{\cos(x)-\cos(y)}}.\label{eq:Chap4-Aux_Retarded_6}\end{eqnarray}

This shows that the \ac{AF} ring has a finite $\tilde{\chi}_{mz;nz}(0)$,
as expected for a gapped chain ($\Delta\sim J/L$); conformal theory
{[}Eq. (\ref{eq:Chap4-ConformalTransformationLaw}){]} could effectively
be used to calculate correlation functions in a different geometry.
Gathering all the constants, we finally reach to the desired result,
\begin{equation}
\tilde{\chi}_{mz;nz}(0)=\mathcal{C}_{mn}\int_{2\pi r/L}^{\pi}d\tau\frac{\tau/\pi-1}{\sqrt{\cos(2\pi r/L)-\cos(\tau)}},\label{eq:Chap4-Suscepitibility_zero_frequency}\end{equation}
with $\mathcal{C}_{mn}=(-1)^{|m-n|}\mathcal{A}/(\sqrt{2}v_{F})$.
Figure~\ref{fig:Chap4-HeisenbergSusceptibility} (Right) shows the
plot of the absolute value of the rescaled response function at zero
frequency (|$\tilde{\chi}_{mz;nz}(0)/\mathcal{C}_{mn}|$) demonstrating
the existence of quasi-perfect \ac{LDE} for a wide range of values
of $r/L$. Note that $\tilde{\chi}_{m\alpha;n\alpha}(0)$ diverges
logarithmically at the origin. Our perturbative approach cannot be
applied unless $J_{p}^{2}\tilde{\chi}_{m\alpha;n\alpha}(0)\ll\Delta\sim J/L$,
and will fail in the thermodynamic limit ($L\to\infty$) for fixed
$r$. The numerical results of Ref. \cite{2006Venuti} show probes
almost completely entangled only for small values of coupling ($J_{p}\sim0.1$),
for a finite chain $L=26$. This value is well estimated by the limit
of validity of our perturbative approach, for $r/L\sim O(1)$, namely,
$J_{p}\ll J/\sqrt{L}$. 

These results strongly suggest that the conditions for \ac{LDE} are
coincident with the conditions for validity of the perturbative approach;
weakly coupled probes get maximally entangled by the effective antiferromagnetic
interaction mediated by the spin chain. We are led then to put forward
the following conjecture:
\begin{conjecture}
The ground states of many-body gapped systems with no symmetry breaking
and dominant long-distance antiferromagnetic correlations mediate
quasi-perfect \ac{LDE} between sufficiently weakly coupled spin-$1/2$
probes.
\end{conjecture}
There is a simple physical picture behind the latter conjecture; if
a spin system has a finite adiabatic susceptibility, then sufficiently
weakly coupled probes will not perturb its ground state very much
(therefore preserving the character of asymptotic correlations). On
the other hand, their perturbation, though very feeble, will be perceived
by each other, being weakly coupled to the bulk. Indeed, they will
necessarily form either a quasi-perfect singlet or a triplet plus
corrections of order $J_{p}^{2}/J^{2}$, for their effective Hamiltonian
must preserve $SU(2)$ symmetry. In perturbation theory, we learned
that whichever state they eventually form depends on the sign of the
adiabatic susceptibility and, therefore, on the nature of the correlations
of the spin system. Although these observations seem quite natural,
a careful study in the next chapter will show that quasi-perfect \ac{LDE}
is a peculiarity of one-dimensional \ac{AF} systems at $T=0$ and
that the general picture is more intricate.

\subsection{Quasi-perfect LDE in the AKLT spin chain at T=0\label{sub:LDE-AKLT} }

Is \ac{LDE} a phenomenon exclusive of systems with long range (or
\emph{quasi}-long range) \ac{AF} order? The conjecture in the previous
section says that even when the correlations decay exponentially we
may still have \ac{LDE} as long as they are antiferromagnetic in
nature and a gap separates the ground state energy from the rest of
the spectrum. Here we consider a different \ac{1D} model of antiferromagnetism:
the \ac{AKLT} spin chain --- a particular case of the spin-$1$ Heisenberg
chain with biquadratic interactions:\begin{equation}
H:=\sum_{i=1}^{L-1}\left[\vec{\boldsymbol{S}}_{i}\cdot\vec{\boldsymbol{S}}_{i+1}+\beta(\vec{\boldsymbol{S}}_{i}\cdot\vec{\boldsymbol{S}}_{i+1})^{2}\right].\label{eq:Chap4-Biquadratic_Heisenberg}\end{equation}
In \ac{1D} the physics of integer and half-integer spins chains differ
very much \cite{1983Haldane,1983Haldane2,Book-Auerbach-1994}; the
latter model has massive excitations in some phases (even in the thermodynamic
limit), a completely different picture than that of spin-$1/2$ isotropic
chains. For $\beta=1/3$ it admits an exact solution known as the
\ac{AKLT} spin chain; a picture of its \ac{GS} is given by the so-called
valence-bond-solid --- each spin-1 is represented by a couple of spins
one-half, as long as the antisymmetric state is projected out. The
valence-bond-solid state is constructed by forming short-ranged singlets
between nearest spin-$1/2$ and then symmetrizing local pairs to get
back $S=1$ states. 

In the thermodynamic limit, the static correlations are very short-ranged
{[}$\xi_{AKLT}=1/\ln(3)\cong0.9${]} \cite{1988Affleck}. For this
reason, we may ask whether two probes are able to get entangled by
interaction mediated by the spin-$1$ chain. We cannot make an exact
computation of the adiabatic susceptibility as for the Heisenberg
model, since the exact dynamical correlations are not known even for
large distances. However, as suggested by Arovas \textit{et}~\textit{al.}~\cite{1988Arovas},
we can apply the single-mode approximation used to deduce the phonon-roton
curve in liquid 4He \cite{1985Girvin,Book-Feynman-1985}, in order
to study the excitations in this model. This is done by assuming that
a excited state at wave vector $q$ is given by\begin{equation}
|q\rangle\equiv S_{q}^{z}|\psi_{0}\rangle=N^{-1/2}\sum_{i}e^{\imath qr_{i}}S_{i}^{z}|\psi_{0}\rangle,\label{eq:Chap4-Single_mode_approximation}\end{equation}
where $|\psi_{0}\rangle$ is the exact \ac{GS} of the \ac{AKLT}
model. Within the single-mode approximation, the dynamical structure
factor defined as \begin{equation}
\mathcal{S}^{\alpha\beta}(q,\omega)=\int dt\int dre^{i\left(\omega t-qr\right)}\langle S_{r}^{\alpha}(t)S_{0}^{\beta}(0)\rangle\label{eq:Chap4-Dynamical_Structure_Factor}\end{equation}
is related with the static structure factor defined as $s^{\alpha\beta}(q)=\langle\psi_{0}|S_{-q}^{\alpha}S_{q}^{\beta}|\psi_{0}\rangle$
in the simple way $\mathcal{S}(q,\omega)\cong s(q)\delta(\omega-\omega_{q})$.
In \cite{1988Arovas} it was shown that, $\omega_{q}=E_{q}-E_{0}=5(5+3\cos q)/27$,
and that $s(q)=(10/27)(1-\cos q)/\omega_{q}$. The knowledge of the
dynamical structure factor allows us to compute the effective couplings
of Eq.~(\ref{eq:Chap4-EffecCouplings}) by inverse Fourier transform.
By inverting Eq. (\ref{eq:Chap4-Dynamical_Structure_Factor}) and
make use of the single-mode approximation (SMA), we can derive the
effective couplings. Indeed, we consider the right-hand side of Eq.~(\ref{eq:Chap4-EffecCouplings})
and make some manipulations using time translation invariance and
rotational symmetry,\begin{eqnarray}
\tilde{\chi}_{m+rz;mz}(0) & = & -\frac{1}{2i}\int_{-\infty}^{\infty}dte^{-0^{+}|t|}sign(t)\left[\langle S_{r}^{z}(t)S_{0}^{z}(0)\rangle-\langle S_{r}^{z}(t)S_{0}^{z}(0)\rangle^{*}\right]\label{eq:Chap4-SPIN1-aux1}\\
 & = & -\text{Im}\left[\int_{-\infty}^{\infty}dte^{-0^{+}|t|}sign(t)\langle S_{r}^{z}(t)S_{0}^{z}(0)\rangle\right]\label{eq:Chap4-SPIN1-aux2}\\
 & = & -\text{Im}\left[\int_{-\infty}^{\infty}dte^{-0^{+}|t|}sign(t)\int\frac{d\omega}{2\pi}\int\frac{dq}{2\pi}e^{-i\left(\omega t-qr\right)}\mathcal{S}^{zz}(q,\omega)\right]\label{eq:Chap4-SPIN1-aux3}\\
 & = & 2\int_{0}^{\infty}dte^{-0^{+}t}\int\frac{d\omega}{2\pi}\sin(\omega t)\int\frac{dq}{2\pi}\cos\left(qr\right)\mathcal{S}^{zz}(q,\omega)\label{eq:Chap4-SPIN1-aux4}\\
 & \underset{SMA}{=} & 2\int_{0}^{\infty}dte^{-0^{+}t}\int\frac{dq}{2\pi}\sin(\omega_{q}t)\cos\left(qr\right)s(q).\label{eq:Chap4-SPIN1-aux5}\end{eqnarray}
The latter expression can be recast into a convenient form by expressing
the static structure factor $s(q)$ as function of the excitation
energy $\omega_{q}$. This yields, \begin{equation}
\tilde{\chi}_{m+rz;mz}(0)=\int_{0}^{\infty}dte^{-0^{+}t}\int_{-\pi}^{\pi}\frac{dq}{2\pi}\cos\left(qr\right)\sin\left(\omega_{q}t\right)\left(a+\frac{b}{\omega_{q}}\right),\label{eq:Chap4-SPIN1-Adiabatic_suscepitibility}\end{equation}
with $a=-2/3$ and $b=80/81$. The integration in the time domain
gives $1/\omega_{q}$, and thus we are left with a one-dimensional
integral,\begin{equation}
\tilde{\chi}_{m+rz;mz}(0)=\int_{-\pi}^{\pi}\frac{dq}{2\pi}\frac{\cos\left(qr\right)}{\omega_{q}}\left(a+\frac{b}{\omega_{q}}\right),\label{eq:Chap4-SPIN1-aux6}\end{equation}
which can be evaluated via contour integration. Indeed, we extend
the integral to the interval $]-\infty,\infty[$ by means of an appropriate
change of variables, $x=\tan\left(q/2\right)$. Let us compute explicitly
the term involving $1/\omega_{q}$. Defining $a_{1}=a/(2\pi*25/27)$
and $a_{2}=a/(2\pi*15/27)$, \begin{equation}
I_{1}:=\int_{-\pi}^{\pi}dq\frac{\cos\left(qr\right)}{a_{1}+a_{2}\cos q}=2\int_{-\infty}^{\infty}dx\frac{\cos\left(2r\arctan x\right)}{\left(1+x^{2}\right)a_{1}+\left(1-x^{2}\right)a_{2}}.\label{eq:Chap4-SPIN1-aux7}\end{equation}
We can invert the relation between $x$ and $q$ given above to get
$q=i\ln\left(1-ix\right)-i\ln\left(1+x\right)$. Under this relation,
the integrand is simplified according to, \begin{equation}
2\cos\left(2r\arctan x\right)\rightarrow\left(\frac{1+ix}{1-ix}\right)^{r}+\left(\frac{1-ix}{1+ix}\right)^{r}.\label{eq:Chap4-SPIN1-aux8}\end{equation}
The conditions for the Jordan lemma are verified and thus we can extend
the integral to the complex plane:\begin{equation}
I_{1}=2\oint dz\frac{\left(\frac{1+iz}{1-iz}\right)^{r}}{\left(1+z^{2}\right)a_{1}+\left(1-z^{2}\right)a_{2}}.\label{eq:Chap4-SPIN1-aux9}\end{equation}
This integral has simple poles at $z=\pm i\sqrt{\left(a_{1}+a_{2}\right)/\left(a_{1}-a_{2}\right)}$
and a pole of order $r$ at $z=-i$. We then conveniently choose to
close the contour in the upper half plane. The residues theorem yields
$I_{1}=2\pi\sqrt{(1-\zeta)/(1+\zeta)}/\sqrt{a_{1}^{2}-a_{2}^{2}}$
with $\zeta=\sqrt{(a_{1}+a_{2})/(a_{1}-a_{2})}$. The calculation
of the remaining integral, \begin{equation}
I_{2}=b\int_{-\pi}^{\pi}\frac{dq}{2\pi}\frac{\cos\left(qr\right)}{\omega_{q}^{2}},\label{eq:Chap4-Aux-I2}\end{equation}
is similar and thus will not be reproduced. Adding both contributions
we arrive at, \begin{equation}
\tilde{\chi}_{m+rz;mz}(0)=\frac{1}{\Delta}(-1)^{r+1}\left(1+\frac{4r}{3}\right)e^{-\frac{r}{\xi_{AKLT}}},\label{eq:Chap4-SPIN1-Xhi_Zero}\end{equation}
where $\Delta=\omega_{\pi}$ is the gap of the chain in units of the
exchange interaction. The sign of the interaction mediated by the
\ac{AKLT} spin chain changes according to the distance between the
probes. This comes from the fact that the static correlations in this
spin chain have a similar alternation. Since the effective coupling
is given by $J_{p}^{2}\tilde{\chi}_{m+rz;mz}(0)$, we conclude that
at $T=0$ the probes get entangled whenever their distance corresponds
to a odd number of sites. 

What happens at $T>0$? The effective interaction vanishes so rapidly
with the distance {[}see Eq.~(\ref{eq:Chap4-SPIN1-Xhi_Zero}){]}
that any finite temperature will \textquotedbl{}thermalize\textquotedbl{}
the probes: their partial state will be a uncorrelated mixed state
in $2\otimes2$. We can give an estimate of the critical temperature
above which no \ac{LDE} is expected; to do so, we restrict ourselves
to temperatures much below the gap (remark that the opposite limit
would necessarily wash out the antiferromagnetic order and thus \ac{LDE}).
In this case, we do not expect real excitations of the spin chain
to be present: only the subspace of states described by $\hat{H}_{\text{eff}}$
{[}Eq.~(\ref{eq:Chap4-Heff_heisenberg}){]} will be populated and
then we may calculate the correlations between the probes using \begin{equation}
\rho_{ab}\simeq\frac{e^{-\beta J_{\text{eff}}\vec{\boldsymbol{\tau}}_{a}\cdot\vec{\boldsymbol{\tau}}_{b}}}{\text{Tr\ensuremath{\left[e^{-\beta J_{\text{eff}}\vec{\boldsymbol{\tau}}_{a}\cdot\vec{\boldsymbol{\tau}}_{b}}\right]}}},\label{eq:Chap4-SPIN1-rho_ab}\end{equation}
with $\beta^{-1}=k_{B}T$. This defines a temperature, $T^{*}\equiv1/(k_{B}\beta^{*})$,
above which entanglement disappears. \ac{PPT} yields: $\beta^{*}J_{mn}\simeq0.27$
where $J_{mn}$ is the effective coupling of probes interacting with
sites $m$ and $n$ and $|m-n|$ is an odd integer. This corresponds
to a very low temperature because $J_{mn}\sim(J_{p}^{2}/\Delta)f_{mn}$
is exponentially suppressed for large distances, $f_{mn}\simeq|m-n|\exp\left(-|m-n|/\xi_{AKLT}\right)$.
Thus, quasi-perfect \ac{LDE} will be present at $T\lll\Delta$, more
precisely when\begin{equation}
k_{B}T^{*}\ll\frac{J_{p}^{2}}{\Delta}re^{-r/\xi_{AKLT}}.\label{eq:Chap4-SPIN1-Cond_Temp_S=1}\end{equation}
This is to be compared to the \ac{AF} half-integer spin chain of
the previous section for which the criterion on temperature for quasi-perfect
\ac{LDE} reads, \begin{equation}
k_{B}T^{*}\ll J_{p}^{2}\tilde{\chi}_{r/L}(0).\label{eq:Chap4-SPIN1-Cond_Temp_S1/2}\end{equation}
The latter is a weaker constraint since $\chi$ is very large for
$r/L<O(1)$ (Fig.~\ref{fig:Chap4-HeisenbergSusceptibility}). We
then conclude that, at realistic temperatures (even if much smaller
than $\Delta/k_{B}$), distant probes will display more entanglement
when weakly coupled to the $S=1/2$ spin chain. This situation should
be rather insensitive to the actual microscopic model of the spin
bus for the fast decay of correlation functions of gapped systems
reflects into the large distance behaviour of the susceptibility $\chi$.
On the other hand, having a large intrinsic gap, the \ac{AKLT} chain
allows to consider larger couplings and yet being well inside perturbation
theory limits. It is in the interplay between a large gap (and thus
the possibility of considering higher $J_{p}$) and a large DC susceptibility
(and thus the possibility of quasi-perfect \ac{LDE}) that a better
performance may be achieved. At temperatures strictly zero both spin
chains will mediate quasi-perfect entanglement for $J_{p}^{2}\ll\Delta/\tilde{\chi}_{m+rz;mz}(0)$,
which is the main result of the present chapter.

\section{Concluding remarks\label{sec:Concluding-remarks}}

In this chapter we have learned that \ac{1D} spin chains are able
to entangle two distant spin-$1/2$ probes which did not interact
directly. At $T=0$ this entanglement is nearly maximal, an attractive
situation for quantum communication and computation. This was done
by implementing an adequate perturbation theory and holds whenever
the following conditions are met:
\begin{enumerate}
\item the bus-probe coupling is small compared to the relevant energy scales
of the many-body system, $J_{p}\ll J$;
\item the effective coupling is small compared to the gap, $J_{p}^{2}\tilde{\chi}_{mz;nz}(0)\ll\Delta$;
\item the system is effectively at zero-temperature, $k_{B}T\ll\Delta$;
\item the system's correlations are asymptotically antiferromagnetic so
that $\tilde{\chi}_{r}(0)>0,$ for large $r$.
\end{enumerate}
We have considered $SU(2)$ systems with full rotational symmetry
for they generally lead to larger amounts of \ac{LDE} even at $T=0$.
As an example of a system without full $SU(2)$ symmetry, think about
the Heisenberg spin-$1/2$ chain in field:\[
H_{0}=J\sum_{i=1}^{L-1}\vec{\boldsymbol{S}}_{i}\cdot\vec{\boldsymbol{S}}_{i+1}+h\sum_{i=1}^{L}S_{i}^{z}.\]
This system entails extra terms in the effective Hamiltonian even
in first order perturbation theory {[}see Eqs.~(\ref{eq:App4-DegPerturbTh-HamProjection})
and (\ref{eq:App4-DegPerturbTh-1stOrder}){]}. Also, the $SU(2)$
invariance of the second-order term in the effective Hamiltonian is
broken:\[
H_{\text{\text{eff}}}(h)=J_{p}^{2}\tilde{\chi}_{xx}(0)\left(\tau_{a}^{x}\otimes\tau_{b}^{x}+\tau_{a}^{y}\otimes\tau_{b}^{y}\right)+J_{p}^{2}\tilde{\chi}_{zz}(0)\tau_{a}^{z}\otimes\tau_{b}^{z}+J_{p}\langle S_{i}^{z}\rangle\left(\tau_{a}^{z}\otimes{\mathbbm{1}}_{b}+{\mathbbm{1}}_{a}\otimes\tau_{b}^{z}\right).\]
The local term above being proportional to $J_{p}$ dominates for
$h$ not to small, yielding aligned (and hence, separable) probes.
The case $h\ll J$ will also lead to less probe entanglement because
the susceptibilities are not isotropic. The present example, although
being a particular case, can be easily extended to other non-rotational
invariant spin chains, confirming our argument regarding the importance
of spin symmetry.

The present chapter focus on $2$ probes that interact with a many-body
system. This treatment, however, can be easily extended to include
$N>2$ probes. In the latter scenario, assuming global rotational
symmetry and homogeneous probes-system couplings, the effective Hamiltonian
{[}Eq.~(\ref{eq:Chap4-Heff_heisenberg}){]} reads instead, \begin{equation}
H^{(abc...)}=J_{p}^{2}\sum_{i<j}\tilde{\chi}_{iz;jz}(0)\vec{\boldsymbol{\tau}}_{i}\cdot\vec{\boldsymbol{\tau}}_{j}.\label{eq:Chap4-Extension_to_many_probes}\end{equation}
Within Hamiltonian (\ref{eq:Chap4-Extension_to_many_probes}) many
multipartite entangled states can be engineered by properly choosing
the many-body system and the probes locations. It would be interesting
to study the possibility of generating \emph{cluster states} --- these
states arise when spins of Ising quantum lattices are initially prepared
in a special state \cite{2001Briegel}; their entanglement is very
robust in the sense that $N/2$ qubits must be measured (for instance,
by the environment) as to completely turn the state separable. Cluster
states are very attractive for quantum computation purposes but are
very difficult to implement in the laboratory. The alternative of
sending probes that would weakly interact with a large system in equilibrium
(like those considered in the present chapter), which does not need
to be \textquotedbl{}reset\textquotedbl{} in a special state, deserves
further study. 

Still regarding the important case of $N=2$ some important questions
where not answered in the present chapter. Namely,
\begin{itemize}
\item What is the specific dependence of entanglement with the relevant
physical parameters? How does it vary with the coupling?
\item What happens in $2$-dimensions where symmetry breaking is known to
exist at $T=0$ for large systems?
\end{itemize}
Both questions will be addressed in the following chapter. Systems
where the gap approaches zero in the thermodynamic limit, accomplish
long-range (or \emph{quasi}-long range) correlations and thus large
susceptibilities, but the perturbative regime requires very weak couplings.
It turns out that increasing the coupling suppresses \ac{LDE} very
generally because of entanglement monogamy (Sec.~\ref{sec:Many-body-systems-as}),
suggesting that the validity of perturbation theory coincides with
the conditions leading to larger \ac{LDE}. Regarding \ac{2D} spin
lattices, we can expect \ac{LDE} to be lessened; models with symmetry
breaking in the thermodynamic limit will lead to a finite sub-lattice
magnetization, which would be perceived as a local magnetic field
by the probes. This should reduce entanglement even in the finite
size scenario for sufficiently large lattices. The next chapter will
disclose some surprises regarding the role of dimensionality in this
problem. 

Finally, we make a remark about the distinct roles of the bath in
the present problem and the opto-mechanical problem of Chapters~\ref{cha:MacroscopicEntang}
and \ref{cha:Stationary-optomechanical-entangl}: there we had two
large systems interacting directly and the continuous monitoring of
a bath resulted in a dilution of entanglement, whereas here we have
two small particles that not interact directly and get highly entangled
via the bath. Indeed, entanglement can arise between two quantum systems
which interact directly, but also can be transferred from highly correlated
many-body ground states to particles that otherwise would be separable.

\chapter{Finite temperature entanglement mediated by 2D antiferromagnets\label{cha:LDE_Finite_Temperature} }

\emph{This chapter is based on the following publication by the author:}
\begin{itemize}
\begin{onehalfspace}
\item Emergence of robust gaps in 2D antiferromagnets via additional spin-$1/2$
probes, AIRES FERREIRA, J. Viana Lopes, and J. M. B. Lopes dos Santos,
submitted to PRL (2009).\end{onehalfspace}

\end{itemize}

\section{Overview\label{sec:Chap5-Overview}}

The complex interplay between many degrees of freedom in solid state
represents a vast laboratory to test the quantum-classical boundary.
In the previous chapter an example of quantum behaviour was studied:
entangled ground states of many-body systems induce quantum correlations
in qubits --- a non-trivial effect that interactions mediated by mesoscopic
(and macroscopic) systems can actually produce more than classical
correlations between external parties. Apart from the fundamental
issues, solid state systems have been exploited for their applications,
especially those related to quantum computing and information processing. 

Solid state quantum computing relies upon the possibility of generating
some kind of entangled state, sufficiently robust not to suffer complete
decoherence from local noise or a global environment. One possibility
is to use the low-lying energy states of a many-body system as qubits,
albeit the decoherence mechanisms are generally very complex and far
away from being completely understood in this case. 

A good chance of maintaining quantum coherence, though, comes about
when the qubits states are separated from the rest of the spectrum
by a large gap; when this happens the effect of temperature, and thus
decoherence, is highly restrained. In this regards, widespread attention
has been paid to carbon-based solid state; for instance, diamond has
been shown to be a realistic candidate to quantum computing at room
temperature: spins in the vicinity of a single nitrogen-vacancy defect
in diamond can be manipulated to the extent of creating entangled
states between electron and nuclear spins that lasts for times as
large as milliseconds, although no simple method to scale up this
system is presently known \cite{2006Childress,2007Corbitt,2008Neumann}. 

Moreover, a considerable body of work has been devoted to systems
of spins-$1/2$ experiencing nearest-neighbor interactions, since
they can be used as models for universal quantum computation meeting
(at least partially) the DiVicenzo's requirements (see Sec.~\ref{sec:Many-body-systems-as}
and also \cite{2002Zhou,2003Benjamin,2006Yung,2007Friesen}). Spin
chains have been shown to be extremely versatile; for instance, they
allow to transfer reliably the state of a single qubit \cite{2003Bose}
and, as shown by \ac{DMRG} simulations \cite{2006Venuti,2007Venuti},
their ground-states are able to mediate an effective long-distance
interaction ultimately entangling distant spin probes. 

The \ac{LDE} phenomenon was the focus of the previous chapter, where
we learned that spin-$1/2$ probes, interacting locally with a large
spin system, can get highly correlated if they interact sufficiently
weakly as to not destroy completely the bulk's ground state: a very
small quantity of entanglement is extracted from the bulk which is
sufficient to force the probes towards a quasi-perfect singlet. Our
results constitute the first analytical support for \ac{LDE} mediated
by ground states of large many-body systems and suggest that, like
in the case of dynamical evolution of two qubits interacting with
a bath of harmonic oscillators \cite{2006Oh}, the role of the bus
gap is crucial. 

We have seen that a finite-size gap (in systems where the gap vanishes
in the thermodynamic limit) is convenient, whereas in systems with
intrinsic massive excitations a large gap results in fast decaying
correlations. The latter reduces the effective coupling, and hence
the robustness of entanglement at $T>0$. The results of Sec.~\ref{sec:A-perturbation-theory}
encourage us to consider more general possibilities and investigate
with more detail the role of the gap and the non-perturbative regime.

\section{Non-perturbative theory from adiabatic continuity\label{sec:Non-perturbative-approach}}

In Sec.~\ref{sec:A-perturbation-theory}, by means of standard condensed
matter methods, we computed the effective Hamiltonian of interaction
between probes that interact weakly with a gapped many-body system.
This was sufficient to conjecture quasi-perfect \ac{LDE} in the limit
of weak coupling, and also to give estimates of critical temperatures
above which probes entanglement would vanish {[}recall Eqs.~(\ref{eq:Chap4-SPIN1-rho_ab})-(\ref{eq:Chap4-SPIN1-Cond_Temp_S1/2}){]}.
However, as we have mentioned, the effective Hamiltonian does not
suffice to compute the density matrix (for reasons that will become
clear in the course of the present section). Here we overcome this
limitation by developing a simple non-perturbative theory describing
very faithfully numerical results for \ac{LDE}. 

The problem of computing the partial state of the probes is complex
even in perturbation theory; efficient finite temperature methods
apply almost exclusively to high-temperatures for any expansion of
the partition function, or density matrix, requires $\beta E\ll1$.
But in this limit, entanglement in systems endowed with a small Hilbert
space, like two spin-$1/2$ probes, is totally destroyed --- high-temperature
entanglement (\emph{i.e.}~above natural energy scales of the system)
can only be achieved in macroscopic systems or in systems where subpurity
is justified, as shown in the first part of this thesis (see Chapters~\ref{cha:MacroscopicEntang}~and~\ref{cha:Stationary-optomechanical-entangl}).
In $2\otimes2$ systems, separability is equivalent to \ac{PPT} (Sec.~\ref{sub:Entanglement_MixedStates})
defining a critical temperature ($T^{*}$) above which probes correlations
become classical: $\beta^{*}J_{ab}\simeq0.27$, where $J_{ab}$ is
the coupling energy between the qubits (Sec.~\ref{sub:LDE-AKLT}). 

We start by considering many-body systems of spins (which we will
designate often by bath) with rotational invariant Hamiltonian%
\footnote{A non-perturbative analytic approach comes with the expense of loss
of generality; whereas in the previous chapter we developed a perturbative
approach valid for any gapped system, here we will develop a theory
that applies to systems with rotational invariance. On the other hand,
as we have argued, rotational invariant buses with antiferromagnetic
order are precisely the systems potentially attaining large amounts
of \ac{LDE}. Thus, our results will apply in most of the interesting
scenarios.%
}, $H_{0}$, and a singlet, non degenerate ground state, $|\psi_{0}\rangle$.
As before, the two probes, $\boldsymbol{\tau}_{a}$ and $\boldsymbol{\tau}_{b}$,
are coupled to the bath by Heisenberg exchange interaction with strength
$J_{p}:=J\alpha$, through sites $A$ and $B$, respectively, \begin{equation}
V=\alpha J\left(\vec{\boldsymbol{S}}_{A}\cdot\vec{\boldsymbol{\tau}}_{a}+\vec{\boldsymbol{S}}_{B}\cdot\vec{\boldsymbol{\tau}}_{b}\right),\label{eq:Chap5-Eq1}\end{equation}
where $J$ denotes a energy scale of the bath (typically its exchange
interaction) and $\alpha$ is a dimensionless parameter. We make the
simple but crucial assumption that there is a one-to-one map of eigenstates
of the uncoupled system ($\alpha=0$) to the eigenstates of the full
Hamiltonian, \emph{i.e.}~we invoke adiabatic continuity \cite{Book-Anderson-1984}.
Hence, we define a canonical transformation between the two basis:\begin{equation}
|\psi_{m}\rangle\otimes|\chi_{\sigma}^{ab}\rangle=e^{-i\hat{S}}|\Psi_{m,\sigma}\rangle,\label{eq:Chap5-Eq2}\end{equation}
where $|\psi_{m}\rangle$ is a bath-only eigenstate, $|\chi_{\sigma}^{ab}\rangle$
a probe state, and $|\Psi_{m,\sigma}\rangle$ an eigenstate for finite
$\alpha$. Note that the generator $\hat{S}$ is an operator acting
on both probe and bath space. This map has important consequences;
the transformed Hamiltonian must have the form of a sum of a probe-only
term ($H_{p}$) with a bath only term ($H_{b}^{'}$), that is\begin{equation}
H_{S}=e^{-i\hat{S}}\left(H_{0}+V\right)e^{i\hat{S}}=H_{p}+H_{b}^{'},\label{eq:Chap5-Eq3}\end{equation}
\begin{figure}[tb]
\noindent \begin{centering}
\includegraphics[width=1\textwidth]{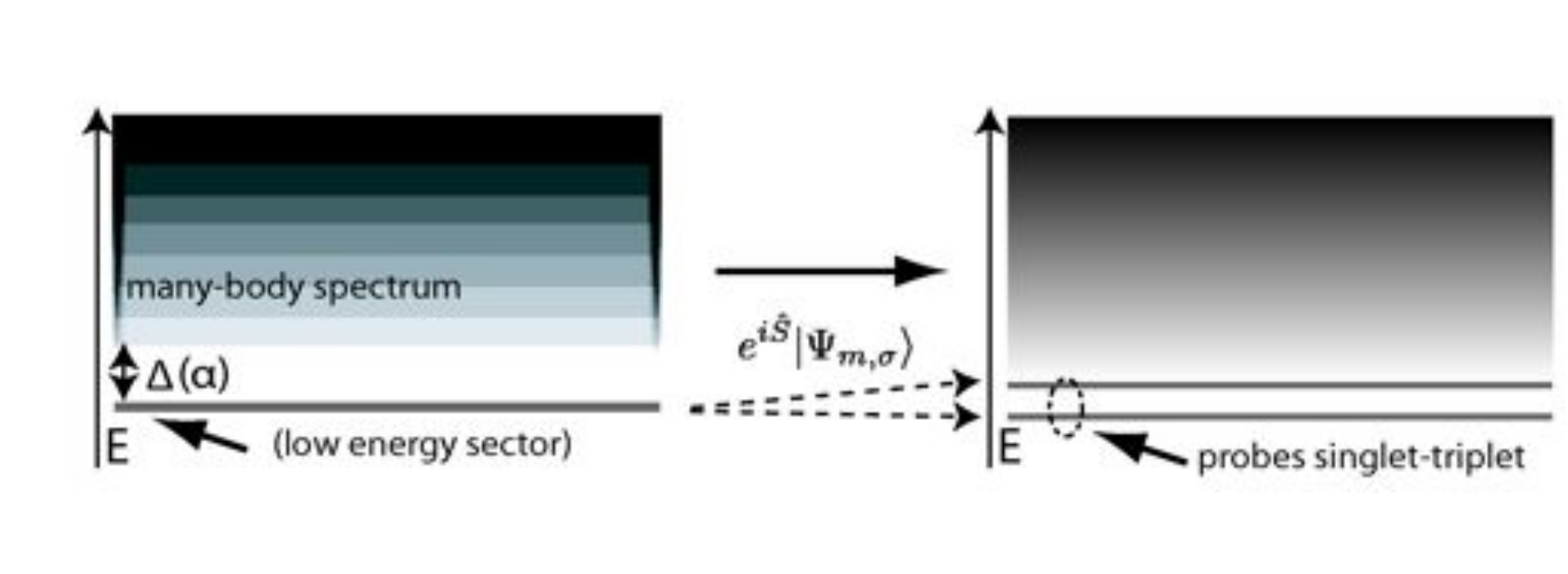}\caption[The canonical transformed spectrum under the robust gap assumption]{\label{fig:Chap5-CanonicalTransf}The schematic of the canonical
transformed many-body spectrum (at right) under the assumptions of
adiabatic continuity and robust gap, $\Delta(\alpha)>0$. If one succeeds
in finding the matrix elements of $\hat{S}$, the low-energy physics
of our problem will be described by an effective Hamiltonian containing
just the probes canonical singlet and triplet states.}

\par\end{centering}

\end{figure}

since the corresponding eigenstates are product states. We now add
the assumption that the lowest lying states, which map to a probe
singlet and probe triplet, \begin{eqnarray}
|\Psi_{0,s}\rangle & = & e^{-i\hat{S}}|\psi_{0}\rangle\otimes|\chi^{s}\rangle,\label{eq:Chap5-Eq4}\\
|\Psi_{0,m}\rangle & = & e^{-i\hat{S}}|\psi_{0}\rangle\otimes|\chi_{m}^{t}\rangle\qquad m=0,\pm1\label{eq:Chap5-Eq5}\end{eqnarray}
are well separated from states which map to excited states of the
bath by a \textquotedbl{}robust gap\textquotedbl{}, $\Delta(\alpha)$.
In other words, we assume that the bath$+$probes system has a low
energy manifold isomorphic to the probe space, well separated in energy
from the remaining energy spectrum. This could seem a strong restriction,
but, on rather general grounds, rotational invariance implies a non-degenerate
singlet of total spin, at least in systems with translational invariance
as implied by Marshall's theorems \cite{Book-Auerbach-1994}. Our
bath$+$probes problem breaks translation invariance (ultimately due
to the probes), but we will see here also the spectrum exhibits a
singlet-triplet low energy sector, well separated from the excited
states. 

The exact density-matrix of the qubits (the partial state of the probes,
$\rho_{ab}$) encodes the full-capabilities of a generic lattice as
a quantum bus (in particular, the possibility of \ac{LDE}):\begin{equation}
\rho_{ab}=\mathcal{Z}_{ab}^{-1}\textrm{Tr}_{i\in\mathcal{L}}\left[e^{-\beta\left(H_{0}+V\right)}\right].\label{eq:PartialState - Def}\end{equation}
The trace is made with respect to the degrees of freedom of the bath,
$\mathcal{L}$, and \begin{equation}
\mathcal{Z}_{ab}=\textrm{Tr}[\exp(-\beta H_{0}-\beta V)]\label{eq:PartitionFunction}\end{equation}
 is the system's partition function. In our case, global $SU(2)$
symmetry implies a very simple form for $\rho_{ab}$,\begin{equation}
\rho_{ab}\propto\exp\left(-\beta J_{ab}(\beta)\boldsymbol{\tau}_{a}\cdot\boldsymbol{\tau}_{b}\right),\label{eq:rho_ab}\end{equation}
where $J_{ab}(\beta)$ is the actual effective coupling of the probes
(not to be confused with the effective coupling in perturbation theory).
As soon as this function is known, bipartite entanglement can be computed
directly $\rho_{ab}$ using the negativity {[}or any other entanglement
monotone, see \emph{e.g.}~Eq.~(\ref{eq:Chap1_Negativity}){]}. The
$SU(2)$ spin symmetry of our problem implies that entanglement will
be given directly by the probe correlation: \begin{equation}
\langle\boldsymbol{\tau}_{a}\cdot\boldsymbol{\tau}_{b}\rangle=\textrm{Tr}\left[\rho_{ab}\boldsymbol{\tau}_{a}\cdot\boldsymbol{\tau}_{b}\right]=\frac{e^{-4\beta J_{ab}(\beta)}-1}{e^{-4\beta J_{ab}(\beta)}+1/3}.\label{eq:tau_atau_b}\end{equation}

\subsection{From renormalized spins to real spins\label{sub:From-renormalized-spins}}

In the previous chapter we approximated the partial state by $\rho_{ab}\propto e^{-\beta\hat{H}_{\text{eff}}}$,
and hence the effective coupling $J_{ab}(\beta)$ was temperature
independent {[}see Eq.~(\ref{eq:Chap4-SPIN1-rho_ab}) and Eq.~(\ref{eq:Chap4-Heff_heisenberg}){]},
\emph{i.e.}~we have assumed%
\footnote{To simplify notation, from now on, the probe's susceptibility {[}Eq.~(\ref{eq:Chap4-RelationBetweenEffectiveCouplingsandXhi}){]}
will be simply denoted by $\tilde{\chi}{}_{ab}^{zz}(0)$.%
} $J_{ab}(\beta)\simeq J_{p}^{2}\tilde{\chi}{}_{ab}^{zz}(0)$ for $\beta\Delta(0)\gg1$.
According to this prescription the probes correlation, $\langle\boldsymbol{\tau}_{a}\cdot\boldsymbol{\tau}_{b}\rangle$,
for instance, would be computed from $\rho_{ab}$ via,\begin{equation}
\langle\boldsymbol{\tau}_{a}\cdot\boldsymbol{\tau}_{b}\rangle=\frac{\textrm{Tr}\left[e^{-\beta H_{\text{eff}}}\left(\boldsymbol{\tau}_{a}\cdot\boldsymbol{\tau}_{b}\right)\right]}{\textrm{Tr}\left[e^{-\beta H_{\text{eff}}}\right]}.\label{eq:Chap5-Approximation}\end{equation}
The procedure $\rho_{ab}\propto\exp\left(-\beta\hat{H}_{\text{eff}}\right)$
only provides a rough estimate to the probes state, even in perturbation
theory ($J_{p}\ll\text{energy scales}$), though. Let us see why using
a quick argument based on the zero temperature limit and \ac{QI}
reasoning.

When the probes are antiferromagnetically correlated, $\tilde{\chi}{}_{ab}^{zz}(0)>0$,
according to the latter approximation, one has, \begin{equation}
\lim_{\beta\rightarrow\infty}\langle\boldsymbol{\tau}_{a}\cdot\boldsymbol{\tau}_{b}\rangle=-3,\label{eq:Chap5-Approximation2}\end{equation}
in which case, their partial state is a perfect singlet with zero
linear entropy: $S_{L}(\rho_{ab})=0$ {[}Eq.~(\ref{eq:Chap1_LinearEntropy}){]}.
This result does not depend on how accurate is our estimate of the
low-temperature limit of $J_{ab}$, for we would obtain the same value
provided that $J_{ab}>0$ is temperature independent. On the other
hand, the probes are also correlated (even if weakly) with the spin
bath by virtue of the local coupling {[}Eq.~(\ref{eq:Chap5-Eq1}){]}:
tracing the degrees of freedom of the bath {[}Eq.~(\ref{eq:PartialState - Def}){]}
then introduces residual entropy in $\rho_{ab}$ , and hence $S_{L}(\rho_{ab})>0$
contradicting Eq.~(\ref{eq:Chap5-Approximation2}).

To investigate the roots of this apparent ambiguity, we compute the
connected correlation between the probes under the adiabatic assumption
and compare it with (\ref{eq:Chap5-Approximation}) {[}obtained from
the approximation, $\rho_{ab}\propto\exp\left(-\beta\hat{H}_{\text{eff}}\right)${]}.
This is achieved by recasting the correlation {[}Eq.~(\ref{eq:tau_atau_b}){]}
into a form involving the effective Hamiltonian $H_{S}=H_{b}^{\prime}+H_{p}$.
To this end, we make use of the canonical transformation {[}Eq.~(\ref{eq:Chap5-Eq2}){]}
to get \begin{equation}
\langle\boldsymbol{\tau}_{a}\cdot\boldsymbol{\tau}_{b}\rangle=\frac{\textrm{Tr}\left[e^{-\beta H_{S}}\left(e^{-i\hat{S}}\boldsymbol{\tau}_{a}\cdot\boldsymbol{\tau}_{b}e^{i\hat{S}}\right)\right]}{\textrm{Tr}\left[e^{-\beta H_{S}}\right]}.\label{eq:Chap5-Eq6}\end{equation}
From our assumption that the lowest energy sector is mapped to a probe
singlet and probe triplet {[}see Eqs.~(\ref{eq:Chap5-Eq4})-(\ref{eq:Chap5-Eq5}){]},
it is clear that the canonical transformed Hamiltonian must be a scalar
in the probes operators, that is $H_{p}\sim\boldsymbol{\tau}_{a}\cdot\boldsymbol{\tau}_{b}$,
hence having the same form we found in perturbation theory {[}$H_{\text{eff}}\sim\boldsymbol{\tau}_{a}\cdot\boldsymbol{\tau}_{b}${]}.
The bottom line comes from the observation that the scalar product
entering in $H_{S}$ (via the term $H_{p}$) is not the same operator
inside brackets {[}in Eq.~(\ref{eq:Chap5-Eq6}){]}, for \begin{equation}
e^{-i\hat{S}}\boldsymbol{\tau}_{a}\cdot\boldsymbol{\tau}_{b}e^{i\hat{S}}=\boldsymbol{\tau}_{a}\cdot\boldsymbol{\tau}_{b}-i\left[\hat{S},\boldsymbol{\tau}_{a}\cdot\boldsymbol{\tau}_{b}\right]+...\label{eq:Chap5-Eq7}\end{equation}
We conclude that, in second order perturbation theory, the correlation
computed from Eq.~(\ref{eq:Chap5-Approximation}) will just yield
a valid approximation to the exact correlation (\ref{eq:tau_atau_b})
in situations where $e^{-i\hat{S}}\boldsymbol{\tau}_{a}\cdot\boldsymbol{\tau}_{b}e^{i\hat{S}}\simeq\boldsymbol{\tau}_{a}\cdot\boldsymbol{\tau}_{b}+O(\alpha^{3})$.
In general, however, $\hat{S}$ (containing bath and probes operators)
will give rise to terms not proportional to $\boldsymbol{\tau}_{a}\cdot\boldsymbol{\tau}_{b}$;
we can anticipate they will reduce the correlation (\ref{eq:Chap5-Approximation2})
at any temperature and henceforth also diminish the \ac{LDE} capability
of generic spin baths. 

In summary, the effective Hamiltonian correctly accounts for the shifts
in energy due to the probes and for the possibility of \ac{LDE},
but, by itself, fails in giving the exact probes partial state; one
must take also in consideration the way the operators change (even
in perturbation theory). Doing so, one finds that, in the previous
chapter, the renormalized spins were taken as being original spins,
which is only an approximation. 

Further insight on the canonical transformation and its relation with
perturbation theory can be found in Appendices~\ref{sec:The-Schrieffer-Wolff-canonical},
\ref{sec:RenormalizationProcedure} and \ref{sec:The-canonical-parameters}.
In what follows we analyze the effect of spin renormalization in the
partial state of the probes.

\subsection{The canonical parameters\label{sub:The-canonical-parameters}}

We learned that the spin operators must be renormalized if one wishes
to get averages corresponding to real spin degrees of freedom (that
is, spin operators have to be conveniently transformed according to
$\hat{S}$). This is consistent with perturbation theory (Appendix~\ref{sec:The-Schrieffer-Wolff-canonical})
for the wave functions also change according to: $|\Psi_{0}^{(S_{2})}\rangle=e^{-iS_{2}}\left|\Psi_{0}\right\rangle ,$
where $S_{2}$ is an appropriate generator in second order perturbation
theory. 

We move gears to the study of the probe correlation (\ref{eq:Chap5-Eq6})
under the robust gap assumption. Also, connections to previous perturbative
results will be made when relevant. We start by introducing new notation
for the renormalized spins; we will denote them as $\boldsymbol{\tau}_{a(b)}^{R}$
to distinguish from the real spin operators {[}those appearing in
Eq.~(\ref{eq:rho_ab}){]}:\begin{equation}
\boldsymbol{\tau}_{a(b)}^{R}:=e^{-i\hat{S}}\boldsymbol{\tau}_{a(b)}e^{i\hat{S}}.\label{eq:Chap5-RenormalizedSpins}\end{equation}
Using symmetry alone, we can relate the scalar product involving real
and renormalized spins. The formal derivation is done in Appendix~\ref{sec:RenormalizationProcedure}
{[}see Eqs.~(\ref{eq:App5-Renormalization-9})-(\ref{eq:tau_tau_can}){]};
here is sufficient to observe that taking averages with respect to
the spin bath \ac{GS}, $\langle\psi_{0}|...|\psi_{0}\rangle:=\langle...\rangle_{\text{bath}}$,
effectively integrates out the bath and yields the low-energy physics
of the probes, when the robust gap assumption is verified {[}see Eqs.~(\ref{eq:Chap5-Eq4})
-(\ref{eq:Chap5-Eq5}) and comments therein{]}. We have, \begin{equation}
\langle e^{-i\hat{S}}\boldsymbol{\tau}_{a}\cdot\boldsymbol{\tau}_{b}e^{i\hat{S}}\rangle_{\text{bath}}=\eta{\mathbbm{1}}_{2\otimes2}+(1-\Phi)\boldsymbol{\tau}_{a}\cdot\boldsymbol{\tau}_{b},\label{eq:Chap5-Example1}\end{equation}
with $\eta=\eta(\alpha)$ and $\Phi=\Phi(\alpha)$ real and bounded.
The reason why other operators do not enter in formula (\ref{eq:Chap5-Example1})
is because the canonical transformation $\hat{S}$ will necessarily
produce rotational invariant probe operators (and there are just two
in $2\otimes2$, namely the identity and the scalar product). 

The probes correlation is obtained by averaging the latter equation.
It is instructive to consider the zero temperature case,\begin{eqnarray}
\langle\boldsymbol{\tau}_{a}\cdot\boldsymbol{\tau}_{b}\rangle_{T=0} & = & \langle\Psi_{0,s}|\boldsymbol{\tau}_{a}\cdot\boldsymbol{\tau}_{b}|\Psi_{0,s}\rangle\label{eq:Chap5-new1}\\
 & = & \langle\chi^{s},\psi_{0}|\boldsymbol{\tau}_{a}^{R}\cdot\boldsymbol{\tau}_{b}^{R}|\psi_{0},\chi^{s}\rangle\label{eq:Chap5-new2}\\
 & = & -3+\eta+3\Phi.\label{eq:Chap5-new3}\end{eqnarray}
The last equality implies the restriction: $4\ge\eta+3\Phi\ge0$.
The scenario of perfect entanglement, $E(\rho_{ab})=1$, requires
$\eta+3\Phi=0.$ Indeed, considering the approximation of the previous
chapter, namely Eq.~(\ref{eq:Chap5-Approximation}), is equivalent
to take $\boldsymbol{\tau}_{a(b)}^{R}\simeq\boldsymbol{\tau}_{a(b)}$,
which results in quasi-perfect \ac{AF} correlations for $T=0$:\begin{equation}
\langle\boldsymbol{\tau}_{a}\cdot\boldsymbol{\tau}_{b}\rangle_{T=0}=\langle\Psi_{0,s}|\boldsymbol{\tau}_{a}\cdot\boldsymbol{\tau}_{b}|\Psi_{0,s}\rangle\simeq\langle\chi^{s},\psi_{0}|\boldsymbol{\tau}_{a}\cdot\boldsymbol{\tau}_{b}|\psi_{0},\chi^{s}\rangle=-3.\label{eq:Chap5-new4}\end{equation}
This approximation is strictly valid when $\eta\simeq\Phi\simeq0$.
We will refer to $\eta$ and $\Phi$ as the canonical parameters (or
canonical corrections); they relate the correlation of the renormalized
spins and the correlation of real spins: \begin{equation}
\langle\boldsymbol{\tau}_{a}\cdot\boldsymbol{\tau}_{b}\rangle=\eta+(1-\Phi)\langle\boldsymbol{\tau}_{a}\cdot\boldsymbol{\tau}_{b}\rangle_{can}.\label{eq:Chap5-RenormalizedAverage}\end{equation}
The term $\langle\boldsymbol{\tau}_{a}\cdot\boldsymbol{\tau}_{b}\rangle_{can}$
is obtained via the Gibbs ensemble constructed directly from $H_{p}$
and equals the correlations obtained via the approximation (\ref{eq:Chap5-Approximation}):
\begin{equation}
\langle\boldsymbol{\tau}_{a}\cdot\boldsymbol{\tau}_{b}\rangle_{\mathrm{can}}=\frac{\textrm{Tr}_{p}\left[e^{-\beta H_{p}}\boldsymbol{\tau}_{a}\cdot\boldsymbol{\tau}_{b}\right]}{\textrm{Tr}_{p}\left[e^{-\beta H_{p}}\right]}=\frac{e^{-\beta J_{can}}-1}{e^{-\beta J_{can}}+1/3},\label{eq:Chap5-J_can}\end{equation}
where we have employed the definition \begin{equation}
J_{can}:=\langle\Psi_{0,m}|H|\Psi_{0,m}\rangle-\langle\Psi_{0,s}|H|\Psi_{0,s}\rangle.\label{eq:Chap5-Example4}\end{equation}
Contrary to the real effective coupling, $J_{ab}(\beta)$, the canonical
coupling, $J_{can}$, is now effectively temperature independent.
Its value in second order theory is $J_{ab}/4$ {[}with $J_{ab}=J_{p}^{2}\tilde{\chi}{}_{ab}^{zz}(0)$,
see for instance Eq.~(\ref{eq:Chap4-Heff_heisenberg}){]}. Equation~(\ref{eq:Chap5-RenormalizedAverage})
is however more general since it does not depend on the particular
perturbation scheme: $J_{can}$ is in fact the gap at any order by
construction {[}see the definition of $H_{S}$, Eq.~(\ref{eq:Chap5-Eq3}):
its energy scale, $4J_{can}$ in our notation, gives directly the
gap between the singlet and triplet low-energy sectors; see also Eq.~(\ref{eq:App5-Renormalization-ExactGap})
and discussion therein{]}. 

Adiabatic continuity then entails a curious result: by measuring the
correlations of the probes at different temperatures we can derive
the many-body gap (in the presence of the probes), which equals the
$T=0$ DC spin response function, $\tilde{\chi}{}_{ab}^{zz}(0)$,
in the weak coupling regime {[}Eq.~(\ref{eq:Chap4-RelationBetweenEffectiveCouplingsandXhi}){]}.

Even in perturbation theory, $\eta$ and $\Phi$ will play a role,
and thus Eq.~(\ref{eq:Chap4-SPIN1-rho_ab}) is only valid if the
canonical corrections are small {[}see Appendix~\ref{sec:RenormalizationProcedure}{]}.
We thus expect quasi-perfect \ac{LDE} to correspond to a very special
scenario, that of $\eta+3\Phi\ll1$. Despite that, as long as the
low-lying states are well protected by the remaining spectrum and
the canonical corrections are not too large, we can still find a good
amount of \ac{LDE}, especially if the probe triplet is not populated
at all, which happens when $k_{B}T\ll J_{can}$. 

Finally, we relate the real effective coupling defined by Eq.~(\ref{eq:rho_ab})
with the canonical parameters. This is accomplished by equalling the
right-hand sides of Eqs.~(\ref{eq:tau_atau_b}) and (\ref{eq:Chap5-RenormalizedAverage})
and solving for $J_{ab}(\beta)$ . We get,\begin{equation}
J_{ab}(\beta)=\frac{1}{4\beta}\ln\left[\frac{3(\Phi-\eta)+(4-3\Phi-\eta)\exp(\beta J_{can})}{4-\Phi+\eta+(\Phi+\eta/3)\exp(\beta J_{can})}\right].\label{eq:Chap5-Jeff as function of Jcan}\end{equation}
We thus have achieved a parameterization of the temperature dependence
of $\rho_{ab}$ as function of three parameters,\begin{equation}
\rho_{ab}(\beta)=\frac{{\mathbbm{1}}_{ab}}{4}+\frac{1}{4}\left(\eta/3+\left(1-\Phi\right)\frac{e^{-\beta J_{can}}-1}{3e^{-\beta J_{can}}+1}\right)\boldsymbol{\tau}_{a}\cdot\boldsymbol{\tau}_{b},\label{eq:Chap5-rho_ab_exacto}\end{equation}
which can be easily computed in perturbation theory (Appendix~\ref{sec:The-canonical-parameters}):
\begin{eqnarray}
J_{can} & \simeq & (2J\alpha)^{2}\tilde{\chi}_{ab}^{zz}(0)\label{eq:Chap5-Can1}\\
\Phi & \simeq & (2J\alpha)^{2}\sum_{m>0}\left(\frac{1}{E_{m}-E_{0}}|\langle\psi_{0}|\left(\boldsymbol{S}_{A}^{z}-\boldsymbol{S}_{B}^{z}\right)|\psi_{m}\rangle|\right)^{2}\label{eq:Chap5-Can2}\\
\eta & \simeq & 0+O(\frac{\alpha^{4}J^{4}}{\Delta^{4}}).\label{eq:Chap5-Can3}\end{eqnarray}
The states of the spin bath are denoted by $|\psi_{m}\rangle$ (with
eigenenergy $E_{m}$ ) and $\Delta:=E_{1}-E_{0}$. The canonical parameters,
for small $\alpha$, can be computed by diagonalizing the spin bath
Hamiltonian, $H_{0}|\psi_{k}\rangle=E_{k}|\psi_{k}\rangle$. This
is however only possible in a few models whose analytical solution
is known (\emph{e.g.}~the \ac{1D} $XY$ model). In general, whether
the canonical parameters describe the correlations of the probes accurately,
for a given spin model, must be investigated by comparing result (\ref{eq:Chap5-RenormalizedAverage})
{[}or equivalently, (\ref{eq:Chap5-Jeff as function of Jcan}){]}
with numerical simulations. We recall that these results will only
describe accurately \ac{LDE} of probes interacting with large lattices
if adiabatic continuity holds and a robust gap is available. The following
section will show an impressive agreement between the canonical theory
and simulations in a large family of \ac{AF} spin systems, which
therefore fulfil these two conditions.

A formal derivation of equations~(\ref{eq:Chap5-Can2})-(\ref{eq:Chap5-Can3})
is found in Appendix~\ref{sec:The-canonical-parameters}; Eq.~(\ref{eq:Chap5-Can1})
is derived in Appendix~\ref{sec:The-Schrieffer-Wolff-canonical}
via the canonical formalism. Finally, the renormalization of the spins
is explained with detail in Appendix~\ref{sec:RenormalizationProcedure}.

\section{LDE in 2D}

Promising advances in the engineering of atomic structures and optical
lattices, where finite spin systems are effectively realized in the
laboratory, encourage the consideration of more general possibilities.
Indeed, we now consider \ac{AF} spin systems, ranging from \ac{1D}
chains to square lattices, and demonstrate the emergence of thermal
probe entanglement. We link this phenomenon to the opening of robust
gaps in the full many-body spectrum by means of the theory of the
previous section. Robust gaps implies negligible thermal occupation
of excited bus states in the entire range of temperatures in which
the probes are entangled; we will see this allows \ac{LDE} at higher
temperatures than previously considered possible. 

As in the previous chapter, we take the probes-bus coupling to be
$SU(2)$ invariant (\emph{i.e.}~Heisenberg type). These interactions
entail universal quantum computation \cite{2000DiVincenzo-b} and
are commonly realized in nature (\emph{e.g.}~in the parent compounds
of copper-oxide high-temperature superconductors, such as the undoped
insulator $La_{2}CuO_{4}$ \cite{Review-Manousakis}; in electronically
coupled quasi- \ac{1D} chains such as $CuGeO_{3}$ \cite{1993Hase};
in the Mott insulating one-dimensional perovskite, $KCuF_{3}$ \cite{1970Hirakawa,1980Satija};
and also more recently in linear chains of $\sim10$ manganese atoms
in engineered structures \cite{2006Hirjibehedin}). 

\begin{figure}[tb]
\noindent \begin{centering}
\includegraphics[width=1\textwidth]{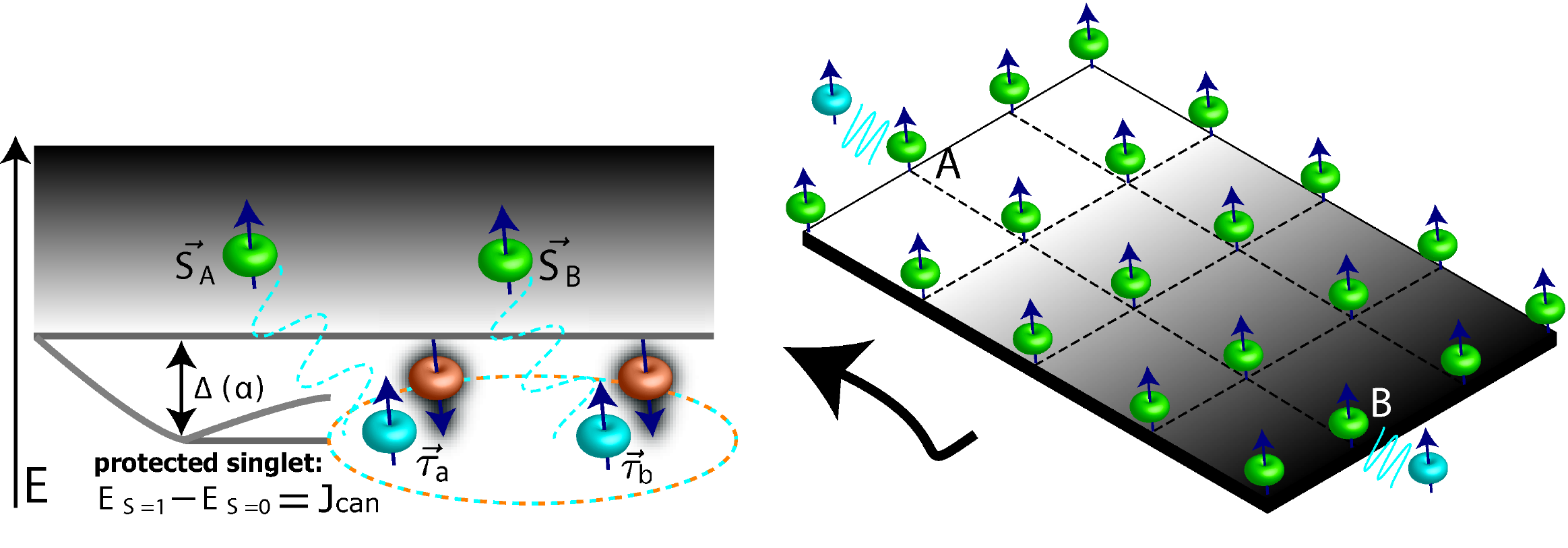}
\par\end{centering}

\caption[Opening of robust gaps in 2D antiferromagnets by extra spin probes]{\label{fig:Chap5-Lattice}Left - Schematic picture of the opening
of a robust gap, $\Delta(\alpha)\gg k_{B}T$, by two spin-$1/2$ probes
that couple locally to the spin bath with arbitrary strength $\alpha J$.
This is observed in all the \ac{AF} lattices considered in the present
section. If adiabatic continuity holds and the singlet is localized
near the additional probes, they will be highly entangled even at
large distances. In this case the singlet has a protection gap (triplet-singlet
energy separation, $J_{can}$) which determines the robustness of
\ac{LDE} regarding temperature. Right - Schematic of the total system:
the probes (blue) interact locally with a \ac{2D} lattice through
sites $A$ and $B$. How does the dimensionality of the spin bath
will change the entanglement picture?}

\end{figure}

Our systems consist of \ac{2D} finite lattices $\mathcal{L}$, with
$N=l\times n_{c}$ spins-$1/2$ and two extra probes, where $l$ is
the number of longitudinal sites and $n_{c}$ stands for the number
of coupled chains, varying from $n_{c}=1$ (spin chain) to $n_{c}=l$
(square lattice), see Fig.~\ref{fig:Chap5-Lattice} for a possible
geometry. The Hamiltonian of the lattice is \begin{equation}
H_{0}=J\sum_{\langle i,j\rangle}\vec{\boldsymbol{S}}_{i}\cdot\vec{\boldsymbol{S}}_{j},\label{eq:Chap5-LatticeHamiltonian}\end{equation}
with $J>0$. The qubit probes interact with the spins at the boundary
of the most central chain (see Fig.~\ref{fig:Chap5-Lattice}) through
an isotropic interaction {[}Eq.~(\ref{eq:Chap5-Eq1}){]}. We expect
a significant change in \ac{LDE} from the common one-dimensional
scenario analyzed in the previous chapter (and numerically in \cite{2006Venuti,2007Venuti}),
as the physics of a \ac{2D} spin bath is very distinct. 

In particular, the 2-leg ladder chain has an Haldane gap \cite{Book-Auerbach-1994}
which should play against a large $J_{ab}$ since very \emph{massive
excitations}, $\Delta\simeq0.504J$, make the correlations die particularly
fast \cite{1994White}. For these systems we are not able to make
exact analytic computations. Thus, we will rely on \ac{QMC} simulations
\cite{ALPS-1,ALPS2,ALPS3,ALPS4,ALPS5} %
\footnote{These simulations were performed with the library \textquotedbl{}looper\textquotedbl{}
from the ALPS (Algorithms and Libraries for Physics Simulations) project. %
} and compare the results with the theoretical prediction {[}Eq.~(\ref{eq:Chap5-Jeff as function of Jcan}){]}. 

\begin{figure}[tb]
\noindent \begin{centering}
\includegraphics[width=1\textwidth]{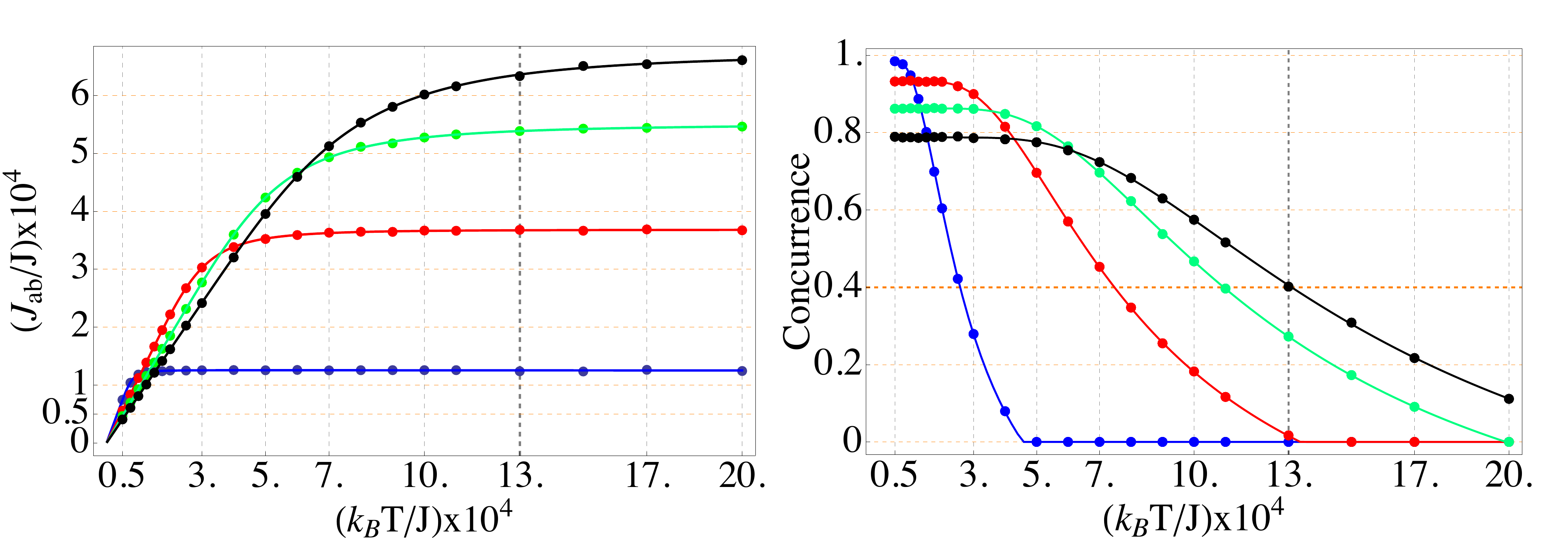}
\par\end{centering}

\caption[Effective coupling as function of temperature for representative lattices]{\label{fig:Jeff_Entang_Temp}The points in the plot (Left) show $J_{ab}$
as function of the temperature from \ac{QMC} simulations for $\alpha=0.05$.
The lines (Right) stand for the fit with the expression given in Eq.~(\ref{eq:Chap5-Jeff as function of Jcan});
with $n_{c}=1$ (blue), $n_{c}=10$ (red), $n_{c}=15$ (blue) and
$n_{c}=20$ (black). All curves saturate for the highest temperatures
{[}Eq.~(\ref{eq:Jeff-high_temp_expan}){]} and therefore entanglement
(Right) vanishes much before our method becomes inaccurate. The agreement
between the \ac{QMC} data is excellent resulting in a average deviation
of $\sim0.1-1\%$ depending on the lattice. }

\end{figure}

We present the results for small probe-bath coupling, $\alpha=0.05$,
before venturing away from perturbation theory. The \ac{1D} scenario
is of special interest as we have a conjecture in this case {[}Eq.~(\ref{eq:Chap4-QuasiPerfectLDE}){]};
our choice of coupling entails, for $L=20$ and $n_{c}=1$: $J_{p}^{2}/\Delta\simeq0.015J$
(with $\Delta$ extracted from the \ac{DMRG} results of \cite{1994White});
well inside perturbation limits and thus our conjecture should hold.
The numerical results indeed show probes almost maximally entangled
(see Fig.~\ref{fig:Jeff_Entang_Temp}), validating quasi-perfect
\ac{LDE} in the \ac{1D} system: the table below shows the canonical
parameters for representative lattices when $\alpha=0.05.$ 

\begin{table}[H]
\noindent \begin{centering}
\begin{tabular}{c||c|c|c|c}
 & Intrinsic (Haldane) gap & The probes gap: $J_{can}$ & $\Phi$ correction & $\eta$ correction\tabularnewline
\hline
\hline 
spin chain & no; $\Delta/J\simeq3.2/L$ {[}{*}{]} & $5.07*10^{-4}J$ & $1.03*10^{-2}$ & $6.23*10^{-4}$\tabularnewline
\hline 
square lattice & no; broken phase for $L\gg1$ & $3.04*10^{-3}J$ & $1.46*10^{-1}$ & $-1.34*10^{-2}$\tabularnewline
\end{tabular}
\par\end{centering}

\caption{The canonical parameters in representative systems for $\alpha=0.05$.
The canonical parameters are calculated by fitting the \ac{QMC} data
for different temperatures with Eq.~(\ref{eq:Chap5-RenormalizedAverage}).
{[}{*}{]} This expression was taken from Ref. \cite{1994White}.\label{tab:Chap5_CanonicalParameters}}

\end{table}

We observe almost perfect \ac{AF} correlations in the \ac{1D} scenario,
since $3\Phi+\eta=O(10^{-2})$ is very small and thus $\langle\boldsymbol{\tau}_{a}\cdot\boldsymbol{\tau}_{b}\rangle\simeq-3$
{[}Eq.~(\ref{eq:Chap5-Approximation2}){]}, and no entanglement mediated
by the $2$-leg and $4$-leg ladders. Regarding these ladders, the
numerical results are well-inside what we could expect: the probe
correlation is nearly zero in the whole temperature range, a consequence
of a very small $J_{can}$ (the fits yield $J_{can}\sim10^{-5}J$)
resulting from an exponentially decaying $\tilde{\chi}_{ab}^{zz}(0)$
(recall discussion of the gapped \ac{AKLT} model, Sec.~\ref{sub:LDE-AKLT}).
Indeed, the probes cannot take advantage of the \ac{AF} coupling
mediated by the system: the probes gap, $J_{can}$, is too small so
that entanglement survives at finite (even though small) temperatures
$T>0$.

The values in the table show a curious property of the \ac{2D} spin
system: a remarkable high $J_{can}$ for a system with such a small
gap. Moreover, the canonical correction $\Phi$ is sufficiently small
as to not delocalize completely the probes singlet. This leads, with
the help of a strong $J_{can}$ , to more robust \ac{LDE} against
temperature! Let us investigate these issues more carefully. 

As an entanglement monotone we adopt the \emph{concurrence} {[}see
Eq.~(\ref{eq:Chap1-EntangFormation}) for definition and Ref.~\cite{1997Wootters}
for a closed expression{]} because it yields an handy expression in
systems with rotational symmetry:\begin{equation}
E_{C}=\max\left[0,|\langle\boldsymbol{\tau}_{a}\cdot\boldsymbol{\tau}_{b}\rangle|/3-\langle\boldsymbol{\tau}_{a}\cdot\boldsymbol{\tau}_{b}\rangle/6-1/2\right]\in[0,1].\label{eq:Chap5-Concurrence}\end{equation}
Using the values of $\langle\boldsymbol{\tau}_{a}\cdot\boldsymbol{\tau}_{b}\rangle$
extracted from the \ac{QMC} simulations, we can compute \ac{LDE}
via the concurrence. For instance, the values of the canonical corrections
in table \ref{tab:Chap5_CanonicalParameters} yield for the spin chain
at zero temperature: $E_{C}\simeq0.985$. In this case the singlet
is localized at the boundary sites (the probes) and thus the mechanism
for \ac{LDE} is optimal; since the singlet is already localized at
the probes, before the canonical transformation, the renormalization
of spins will not change much the density matrix given by the old
expression of Eq.~(\ref{eq:Chap4-SPIN1-rho_ab}). 

\begin{figure}[tb]
\begin{centering}
\includegraphics[width=0.6\textwidth]{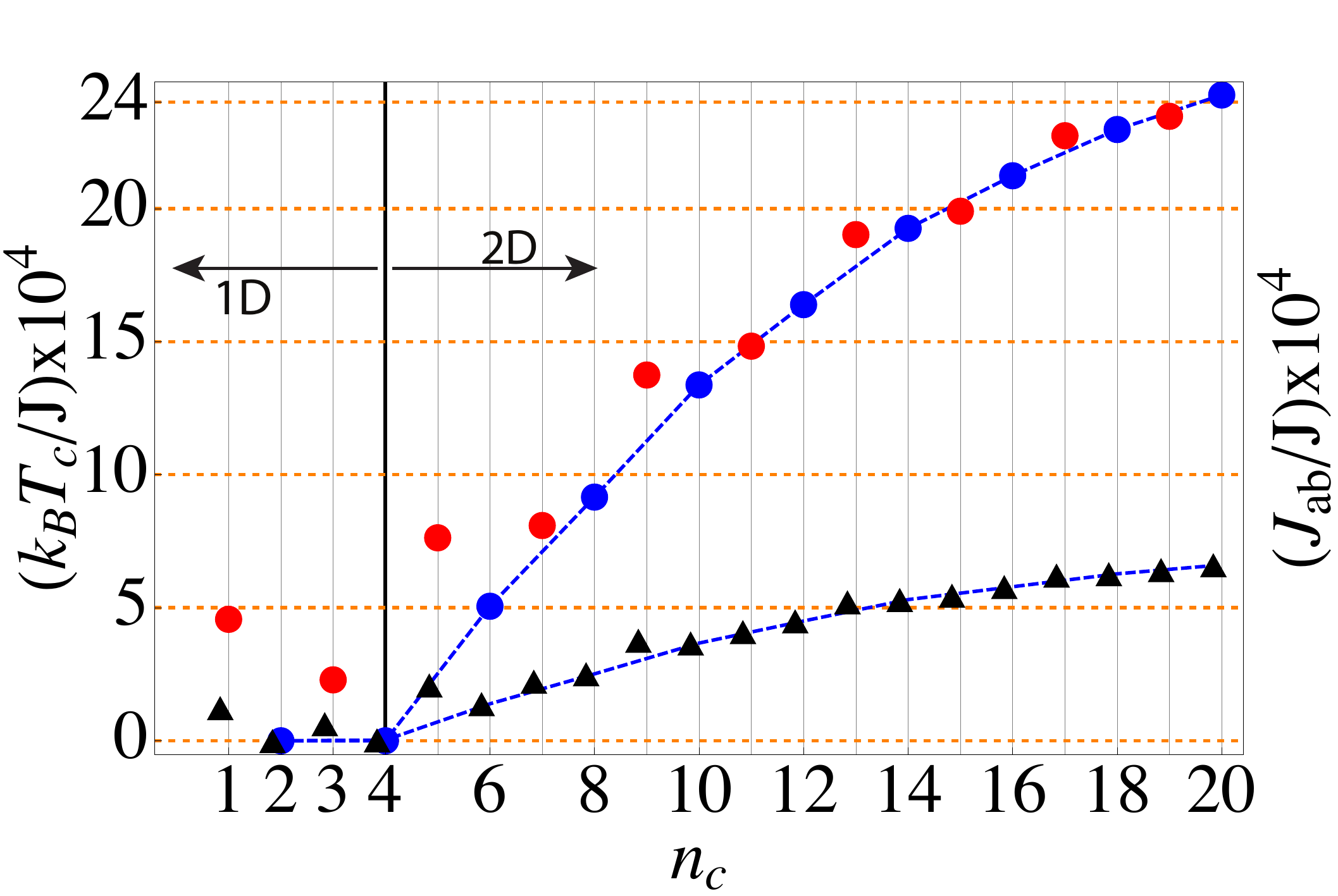}
\par\end{centering}

\caption[Effective coupling between two spin probes mediated by 2D antiferromagnets]{\label{fig:RobustnessGap}Black triangles: effective coupling, $J_{ab},$
at a distance $l=20$, as a function of $n_{c}$, number of transverse
chains, and $k_{B}T=2\times10^{-3}J$ ; blue and red dots: critical
temperature above which \ac{LDE} vanishes. The error bars from \ac{QMC}
cannot be seen as they are typically below $1$\%.}

\end{figure}

\begin{figure}[tb]
\noindent \begin{centering}
\includegraphics[width=1\textwidth]{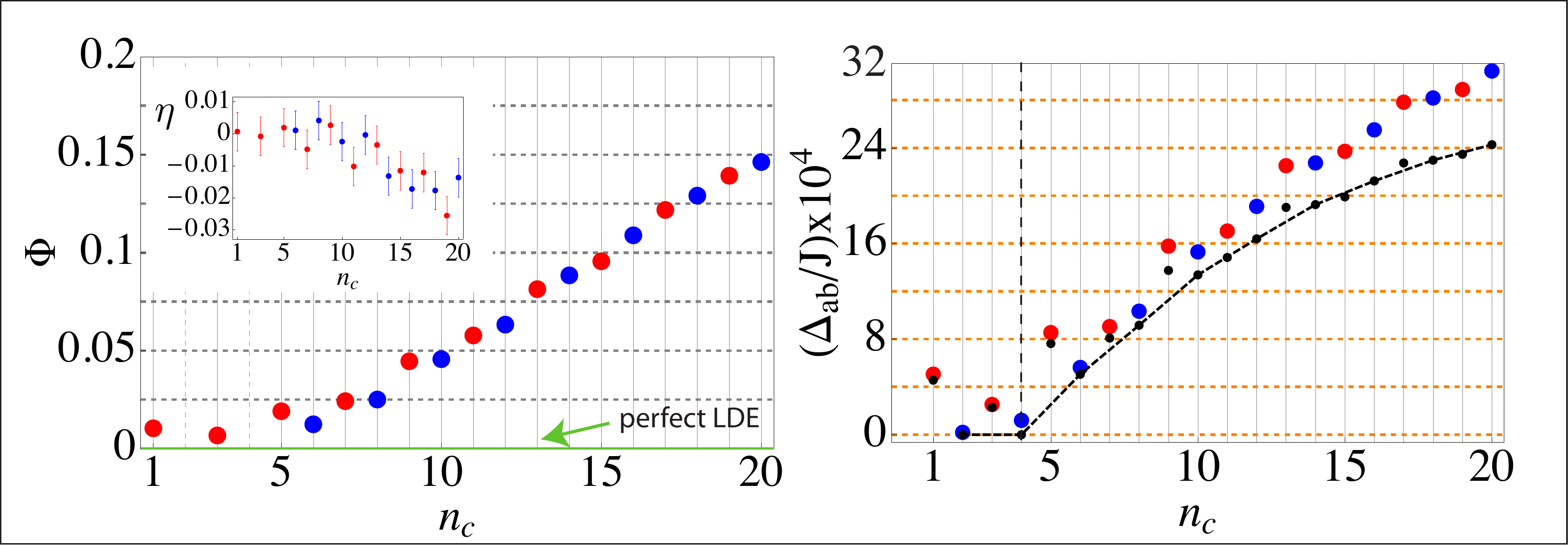}
\par\end{centering}

\caption[Canonical parameters for the entire family of 2D spin lattices]{\label{fig:Phi}Left - Canonical correction $\Phi$ for the same
family of spin lattices of Fig.~\ref{fig:RobustnessGap}. The inset
shows that the $\eta$ correction is negligible compared to $\Phi$;
the bars represent an estimate of the error due to \ac{QMC} fluctuations.
The $n_{c}=2$ and $n_{c}=4$ systems are not represented since the
data does not provide reliable values for canonical corrections. Right
- The singlet protection gap for finite probe coupling ($\alpha=0.05$)
in blue and red dots. The critical temperature is represented (small
black dots) for comparison.}

\end{figure}

We represent the most important physical parameters for the entire
family of lattices in Figure~\ref{fig:RobustnessGap} , namely $J_{ab}$
(for the highest temperature considered in the simulations) and the
critical temperature above which the probes get disentangled. The
latter is the probes gap apart from small corrections (see Fig.~\ref{fig:Phi}).
The exact singlet-triplet gap $(\Delta_{ab}=J_{can}$) as well as
the other canonical parameters are found in Figure~\ref{fig:Phi}
(right). These plots show a clear enhancement of the ability of the
antiferromagnet to generate long-range effective interactions among
distant probes as one reaches the square-lattice. 

Also, a wiggly behavior up to $n_{c}=4$ and a transition for $n_{c}>4$
are observed: the increase of the protection gap $J_{can}$ (and also
$J_{ab}(\beta)$) becomes smooth and the Haldane finite-size gap,
very strong for $n_{c}=\left\{ 2,4\right\} $, gets suppressed ---
the $2D$ physics is reached monotonously as the bath gap disappears. 

We expect the ground state of $2D$ antiferromagnets to reduce substantially
the \ac{LDE} due to the symmetry breaking at $T=0$, for large lattices
\cite{Review-Manousakis,Book-Auerbach-1994}; the finite sub-lattice
magnetization should reduce the amount of genuine quantum correlations
shared by the probes. This is borne out by the results of the \ac{QMC}
simulations, shown in Fig.~\ref{fig:Jeff_Entang_Temp}, where $J_{ab}$
(and hence entanglement) is found to decrease at low temperatures,
when the number of chains increase. Nevertheless, at higher temperatures,
the opposite occurs, $J_{ab}$ increases with $n_{c}$; this reflects
the increase of the probes protection gap, $J_{can}$, for it sets
the temperature scale at which entanglement vanishes.

Having shown the \ac{QMC} results for $20$ spin lattices, we now
compare them with Eq.~(\ref{eq:Chap5-Jeff as function of Jcan})
for several temperatures and compute the entanglement via the concurrence
expression {[}Eq.~(\ref{eq:Chap5-Concurrence}){]}: Fig.~\ref{fig:Jeff_Entang_Temp}
shows a perfect fit to the data. For sake of clarity, we have presented
the agreement just for $4$ lattices although all them show the same
degree of accuracy. The observed linear dependence of $J_{ab}$ with
the temperature for $T\rightarrow0$ is easily understood: a zero
temperature (finite) entanglement below the maximum value of $1$
requires $J_{ab}/T\to\mathrm{constant}$. This constant can be derived
from Eq.~(\ref{eq:Chap5-Jeff as function of Jcan}), yielding, \begin{equation}
\beta J_{ab}\underset{T\rightarrow0}{\rightarrow}\frac{1}{4}\ln\left[\frac{4-3\Phi-\eta}{\Phi+\eta/3}\right].\label{eq:Chap5-ZeroTemperatureJeff}\end{equation}
Thus, the \emph{canonical corrections} ($\Phi$ and $\eta$) determine
the low-temperature physics of the probes. From this expression it
is clear that the quasi-perfect \ac{LDE} phenomenon reported in the
previous chapter can only happen for $3\Phi+\eta\simeq0$, when $\beta J_{ab}$
is very large and thus, according to Eq. (\ref{eq:tau_atau_b}), $\langle\boldsymbol{\tau}_{a}\cdot\boldsymbol{\tau}_{b}\rangle_{T=0}\simeq-3$
{[}see also Eq.~(\ref{eq:Chap5-new3}){]}. This is a very special
scenario which happens in \ac{1D} antiferromagnets and also in dimerized
chains (see \cite{2007Venuti-2} for an analytical treatment in the
latter scenario). The present results reveal that quasi-perfect \ac{LDE}
is also possible in the $3$-ladder chain $3\Phi+\eta\simeq0.007$.
As soon as we approach the \ac{2D} scenario, the $\Phi$ correction
gets larger (see Fig.~\ref{fig:Phi}) and a fraction of the entanglement
is lost. 

In all our simulations, the value of $\eta$ is negligible (a careful
inspection shows that the fits we present are virtually indistinguishable
from the fits with $\eta=0$ up to $\alpha\simeq0.1$). In \ac{2D}
, the correction $\Phi$ will be appreciable (Fig. \ref{fig:Phi})
and the singlet will be only partially localized at the probes. Indeed,
$\Phi$ measures the \textquotedbl{}delocalization rate\textquotedbl{}. 

The $n_{c}=3$ spin system has the best singlet locatization rate
towards the probes, $\Phi\simeq0.007$, of all systems simulated%
\footnote{Curiously, this behaviour is not altered by varying the coupling,
in the entire range we have simulated: $\alpha\in[0.05,0.2]$. In
fact, for large $\alpha$, namely $\alpha=0.2$, the discrepancy between
the $n_{c}=1$ and $n_{c}=3$ lattices is quite significant: $\Phi\simeq0.238$
against $\Phi\simeq0.079$, respectively.%
} (see Fig. \ref{fig:Phi}), yielding $E_{C}\simeq0.99$ --- note that
for an odd number of coupled chains, the case of $n_{c}=3$ is peculiar:
increasing the number of chains increases the protection gap and $\Phi$,
with the exception of the $n_{c}=1\rightarrow n_{c}=3$ transition,
where the entanglement becomes less robust regarding temperature ($J_{can}$
decreases) and more efficient at $T=0$ ($\Phi$ also decreases).

For the highest temperatures simulated, the effective coupling saturates
(Fig.~\ref{fig:Jeff_Entang_Temp}) to a constant value, $J_{can}(1-\Phi)/4$,
when $\eta$ is negligible (\emph{i.e.}~not far away from the perturbation
limit), suffering a slightly change with temperature otherwise, \begin{equation}
J_{ab}(T)\simeq\frac{J_{can}}{4}\left(1-\Phi\right)-\frac{k_{B}T}{12}\eta+O(\frac{J_{can}^{2}\Phi}{k_{B}T}).\label{eq:Jeff-high_temp_expan}\end{equation}
Entanglement between the probes, and thus \ac{LDE}, will survive
up to $\beta J_{ab}(\beta)\simeq0.27$ (Sec.~\ref{sub:Entanglement_MixedStates}).
An estimate of the critical temperature is obtained by noting that
$J_{ab}(T)$ has already saturated when the concurrence vanishes (see
Fig.~\ref{fig:Jeff_Entang_Temp}). Indeed, using the equation above
we get,\begin{equation}
k_{B}T^{*}\simeq0.93J_{can}(1-\Phi).\label{eq:Chap5-CriticalTemp}\end{equation}
This agrees with the numerical results within $1\%$ --- in fact,
Fig.~\ref{fig:Phi} shows a mismatch between $k_{B}T^{*}$ and the
gap $J_{can}$ growing with $\Phi$. Eq.~(\ref{eq:Chap5-CriticalTemp})
resembles the previous result for the spin-$1/2$ \ac{AF} Heisenberg
ring {[}Eq.~(\ref{eq:Chap4-SPIN1-Cond_Temp_S1/2}){]} by noting that
$J_{can}$ is proportional to $\alpha^{2}\tilde{\chi}_{ab}^{zz}(0)$
in the perturbative regime {[}Eq.~(\ref{eq:Chap5-Can1}){]}. Here,
however, we have the correction introduced by the spin renormalization,
namely the canonical correction $\Phi$. 

The square lattice is the system with best thermal robustness regarding
\ac{LDE}, despite its appreciable delocalization rate, $\Phi\simeq0.146$;
a fact explained by the emergence of a large singlet-triplet gap,
$J_{can}$, which is about $6$ times the protection gap of the single
spin chain. The regime of high temperatures, $T\gtrsim\Delta(\alpha=0)$,
is not described by Eq.~(\ref{eq:Chap5-Jeff as function of Jcan})
anymore, which requires zero thermal occupancy of excited states of
the spin bath, a crucial assumption of our analytical modelling. However,
according to our estimate {[}Eq.~(\ref{eq:Chap5-CriticalTemp}){]},
no \ac{LDE} is to be expected in this scenario, and hence $J_{ab}(\beta)$
should decrease with the temperature, at some point, and eventually
drop to zero, yielding probes totally uncorrelated. 

\begin{figure}[tb]
\noindent \begin{centering}
\includegraphics[width=1\textwidth]{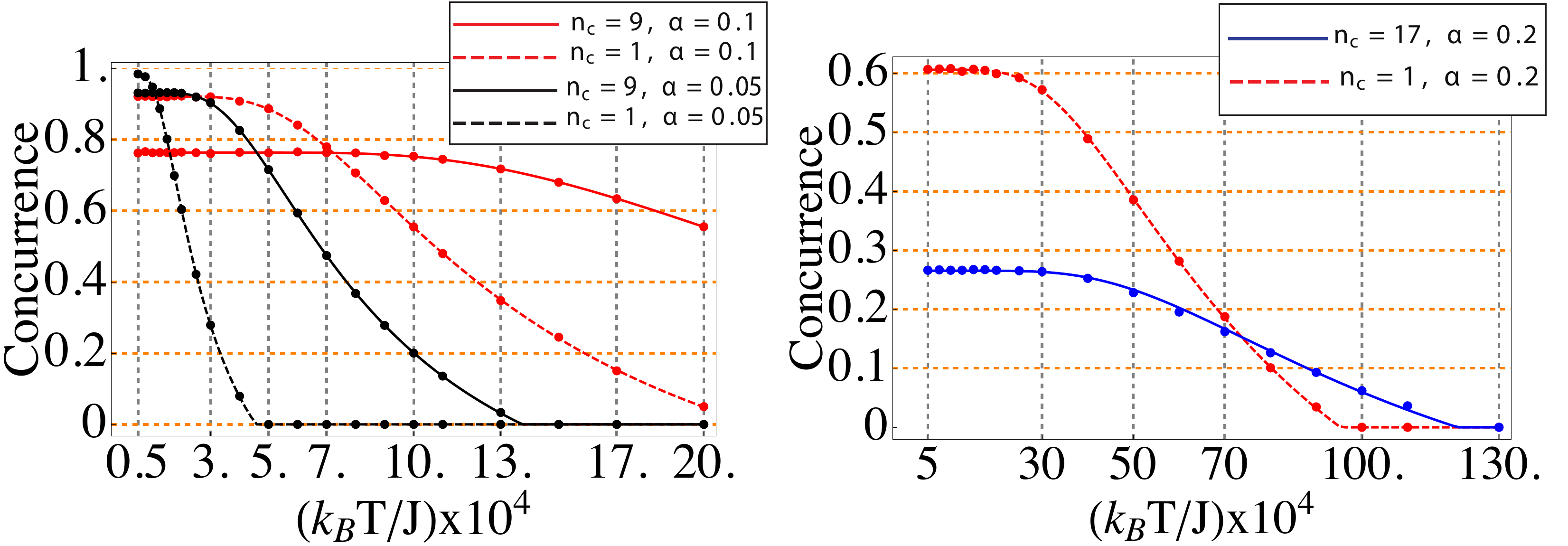}
\par\end{centering}

\caption[Quantum correlations mediated by 2D lattices at strong coupling]{\label{fig:Entang_Various}Concurrence, an entanglement monotone
for qubits, as function of temperature for different lattices and
couplings. Lattices supporting more entanglement at strictly $T=0$
have worse performance at higher temperatures. }

\end{figure}

Figures~\ref{fig:RobustnessGap} and \ref{fig:Phi} deal with relatively
small probe-bath coupling, but the results presented so far are more
general. For instance, choosing a sufficiently large $\alpha$ to
strongly suppress the zero temperature entanglement, via partial frustration
among the neighborhood of the bulk spins connected with the probes,
we again find an excellent agreement with Eq.~(\ref{eq:Chap5-Jeff as function of Jcan}).
For intermediate probe-bath coupling, $\alpha=0.1$ and $\alpha=0.2$,
the measured concurrence is fitted with an expression derived from
Eq.~(\ref{eq:Chap5-RenormalizedAverage}), as shown in Fig.~\ref{fig:Entang_Various}.
These results show that in all our measured systems, the condition
of a robust gap is verified. This is surprising, particularly in the
case of large lattices, which has a gap much smaller than $\alpha J$;
one would not expect, in this situation, the appearance of a well
protected singlet. The emergence of the robust gap has to be \emph{attributed}
\emph{to the coupling of the probes}: the lowest singlet and triplet
are pulled down from the rest of the spectrum, allowing a complete
description of entanglement only in terms of these two energy levels. 

On the other hand, whereas the strong coupling to the spin bath reduces
the zero temperature entanglement, it also allows a larger split between
the singlet and triplet, leading to entangled probes at much higher
temperatures. Typically, exchange interactions in antiferromagnets
can be of the order of $0.1\mathrm{\, eV}$, resulting in an effective
coupling of the order of $0.3\,\mathrm{meV}$ for the square lattice
($l=20)$ at temperature $\sim12\,\mathrm{K}$ and $\alpha=0.2$.
This is to be compared with the value of $0.01-0.1\mathrm{\, me}\mathrm{V}$
achievable in quantum dot spins \cite{1998Loss,2000DiVincenzo} although
decoherence effects in spin lattices can lessen this difference. 

Regarding the quantum-classical transition in these systems, we see
that the critical temperature (above which the correlations shared
by the probes are completely classical) can be increased by a factor
of $20$ from a weakly coupled spin chain ($\alpha=0.05)$ to an intermediate
coupled ($\alpha=0.2)$ \ac{2D} lattice, entanglement surviving up
to $k_{B}T\simeq1.2\times10^{-2}J$ (Fig.~\ref{fig:Entang_Various})
--- an appreciable enhancement of the thermal robustness of such correlations.

\section{Concluding remarks}

In the present chapter we have answered the most important questions
left open previously, namely the effect of a larger probe-bath coupling
and the performance of \ac{2D} lattices towards long-distance entanglement.
We have studied a family of $20$ spin systems serving as quantum
baths, including a single spin chain and a square lattice, and considered
the effect of weak and moderate probe-bath couplings. We found an
increase of the thermal robustness of probe entanglement in \ac{2D}
systems, due to the emergence of robust gaps. This was achieved by
combining \ac{QMC} simulations in large systems and an analytical
model derived from the Schrieffer-Wolff canonical transformation formalism.
We concluded that the canonical parameters give an adequate parametrization
of correlations (and hence entanglement). They are:
\begin{enumerate}
\item the probes singlet-triplet gap, $J_{can}.$ In perturbation theory,
this equals the effective coupling, $J_{\text{eff}}$ --- a single
parameter, which in Chapter~\ref{cha:LDE_Via_GS_GappedSpinChains}
led us to the conjecture of quasi-perfect \ac{LDE} in \ac{1D} systems
(where no symmetry breaking exists). However, to achieve a correct
parametrization of the probe correlation,  two new parameters must
be considered, namely:
\item the \textquotedbl{}singlet delocalization rate\textquotedbl{}, $\Phi$,
which provides a crucial correction to \ac{LDE}, is the main factor
leading to the real effective coupling, $J_{ab}(\beta)$. In particular,
we have seen that $\Phi$ is very small in \ac{1D} validating the
earlier conjecture on quasi-perfect \ac{LDE};
\item the canonical correction $\eta$ --- a constant contribution to the
probes correlator {[}see Eq.~(\ref{eq:Chap5-RenormalizedAverage}){]},
which does not play a role for small probe-bath coupling. 
\end{enumerate}
In the concluding remarks of the previous chapter, we anticipated
that symmetry breaking, a phenomenon occurring for \ac{2D} systems
in the thermodynamic limit, would decrease the amount of quantum correlations
even in the finite size scenario, perhaps completely destroying \ac{LDE}.
Here, we have seen that, indeed, the $T=0$ entanglement considerably
decreases in the square lattice, but that, at the same time, a robust
gap emerges, allowing moderate entanglement at higher temperatures
than previously thought to be possible. 

Although a numerical study had already been performed before in Ref.~\cite{2007Venuti},
here we measured directly the correlator $\langle\boldsymbol{\tau}_{a}\cdot\boldsymbol{\tau}_{b}\rangle$
by \ac{QMC}; have not assumed a low-lying spectrum consisting of
a singlet and triplet states, but rather confirmed it. Moreover, here
we have gone beyond the \ac{1D} scenario by considering a \ac{2D}
family of antiferromagnets. We demonstrated that entanglement mediated
in \ac{2D} is more efficient, at realistic temperatures, and identified
the opening of protected singlet-triplet states (well separated by
the remaining spectrum, Fig.~\ref{fig:Chap5-Lattice}) as the main
cause. Our analytical model provides reliable fits up to $\alpha=0.2$,
meaning that robust gaps, $\Delta(\alpha)>0$, separate the low-lying
energy sector from excited states, quite generally in $SU(2)$ spin
lattices spin systems.

These results entail that, for two spin-$1/2$ probes interacting
with large many body systems, the quantum-classical boundary is very
sensitive to the particular model describing the underlying effective
physics of such systems, and also to the dimensionality. We have seen
a curious interplay  between the capacity towards \ac{LDE} at strictly
zero temperature and finite (realistic) temperatures. \ac{2D} systems
offer the possibility to reach quite large long-distance effective
couplings, boosting entanglement, at temperature as high as $k_{B}T\sim\frac{1}{100}J$,
where $J$ is the energy scale of the system's exchange interaction.

\chapter{Final Remarks\label{cha:Conclusions} }

In this thesis we have applied Quantum Information tools to investigate
the characteristics of the quantum-classical boundary in two distinct
scenarios:

--- a thermalized moveable mirror, in a cavity geometry, interacting
via radiation-pressure with confined photons;

--- a family of antiferromagnetic spin systems, with variable geometry,
interacting with two-level systems.

Our motivation stemmed from promising developments in experimental
physics, paving the way for the demonstration of genuine quantum effects
beyond the microscopic domain. 

While interference was observed in large molecules, it has been very
difficult to overcome this barrier and even seek for similar phenomena
in different contexts. The main obstacle is the existence of many
channels of decoherence in systems endowed with large effective Hilbert
spaces, for their internal degrees of freedom can be activated at
any time by the environment, leading to dissipation and lost of coherence.
A few notable expections do exist: in diamond defects, for instance,
spins can exploit the large electronic gaps of these materials to
maintain coherence for long times, or to get entangled with nuclear
spins. Quite generally, however, the observed quantum effects fade
away when the complexity increases. 

The contribution of this thesis to the debate in the field was to
develop methods based on the entanglement approach, to investigate
the possibility of establishing quantum correlations at finite temperature
in complex systems. The major conclusion emerging from these methods
is that our bipartite systems, albeit very different energy scales,
can attain entanglement at thermal equilibrium. 

In the opto-mechanical scenario this arises due to the particularly
robust radiation-pressure interaction and to the possibility of feeding
continuously the cavity with many photons. In the two spins-$1/2$
interacting with large spin lattices, the possibility of high-quality
entanglement, at moderate spin-lattice coupling, is due to the emergence
of a protected singlet well separated from excited states.

The exact place where the quantum-to-classical transition takes place
can be manipulated, within the limits imposed by Quantum Mechanics:

--- by adding a coherent driving source, and considering non-zero
source-cavity detuning, which results in stationary cavity-mirror
entanglement at higher temperatures, in the opto-mechanical problem; 

--- by considering $SU(2)$ symmetric interactions and increasing
the dimensionality of the bulk spin system, leading to entanglement
at higher temperatures, in the problem of entanglement mediation by
spin systems.

These studies are far from being exhaustive, and many questions are
left open. Some possible lines of research have been mentioned throughout
this monograph. Here we add two more regarding many-body systems:
how does the \textquotedbl{}robust gap\textquotedbl{} picture changes
by considering dimerization and frustration in the spin lattice? These
mechanisms are common in magnetic compounds, and its effects on the
generation of long-range correlations deserves further study. Also,
we believe that the inspection of three-body correlations of bulk
spins, via an entanglement approach, can give further insight on the
properties of many-body systems, in the same manner that the study
of entanglement entropy between regions in spin chains led to a better
understanding of the efficiency of numerical methods. Up to our knowledge
this is an unexplored field and may reveal nice surprises regarding
the phases of matter. 

Regarding entanglement between macroscopic systems, at the moment,
physicists are approaching the first experimental demonstrations of
quantum effects in macroscopic mechanical oscillators, and hence the
results of the present thesis may be tested in the laboratory in the
near future. It is thus important to consider feasible generalizations
of our setup endeavoring to obtain larger amounts of entanglement.
One possible route is to manipulate the cavity field statistics by
changing the driving pump characteristics --- there is plenty of room
for entanglement in the cavity-mirror Hilbert space, as we have seen
in this monograph. 

In summary, we showed that entanglement can in principle exist between
two quantum systems which interact directly, despite one of them being
massive, and it can also be transferred from macroscopic systems to
particles that otherwise would share classical correlations. These
results suggest that macroscopic quantum behavior beyond the microscopic
world is not restrict to collective phenomena, such as Bose-Einstein
condensation, but can arise between two distinct systems, as long
as a sufficient control of decoherence is achieved.

\appendix

\chapter{Appendices for chapter~\ref{cha:MacroscopicEntang}}

\section{The standard quantum limit\label{sec:App-Standard_Quantum_Limit}}

This appendix will be useful for the reader new to quantum measurements.
It attempts to make a brief outline of single quantum measurements
when applied to macroscopic systems. This section will be based on
reviews and text books by Braginsky who greatly contributed to the
field of high precision measurements \cite{1980Braginsky,Book-Braginsky-1992,Review-Braginsky-1996}. 

The \ac{SQL} can be defined as the ultimate precision an experimentalist
has to achieve if he wishes to probe quantum effects of macroscopic
objects. As pointed out many times by Braginsky, it is clear that
a macroscopic body will behave the more quantum mechanically the more
precise we measure it. On the other hand, the more precise a physical
quantity, such as the position, is measured the more disturbed the
system being measured gets. It is in the interplay between the maximum
possible accuracy (in order to probe the quantum world) and the uncertainty
principle that the \ac{SQL} arises. 

Following \cite{Book-Braginsky-1992} let us derive the \ac{SQL}
for an harmonic oscillator (representing, for instance, a single-mode
of a mechanical oscillator with natural frequency $\omega$ and mass
$m$). The Hamiltonian of the harmonic oscillator with generalized
coordinates $Q$ and $P$, satisfying $\left[Q,P\right]=i\hbar$,
reads:\begin{equation}
H=\frac{P^{2}}{2m}+\frac{1}{2}m\omega^{2}Q^{2}.\label{eq:App1-SQL-HamiltonianHO}\end{equation}
The Heisenberg equations of motion produce the following evolution
for the generalized position:\begin{equation}
Q(t)=q_{1}\cos\left(\omega t\right)+q_{2}\sin\left(\omega t\right),\label{eq:App1-SQL-Q(t)}\end{equation}
where $q_{1}=Q(0)$ and $q_{2}=P(0)/\left(m\omega\right)$ are the
initial values of the mode's quadratures. The Heisenberg uncertainty
relation, $\Delta P(0)\Delta Q(0)\ge\hbar/2$ {[}with $\Delta O$
meaning the usual root-mean-square deviation of the mean value of
operator $O$, \emph{i.e.}~$\Delta O=\sqrt{\langle\left(O-\langle O\rangle\right)^{2}\rangle}${]}
entails a similar relation for the quadratures, namely $\Delta q_{1}\Delta q_{2}\ge\hbar/\left(2m\omega\right)$.
In this scenario, the \ac{SQL} is defined as the best precision an
experimentalist may achieve if he/she wishes to monitor $Q(t)$ with
a precision which is time-independent. Indeed, according to Eq.~(\ref{eq:App1-SQL-Q(t)})
this implies that the experiment must be designed as to measure both
mode quadratures, $q_{1}$ and $q_{2}$, with equal precisions: $\Delta q_{1}=\Delta q_{2}$.
The \ac{SQL} therefore reads,\begin{equation}
\Delta Q_{SQL}=\Delta q_{1}=\Delta q_{2}=\sqrt{\frac{\hbar}{2m\omega}},\label{eq:App1-SQL-SQL}\end{equation}
and equals the \ac{ZPF}. The \ac{SQL} is then the best possible
time-independent accuracy we can reach in monitoring the position
of the mechanical oscillator. When the oscillator is in a heat bath,
the \ac{SQL} depends crucially on the measurement time (in general
will depend on the specific measurement apparatus) and again, due
to quantum fluctuations, cannot be made arbitrarily small for two
canonical conjugate variables. It is possible, however, to overcome
the standard quantum limit in some cases by performing a \ac{QND}
measurement. Historically, the first example of such a measurement
is based precisely on the same opto-mechanical system of Chapters~\ref{cha:MacroscopicEntang}
and \ref{cha:Stationary-optomechanical-entangl}, \emph{i.e.}~by
measuring the radiation pressure we can measure the energy of the
resonator with boundless sensitivity. The position of a particle cannot
be measured in a \ac{QND} way; in fact only integrals of motion,
such as the number of photons in the radiation pressure interaction
{[}Eq.~(\ref{eq:Chap2-Hamiltonian}){]} admit such high sensitivity
(the reader is referred to \cite{Review-Braginsky-1996,1980Braginsky}
for more details on \ac{QND} measurements). 

The \ac{SQL} for the oscillator's energy can be obtained in a similar
way by recalling that the energy is related to the amplitude $A=\sqrt{q_{1}^{2}+q_{2}^{2}}$
of the oscillations by \begin{equation}
E=\frac{1}{2}m\omega^{2}A^{2}+\frac{1}{2}\hbar\omega=\hbar\omega\left(n+1/2\right),\label{eq:App1-SQL-Aux}\end{equation}
where $n$ stands for the number of quanta in the oscillator. From
a continuous measurement of $Q(t)$, the amplitude can be obtained
with the same precision than the quadratures: $\Delta A=\Delta q_{1(2)}$,
which results in an accuracy for the energy of $\Delta E\simeq m\omega^{2}A\Delta A$,
yielding the following \ac{SQL}:\begin{equation}
\Delta E_{SQL}=\hbar\omega_{m}\sqrt{n},\label{eq:App1-SQL-MechanE_SQL}\end{equation}
The latter expression is only valid for large number of quanta as
$\Delta A\ll A$ (necessary for the validity of the expression given
above for $\Delta E$) implies $n\gg1$. In order to observe the mechanical
oscillator up to such accuracy, we have to overcome all the classical
sources of noise and especially temperature. This imposes hard constraints
on the experiment as one needs to guarantee,\begin{equation}
k_{B}T\lesssim\frac{\hbar\omega_{m}}{2},\label{eq:App1-SQL-TempCriterion1}\end{equation}
in order to actually observe the macroscopic object behaving quantum
mechanically. In fact, the rigid constraint (\ref{eq:App1-SQL-TempCriterion1})
applies only when the measuring time, $\tau$, is much larger than
a typical mechanical oscillator's relaxation time, $\tau_{relax}$
(in a dissipative environment this is the inverse of the damping constant).
For $\tau\ll\tau_{relax}$, it can be shown that the relevant criterion
is instead, \begin{equation}
k_{B}T\lesssim\frac{\tau_{relax}}{\tau}*\frac{\hbar\omega_{m}}{2}.\label{eq:App1-SQL-TempCriterion2}\end{equation}
The latter can be easily satisfied at Helium liquid temperatures both
in mechanical oscillators and \ac{EM} optical cavities and represents
a much less harder constraint to an experimentalist aiming to test
the quantumness of large systems. The criterion (\ref{eq:App1-SQL-TempCriterion2})
was derived early \cite{1967Braginsky} in the context of gravitational-waves
detection and it has been rediscovered several times afterwards (for
instance, in the rigorous derivation of the decoherence suffered by
a quantum particle in a heat bath by Caldeira and Leggett \cite{1983Caldeira}).

\section{Coherent states of bosons\label{sec:App-CoherentStates}}

Bosonic coherent states are very familiar in Quantum Optics as they
describe the statistical properties of coherent sources such as the
\ac{LASER} light and preserve their coherent character when interacting
with linear optical elements \cite{1963Glauber,1969Glauber,Book-Mandel-1995,Book-Schleich-2001,Book-Leonhardt-1998}.
Also, they are very useful in other branches of physics for their
mathematical properties. In this short appendix, we review the properties
of coherent states in the basis of the derivations of Chap.~\ref{cha:MacroscopicEntang}.
The bosonic coherent state, $|\alpha\rangle$, is defined as the eigenstate
of the annihilation operator, $i.e.$\begin{equation}
a|\alpha\rangle=\alpha|\alpha\rangle.\label{eq:App2-Coh_States-Definition}\end{equation}
Inserting an expansion in the Fock basis, $\left\{ |n\rangle\right\} $(with
$n\in\mathbb{N}_{0}$), in both sides of the above equation, we get
a recurrence relation whose solution is,\begin{equation}
|\alpha\rangle=e^{-|\alpha|^{2}/2}\sum_{n\in\mathbb{N}_{0}}\frac{\alpha^{n}}{\sqrt{n!}}|n\rangle.\label{eq:App2-Coh_States-FockExp}\end{equation}
Using the Campbell-Baker-Hausdorff formula one readily concludes that
the displacement operator {[}Eq.~(\ref{eq:Chap1_DisplacementOperator}){]}
is the suitable operator form of the latter equation, that is,\begin{equation}
|\alpha\rangle=D(\alpha)|0\rangle=e^{\alpha a^{\dagger}-\alpha^{*}a}|0\rangle.\label{eq:App2-Coh_States-DisplacementOperator}\end{equation}
The box below summarizes the main properties of the displacement operator
and the coherent-states. All of them can be easily proved from the
definition (\ref{eq:App2-Coh_States-Definition}) , equations~(\ref{eq:App2-Coh_States-FockExp})
and (\ref{eq:App2-Coh_States-DisplacementOperator}).

\framebox{\begin{minipage}[t]{1\columnwidth}%
\begin{enumerate}
\item In the Heisenberg picture the operators are displaced as to preserve
(\ref{eq:App2-Coh_States-Definition}),\begin{equation}
D^{\dagger}(\alpha)aD(\alpha)=a+\alpha.\label{eq:App2-Coh_States-Operators_get_displaced}\end{equation}

\item The consecutive action of displacement operators displaces the vacuum,
\begin{equation}
D(\alpha)D(\beta)=e^{i\text{Im}(\alpha\beta^{*})}D(\alpha+\beta).\label{eq:App2-Coh_States-ConsecutiveActions}\end{equation}

\item Although the coherent states are not orthogonal,\begin{equation}
\langle\alpha|\beta\rangle=e^{-\frac{|\alpha|^{2}+|\beta|^{2}}{2}+\alpha^{*}\beta}\label{eq:App2-Coh_States-Overlap}\end{equation}

\item they form an overcomplete basis of the Hilbert space,\begin{equation}
\int_{\mathbb{C}}\frac{d^{2}\alpha}{\pi}|\alpha\rangle\langle\alpha|=\sum_{n\in\mathbb{N}_{0}}|n\rangle\langle n|={\mathbbm{1}}.\label{eq:App2-Coh_States-Overcompletness}\end{equation}

\end{enumerate}
\end{minipage}}

\smallskip{}
The latter feature turns out to be very useful in calculating partial
states {[}Eq.~(\ref{eq:Chap1_PartialStates}){]} and it has been
widely used in the derivations of Chap.~\ref{cha:MacroscopicEntang}.
Let us derive the partial state of a confined \ac{EM} field, $\rho_{cav}(t)$,
for the opto-mechanical problem analyzed in this monograph {[}Eqs.~(\ref{eq:Chap2-PartialStates1})
and (\ref{eq:Chap2-PartialStates2}){]} by means of bosonic coherent
states. 

We wish to compute $\rho_{cav}(t)=\rho_{nm}(t)|n\rangle\langle m|$
for the initial state (\ref{eq:Chap2-InitialState}). From Eq.~(\ref{eq:Chap2-InitialCoherentStateEv})
and defining $\tilde{\Phi}_{nm}:=\Phi_{nm00}$ we obtain, \begin{equation}
\rho_{nm}(t)=\tilde{\Phi}_{nm}\text{Tr}_{m}\left[\int_{\mathbb{C}}\frac{d^{2}z}{\bar{n}\pi}|z+\eta(t)kn\rangle\langle z+\eta(t)km|\right].\label{eq:App2-Coh_States-Example1}\end{equation}
Now we insert the resolution of identity as give previously (\ref{eq:App2-Coh_States-Overcompletness})
and compute the corresponding coherent state overlaps (\ref{eq:App2-Coh_States-Overlap}),\begin{eqnarray}
\rho_{nm}(t) & = & \tilde{\Phi}_{nm}\int\frac{d^{2}w}{\pi}\int\frac{d^{2}z}{\bar{n}\pi}e^{-|w|^{2}-|z|^{2}/\bar{n}-|z+\eta(t)kn|^{2}/2}*\label{eq:App2-Coh_States-Example2}\\
 &  & *e^{-|z+\eta(t)km|^{2}/2+w^{*}(z+\eta(t)kn)+(z^{*}+\eta(t)^{*}km)w}\end{eqnarray}
The integrals in the above equation can be solved by an useful identity%
\footnote{This identity is straightforwardly derived by expanding the exponential
in the integrand in Taylor Series around $B=0$ and $C=0$, integrate
each of the terms of the expansion and make the resummation at the
end.%
},\begin{equation}
\int_{\mathbb{C}}\frac{d^{2}z}{\pi}e^{-A|z|^{2}+Bz+Cz^{*}}=\frac{1}{A}e^{\frac{BC}{A}},\label{eq:App2-Coh_States-Int_Identity}\end{equation}
valid for $B,C\in\mathbb{C}$ and $\text{Re}\: A>0$. After integration
in the all complex plane defined by $w$ and $z$, we recover the
result of Chap.~\ref{cha:MacroscopicEntang} {[}Eq.~(\ref{eq:Chap2-PartialStates1}){]},\begin{equation}
\rho_{nm}(t)=e^{-|k\eta(t)|^{2}(n-m)^{2}(2+\bar{n})/4}.\label{eq:App2-Coh-States-Example3}\end{equation}
It is curious to observe that although the formal treatment of Chap.~\ref{cha:MacroscopicEntang}
did not include the issue of decoherence, the cavity field perceives
the mirror as an effective thermal bath. This can be seen in the strong
suppression of the off-diagonal elements when $|m-n|\gg1$. The same
happens with the coherence between different positions (with separation
$\Delta x$) of a particle interacting with a bath. The off-diagonal
elements of the density matrix are rapidly suppressed with a rate
proportional to $\Delta x^{2}$ \cite{Book-Zurek-1996}. Naturally,
this analogy with the universal phenomenon of decoherence cannot be
pushed too far. Recall that (\ref{eq:App2-Coh-States-Example3}) was
derived by unitary evolution of the compound system; the entropy flows
back and forward from the mirror to the cavity field without any dilution
with an active environment.

\section{Unitary evolution of the opto-mechanical density matrix\label{sec:App-DensityMatrix}}

In Chapter~\ref{cha:MacroscopicEntang} we consider the interaction
of a moveable mirror (with mass $m$ and natural frequency $\omega_{m}$),
placed at one end of a Fabry-Perot cavity, and a single-mode of the
intra-cavity \ac{EM} field (with frequency $\omega_{c}$). In the
adiabatic limit, $\omega_{c}\gg\omega_{m}$, this interaction is well-described
by the radiation-pressure Hamiltonian {[}Eq.~(\ref{eq:Chap2-Hamiltonian}){]}.
Our idealized physical situation considers a completely thermalized
mirror, with thermal occupation $\bar{n},$ which is put in contact
with a cavity prepared in a well-defined coherent state of the \ac{EM}
field, with complex amplitude $\alpha$, \begin{equation}
\rho(t_{0})=\frac{1}{\bar{n}}\int_{\mathbb{C}}\frac{d^{2}z}{\pi}e^{-|z|^{2}/\bar{n}}|\alpha\rangle\langle\alpha|\otimes|z\rangle\langle z|,\label{eq:App2-DensityM-1}\end{equation}
which is taken to evolve unitarily according to the evolution operator
$U(t)$ {[}Eq.~(\ref{eq:Chap2-EvolutionOperator}){]}:\begin{equation}
\rho(t)=e^{-i\omega_{c}a^{\dagger}at}e^{ik^{2}(a^{\dagger}a)^{2}\Lambda(t)}D_{m}(\eta(t)ka^{\dagger}a)e^{-i\omega_{m}b^{\dagger}bt}\rho(t_{0})e^{i\omega_{m}b^{\dagger}bt}D_{m}^{\dagger}(\eta(t)ka^{\dagger}a)e^{-ik^{2}(a^{\dagger}a)^{2}\Lambda(t)}e^{i\omega_{c}a^{\dagger}at}\label{eq:App2-DensityM-2}\end{equation}
The effect of the free evolution of the mirror, $U_{b}(t)=\exp\left(-i\omega_{m}b^{\dagger}bt\right)$,
in the coherent states $|z\rangle$ is to rotate the amplitude $z$
in the phase space: \begin{equation}
U_{b}(t)|z\rangle=|e^{-i\omega_{m}t}z\rangle,\label{eq:App2-DensityM-3}\end{equation}
This has no effect when performing the integration in $z$ to get
the thermal state of the mirror {[}Eq.~(\ref{eq:App2-DensityM-1}){]};
\begin{eqnarray*}
 &  & \int_{\mathbb{C}}\frac{d^{2}z}{\pi}e^{-|z|^{2}/\bar{n}}e^{-i\omega_{m}b^{\dagger}bt}|z\rangle\langle z|e^{i\omega_{m}b^{\dagger}bt}\\
 & = & \int_{\mathbb{C}}\frac{dzdz^{*}}{\pi}e^{-|z|/\bar{n}}|e^{-i\omega_{m}t}z\rangle\langle ze^{-i\omega_{m}t}|\\
 & \underset{\omega\leftarrow e^{-i\omega_{m}t}z}{=} & \int_{\mathbb{C}}\frac{d\omega d\omega^{*}}{\pi}e^{-|\omega|/\bar{n}}|\omega\rangle\langle\omega|.\end{eqnarray*}
Hence, without loss of rigour, we can write:\begin{eqnarray}
\rho(t) & = & e^{-i\omega_{c}a^{\dagger}at}e^{ik^{2}(a^{\dagger}a)^{2}\Lambda(t)}D_{m}(\eta(t)ka^{\dagger}a)\rho(t_{0})D_{m}^{\dagger}(\eta(t)ka^{\dagger}a)e^{-ik^{2}(a^{\dagger}a)^{2}\Lambda(t)}e^{i\omega_{c}a^{\dagger}at}\label{eq:App2-DensityM-4}\\
 & = & \frac{1}{\bar{n}}\sum_{n,m}\Theta_{nm}\int_{\mathbb{C}}\frac{d^{2}z}{\pi}e^{-|z|^{2}/\bar{n}}U(t)|n\rangle\langle m|\otimes|z\rangle\langle z|U^{\dagger}(t).\label{eq:App2-DensityM-5}\end{eqnarray}
The last line follows from expanding the coherent state of the cavity
field in the Fock basis {[}Eq.~(\ref{eq:App2-Coh_States-FockExp}){]}
and \begin{equation}
\Theta_{nm}=\frac{\alpha^{n}\left(\alpha^{*}\right)^{m}}{\sqrt{n!m!}}e^{-|\alpha|^{2}}.\label{eq:App2-DensityM-6}\end{equation}
It is useful to write explicitly the action of $U(t)$ on a generic
separable state of the form $|n\rangle\otimes|z\rangle$, with $n\in\mathbb{N}$
and $z\in\mathbb{C}$, before proceeding with the calculation {[}compare
with Eq.~(\ref{eq:Chap2-Aux}){]}:\begin{equation}
\left(e^{-i\omega_{c}a^{\dagger}at}e^{ik^{2}(a^{\dagger}a)^{2}\Lambda(t)}D_{m}(\eta(t)ka^{\dagger}a)\right)|n\rangle\otimes|z\rangle=e^{-i\phi_{n}(t)}|n\rangle\otimes|z+kn\eta(t)\rangle,\label{eq:App2-DensiityM-7}\end{equation}
where $\phi_{n}(t):=-i\omega_{c}n+ik^{2}n^{2}\Lambda(t)$. Indeed,
the integrand in Eq.~(\ref{eq:App2-DensityM-5}) can be written in
a more convenient form;\begin{equation}
\rho(t)=\frac{1}{\bar{n}}\sum_{n,m}\Theta_{nm}\int_{\mathbb{C}}\frac{d^{2}z}{\pi}e^{-|z|^{2}/\bar{n}}e^{-i\left(\phi_{n}(t)-\phi_{m}(t)\right)}|n\rangle\langle m|\otimes|z+kn\eta(t)\rangle\langle z+km\eta(t)|.\label{eq:App2-DensityM-8}\end{equation}
The next step is to express the coherent states (on the righ-hand
side of the latter equation) in the Fock basis {[}Eq.~(\ref{eq:App2-Coh_States-FockExp}){]}:\begin{equation}
\rho(t)=\frac{1}{\bar{n}}\sum_{n,m}\Phi_{nm\mu\nu}(t)\int_{\mathbb{C}}\frac{d^{2}z}{\pi}e^{-K_{nm}(z)}F(z)^{\mu}F_{m}^{*}(z)^{\nu}|n\rangle\langle m|\otimes|\mu\rangle\langle\nu|.\label{eq:App2-DensityM-9}\end{equation}
We have defined $\Phi_{nm\mu\nu}(t):=\Theta_{nm}e^{-i\left(\phi_{n}(t)-\phi_{m}(t)\right)}/\sqrt{\mu!\nu!}$
as to meet the notation of Chapter~\ref{cha:MacroscopicEntang};
the definitions of $K_{nm}(z)$ and $F_{n}(z)$ read:\begin{eqnarray}
F_{n}(z) & := & z+kn\eta(t)\label{eq:App2-DensityM-10}\\
K_{nm}(z) & := & |F_{n}(z)|^{2}/2+|F_{m}(z)|^{2}/2+|z|^{2}/\overline{n}.\label{eq:App2-DensityM-11}\end{eqnarray}
The density matrix elements $\rho_{nm\mu\nu}(t)$ will now be worked
out more explicitly. Let \begin{equation}
\sigma_{nm\mu\nu}(t)=\left(\bar{n}/\Phi_{nm\mu\nu}(t)\right)\rho_{nm\mu\nu}(t),\label{eq:App2-DensityM-12}\end{equation}
hence, from Eq.~(\ref{eq:App2-DensityM-9}), we have\begin{equation}
\sigma_{nm\mu\nu}(t)=\int_{\mathbb{C}}\frac{d^{2}z}{\pi}e^{-K_{nm}(z)}F_{n}(z)^{\mu}F_{m}^{*}(z)^{\nu}\label{eq:App2-DensityM-13}\end{equation}
The exponential in the integrand reads\begin{equation}
e^{-K_{nm}(z)}=\exp\left[-k^{2}|\eta(t)|^{2}\frac{n^{2}+m^{2}}{2}-z\frac{k}{2}\eta(-t)\left(n+m\right)-z^{*}\frac{k}{2}\eta(t)\left(n+m\right)-|z|^{2}/x\right].\label{eq:App2-DensityM-14}\end{equation}
In the last line we have used the fact that $\eta(t)=\eta(-t)$ {[}see
Eq.~(\ref{eq:Chap2-EvolutionOperator}) and comments therein{]} and
defined $x:=\bar{n}/\left(\bar{n}+1\right)$. In order to perform
the integration in the variables $z$ and $z^{*}$ we add two terms
to $K_{nm}(z)$ (which we set to zero at the end of calculation):\begin{equation}
K_{nm}(z)\rightarrow J_{nm}(z,a,b):=K_{nm}(z)+a\left(z+kn\eta(t)\right)+b\left(z^{*}+km\eta(-t)\right).\label{eq:App2-DensityM-15}\end{equation}
The density matrix elements can be written by taking the correct number
of derivatives with respect to $a$ and $b$ (the same method used
to evaluate path integrals):\begin{equation}
\sigma_{nm\mu\nu}(t)=\left[\int_{\mathbb{C}}\frac{d^{2}z}{\pi}\partial_{a}^{\mu}\partial_{b}^{\upsilon}\exp\left(-J_{nm}(z,a,b)\right)\right]_{(a=0,b=0)}.\label{eq:App2-DensityM-16}\end{equation}
We now evaluate explicitly the integral by exchanging the partial
derivatives and the integral sign. First we evaluate the following
function,\begin{eqnarray}
I(a,b) & := & \int_{\mathbb{C}}\frac{d^{2}z}{\pi}\exp\left(-J_{nm}(z,a,b)\right)\label{eq:App2-DensityM-17}\\
 & = & \int_{\mathbb{C}}\frac{d^{2}z}{\pi}\exp\left\{ -|z|^{2}/x+zG_{1}(a,-t)+z^{*}G_{1}(b,t)\right\} e^{R(a,b)},\label{eq:App2-DensityM-18}\end{eqnarray}
with $G_{1}(X,t)=X-\eta(t)\left(n+m\right)/2$ and \begin{equation}
R(a,b):=-k^{2}|\eta(t)|^{2}\frac{n^{2}+m^{2}}{2}+akn\eta(t)+bkm\eta(-t).\label{eq:App2-DensityM-19}\end{equation}
With the integral written in the above form {[}Eq.~(\ref{eq:App2-DensityM-18}){]}
we can apply directly formula (\ref{eq:App2-Coh_States-Int_Identity})
to obtain:\begin{equation}
I(a,b)=x\exp\left[G_{1}(a,-t)G_{1}(b,t)/x\right]e^{R(a,b)}.\label{eq:App2-DensityM-20}\end{equation}
Finally, the equation given in Chapter~\ref{cha:MacroscopicEntang}
{[}namely, Eq.~(\ref{eq:Chap2-DensityMatrixTOTAL}){]} follows immediately
from equations~(\ref{eq:App2-DensityM-12}), (\ref{eq:App2-DensityM-16})
and (\ref{eq:App2-DensityM-20}):\begin{equation}
\rho_{nm\mu\nu}(t)=\frac{\Phi_{nm\mu\nu}(t)}{\bar{n}+1}\left\{ \partial_{a}^{\mu}\partial_{b}^{\upsilon}\exp\left[G_{1}(a,-t)G_{1}(b,t)/x+R(a,b)\right]\right\} _{(a=0,b=0)}.\label{eq:App2-DensityM-21}\end{equation}
The above expression is an explicit formula for the complete density
matrix of the system for all times and it is expected to be a good
description of the opto-mechanical system in the conditions described
in Chapter~\ref{cha:MacroscopicEntang}, namely in the adiabatic
regime and for perfect reflecting Fabry-Perot end mirrors.

\section{Entropy, mutual information and entropic inequalities \label{sec:App-Entropy}}

In this appendix we introduce the mutual information and the Araki-Lieb
inequality which is in the basis of the entropic inequality used by
the authors in Chapter~\ref{cha:MacroscopicEntang}. We first summarize
the main properties of the von Neumann entropy {[}Eq.~(\ref{eq:Chap1_VonNeumann}){]}; 
\begin{itemize}
\item it is zero if and only if $\rho$ is a pure state, \emph{i.e. \begin{equation}
S(|\psi\rangle\langle\psi|)=0,\label{eq:App1-Entropy-Prop1}\end{equation}
}
\item if $d$ is the dimension of the Hilbert space, then it is maximal
when $\rho$ is the maximally mixed state $\rho_{d}=\frac{1}{d}{\mathbbm{1}}$,
\begin{equation}
\underset{\rho}{\max}\left\{ S(\rho)\right\} =S(\rho_{d})=\ln d,\label{eq:App1-Entropy-Prop2}\end{equation}

\item it is invariant under global unitary transformations $U$: \begin{equation}
S(\rho)=S(U\rho U^{\dagger}),\label{eq:App1-Entropy-Prop3}\end{equation}

\item is a concave function, that is, for $\lambda_{i}>0$ and the ensemble
of density matrices $\left\{ \rho_{i}\right\} $,\begin{equation}
S(\sum_{i}\lambda_{i}\rho_{i})\ge\sum_{i}\lambda_{i}\rho_{i},\label{eq:App1-Entropy-Prop4}\end{equation}

\item is additive, that is, for any two density matrices, $\rho_{A}$ and
$\rho_{B}$, we have,\begin{equation}
S(\rho_{A}\otimes\rho_{B})=S(\rho_{A})+S(\rho_{B}).\label{eq:App1-Entropy-Prop5}\end{equation}

\end{itemize}
With the exception of the concavity {[}Eq.~(\ref{eq:App1-Entropy-Prop4}){]}
(please refer to \cite{1978Wehrl} for a derivation) all the other
properties follow trivially from the definition. Let us, for instance,
prove the additivity of entropy (\ref{eq:App1-Entropy-Prop5}). We
first note that the trace of a matrix is invariant under change of
basis, and conveniently choose the basis where $\rho_{A}$ and $\rho_{B}$
are diagonal and denote their eigenvalues by $a_{1}...a_{d}$ and
$b_{1}...b_{d}$, respectively. Let $U:=U_{A}\otimes U_{B}$ be the
unitary matrix bringing $\rho_{A}\otimes\rho_{B}$ to its diagonal
form. Thus,\[
U\rho_{A}\otimes\rho_{B}U^{\dagger}=\bigoplus_{i=1}^{d}a_{i}\otimes\bigoplus_{i=1}^{d}b_{i}.\]
and hence,\begin{eqnarray*}
S(\rho_{A}\otimes\rho_{B}) & = & \sum_{i,j=1}^{d}a_{i}b_{j}\ln\left(a_{i}b_{j}\right)=\sum_{i,j=1}^{d}a_{i}b_{j}\left(\ln a_{i}+\ln b_{j}\right)\\
 & = & \sum_{i}\left(a_{i}\ln a_{i}+b_{i}\ln b_{i}\right)\\
 & = & S(\rho_{A})+S(\rho_{B}).\end{eqnarray*}
From the above expression we conclude that, analogously to classical
statistical mechanics, the following inequalities hold,\begin{eqnarray}
S(\rho_{A}\otimes\rho_{B}) & \ge & S(\rho_{A})\label{eq:App1-Additivity1}\\
S(\rho_{A}\otimes\rho_{B}) & \ge & S(\rho_{B}).\label{eq:App1-Additivity2}\end{eqnarray}
This is in agreement with the classical picture of the whole at least
as entropic as its parts. However, we know from our earlier discussion
on \ac{EPR} correlations {[}Sec.~\ref{sec:C-versus-Q}{]} that the
quantum world can easily violate the above inequalities, when the
state is non separable {[}see for instance Eq.~(\ref{eq:Chap1_partial_sates_singlet}){]},
being the compound state well-defined but not its parts. Araki and
Lieb showed that a \textquotedbl{}triangle inequality\textquotedbl{}
holds for bipartite quantum states \cite{1970Araki-Lieb}, \begin{equation}
|S(\rho_{A})-S(\rho_{B})|\le S(\rho_{AB})\le S(\rho_{A})+S(\rho_{B}).\label{eq:App1_Triangle_Ineq}\end{equation}
The first inequality encompasses all classes of bipartite states and
differs significantly from the classical analogue based on the Shannon
entropy {[}Eq.~(\ref{eq:Chap1_Shannon Entropy}){]},\begin{equation}
H(X,Y)\ge\max\left\{ H(X),H(Y)\right\} .\label{eq:App1-Classical_bound_entropy}\end{equation}
We can take advantage of the above discrepancy between the entropic
content shared by two random classical variables and that of quantum
states in order to derive an entropic witness of \textquotedbl{}quantumness\textquotedbl{}
for bipartite states. For that end, we use the mutual information
which quantifies the total amount of correlations in a given state,
\begin{equation}
I(\rho_{AB}):=S(\rho_{A})+S(\rho_{B})-S(\rho_{AB}).\label{eq:App1-Mutual_Info}\end{equation}
In general the above quantity does not discriminate between purely
classical and genuine quantum correlations, but still has some desirable
properties; i) it is zero for separable states of the form $\rho_{A}\otimes\rho_{B}$;
ii) it is maximal for maximally entangled pure states (\emph{e.g.}~the
singlet {[}Eq.~(\ref{eq:Chap1_SingletState}){]}); iii) it meets
a natural upper bound within the framework of classical random variables
(or canonical variables in statistical mechanics). The latter turns
out to be a very useful result when one is not able to compute the
exact entanglement of a mixed bipartite state. In order to expose
more clearly this idea, we define the normalized mutual information
as,\begin{equation}
\mathit{\mathcal{I}}(\rho_{AB}):=\frac{I(\rho_{AB})}{S(\rho_{A})+S(\rho_{B})}.\label{eq:App1-Normalized_mutual_information}\end{equation}
It should be clear that this definition holds only when $S(\rho_{A})+S(\rho_{B})>0$.
This of course excludes separable pure states but, nevertheless, it
holds for the most relevant cases, \emph{i.e.} when the partial states
have some entropy (see Sections~\ref{sec:C-versus-Q} and \ref{sec:Entang-th}).
From inequality (\ref{eq:App1-Classical_bound_entropy}) we derive
the upper bound for classical random variables,\begin{equation}
\mathcal{I}_{c}\le\frac{1}{2}.\label{eq:App1-Upper_bound_classical}\end{equation}
The above result has a clear and important interpretation: the correlations
shared by classical random variables cannot be as strong (when properly
normalized by the entropy of the subsystems) as their quantum counterpart.
Thus, violation of the upper bound (\ref{eq:App1-Upper_bound_classical})
is and indicator of quantumness. On the other hand, however, it is
not guaranteed that a large normalized mutual information expresses
a large amount of entanglement. For instance, for every pure state
of two two-level systems the normalized mutual information is always
maximal even if the parties share little entanglement, that is, for
$\theta\in]0,\pi/2[$ we have,\[
\mathcal{I}(\cos\theta|\uparrow,\uparrow\rangle+\sin\theta|\downarrow,\downarrow\rangle)=1,\]
although entanglement vanishes when $\cos\theta\rightarrow0$ or $\sin\theta\rightarrow0$.

\chapter{Appendices for chapter~\ref{cha:Stationary-optomechanical-entangl}}

\section{An outline of optical cavities\label{sec:App-Cavities}}

\begin{figure}[tph]
\noindent \begin{centering}
\begin{tabular}{ccc}
\multicolumn{1}{c}{\includegraphics[scale=0.65]{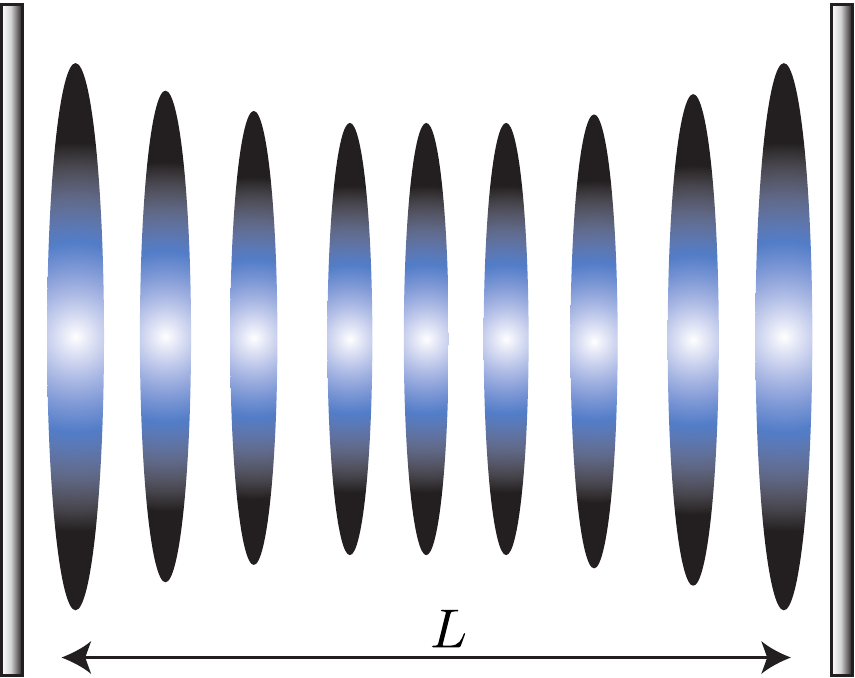}} & \,\, & \includegraphics[width=0.45\textwidth]{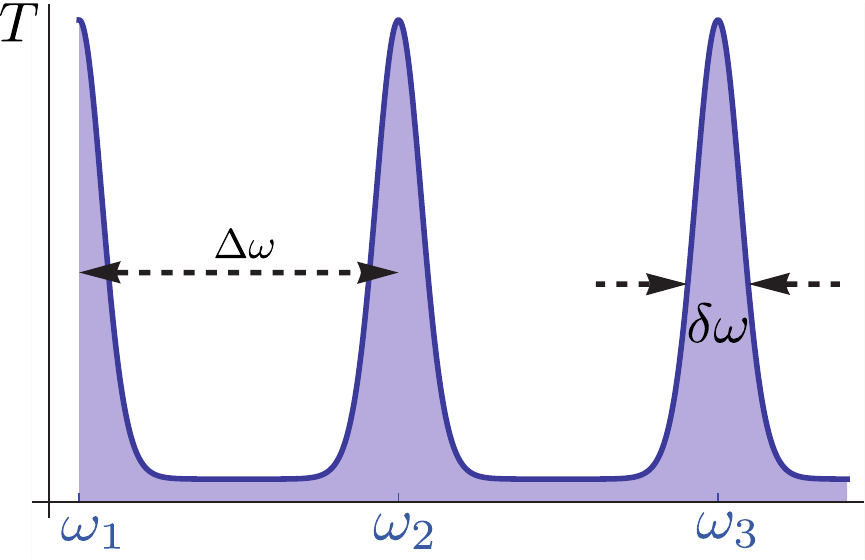}\tabularnewline
\end{tabular}
\par\end{centering}

\caption[The transmissivity mode spectrum of a Fabry-Perot cavity]{\label{fig:AppB_ModeSpectrum}The transmissivity mode spectrum of
a Fabry-Perot cavity (\emph{i.e}.~a linear optical cavity with flat
end-mirrors) of size $L$ and reflectivity $R<1$ has characteristic
lineshapes, frequency separations and other properties controlling
their operation. The most important features of an optical cavity
are the free-spectral range ($\Delta\omega=2\pi c/L$) and the FWHM
($\delta\omega$). The peaks get sharper as one approaches the limit
of a perfect reflective mirror $R=0$. The interplay between the resonant
enhancement, the free-spectral range, and the FWHM will determine
the finesse and the quality factor of the cavity.}

\end{figure}

In this appendix we derive the mode spectrum of a generic Fabry-Perot
optical resonator and afterwards discuss the relevant physical parameters
driving the operation of an optical cavity. A Fabry-Perot cavity (also
known as Fabry-Perot interferometer) is made of two parallel highly
reflecting flat mirrors {[}see Fig.~(\ref{fig:AppB_ModeSpectrum}){]}
and is widely used for controlling and measuring the state of light.
Its mode spectrum is characterized by looking at the transmittance
for different wavelengths. Let us consider a Fabry-Perot cavity made
of equals mirrors placed in vacuum (with real transmission and reflection
coefficients $t$ and $r$, respectively, obeying $r^{2}+t^{2}=1$)
. A classical \ac{EM} wave with amplitude $\boldsymbol{\mathcal{E}}_{i}$
and wave number $k$ will be partially transmitted into the cavity
through mirror $1$ (\emph{e.g.}~the one at left). We make the useful
definitions; $\boldsymbol{\mathcal{E}}_{r}$ and $\boldsymbol{\mathcal{E}}{}_{t}$
are the outer-cavity amplitudes leaving the resonator through mirror
$1$ and $2$, respectively. Also, we define the intra-cavity field
leaving mirror $1$ and mirror $2$ by $\boldsymbol{\mathcal{E}}_{1}$
and $\boldsymbol{\mathcal{E}}_{2}$ , respectively. Then we can write,
$\boldsymbol{\mathcal{E}}_{1}=t\boldsymbol{\mathcal{E}}_{i}-r\boldsymbol{\mathcal{E}}_{2}$,
where the minus sign must be included in order to guarantee conservation
of energy flux in mirror $1$:\begin{equation}
|\boldsymbol{\mathcal{E}}_{1}|^{2}+|\boldsymbol{\mathcal{E}}_{r}|^{2}=|\boldsymbol{\mathcal{E}}_{i}|^{2}+|\boldsymbol{\mathcal{E}}_{2}|^{2},\label{eq:AppB1-flux_conservation}\end{equation}
The relation between the fields read,\begin{eqnarray}
\boldsymbol{\mathcal{E}}_{t} & = & t\boldsymbol{\mathcal{E}}_{1}e^{ikL},\label{eq:AppB1-transmited wave}\\
\boldsymbol{\mathcal{E}}_{2} & = & -r\boldsymbol{\mathcal{E}}_{1}e^{2ikL},\label{eq:AppB1-intracavity reflected wave}\end{eqnarray}
where the origin of the phases have conveniently be chosen at mirror
$1$. The transmittance is defined as,\begin{equation}
T:=\left|\frac{\boldsymbol{\mathcal{E}}_{t}}{\boldsymbol{\mathcal{E}}_{i}}\right|^{2}.\label{eq:AppB1-Transmittance}\end{equation}
Using the equations above it is straightforward to derive,\begin{equation}
T=\frac{1}{1+\frac{4R}{(1-R)^{2}}\sin\left(kL\right)^{2}}.\label{eq:AppB1-Transmittance_calculated}\end{equation}

The transmittance is maximum for $kL=n\pi$ (with $n\in\mathbb{N}_{0}$)
even if the mirror's reflectance $R:=r^{2}$ is arbitrarily close
to one. This defines the resonant frequencies of the cavity\begin{equation}
\omega_{n}=\frac{\pi c}{L}n.\label{eq:AppB1-ResonantFrequencies}\end{equation}
The function $4R/(1-R)^{2}$ in Eq.~(\ref{eq:AppB1-Transmittance_calculated})
measures the resolution of the peaks in the spectrum. It is therefore
the most important physical parameter driving the operation of an
optical cavity and defines the so-called \emph{optical finesse, $\mathcal{F}:=\pi\sqrt{R}/(1-R)$},
which equals the ratio between the frequency intervals and the full-width
at half-maximum (FWHM) {[}Fig.~(\ref{fig:AppB_ModeSpectrum}){]}\emph{.
}Hence, a Fabry-Perot cavity is fully characterized by,
\begin{enumerate}
\item the free-spectral range, $\Delta\omega_{c}$. This is a simply function
of the cavity size, $\Delta\omega_{c}=\pi c/L$, and measures the
distance between two resonant frequencies; 
\item the FWHM, $\delta\omega_{c}$. This depends on the reflectivity of
both mirrors and it measures how many modes are actually contributing
to the real mode spectrum of the cavity (see Fig.~\ref{fig:AppB_ModeSpectrum}).
\end{enumerate}
The relevant physical quantities describing how adequate is a cavity
for a given experiment will depend on the interplay between the free-spectral
range and the FWHM (Fig.~\ref{fig:AppB_ModeSpectrum}). They are,
\begin{enumerate}
\item the $Q$-factor. This is defined in the same manner as in electronics;
the ratio between the frequency at which the field oscillates and
the rate at which it dissipates its energy. It is an important measure
as it tell us about the life-time of resonant photons $\tau=Q/\omega_{c}$,
and thus the reliability of a cavity as a stable amplifier of the
radiation-pressure mechanism. It can be expressed as,\begin{equation}
Q=\frac{\omega_{c}}{\delta\omega_{c}},\label{eq:AppB0-QualityFactor}\end{equation}
and depends on the frequency of the photons\emph{;}
\item the optical-finesse, $\mathcal{F}$. It measures the resolution of
the peaks separation in the mode spectrum,\begin{equation}
\mathcal{F}=\frac{\Delta\omega_{c}}{\delta\omega_{c}}.\label{eq:AppB1-OpticalFinesse}\end{equation}
For a Fabry-Perot microcavity, the free-spectral range is similar
to the cavity mode frequency and thus the optical finesse and the
$Q$-factor are essential the same. The finesse can be determined
by the ring-down time of the cavity, $\tau$. The latter can be measured
using an avalanche photodiode counting the individual photons leaking
out the cavity as function of time \cite{2006Kleckner},\begin{equation}
\mathit{\mathcal{F}}=\frac{\pi c}{L}\tau.\label{eq:AppB1-Finesse}\end{equation}

\end{enumerate}

\section{The input-output theory\label{sec:App-The-input-output-theory}}

This appendix is concerned with the theory of Gardiner and Collett
and aims to give a derivation of the input-output relations for optical
cavities \cite{1984Collett,1985Gardiner}. We consider an optical
cavity with fixed length $L$, and a single-mode (with frequency $\omega_{c}$)
interacting with an input field. The full Hamiltonian of the system
reads, \begin{equation}
H=\hbar\omega_{c}a^{\dagger}a+H_{bath}+H_{int}.\label{eq:AppB1-TotalHamiltonian}\end{equation}
The term $H_{bath}$ is the Hamiltonian of the external fields and
$H_{int}$ describes the interaction of these fields with the mode
of interest, $a$. The standard assumptions of Quantum Optics are
made, namely,
\begin{enumerate}
\item the interaction between the bath and the cavity is chosen to be linear
in the bath operators;
\item an approximated form of $H_{int}$ is considered (the rotating-wave
approximation);
\item the bath is made of many independent harmonic oscillators,\begin{equation}
H_{bath}=\int_{0}^{\infty}d\omega\hbar\omega b^{\dagger}(\omega)b(\omega),\label{eq:AppB1-H_bath}\end{equation}
and the coupling constant (in $H_{int}$) is made frequency-independent
(the so-called Markovian approximation).
\end{enumerate}
With these premises a Langevin equation describing the dissipative
dynamics of the cavity mode can be derived as follows. We start by
writing the interaction term in the rotating-wave approximation%
\footnote{The motivation behind the rotating-wave approximation is the following;
terms like $a^{\dagger}b^{\dagger}$ (corresponding to emission with
excitation) and $ab$ (corresponding to absortion with de-excitation)
oscillate at very high frequencies --- in the Heisenberg picture $(ab)(t)\sim\exp\left[i(\omega_{c}+\omega)t\right]$---
and thus its contribution to the dynamics is negligible. %
},\begin{equation}
H_{int}=\imath\hbar\int_{0}^{\infty}d\omega\kappa(\omega)\left(ab^{\dagger}(\omega)-a^{\dagger}b(\omega)\right),\label{eq:AppB1-H_int}\end{equation}
where $\kappa(\omega)$ is the coupling between the modes. The only
contribution to time averages of the operators will come from frequencies
near $\omega_{c}$, and thus we extend the integration limit in Eq.~(\ref{eq:AppB1-H_int})
which is consistent with the rotating-wave approximation. The dynamics
of $a$ and $b(\omega)$ will follow the Heisenberg equation of motion,
$i\hbar\dot{a}(\dot{b})=\left[a(b),H\right],$ yielding,\begin{eqnarray}
\dot{a} & = & -i\omega_{c}a-\int_{-\infty}^{\infty}d\omega\kappa(\omega)b(\omega),\label{eq:AppB1-HeisenbergEqs1}\\
\dot{b}(\omega) & = & -i\omega b(\omega)+\kappa(\omega)a.\label{eq:AppB1-HeisenbergEqs2}\end{eqnarray}
The integration of the above equations is straightforward. Defining
the initial and final times by subscripts $i$ and $f$, respectively,
and the under the Markovian assumption, $\kappa^{2}(\omega)=\gamma/2\pi$,
we have,\begin{equation}
b(\omega)=e^{-i\omega(t-t_{i})}b_{i}(\omega)+\sqrt{\frac{\gamma}{2\pi}}\int_{t_{i}}^{t}d\tau e^{-i\omega(t-\tau)}a(\tau),\label{eq:AppB1-Aux1}\end{equation}
when integrating Eq.~(\ref{eq:AppB1-HeisenbergEqs2}) for $t>t_{i}$.
The initial time, $t_{i}$, should be interpreted a remote time in
the past when no wave packet has reached the cavity; while the first
term is just the free evolution of the bath modes, the second represents
the waves radiated by the cavity. A similar equation holds for $t>t_{f}$
, namely, \begin{equation}
b(\omega)=e^{-i\omega(t-t_{f})}b_{f}(\omega)-\sqrt{\frac{\gamma}{2\pi}}\int_{t}^{t_{f}}d\tau e^{-i\omega(t-\tau)}a(\tau),\label{eq:AppB1-Aux2}\end{equation}
The cavity mode will evolve according to,\begin{equation}
\dot{a}=-i\omega_{c}a-\sqrt{\frac{\gamma}{2\pi}}\int_{-\infty}^{\infty}d\omega e^{-i\omega(t-t_{i})}b_{i}(\omega)\underset{-(\gamma/2)a(t)}{\underbrace{-\frac{\gamma}{2\pi}\int_{-\infty}^{\infty}d\omega\int_{t_{i}}^{t}d\tau e^{-i\omega(t-\tau)}a(\tau)}}.\label{eq:AppB1-Aux3}\end{equation}
The parameter $\gamma$ clearly plays the role of the damping frequency
of the cavity due to partial reflectivity of the mirror.%
\begin{figure}[tph]
\noindent \begin{centering}
\includegraphics[width=0.5\textwidth]{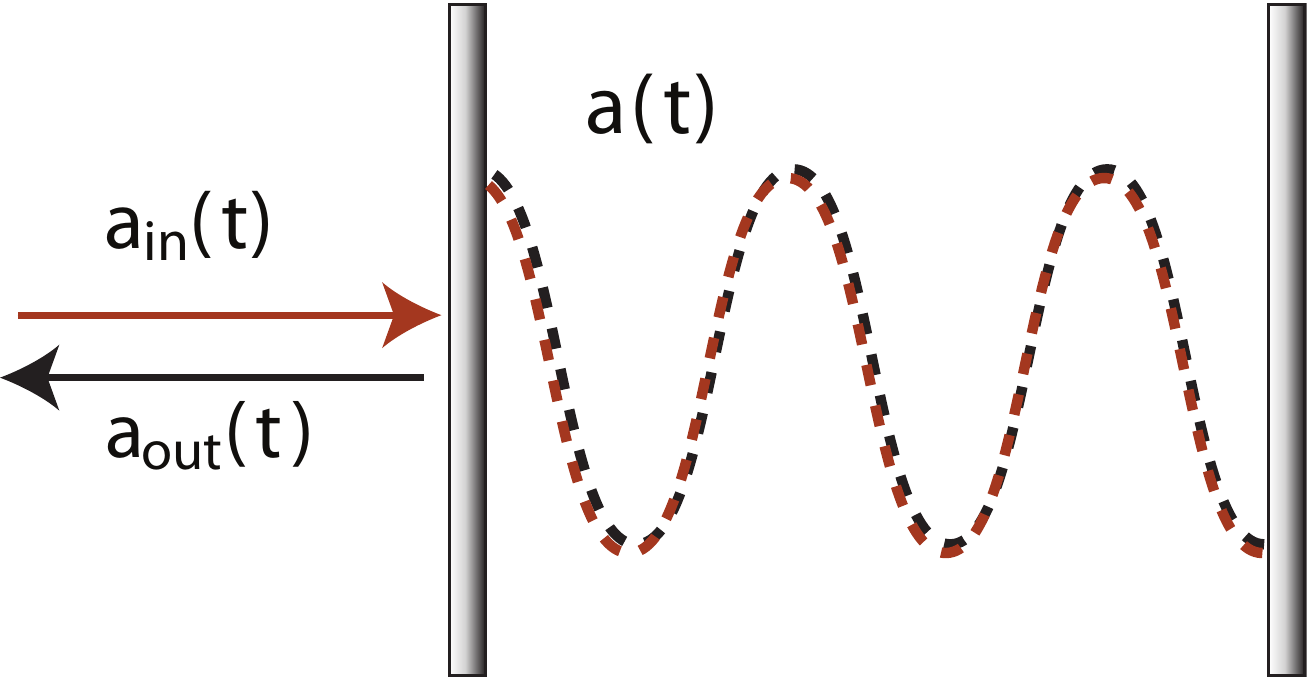}
\par\end{centering}

\caption[Schematic picture of the input, output and intra-field for a Fabry
Perot cavity]{The input field, $a_{in}$, couples to a cavity with partial reflecting
mirrors ($R<1$). The relation between this field, the cavity field
and the field leaking from the cavity, $a_{out}$, is given by Eq.~(\ref{eq:AppB1-Relation_ain_aout}).}

\end{figure}
Now we make the crucial step of the derivation by defining the input-output
field operators,\begin{eqnarray}
a_{in}(t) & = & \frac{1}{2\pi}\int_{-\infty}^{\infty}d\omega e^{-i\omega(t-t_{i})}b_{i}(\omega),\label{eq:AppB1-a_in}\\
a_{out}(t) & = & \frac{1}{2\pi}\int_{-\infty}^{\infty}d\omega e^{-i\omega(t-t_{f})}b_{f}(\omega).\label{eq:AppB1-a_out}\end{eqnarray}
In a open-system like a lossy cavity the input, output and intra-cavity
fields are not essentially different one from each others, but the
formal separation in the above equations is justified for high-quality
cavities where an effective model of spontaneous emission processes
(where a bath boson is created at the expense of a cavity boson) describes
well the physics. These operators satisfy the canonical commutation
relations,\begin{equation}
\left[a_{in}(t),a_{in}^{\dagger}(t^{\prime})\right]=\left[a_{out}(t),a_{out}^{\dagger}(t^{\prime})\right]=\delta(t-t^{\prime}).\label{eq:AppB1-commutation_relations}\end{equation}
The Langevin equation containing the dynamics of the cavity mode reads,\begin{equation}
\dot{a}=-i\omega_{c}a-\frac{\gamma}{2}a-\sqrt{\gamma}a_{in}(t).\label{eq:AppB1-LangevinEquation}\end{equation}
We can easily compute any averages containing the cavity operator,
$a$, if the statistical properties to the input field, $a_{in}$,
are known. For this fact, the latter equation is central for the discussion
of Chap.~\ref{cha:Stationary-optomechanical-entangl}. There we focus
on a cavity which is continuously fed by a pumping \ac{LASER} and
therefore an adequate operator equation relating the flux of energy
entering and exiting the cavity is compulsory. A similar equation
to (\ref{eq:AppB1-LangevinEquation}) can be obtained that depends
on $a_{out}$,\begin{equation}
\dot{a}=-i\omega_{c}a+\frac{\gamma}{2}a-\sqrt{\gamma}a_{out}(t).\label{eq:AppB1-LangEq_out}\end{equation}
The latter is not as useful as Eq.~(\ref{eq:AppB1-LangevinEquation})
as it depends on unknown boundary conditions ($b_{f}$), but together
with the Langevin equation {[}Eq.~(\ref{eq:AppB1-LangevinEquation}){]}
it provides an important input-output relation in Quantum Optics,
\begin{equation}
a_{out}(t)=a_{in}(t)+\sqrt{\gamma}a(t).\label{eq:AppB1-Relation_ain_aout}\end{equation}
Assuming the bath to be initially in a thermal state with temperature
$T$, in the rotating-wave approximation, the statistical properties
of the input-field assume a simple form \cite{Book-Gardiner-2004}:\begin{eqnarray}
\langle a_{in}(t)\rangle & = & 0,\label{eq:AppB1-Averages_a_in0}\\
\langle a_{in}^{\dagger}(t)a_{in}(t^{\prime})\rangle & \simeq & \bar{n}(\omega_{c})\delta(t-t^{\prime}),\label{eq:AppB1-Averages_a_in_00}\\
\langle a_{in}(t)a_{in}^{\dagger}(t^{\prime})\rangle & \simeq & \left(\bar{n}(\omega_{c})+1\right)\delta(t-t^{\prime}),\label{eq:AppB1-Averages_a_in}\end{eqnarray}
where $\bar{n}(\omega_{c})$ represents the Bose occupation number
at temperature $T$ and frequency $\omega_{c}$. These relations allow
us to compute the thermal averages of cavity operators evolving according
to the Langevin equation {[}Eq.~(\ref{eq:AppB1-LangevinEquation}){]}
(or the Langevin equation of Sec. \ref{sec:The-dynamics} in which
the cavity field also interacts with a moveable mirror). To this end,
one can for instance take the Fourier transform of Eq.~(\ref{eq:AppB1-LangevinEquation})
to get:\begin{equation}
(\mathcal{F}a)[\omega]=\frac{-\sqrt{\gamma}}{i(\omega_{c}-\omega)+\gamma/2}(\mathcal{F}a_{in})[\omega]=-\sqrt{\gamma}\chi_{c}[\omega](\mathcal{F}a_{in})[\omega],\label{eq:AppB1-FT-a}\end{equation}
where $\chi_{c}[\omega]$ is the susceptibility of the cavity. This
together with the time statistics for $a_{in}$ allow us to extract
easily any time-dependent average of cavity operators.

\section{The quantum Brownian motion\label{sec:App-QuantumBrownianMotion} }

\textbf{\large Quantum Langevin equations (brief outline)}\smallskip{}

The original derivation of the quantum Langevin equation has more
than $20$ years and is due to Benguria and Kac \cite{1981Benguria,1988Ford}.
Here we outline a more recent derivation by Giovannetti and Vitali
\cite{2001Giovannetti}, and briefly discuss the differences between
classical and quantum Brownian motion. For a complete survey into
this subject and related topics (the master equation and phase-space
methods) the reader may consult the excellent book on quantum noise
by Gardiner \cite{Book-Gardiner-2004}. 

We consider a particle with Hamiltonian $H_{0}$ interacting with
a reservoir made of $N$ independent harmonic oscillators with frequencies
$\omega_{j}$ and couplings $k_{j}$ (with $j=1,...,N)$, whose rescaled
canonical coordinates read $Q_{j}$ and $P_{j}$. The interaction
is chosen such that the bath is sensitive to displacements of the
position of the particle, $q$. After an appropriate canonical transformation,
the total Hamiltonian of the system reads \cite{Book-Gardiner-2004}\begin{equation}
H=H_{0}+\frac{1}{2}\sum_{j=1}^{N}\left[\left(P_{j}-k_{j}q\right)^{2}+\omega_{j}^{2}Q_{j}^{2}\right].\label{eq:AppB3-TotalH}\end{equation}
The reservoir annihilation (and creation) operators are defined in
the usual way,\begin{equation}
a_{j}=\frac{1}{\sqrt{2\hbar\omega_{j}}}\left(\omega_{j}Q_{j}+iP_{j}\right).\label{eq:AppB3-a_op}\end{equation}
The Heisenberg equations of motion for the particle operators are
obtained from (\ref{eq:AppB3-TotalH}), and their integration from
the initial time $t_{0}$ resemble the derivation presented in Sec.~(\ref{sec:App-The-input-output-theory}),
but the final expression is somewhat cumbersome and will not be displayed
in full generality. Here we present the solution for a bath spectrum
approximated by a continuum ($N\rightarrow\infty$). The continuous
limit is taken according to,\begin{equation}
\sum_{j=1}^{N}k_{j}^{2}(...)\rightarrow\int_{0}^{\infty}k^{2}(\omega)\frac{dn}{d\omega}(...)\underset{\text{Markovian Assumption}}{\equiv}\frac{2\eta}{\pi}\int_{0}^{\infty}d\omega(...),\label{eq:AppB3_1stMarkov}\end{equation}
where $n^{\prime}(\omega)$ is the oscillator density and $\eta$
is the friction coefficient. The ideal situation occurs when $k^{2}(\omega)n^{\prime}(\omega)=\text{constant}$
--- the so-called Markovian approximation (recall that in classical
dynamics this approximation leads to a stochastic Langevin equation
with an extra random force term without memory). This prescription
introduces the mechanical damping for the particle via $\eta$ and
assumes a very large cut-off compared to typical frequencies $\Omega_{cutoff}\rightarrow\infty$
reflecting the extremely fast dynamics of a large reservoir. The interested
reader is referred to \cite{2001Giovannetti} for a detailed an rigorous
calculation. The quantum Langevin equations resemble their classical
version: \begin{eqnarray}
\dot{q}(t) & = & \frac{p(t)}{m},\label{eq:AppB3-Langevin1}\\
\dot{p}(t) & = & \frac{i}{\hbar}\left[H_{0},p\right]-\eta\frac{p}{m}+\boldsymbol{\xi}(t).\label{eq:AppB3-Langevin2}\end{eqnarray}
but now the noise, $\boldsymbol{\xi}$, is an hermitian operator obeying
the following commutation relation,\begin{equation}
\left[\boldsymbol{\xi}(t),\boldsymbol{\xi}(t^{\prime})\right]=2i\eta\hbar\frac{d}{dt}\delta(t-t^{\prime}).\label{eq:AppB3-LangevinCommutation}\end{equation}
Although a Markov assumption has been made {[}Eqs.~(\ref{eq:AppB3-Langevin1})
and (\ref{eq:AppB3-Langevin2}){]} the differential equations do not
entail a Markovian process. This fact, unfamiliar to classical statistical
mechanics, emerge for the physics also depends on the state vector
and this introduces a non-zero correlation time in a genuine quantum
stochastic process. The auto-correlation function of the quantum random
force reads \cite{Book-Gardiner-2004},\begin{equation}
\left\langle \boldsymbol{\xi}(t)\boldsymbol{\xi}(t')\right\rangle =\hbar\eta\int_{-\infty}^{\infty}\frac{d\omega}{2\pi}e^{-i\omega(t-t')}\omega\left[\coth\left(\frac{\hbar\omega}{2k_{B}T}\right)+1\right].\label{eq:AppB3-Correlation}\end{equation}
Clearly a genuine quantum Brownian motion is not a Markovian process
in general. It is well-known that the quantum harmonic oscillator
finds anomalies at very small temperature and/or strong damping; \emph{e.g.}~power-decay
of expectation values of correlation functions and strong squeezing
of the position and momentum uncertainties, a clear signature of quantumness
\cite{1985Haake}. At which extent we should expect non-Markovian
physics will depend on the balance of the thermal correlation time,
$\tau_{T}=\hbar/(2\pi k_{B}T$), and the time-scale of a given physical
process. The classical limit is obtained setting $\hbar\rightarrow0$
in the auto-correlation function%
\footnote{In fact this limit must be interpreted with some care, since for any
finite $\hbar$ the integrand in Eq.~(\ref{eq:AppB3-Correlation})
diverges. The origin of the problem its in the \textquotedbl{}first
Markovian approximation\textquotedbl{}, namely $k^{2}(\omega)n^{\prime}(\omega)=\text{constant}$.
In a realistic scenario this function is not constant but falls off
at very high frequencies. The result {[}Eq. (\ref{eq:AppB3-CorrelHighTemp}){]}
is thus the wide bandwidth limit of the classical limit.%
},\begin{equation}
\left\langle \boldsymbol{\xi}(t)\boldsymbol{\xi}(t')\right\rangle \underset{\hbar\rightarrow0}{=}2\eta k_{B}T\delta(t-t^{\prime}),\label{eq:AppB3-CorrelHighTemp}\end{equation}
which is the familiar result of the classical Brownian motion. It
is pedagogical to review the case of zero temperature of a free particle,
since it entails the ultimate quantum Brownian motion as no thermal
noise plays and only the vacuum fluctuations drive the system. The
Hamiltonian of the particle reads $H_{0}=p^{2}/2m$, and the position
equation of motion {[}Eq.~(\ref{eq:AppB3-Langevin1}){]} can be immediately
solved,\begin{equation}
q(t)=q(0)+\frac{p(0)}{\eta}\left(1-e^{-\eta t/m}\right)+\frac{1}{m}\int_{0}^{t}d\tau\int_{0}^{\tau}dse^{-\eta(\tau-s)/m}\boldsymbol{\xi}(\tau).\label{eq:AppB3-motion_of_free_Particle}\end{equation}
The mean square displacement, $\left\langle [q(t_{2})-q(t_{1})]^{2}\right\rangle $,
can be calculated from the above expression and Eq.~(\ref{eq:AppB3-Correlation})
(with $\omega\coth\left(\hbar\omega/2k_{B}T\right)\rightarrow|\omega|$)
by making two assumptions; i) $t_{1}$ and $t_{2}$ are very large
so that transient terms vanish and ii) a frequency cutoff is introduced
in the divergent terms in the integrand \cite{Book-Gardiner-2004}.
Here, we just state the result,\begin{equation}
\left\langle [q(t_{2})-q(t_{1})]^{2}\right\rangle \underset{t_{2}-t_{1}\gg m/\eta}{\longrightarrow}\frac{2\hbar}{\eta\pi}\ln\left(\frac{\eta\left|t_{2}-t_{1}\right|}{m}\right).\label{eq:AppB3-ZeroTempBrownianMotion}\end{equation}
The result (\ref{eq:AppB3-ZeroTempBrownianMotion}) is remarkable;
the \ac{ZPF} of the bath oscillators induce a genuine quantum Brownian
motion which takes place much slower than its thermal classical version,\begin{equation}
\left\langle [q(t_{2})-q(t_{1})]^{2}\right\rangle \thickapprox\frac{2k_{B}T}{\eta}\left|t_{2}-t_{1}\right|.\label{eq:AppB3-ClassicalRandomWalk}\end{equation}
The interested reader is referred to the excellent book on quantum
noise by Gardiner and Zoller for more insight into the distinctive
features of quantum Brownian processes \cite{Book-Gardiner-2004}.
\smallskip{}

\textbf{\large The opto-mechanical regime for Brownian motion}{\large \par}

\smallskip{}

What is the physical regime an experimentalist will face in a opto-mechanical
experiment with a massive mechanical oscillator? In order to answer
this question we recall the definition of the quality-factor, $\mathcal{Q}$,
of a mechanical oscillator. The quality factor is defined analogously
as for a resonator cavity {[}Eq. (\ref{eq:AppB0-QualityFactor}){]},
\emph{i.e.}~as the ratio between the natural frequency of the oscillator
and the rate at which it dissipates its energy,\begin{equation}
\mathcal{Q}:=\frac{m\omega_{m}}{\eta}=\frac{\omega_{m}}{\gamma_{m}}.\label{eq:AppB3-Qfactor}\end{equation}
It should be clear that the mechanical oscillator (\emph{e.g.} a free-standing
mirror, a mirror attached to a cantilever, \emph{etc.}) must dissipate
its energy to the bath very slowly compared to its own dynamics, if
one wishes to observe \emph{bona fide }quantum phenomena. Indeed,
the experimentalist must use high-quality mirrors, $\mathcal{Q}\gg1$.
Also, one must deal with the limit of high-temperatures, for we have
$k_{B}T\gg\hbar\omega_{m}$ even at cryogenic temperatures. The latter
condition poses a serious problem in reaching the realm of quantum
effects. In fact, the large number of phononic excitations is responsible
for making the famous Penrose proposal \cite{Book-Penrose-1986,2003Marshall}
impossible to realize even with state-of-the-art cooling methods ---
see the discussion of \cite{2005Bassi,2006Bernard}. 

Notwithstanding, we found a way to circumscribe the effect of high-temperature
in opto-mechanical systems and recover quantum effects beyond the
low temperature regime. This is accomplish by preparing the cavity
field with a sufficiently large number of coherent photons as conjectured
in Sec. \ref{sub:Sec2.2.3-Mac_therm_entang} and confirmed, for a
realistic scenario of a driven open system, in Sec.~\ref{sec:The-dynamics}
by showing that the effective opto-mechanical coupling {[}Eq.~(\ref{eq:Chap3-EffectiveOptoMechanicalCoupling}){]}
is proportional to the square root of the number of intra-cavity photons.
As a consequence, entanglement may be observed at temperatures much
higher than than the mirror's ground state (Sec.~\ref{sec:Approaching-stationary-entanglement}).
This clearly makes the life much easier to the experimentalist aiming
to observe genuine quantum effects on macroscopic oscillators.

In what follows, and to make a more clear connection between Sec.~\ref{sec:The-dynamics}
and the Brownian motion, we re-define the position and momentum operators
to be dimensionless {[}Eq.~(\ref{eq:Chap3_renormaliz}){]}. As a
consequence the auto-correlation function of the noise operator acquires
dimensions of frequency and its high-temperature limit reads \cite{Book-Breuer-2002}\cite{1981Benguria}, 

\begin{equation}
\left\langle \boldsymbol{\xi}(t)\boldsymbol{\xi}(t')\right\rangle \underset{\beta\hbar\omega_{m}\ll1}{=}\gamma_{m}\left(2\bar{n}+1\right)\delta(t-t^{\prime}),\label{eq:AppB3-CorrelationNoiseHighTHighQ}\end{equation}
where $\bar{n}$ is the Bose occupation number of phonons for the
mechanical frequency $\omega_{m}$ at temperature $T$. This expression
coincides with the classical limit {[}recall that for $\beta\hbar\omega_{m}\ll1$
we have $\bar{n}\simeq k_{B}T/\hbar\omega_{m}$ and the above expression
reduces to Eq.~(\ref{eq:AppB3-CorrelHighTemp}) with $m\rightarrow1/\hbar\omega_{m}${]}
of the quantum Brownian motion and holds for weak mirror-environment
coupling (\emph{i.e.}~$\mathcal{Q}\gg1$). Experimentally, the quality-factor
can be as high as $10^{4}$ for low free-standing mirrors (with $m\simeq400\:\mathrm{ng}$)
\cite{2006Gigan} and may be improved with state-of-the-art microfabrication
techniques; in fact, very recently, the Vienna group has demonstrated
a micromechanical resonator with $\mathcal{Q}\approx30000$ and $m\simeq43\:\mathrm{ng}$
operating at $T=5\:\mathrm{K}$ \cite{2009Groblacher}. This makes
the Markovian assumption very accurate in realistic scenarios.

\section{Equation of motion for the covariance matrix\label{sec:CovarianceMatrix}}

Here we derive the equation of motion for the covariance matrix associated
with the dynamical system:\begin{equation}
\dot{X}(t)=A(t)X(t)+Y(t).\label{eq:App-B4-Eq1}\end{equation}
An example of a time-dependent first order inhomogeneous equations
is found in Chapter~\ref{cha:Stationary-optomechanical-entangl}.
For the sake of generality, in this appendix, we will think of $X(t)$
as being a generic $d$-dimensional vector with components $\left(X_{1}(t),...,X_{d}(t)\right)$,
where each $X_{i}(t)$ is the Heisenberg representation for the quantum
operator $X_{i}$; the $d\times d$ matrix $A$ will depend on time,
in general, and $Y(t)$ is a generic vector of operators $Y(t)=\left(Y_{1}(t),...,Y_{d}(t)\right)$.
The formal solution of (\ref{eq:App-B4-Eq1}) reads;\begin{equation}
X(t)=M(t)X(0)+\int_{0}^{t}dsM(s)Y(t-s),\label{eq:App-B4-Eq2}\end{equation}
where $M(t)$ denotes the principal matrix solution of the homogeneous
system $\dot{M}(t)=A(t)M(t)$ with $M(0)={\mathbbm{1}}_{d}$. Recall
that for the opto-mechanical linearized equations of motion $A$ does
not depend on time and hence $M(t)=\exp\left(At\right)$ {[}Eq.~(\ref{eq:Chap3_LangevinEqsCompactForm}){]};
in this case, $M(t)=\exp\left(At\right)$. The covariance matrix associated
with the operators $\left\{ X_{i}(t)\right\} $ is defined in the
usual way, \begin{equation}
V_{ij}(t)=\frac{1}{2}\langle\left\{ X_{i}(t),X_{j}(t)\right\} \rangle,\label{eq:App-B4-Eq3}\end{equation}
where $\left\{ A,B\right\} =[A,B]_{+}$ is the anti-commutator. To
find the equation of motion for $V(t)$ we take the derivative of
the latter equation:\begin{eqnarray}
\dot{V}_{ij}(t) & = & \frac{1}{2}\langle\left\{ \dot{X}_{i}(t),X_{j}(t)\right\} +\left\{ X_{i}(t),\dot{X}_{j}(t)\right\} \rangle\label{eq:App-B4-Eq4}\\
 & = & \frac{1}{2}\sum_{k=1}^{d}\langle\left\{ A_{ik}(t)X_{k}(t)+Y_{i}(t),X_{j}(t)\right\} +\left\{ X_{i},A_{jk}(t)X_{k}(t)+Y_{j}(t)\right\} \rangle.\label{eq:App-B4-Eq5}\end{eqnarray}
The last equality was obtained inserting the lef-hand side of Eq.~(\ref{eq:App-B4-Eq1})
in $\dot{X}_{i(j)}(t)$. We now drop the explict time depence and
adopt the summation convention for repeated indexes as to ease the
notation. Indeed, we can recast the above formulas into the form\begin{equation}
\dot{V}_{ij}=\langle A_{ik}\underset{V_{kj}}{\underbrace{\frac{1}{2}\left\{ X_{k},X_{j}\right\} }}+A_{jk}\underset{V_{ik}}{\underbrace{\frac{1}{2}\left\{ X_{i},X_{k}\right\} }}\rangle+\frac{1}{2}\langle\left\{ Y_{i},X_{j}\right\} +\left\{ X_{i},Y_{j}\right\} \rangle.\label{eq:App-B4-Eq6}\end{equation}
It is useful to define the symmetric matrix,\begin{equation}
W_{ij}:=\frac{1}{2}\langle\left\{ Y_{i},X_{j}\right\} +\left\{ X_{i},Y_{j}\right\} \rangle.\label{eq:App-B4-Eq7}\end{equation}
We then get the general form for the equation of motion of the covariance
matrix:\begin{equation}
\dot{V}(t)=A.V(t)+V(t).A^{T}+W.\label{eq:App-B4-Eq8}\end{equation}
In order to solve for the above dynamical system, we have to resort
to some particular case. Indeed, we focus on those cases equivalent
to the case studied in this monograph (Sec.~\ref{sec:The-dynamics})
and solve for the stationary solution, \emph{i.e.}~we consider the
only non-vanishing averages containing \textquotedbl{}noise operators\textquotedbl{}
$Y_{i}$ to be $\langle Y_{i}Y_{j}\rangle$, and \begin{equation}
\langle Y_{i}(t)\rangle=0.\label{eq:App-B4-Eq9}\end{equation}
This should not be thought as restrictive in the context of quantum
open systems as single averages of noise operators vanish in quite
general grounds. We now introduce the formal solution of $X_{i}(t)$
in $W$ to get:\begin{eqnarray}
W_{ij}(t) & = & \frac{1}{2}\langle\left\{ Y_{i}(t),M_{jk}(t)X_{k}(0)+\int_{0}^{t}dsM_{jk}(s)Y_{k}(t-s)\right\} +\nonumber \\
 &  & +\left\{ M_{ik}(t)X_{k}(0)+\int_{0}^{t}dsM_{ik}(s)Y_{k}(t-s),Y_{j}(t)\right\} \rangle\label{eq:App-B4-Eq10}\\
 & = & \frac{1}{2}\langle\int_{0}^{t}dsM_{jk}(s)\left\{ Y_{i}(t),Y_{k}(t-s)\right\} +\int_{0}^{t}dsM_{ik}(s)\left\{ Y_{k}(t-s),Y_{j}(t)\right\} \rangle.\label{eq:App-B4-Eq11}\end{eqnarray}
The existence of a stable solution depends on the nature of the principal
matrix solution $M(t)$; in order to keep going we assume $A$ to
be time-independent and $A<0$ so that $M(t)=\exp\left(At\right)\rightarrow{\mathbbm{1}}_{d}$
when $t\rightarrow\infty$. In practice we do not actually need to
solve for the eigenvalues of $A$; applying the Routh-Hurwitz criterion
is sufficient \cite{1987DeJesus}. In the asymptotic regime, and for
a stable system, we have, $\dot{V}(\infty)=0$, and hence,\begin{equation}
A.V(\infty)+V(\infty).A^{T}=-W(\infty),\label{eq:App-B4-Eq12}\end{equation}
Further progress is obtained by simplifying Eq.~(\ref{eq:App-B4-Eq11})
by considering Markovian delta-correlated noise; to this end we introduce
the matrix of stationary noise correlation functions:\begin{eqnarray}
\Phi_{ij}(s-s') & = & \frac{1}{2}\langle\left\{ Y_{i}(s),Y_{j}(s')\right\} \rangle,\label{eq:App-B4-Eq13}\\
\Phi_{ij} & := & D_{ij}\delta(s-s').\label{eq:App-B4-Eq14}\end{eqnarray}
This entails the following simplification, \begin{equation}
W_{ij}(t)\underset{\text{Markovian}}{=}\langle\int_{0}^{t}dsM_{jk}(s)D_{ik}\delta(s)+\int_{0}^{t}dsM_{ik}(s)D_{kj}(s)\rangle.\label{eq:App-B4-Eq15}\end{equation}
The asymptotic limit ($t\rightarrow\infty$) yields%
\footnote{The justification to extend the integral to $-\infty$ (with factor
$1/2$ to compensate) stems from $\Phi_{ij}(s,s^{\prime})$ being
a even function of the time difference $s-s^{\prime}$, that is, $\Phi_{ij}(s-s^{\prime})=\Phi_{ij}(s^{\prime}-s)$.%
} $W_{ij}(\infty)=D_{ij}$. We thus arrive at the following equation
for the steady-state correlation matrix: \begin{equation}
A.V(\infty)+V(\infty).A^{T}=-D(\infty),\label{eq:App-B5-Eq15}\end{equation}
which is a linear algebraic equation for $V$ and can be straightforwardly
solved.

\chapter{Appendices for chapter~\ref{cha:LDE_Via_GS_GappedSpinChains}}

\section{Degenerate perturbation theory\label{sec:App-Degenerate-perturbation-theory}}

In the problems of Chapters~\ref{cha:LDE_Via_GS_GappedSpinChains}
and \ref{cha:LDE_Finite_Temperature} we have an enlargement of the
Hilbert space of a many-body system with Hamiltonian $H_{0}$ due
to the introduction of extra quantum systems (\emph{i.e.}~probes).
In general grounds, the system-probes interaction is described by
Eq.~(\ref{eq:Chap4-Hamiltonian_Probes_System}), namely\begin{equation}
V=\sum_{\alpha=1}^{p}\left(\gamma_{\alpha}^{a}O_{m}^{\alpha}\otimes A^{\alpha}\otimes{\mathbbm{1}}_{b}+\gamma_{\beta}^{b}O_{n}^{\alpha}\otimes{\mathbbm{1}}_{a}\otimes B^{\alpha}\right),\label{eq:App4-DegPerturbTh-Interaction}\end{equation}
where $A$ and $B$ are generic operators of two probes, $a$ and
$b$, respectively, and $O$ denote system's operators. The Hamiltonian
of the full system reads, \begin{equation}
H=H_{0}+V.\label{eq:App4-DegPerturbTh-TotalHam}\end{equation}
The many-body system Hamiltonian has the following spectrum,\begin{equation}
H_{0}|\psi_{k}\rangle=E_{k}|\psi_{k}\rangle.\label{eq:App4-DegPerturbTh-e1}\end{equation}
When the couplings vanish, $\gamma_{\alpha}^{a(b)}=0$, the full system
becomes degenerate as any quantum configuration of the probes contributes
with the same energy. In this case,\begin{equation}
H(\gamma=0)|\Psi_{k}\rangle=E_{k}|\Psi_{k}\rangle,\label{eq:App4-DegPerturbTh-e2}\end{equation}
where $|\Psi_{k}\rangle$ has degeneracy that equals the Hilbert dimension
of the probes and hence the projectors onto the eigenstates obey,\begin{equation}
{\cal P}_{k}=|\psi_{k}\rangle\langle\psi_{k}|\otimes{\mathbbm{1}}_{a}\otimes{\mathbbm{1}}_{b}.\label{eq:App4-DegPerturbTh-e3}\end{equation}
In general this degeneracy is lifted when the coupling to the probes
is turned on, $|\gamma|>0$. If this coupling is not too strong then
degenerate perturbation theory will account for the necessary correction
to the energy (and eigenstates) of the system. Since we are only interested
in the physics of the probes, such as their correlations, we may get
a general description of the problem by integrating out the degrees
of freedom of the many-body system. This corresponds to projecting
the Hamiltonian into the probes's subspace. 

Here we derive equations~(\ref{eq:Chap4-Aux}) and (\ref{eq:Chap4-Aux2})
via the formalism of degenerate perturbation theory; a more general
method --- the Schrieffer-Wolff canonical transformation --- will
be introduced later (Appendix~\ref{sec:The-Schrieffer-Wolff-canonical}).
The procedure leading to the integration of the many-body degrees
of freedom reads,\begin{equation}
H=H_{0}+V\rightarrow H^{(ab)}:=\mathcal{P}_{0}\left(H_{0}+W^{(1)}+W^{(2)}+...\right)\mathcal{P}_{0},\label{eq:App4-DegPerturbTh-HamProjection}\end{equation}
where $H_{0}+W^{(1)}+W^{(2)}+...$ stands for an adequate perturbation
series of $H$ and $H^{(ab)}$ the corresponding effective Hamiltonian
of the probes. Please remark that $\mathcal{P}_{0}$ commutes with
any operator with support in the Hilbert space of the probes; indeed,
Eq.~(\ref{eq:App4-DegPerturbTh-HamProjection}) corresponds to a
ground state average. In what follows, for the sake of generality,
we specify neither the nature of the probes nor the type of coupling
to the many-body system.

\textbf{\large Generic Formalism}{\large \par}

We rewrite Eq.~(\ref{eq:App4-DegPerturbTh-TotalHam}) as $H(\alpha)=H_{0}+\alpha V$
where $\alpha$ is a dimensionless parameter which we suppose sufficiently
small so that near $\alpha=0$ the energy eigenstates are differentiable
functions of $\alpha$. It is convenient to write the projector onto
the degenerate ground state as:\[
{\cal P}_{0}=\sum_{n=1}^{d}|\Psi_{0,n}\rangle\langle\Psi_{0,n}|,\]
where $\left\{ |\Psi_{0,n}\rangle\right\} $, with dimension $d$,
spans the degenerate ground state wave functions. An approximation
to the ground state energy, $E_{0,n}$, can be obtained through a
Taylor expansion of $E_{0,n}(\alpha)$, namely,\begin{equation}
E_{0,n}(\alpha)=E_{n}^{(0)}+\alpha E_{n}^{(1)}+\alpha^{2}E_{n}^{(2)}...,\label{eq:eq:App4-DegPerturbTh-EnTaylor}\end{equation}
where we have defined $E_{n}^{(0)}:=E_{0,n}(0)$, $E_{n}^{(1)}=\dot{E}_{0,n}(0)$
and $E_{n}^{(2)}=\ddot{E}_{0,n}(0)/2$ (overdot denotes $\partial_{\alpha}$
). By taking derivatives to the Schr\"{o}dinger equation we get:\begin{equation}
(H-E_{0,n})|\Psi_{0,n}\rangle=0\Rightarrow\begin{cases}
\left(\dot{H}-E_{n}^{(1)}\right)|\Psi_{0,n}\rangle+\left(H-E_{n}^{(0)}\right)|\dot{\Psi}_{0,n}\rangle=0\\
\left(H-E_{n}^{(0)}\right)\ddot{|\Psi_{0,n}}\rangle-2E_{n}^{(2)}|\Psi_{0,n}\rangle+2\left(\dot{H}-E_{n}^{(1)}\right)|\dot{\Psi}_{0,n}\rangle=0\end{cases},\label{eq:App4-DegPerturbTh-HamTaylor}\end{equation}
where all kets and operators are evaluated at $\alpha=0$. We simplify
the notation by making the following identifications: $H(0)=H_{0}$
and $\dot{H}=V$ to get the following set of equations,\begin{eqnarray}
\left(H_{0}-E_{n}^{(0)}\right)|\Psi_{0,n}\rangle & = & 0,\label{eq:App4-DegPerturbTh-HamTaylor1}\\
\left(V-E_{n}^{(1)}\right)|\Psi_{0,n}\rangle+\left(H_{0}-E_{n}^{(0)}\right)|\dot{\Psi}_{0,n}\rangle & = & 0,\label{eq:App4-DegPerturbTh-HamTaylor2}\end{eqnarray}
\begin{equation}
-2E_{n}^{(2)}|\Psi_{0,n}\rangle+2\left(V-E_{n}^{(1)}\right)|\dot{\Psi}_{0,n}\rangle+\left(H_{0}-E_{n}^{(0)}\right)\ddot{|\Psi}_{0,n}\rangle=0.\label{eq:App4-DegPerturbTh-HamTaylor3}\end{equation}
The first-order shift to the ground state energy is obtained acting
with the projector onto the subspace of states with energy $E_{0,n}$
when $\alpha=0$,~\emph{i.e. }with $\mathcal{P}_{0}$, on Eq.~(\ref{eq:App4-DegPerturbTh-HamTaylor2}),\begin{eqnarray}
E_{n}^{(1)} & = & \langle\Psi_{0,n}|V|\Psi_{0,n}\rangle,\label{eq:App4-DegPerturbTh-Secular}\\
\mathcal{P}_{0}V|\Psi_{0,n}\rangle & = & E_{n}^{(1)}|\Psi_{0,n}\rangle.\end{eqnarray}
The latter equation tell us that in the limit $\alpha\rightarrow0$
the eigenstate $|\Psi_{0,n}(\alpha)\rangle$ of $H(\alpha)$ is also
an eigenstate of $\mathcal{P}_{0}V$. This equation can be written
in matrix form by introducing the resolution of the identity and making
the inner product with an unperturbed state $\langle\psi_{p}|$,\begin{equation}
\sum_{q}\langle\psi_{p}|V|\psi_{q}\rangle\langle\psi_{q}|\Psi_{0,n}\rangle=E_{n}^{(1)}\langle\psi_{p}|\Psi_{0,n}\rangle,\label{eq:App4-DegPerturbTh-Aux1}\end{equation}
where $\{|\psi_{q}\rangle\}$ is the unperturbed energy eigenbasis
spanning the degenerate subspace with energy $E_{n}^{(0)}$. The first-order
contribution in Eq.~(\ref{eq:App4-DegPerturbTh-HamProjection}) therefore
reads \begin{equation}
W^{(1)}=V.\label{eq:App4-DegPerturbTh-1stOrder}\end{equation}
The derivation of $W^{(2)}$ can be carried out as follows; define
the projector onto the states with unperturbed energy $E_{n}^{(0)}$
that at the same time are eigenvectors of $\mathcal{P}_{0}V$ with
eigenvalue $E_{n}^{(1)}$, \emph{i.e.}%
\footnote{From Eq.~(\ref{eq:App4-DegPerturbTh-Secular}) and the definition
of $\mathcal{P}_{n}^{\prime}$ we can write $\mathcal{P}_{0}V$ as,\begin{equation}
\mathcal{P}_{0}V=E_{n}^{(1)}\mathcal{P}_{n}^{\prime}+\sum_{k}\epsilon^{k}Q^{k},\label{eq:App4-DegPerturbTh-Aux4}\end{equation}
where $\{\epsilon^{k}\}$ stand for the eigenvalues of $\mathcal{P}_{0}V$
other than $E_{n}^{(1)}$ and $Q^{k}$ the projectors onto the respective
eigenvectors. Applying $\mathcal{P}_{n}^{\prime}$ to the right of
Eq.~(\ref{eq:App4-DegPerturbTh-Aux4}) and using the fact that the
vectors forming each $Q^{k}$ must be orthogonal to $|\Psi_{0,n}\rangle$,
one gets Eq.~(\ref{eq:App4-DegPerturbTh-Aux3}). %
}\begin{eqnarray}
\mathcal{P}_{n}^{\prime}|\Psi_{0,n}\rangle & = & |\Psi_{0,n}\rangle,\label{eq:App4-DegPerturbTh-Aux2}\\
\mathcal{P}_{n}^{\prime}\left(V-E_{n}^{(1)}\right)\mathcal{P}_{0} & = & 0.\label{eq:App4-DegPerturbTh-Aux3}\end{eqnarray}
We act with $\mathcal{P}_{n}^{\prime}$ on Eq.~(\ref{eq:App4-DegPerturbTh-HamTaylor3}):
$\mathcal{P}_{n}^{\prime}\left(V-E_{n}^{(1)}\right)|\dot{\Psi}_{0,n}\rangle=E_{n}^{(2)}|\Psi_{0,n}\rangle$,
and use relation (\ref{eq:App4-DegPerturbTh-Aux3}) to get \begin{equation}
\mathcal{P}_{n}^{\prime}\left(V-E_{n}^{(1)}\right)\left(1-\mathcal{P}_{0}\right)|\dot{\Psi}_{0,n}\rangle=E_{n}^{(2)}|\Psi_{0,n}\rangle.\label{eq:App4-DegPerturbTh-Aux5}\end{equation}
All we have to do now is to recast $|\dot{\Psi}_{0,n}\rangle$ into
a suitable form; this can be done acting with $\left(1-\mathcal{P}_{0}\right)$
on (\ref{eq:App4-DegPerturbTh-HamTaylor2}), yielding 

\begin{equation}
\left(1-\mathcal{P}_{0}\right)|\dot{\Psi}_{0,n}\rangle=\frac{1}{E_{n}^{(0)}-H_{0}}\left(1-\mathcal{P}_{0}\right)V|\Psi_{0,n}\rangle.\label{eq:App4-DegPerturbTh-1stOrderCorrState}\end{equation}
Eq.~(\ref{eq:App4-DegPerturbTh-Aux5}) now reads:\begin{equation}
\mathcal{P}_{n}^{\prime}V\frac{1}{E_{n}^{(0)}-H_{0}}\left(1-\mathcal{P}_{0}\right)V|\Psi_{0,n}\rangle=E_{n}^{(2)}|\Psi_{0,n}\rangle.\label{eq:App4-DegPerturbTh-Aux6}\end{equation}
In problem of the probes interacting with a spin bath, one has $E_{n}^{0}=E_{0}$,
with $E_{0}$ standing for the spin bath \ac{GS} energy assumed non-degenerate
{[}Eq.~(\ref{eq:App4-DegPerturbTh-e3}){]}. Hence, we drop the subscript
$n$ from now on. Finally, by noticing that $1-\mathcal{P}_{0}=\sum_{k>0}\mathcal{P}_{k}$
, and applying the same procedure leading to Eq.~(\ref{eq:App4-DegPerturbTh-Aux1}),
we get \begin{equation}
\sum_{q}\sum_{k>0}\left[\frac{\langle\psi_{p}|V|\psi_{k}\rangle\langle\psi_{k}|V|\psi_{q}\rangle}{E^{(0)}-E_{k}}\right]\langle\psi_{q}|\Psi_{0}\rangle=E^{(2)}\langle\psi_{p}|\Psi_{0}\rangle,\label{eq:App4-DegPerturbTh-SecularOrder2}\end{equation}
from which the expression for the second-order term in Eq.~(\ref{eq:App4-DegPerturbTh-HamProjection})
can be read out:\begin{equation}
W^{(2)}=-\sum_{k>0}V\frac{\mathcal{P}_{k}}{E_{k}-E^{(0)}}V.\label{eq:App4-DegPerturbTh-2stOrder}\end{equation}
According to Eq.~(\ref{eq:App4-DegPerturbTh-HamProjection}) the
effective Hamiltonian is obtained by adding up the zero and first-order
contributions to the latter equation and computing their ground state
average $\langle...\rangle_{0}=\mathcal{P}_{0}...\mathcal{P}_{0}$,\begin{equation}
H^{(ab)}=\left(E_{0}+\langle V\rangle_{0}-\sum_{k>0}\frac{1}{E_{k}-E^{(0)}}\langle\bar{V}\mathcal{P}_{k}\bar{V}\rangle_{0}+...\right)\mathcal{P}_{0},\label{eq:App4-DegPerturbTh-H(ab)}\end{equation}
where we have introduced the operator $\bar{V}=V-\langle V\rangle_{0}$
and use the fact that $\langle\mathcal{P}_{k}\rangle_{0}=\langle V\mathcal{P}_{k}\rangle_{0}=0$
when $k\neq0$. 

\textbf{\large Local coupling to the many-body system}{\large \par}

The treatment above is totally general and applies for any kind of
perturbation $V$ as long as the many-body system is gapped. Now we
specialize to the case where the probes interact locally with the
many-body bulk. To this end we follow a similar procedure used to
express cross sections of scattering by many-body systems in terms
of its correlation functions \cite{Book-Squires-1978}. Indeed, we
start by expressing the denominator in the second-order term of Eq.~(\ref{eq:App4-DegPerturbTh-H(ab)})
as: 

\begin{equation}
H_{NL}^{(ab)}=-\int_{-\infty}^{+\infty}\frac{dE}{E}\sum_{k>0}\langle\psi_{0}|\bar{V}\mathcal{P}_{k}\bar{V}|\psi_{0}\rangle\delta(E-E_{k}+E_{0}),\label{eq:App4-DegPerturbTh-Manipulations1}\end{equation}
and use the integral representation of the Dirac delta function, \begin{equation}
\delta(E)=(2\pi)^{-1}\int_{-\infty}^{+\infty}dte^{iEt},\label{eq:App4-DegPerturbTh-DiracDeltaRepresentation}\end{equation}
to make the following manipulation,\begin{equation}
\langle\psi_{0}|\bar{V}\mathcal{P}_{k}\bar{V}|\psi_{0}\rangle\delta(E-E_{k}+E_{0})\rightarrow...\langle\psi_{0}|e^{iE_{0}t}\bar{V}e^{iE_{k}t}\mathcal{P}_{k}\bar{V}|\psi_{0}\rangle\label{eq:App4-DegPerturbTh-Manipulations2}\end{equation}
which conveniently introduces the evolution of the operators in the
Heisenberg representation for the many-body system: \begin{equation}
e^{iE_{0}t}\bar{V}e^{iE_{k}t}=\bar{V}(t)\label{eq:App4-DegPerturbTh-HeisenbergPic}\end{equation}
Finally, because $\langle\bar{V}\rangle_{0}=0$ by definition we can
include the term $k=0$ in the sum over $k$ to our advantage, to
get:\begin{equation}
H_{NL}^{(ab)}=-\frac{1}{2\pi}\int_{-\infty}^{+\infty}\frac{dE}{E}\int_{-\infty}^{+\infty}dt\langle\bar{V}(t)\bar{V}\rangle_{0}e^{iEt}.\label{eqApp4-DegPerturbTh-FinalEquation}\end{equation}

\section{The response function and time correlation functions\label{sec:Time-correlation-functions}}

In this appendix we derive the relation between the adiabatic susceptibility
at zero frequency and time-dependent correlation functions {[}Eq.~(\ref{eq:Chap4-RelationBetweenEffectiveCouplingsandXhi}){]}.
In order to do so we use the spectral and Lehman representations.
We assume time translation invariance and introduce the following
notation,\begin{equation}
S_{AB}(t_{1},t_{2})=\langle A(t_{1})B(t_{2})\rangle=\langle A(t_{1}-t_{2})B(0)\rangle:=S_{AB}(t_{1}-t_{2}),\label{eq:App4-Relation-1}\end{equation}
and $O(t)=e^{iHt}Oe^{-iHt}$ is the Heisenberg representation for
operator $O$. With this notation it is sufficient to show that the
following relation holds\begin{equation}
\tilde{\chi}_{AB}(0)=\frac{i}{2}\int_{-\infty}^{\infty}dt\left(S_{AB}(t)+S_{BA}(t)\right)sign(t)e^{-0^{+}|t|},\label{eq:App4-Relation-1b}\end{equation}

where $\tilde{\chi}_{AB}(0)$ is the adiabatic susceptibility (or
response function) at zero frequency; then, Eq.~(\ref{eq:Chap4-RelationBetweenEffectiveCouplingsandXhi})
immediately follows as it can be seen by direct inspection. To this
purpose, we introduce (for later convenience) the \emph{spectral function
}$\phi_{AB}$:\begin{equation}
\phi_{AB}(t)=S_{AB}(t)\mp S_{BA}(-t),\label{eq:App4-Relation-2}\end{equation}
where the $\mp$ applies when the operators $A$($B)$ are boson(fermion)-like
operators. Following the tradition, we define the following Green
functions\begin{eqnarray}
G_{AB}^{R}(t) & := & -i\langle\left[A(t),B(0)\right]_{\mp}\rangle\theta(t)\qquad\text{(retarded)}\label{eq:App4-Relation-3}\\
G_{AB}^{A}(t) & := & i\langle\left[A(t),B(0)\right]_{\mp}\rangle\theta(-t)\qquad\text{(advanced)},\label{eq:App4-Relation-4}\end{eqnarray}
and recall that the response function $\chi_{AB}(t)$ --- the central
object of linear response theory --- is basically given by the retarded
Green function,\begin{equation}
\chi_{AB}(t)=-G_{AB}^{R}(t).\label{eq:App4-Relation-5}\end{equation}
The Lehman representation of correlation functions is obtained inserting
the resolution of identity ${\mathbbm{1}}=\sum_{n}|n\rangle\langle n|$,
where $\left\{ |n\rangle\right\} $ is a complete set of energy eigenstates:\begin{equation}
S_{AB}(t)=\frac{1}{\mathcal{Z}}\sum_{n,m}e^{-\beta E_{n}}\langle n|A|m\rangle\langle m|B|n\rangle e^{i(E_{n}-E_{m})t},\label{eq:App4-Relation-7}\end{equation}
whose Fourier transform $\tilde{S}_{AB}(\omega)$ reads \begin{equation}
\tilde{S}_{AB}(\omega)=\frac{2\pi}{\mathcal{Z}}\sum_{n,m}e^{-\beta E_{n}}\langle n|A|m\rangle\langle m|B|n\rangle\delta\left(\omega-E_{m}+E_{n}\right),\label{eq:App4-Relation-8}\end{equation}
obeying the \emph{detailed balance condition}:\begin{equation}
\tilde{S}_{AB}(\omega)=e^{\beta\omega}S_{BA}(-\omega).\label{eq:App4-Relation-9}\end{equation}
The latter can be easily proven by noticing that because of\emph{
}the delta function we can make the substitution: $E_{n}\rightarrow E_{m}-\omega$,
in Eq.~(\ref{eq:App4-Relation-8}). Using the detailed balance condition
and the definition of spectral function we arrive at the important
relations,\begin{eqnarray}
\tilde{\phi}_{AB}(\omega) & = & \tilde{S}_{AB}(\omega)\left(1\mp e^{-\beta\omega}\right)\label{eq:App4-Relation-10}\\
 & = & \tilde{S}_{BA}(-\omega)\left(e^{-\beta\omega}\mp1\right).\label{eq:App4-Relation-11}\end{eqnarray}
We are interested in the limit of zero temperature;\begin{equation}
\tilde{\phi}_{AB}(\omega)=\begin{cases}
\tilde{S}_{AB}(\omega) & ,\omega>0\\
0 & ,\omega<0.\end{cases}\qquad T=0\label{eq:App4-Relation-12}\end{equation}
The second line comes from Eq.~(\ref{eq:App4-Relation-8}) as $\beta\rightarrow\infty$
implies a single contribution to the sum, namely $n=0$, and thus
$\tilde{S}_{AB}(\omega)\rightarrow0$ for $\omega<0$, since all states
have $E_{m}>E_{0}$. Let us introduce the spectral representation
of the retarded Green function (analogous formulas holds for the advanced
Green function); from the definitions (\ref{eq:App4-Relation-2})-(\ref{eq:App4-Relation-3})
we have\begin{equation}
iG_{AB}^{R}(t)=\phi_{AB}(t)\theta(t),\label{eq:App4-Relation-13}\end{equation}
from which we expect analytical behavior in the upper half of the
complex plane as long as $\phi_{AB}(t)$ does not grow as an exponential
when $t\rightarrow\infty$,\begin{equation}
i\tilde{G}_{AB}^{R}(z)=\int_{0}^{\infty}dt\phi_{AB}(t)e^{izt}\qquad,\text{Im}z>0.\label{eq:App4-Relation-14}\end{equation}
We now introduce the Fourier representation of the spectral function
to write the Laplace transform as,\begin{equation}
i\tilde{G}_{AB}^{R}(z)=\int_{-\infty}^{\infty}\frac{d\omega}{2\pi i}\frac{\tilde{\phi}_{AB}(\omega)}{z-\omega}\qquad,\text{Im}z>0.\label{eq:App4-Relation-15}\end{equation}
We know take advantage of the special form of $\phi_{AB}(\omega)$
at zero temperature {[}Eq.~(\ref{eq:App4-Relation-12}){]} to write
the Fourier transform {[}obtained via analytical continuation to the
real axis ($z\rightarrow\omega+i0^{+}$) of the Laplace transform{]}:\begin{eqnarray}
i\tilde{G}_{AB}^{R}(\omega) & = & \int_{-\infty}^{\infty}\frac{dx}{2\pi i}\frac{\tilde{\phi}_{AB}(x)}{\omega-x+0^{+}}\label{eq:App4-Relation-16}\\
 & = & \text{PV}\int\frac{dx}{2\pi i}\frac{\tilde{\phi}_{AB}(x)}{\omega-x}-\frac{1}{2}\tilde{\phi}_{AB}(\omega),\label{eq:App4-Relation-17}\end{eqnarray}
where PV denotes the Cauchy principal value. The detailed balance
condition at $T=0$ yields\begin{equation}
i\tilde{G}_{AB}^{R}(\omega)=\text{PV}\int_{0^{+}}^{\infty}\frac{dx}{2\pi i}\left(\frac{\tilde{S}_{AB}(x)}{\omega-x}\mp\frac{\tilde{S}_{BA}(x)}{\omega+x}\right)-\frac{1}{2}\left(S_{AB}(\omega)\mp S_{BA}(-\omega)\right).\label{eq:App4-Relation-18}\end{equation}
If the system has an unique ground state and a gap then it is clear
from the Lehman representation of the correlations {[}Eq.~(\ref{eq:App4-Relation-8}){]}
that in the limit $T\rightarrow0$ the only term surviving for $\omega=0$
is\begin{equation}
\tilde{S}_{AB}(0)=\tilde{S}_{BA}(0)=2\pi\langle A\rangle_{0}\langle B\rangle_{0}\delta(\omega).\label{eq:App4-Relation-19}\end{equation}
The zero frequency response for boson-like operators, $\tilde{\chi}_{AB}(0)=-\tilde{G}_{AB}^{R}(0)$,
is therefore given by,\begin{equation}
\tilde{\chi}_{AB}(0)=\text{PV}\int_{0^{+}}^{\infty}\frac{dx}{2\pi}\left(\frac{\tilde{S}_{AB}(x)}{x}+\frac{\tilde{S}_{BA}(x)}{x}\right)\qquad,T=0.\label{eq:App4-Relation-20}\end{equation}
In order to obtain relation (\ref{eq:App4-Relation-1b}) we must relate
the integral above to an integral in the time domain,\begin{eqnarray}
\int_{0}^{\infty}dtS_{AB}(t)e^{i(\omega+i0^{+})t} & = & -\int_{-\infty}^{\infty}\frac{dx}{2\pi i}\frac{\tilde{S}_{AB}(x)}{\omega-x+i0^{+}}\label{eq:App4-Relation21}\\
\int_{-\infty}^{0}dtS_{AB}(t)e^{i(\omega-i0^{+})t} & = & \int_{-\infty}^{\infty}\frac{dx}{2\pi i}\frac{\tilde{S}_{AB}(x)}{\omega-x-i0^{+}},\label{eq:App4-Relation22}\end{eqnarray}
where $0^{+}$ assures convergence. We now perform the following manipulation,\begin{equation}
\int_{-\infty}^{\infty}\frac{dx}{2\pi i}\frac{\tilde{S}_{AB}(x)}{\omega-x\pm i0^{+}}=\text{PV}\int_{-\infty}^{\infty}\frac{dx}{2\pi i}\frac{\tilde{S}_{AB}(x)}{\omega-x}\mp\frac{\tilde{S}_{AB}(\omega)}{2}.\label{eq:App4-Relation-23}\end{equation}
Finally we add up contributions (\ref{eq:App4-Relation21}) and (\ref{eq:App4-Relation22})
in the light of the latter relation to get,\begin{equation}
\int_{-\infty}^{\infty}dtsign(t)S_{AB}(t)e^{-0^{+}|t|}=\text{PV}\int_{-\infty}^{\infty}\frac{dx}{\pi i}\frac{\tilde{S}_{AB}(x)}{x}\qquad,\omega=0,\label{eq:App4-Relation-24}\end{equation}
from which we prove the desired result {[}Eq.~(\ref{eq:App4-Relation-1b}){]}
and thus Eq.~(\ref{eq:Chap4-RelationBetweenEffectiveCouplingsandXhi}).

\section{Analytic continuation of Green's functions \label{sec:Analytic-continuation}}

It is possible to relate time ordered Green functions (with real or
imaginary time) to the retarded Green's function for bosons $G_{AB}^{R}(t$)
by making a proper analytic continuation into the whole complex plane.
These relations can be useful when some method is available that is
easier to compute one of them. For instance, functional integral methods
allow to obtain time-ordered correlation functions in a consistent
way, although, at the end, we are always interested in physical quantities
such as the adiabatic susceptibility and these correspond to retarded
Green's functions {[}Eq.~(\ref{eq:App4-Relation-5}){]} not to ordered
functions. Analytic continuation is a powerful method that makes the
bridge between these two kinds of Green's functions. They can take
place in the frequency domain (commonly seen in Condensed Matter)
or in the time domain (this will be used to compute the retarded Green's
function on a cylinder in Appendix~\ref{sec:Analytic-continuation}).
This appendix outlines both methods.

Denoting real time by $t$, imaginary time (or temperature variable)
by $\sigma$ and the time ordering operator by $T$, these Green's
functions read:\begin{eqnarray}
G_{AB}^{R}(t) & := & -i\langle\left[A(t),B(0)\right]\rangle\theta(t)\qquad\text{(retarded Green function)}\label{eq:App4-Analytic-1}\\
G_{AB}^{T}(t) & := & \langle T_{t}\left[A(t)B(0)\right]\rangle\qquad\text{(time-ordered Green function)}\label{eq:App4-Analytic-2}\\
G_{AB}(\tau) & := & \langle T_{\sigma}\left[A(\sigma)B(0)\right]\rangle\qquad\text{(Matsubara Green function)},\label{eq:App4-Analytic-3}\end{eqnarray}
where the action of the time ordering operator is to take the operators
defined at later time to the left\emph{:\begin{eqnarray}
T_{t}[A(t)B(0)] & := & \theta(t)A(t)B(0)+\theta(-t)B(0)A(t),\label{eq:App4-Analytic-4}\\
T_{\sigma}[A(\sigma)B(0)] & := & \theta(\sigma)A(\sigma)B(0)+\theta(-\sigma)B(0)A(\sigma),\label{eq:App4-Analytic-4b}\end{eqnarray}
}and all the operators are written in the Heisenberg representation,
namely, $A(t)=e^{iHt}Ae^{-iHt}$ and $A(\sigma)=e^{\sigma H}Ae^{-\sigma H}$. 

\textbf{\large Analytic continuation (frequency domain)}{\large \par}

We wish to relate the Matsubara Green's function to the retarded Green's
function in the frequency domain. To this end we express $G_{AB}(\tau)$
in the Lehman representation {[}see in Eq.~(\ref{eq:App4-Relation-7}){]}:\begin{equation}
\langle A(\sigma)B(0)\rangle=\frac{1}{\mathcal{Z}}\sum_{n,m}e^{-\beta E_{m}}\langle m|A|n\rangle\langle n|B|m\rangle e^{\sigma(E_{m}-E_{n})},\label{eq:App4-Analytic-5}\end{equation}
\begin{equation}
\langle B(0)A(\sigma)\rangle=\frac{1}{\mathcal{Z}}\sum_{n,m}e^{-\beta E_{n}}\langle n|B|m\rangle\langle m|A|n\rangle e^{\sigma(E_{m}-E_{n})}.\label{eq:App4-Analytic-6}\end{equation}
These expressions imply that the Matsubara Green's function (or temperature
Green's function) is periodic as:\begin{equation}
G_{AB}(\sigma)=G_{AB}(\sigma-\beta),\label{eq:App4-Analytic-7}\end{equation}
for $\sigma\in]0,\beta[$. Therefore one should be able to expand
$G_{AB}(\sigma)$ in Fourier series in the interval $0\le\sigma\le\beta$:\begin{equation}
G_{AB}(\sigma)=\sum_{\mu}e^{2\pi i\mu\sigma/\beta}\tilde{G}_{AB}(\mu),\label{eq:App4-Analytic-8}\end{equation}
where $\mu=0,\pm1,\pm2,...$. The Fourier coefficients read\begin{eqnarray}
\tilde{G}_{AB}(\mu) & = & \frac{1}{\beta}\int_{0}^{\infty}d\sigma e^{-2\pi i\mu\sigma/\beta}G_{AB}(\sigma)\nonumber \\
 & = & \frac{1}{\beta}\int_{0}^{\infty}d\sigma e^{-2\pi i\mu\sigma/\beta}A(\sigma)B(0)\nonumber \\
 & = & \frac{1}{\beta\mathcal{Z}}\int_{0}^{\infty}d\sigma e^{-2\pi i\mu\sigma/\beta}\sum_{n,m}e^{-\beta E_{m}}\langle m|A|n\rangle\langle n|B|m\rangle e^{\sigma(E_{m}-E_{n})}\nonumber \\
 & = & \frac{1}{\beta\mathcal{Z}}\sum_{n,m}\langle m|A|n\rangle\langle n|B|m\rangle\frac{e^{-\beta E_{n}}-e^{-\beta E_{m}}}{E_{m}-E_{n}-i\bar{\mu}},\label{eq:App4-Analytic-9}\end{eqnarray}
where we have defined $\bar{u}=2\pi\mu/\beta$. Comparing the latter
equation with the Fourier transform $\tilde{S}_{AB}(\omega)$ of the
correlation function $\langle A(t)B(0)\rangle$ {[}Eq.~(\ref{eq:App4-Relation-8}){]}
we immediately conclude,\begin{eqnarray}
\tilde{G}_{AB}(\bar{\mu}) & = & \frac{1}{\beta}\int_{-\infty}^{\infty}\frac{dx}{2\pi}\frac{\tilde{S}_{AB}(x)-\tilde{S}_{BA}(x)}{i\bar{\mu}+x},\label{eq:App4-Analytic-10}\\
 & = & \frac{1}{\beta}\int_{-\infty}^{\infty}\frac{dx}{2\pi}\frac{\tilde{\phi}_{AB}(x)}{i\bar{\mu}+x}.\label{eq:App4-Analytic-11}\end{eqnarray}
The last equality was obtained via the definition of spectral function
{[}Eq.~(\ref{eq:App4-Relation-2}){]} and can be simplified using
the detailed balance condition {[}Eq.~(\ref{eq:App4-Relation-10}){]}:\begin{equation}
\tilde{G}_{AB}(\bar{\mu})=\frac{1}{\beta}\int_{-\infty}^{\infty}\frac{dx}{2\pi}\frac{\tilde{S}_{AB}(x)}{i\bar{\mu}+x}\left(1-e^{-\beta x}\right),\label{eq:App4-Analytic-12}\end{equation}
showing that the Fourier coefficients can be computed from a single
correlation function. Comparing (\ref{eq:App4-Analytic-11}) with
the Fourier transform of the retarded Green's function {[}Eq.~(\ref{eq:App4-Relation-16}){]}
we get the desired relation:\begin{equation}
\tilde{G}_{AB}^{R}(\omega)=-\beta G_{AB}(\bar{\mu}\rightarrow i\omega-0^{+}).\label{eq:App4-Analytic-13}\end{equation}
The formal procedure $\bar{\mu}\rightarrow i\omega-0^{+}$ amounts
to an analytic continuation for $\tilde{G}_{AB}(\bar{\mu})$ is defined
only at a discrete set of points in the complex plane, namely, $\bar{\mu}=0,\pm2\pi/\beta,\pm4\pi/\beta,...$.

\textbf{\large Analytic continuation (time domain)}{\large \par}

In the time domain the relation between the time-ordered (\ref{eq:App4-Analytic-2})
and retarded (\ref{eq:App4-Analytic-1}) Green's functions can be
obtained with little algebra: \begin{equation}
G_{AB}^{R}(t)=i\left[G_{AB}^{T}(t)-\left(G_{B^{\dagger}A^{\dagger}}^{T}(-t)\right)^{*}\right]\theta(t).\label{eq:App14-Analytic-14}\end{equation}
It is useful to consider the particular case of Hermitian operators
(\emph{i.e.}~$A^{\dagger}=A$ and $B^{\dagger}=B$). In this case
we get a simpler expression,\begin{equation}
G_{AB}^{R}(t)=-2\text{Im}\left[G_{AB}^{T}(t)\right]\theta(t).\label{eq:App14-Analytic-15}\end{equation}
Since one usually works with imaginary time (\emph{e.g.}~when computing
correlations from finite temperature path integral methods), it is
convenient to express $G_{AB}^{R}(t)$ as function of the Matsubara's
Green function. This is done via Wick rotation according to $\sigma\rightarrow it+0^{+}sign(t)$,
\begin{equation}
G_{AB}^{R}(t)=-2\text{Im}\left[G_{AB}\left(\sigma\rightarrow it+0^{+}sign(t)\right)\right]\theta(t).\label{eq:App14-Analytic-16}\end{equation}

\section{Time-ordered Green's function from conformal mapping\label{sec:Correlations-from-conformal}}

The method of relating the physics of different geometries from conformal
invariance was originally developed by Cardy \cite{1984Cardy,Review-Cardy-2008}.
Here we apply this method to our problem, namely that of computing
the time-ordered Green's function $G_{AB}^{T}(t)$ for the finite
\ac{AF} spin chain with $SU(2)$ symmetric Heisenberg interactions.
First we extract the asymptotic Matsubara function of the \ac{AF}
Heisenberg ring from the correlations of the infinite chain and, at
the end, perform an analytic continuation to real time in order to
get $G_{AB}^{T}(t)$ .

\begin{figure}[tph]
\noindent \begin{centering}
\includegraphics[width=0.8\columnwidth]{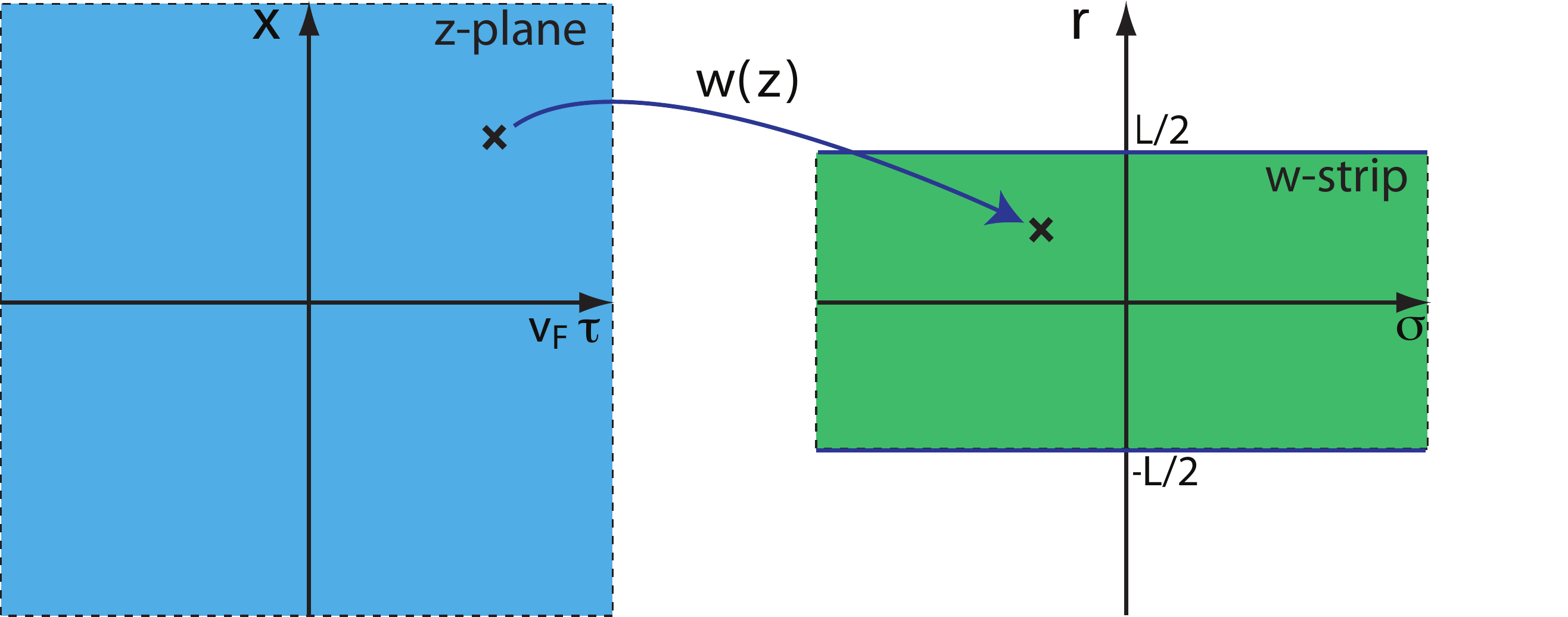}
\par\end{centering}

\caption[Conformal compactification of the plane]{\label{fig:App4-The-conformal-transformation}The conformal transformation
$w(z)=\sigma+\imath r$ maps every point $(v_{F}\tau,x)$ in the plane
into the strip geometry ($\sigma\in]-\infty,\infty[$, $r\in[-L/2,L/2]$)
with periodic boundary conditions along the $r$ direction. The plane
is effectively compactified acquiring the topology of a cylinder.}

\end{figure}

The crucial point is to find a proper analytic mapping between the
critical theory (which is defined in the whole plane: $1+1$ space-time)
and the theory defined in the ring (which including time corresponds
to a strip with boundary conditions along the spatial direction).
The analytic mapping performing such mapping is\begin{equation}
w=\frac{L}{2\pi}\ln z\equiv\sigma+\imath r,\label{eq:App4-Correl_from_Conf-1}\end{equation}
where we have taken the primary branch of the log-function and $z$
is a complex coordinate in the plane $z=v_{F}\tau+\imath x$. The
imaginary part of $\ln z$ lies in the interval $[-\pi,\pi]$ whereas
its real part can take any value, thus achieving the desired mapping
(see Fig.~\ref{fig:App4-The-conformal-transformation}). 

Afterwards, all one has to do is to apply the transformation law for
conformal invariant systems. Indeed, we begin by writing the Matsubara
Green's function of the \ac{1D} antiferromagnet {[}see \cite{Book-Giamarchi-2004,Book-Sachdev-2004}
for a derivation and also Eq.~(\ref{eq:Chap4-MatsubaraGreenFunc}){]}
in an appropriate form, \begin{equation}
\langle M(z)M(0)\rangle_{\infty}\sim\frac{\mathcal{A}}{\left|x-iv_{F}\tau\right|}=\frac{\mathcal{A}}{\sqrt{\left(z\bar{z}\right)}}.\label{eq:App4-Correl_from_Conf-2}\end{equation}
We have used the notation $G(x,\tau)=\langle\hat{T}_{\tau}M(x,\tau)M(0,0)\rangle:=\langle M(z)M(0)\rangle$,
where $M(x,\tau)=e^{\tau H}M(x)e^{-\tau H}$ denotes the staggered
magnetization in the Heisenberg representation with imaginary time.
The theory is critical with conformal weight $\mu=1/2$ {[}Eq.~(\ref{eq:Chap4-CriticalTwoPoint}){]}
and therefore conformal invariance implies the following transformation
law {[}Eq.~(\ref{eq:Chap4-ConformalTransformationLaw}){]}: \begin{equation}
\langle M(w)M(0)\rangle_{strip}=\left|\frac{\partial w}{\partial z}(z)\frac{\partial w}{\partial z}(0)\right|^{-1/2}\langle M(w)M(0)\rangle_{\infty},\label{eq:App4-Correl_from_Conf-3}\end{equation}
with $\langle M(w)M(0)\rangle_{strip}$ being the Matsubara Green's
function defined in the strip geometry corresponding to a finite chain
with periodic boundary conditions. Denoting the time variable by $u$
and the space variable by $r$, a simple calculation yields Eq.~(\ref{eq:Chap4-Matsubara_in_Cylinder}),
namely, \begin{equation}
\langle M(w)M(0)\rangle_{strip}=\frac{2\pi}{\sqrt{2}L}\frac{\mathcal{A}}{\sqrt{\cosh\left(\frac{2\pi\sigma}{L}\right)-\cos\left(\frac{2\pi r}{L}\right)}}.\label{eq:App4-Corr_from_Conf-4}\end{equation}

\begin{figure}[tph]
\noindent \begin{centering}
\includegraphics[width=0.5\columnwidth]{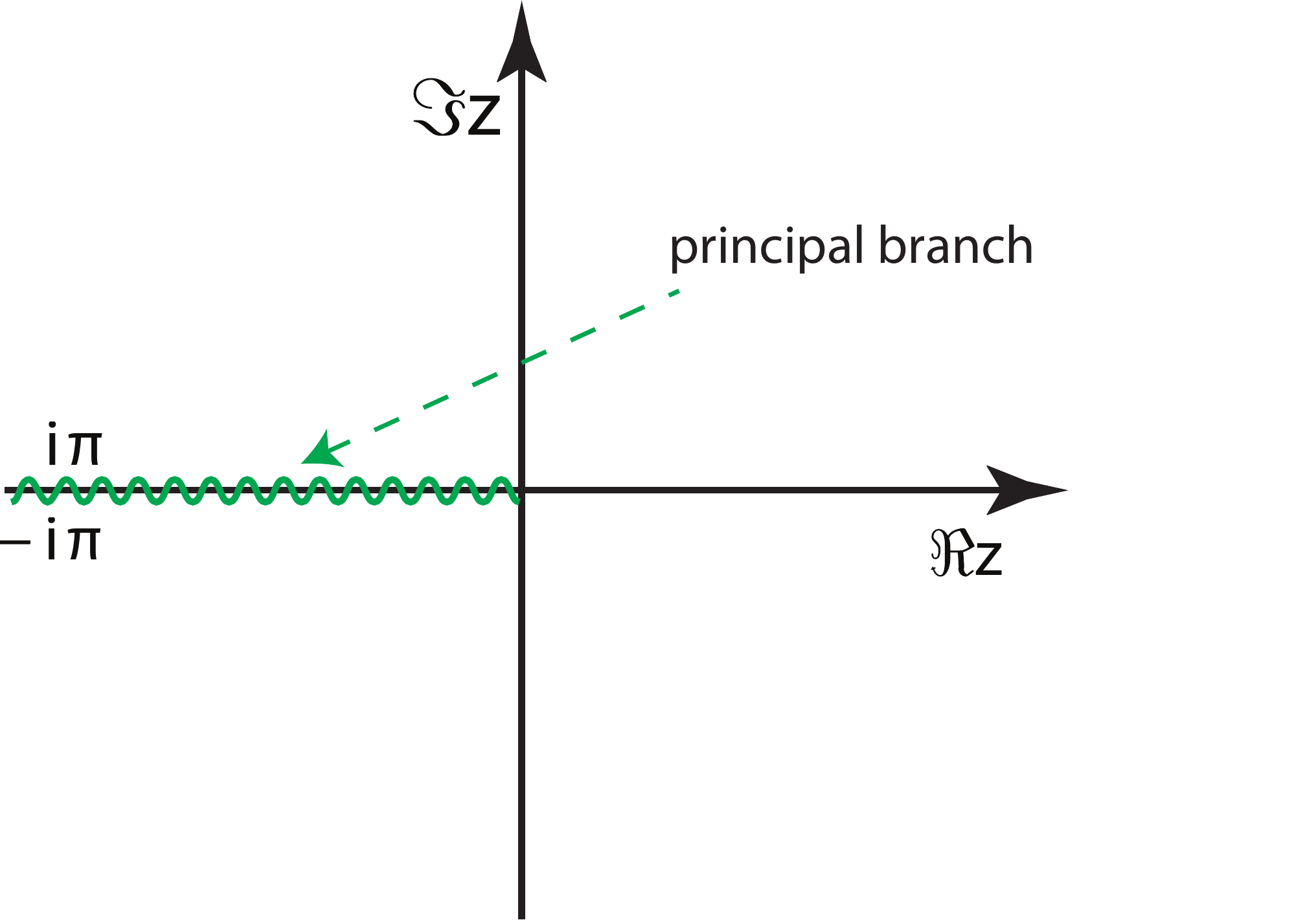}
\par\end{centering}

\caption[The principal branch cut of the logarithm function]{\label{fig:App4-PB_Log}The principal branch cut of the logarithm
function: $\mathbb{C}\setminus\left\{ x+iy|x\le0\wedge y=0\right\} $. }

\end{figure}

Some comments are in order; the latter expression is periodic in the
spatial coordinate ($r$) in accordance with the choice we made in
the space-time labelling {[}Eq.~(\ref{eq:App4-Correl_from_Conf-1}){]}.
The analytic form of the non-universal amplitude $\mathcal{A}$ arising
from the bosonization of the Hamiltonian is unknown, but Eq.~(\ref{eq:App4-Corr_from_Conf-4})
has all the information we need to prove quasi-perfect \ac{LDE} (Sec.~\ref{sub:LDE_HeisenbergAF}).
We also remark that the finite temperature correlator of the infinite
chain can be obtained via conformal invariance if instead of make
a compactification of the spatial variable {[}Eq.~(\ref{eq:App4-Correl_from_Conf-1}){]}
we do it in time. This amounts to make the analytic continuation $L\rightarrow i\beta$,
since now it is the temperature that takes values within a finite
range, one gets:\begin{equation}
\langle M(w)M(0)\rangle_{\beta}=\frac{2\pi}{\sqrt{2}\beta}\frac{\mathcal{A}}{\sqrt{\cosh\left(\frac{2\pi r}{\beta}\right)-\cos\left(\frac{2\pi\sigma}{\beta}\right)}},\label{eq:App4-Correl_from_Conf-5}\end{equation}
recovering the familiar result of statistical mechanics of a exponentially
decaying correlation function $\sim e^{-\pi r/\beta}$ with correlation
length inversely proportional to the temperature $\xi=\beta/\pi$
in the limit $r\gg\beta$. 

We now compute $G^{R}(x,t)$ from Matsubara's Green's function {[}Eq.~(\ref{eq:App4-Corr_from_Conf-4}){]}
via the procedure discussed in the previous Appendix {[}see Eqs.~(\ref{eq:App14-Analytic-15})-(\ref{eq:App14-Analytic-16}){]}.
First, we need the time-ordered Green's function, $G^{T}(x,t)=\langle\hat{T}_{t}M(x,t)M(0,0)\rangle$.
To this end, we Wick rotate the imaginary time variable $\sigma\rightarrow it+0^{+}sign(t)$.
In accordance, we must perform the replacement \begin{eqnarray}
\cosh\left(\frac{2\pi\sigma}{L}\right) & \rightarrow & \cosh\left[\frac{2\pi}{L}\left(it+0^{+}sign(t)\right)\right]\label{eq:App4-Correl_from_Conf-6a}\\
 & = & \cos\left(\frac{2\pi t}{L}\right)-i0^{+}sign(t)\sin\left(\frac{2\pi t}{L}\right)\label{eq:App4-Correl_from_Conf-6}\end{eqnarray}
in Eq.~(\ref{eq:App4-Corr_from_Conf-4}) and take the imaginary part.
It is convenient to recast $G^{R}(x,t)$ into the form \begin{equation}
G^{R}(x,t)=\sqrt{2}\pi(\mathcal{A}/L)e^{-1/2\ln K(t)},\label{eq:App4-Trick}\end{equation}
where we have considered the branch cut of the square-root-function
to be the principal branch cut of the log-function, \emph{i.e.} $\mathbb{C}\setminus\left\{ x+iy|x\le0\wedge y=0\right\} $
(see Fig.~\ref{fig:App4-PB_Log}). With these definitions,  \begin{equation}
K(t)=\cos\left(\frac{2\pi t}{L}\right)-i0^{+}sign(t)\sin\left(\frac{2\pi t}{L}\right)-\cos\left(\frac{2\pi r}{L}\right).\label{eq:App4-Correl_from_Conf-7}\end{equation}
Hence, $G^{T}(x,t)$ has a imaginary part if and only if $\cos\left(2\pi t/L\right)<\cos\left(2\pi r/L\right)$:\begin{eqnarray}
\ln K(t) & = & \ln\left[\left|\cos\left(\frac{2\pi t}{L}\right)-\cos\left(\frac{2\pi r}{L}\right)\right|\right]+\nonumber \\
 &  & +i\pi sign\left[\sin\left(\frac{2\pi t}{L}\right)\right]\theta\left[\cos\left(\frac{2\pi t}{L}\right)-\cos\left(\frac{2\pi r}{L}\right)\right].\label{eq:App4-Correl_from_Conf-8}\end{eqnarray}
From the latter expression and Eq.~(\ref{eq:App14-Analytic-15})
we finally get the retarded Green's function: \begin{equation}
G^{R}(x,t)=-\frac{2\pi}{\sqrt{2}L}\frac{\theta\left[\cos\left(\frac{2\pi t}{L}\right)-\cos\left(\frac{2\pi r}{L}\right)\right]}{\sqrt{\cos\left(\frac{2\pi t}{L}\right)-\cos\left(\frac{2\pi r}{L}\right)}}sign\left[\sin\left(\frac{2\pi t}{L}\right)\right]\theta(t).\label{eq:App4-Correl_from_Conf-9}\end{equation}

\chapter{Appendices for chapter~\ref{cha:LDE_Finite_Temperature}}

\section{The Schrieffer-Wolff canonical transformation\label{sec:The-Schrieffer-Wolff-canonical}}

In Appendix~\ref{sec:App-Degenerate-perturbation-theory} the effective
Hamiltonian between two probes interacting with a larger system was
derived via degenerate perturbation theory. Here we will derive the
same result via a more powerful method, namely the Schrieffer-Wolff
canonical transformation \cite{1966Schrieffer}. This method will
allow us to take in consideration the effect of the probes renormalization
and open the door to compute their partial state in perturbation theory
(Appendix~\ref{sec:RenormalizationProcedure}):

We start by recalling the Hamiltonian of the full system:\begin{equation}
H=H_{0}+V,\label{eq:Full Hamiltonian}\end{equation}
where $H_{0}$ is the many-body Hamiltonian and $V$ describes an
interaction. In Chapters~\ref{cha:LDE_Via_GS_GappedSpinChains} and
\ref{cha:LDE_Finite_Temperature} the interaction corresponds to two
probes, $a$ and $b$, coupling locally to the many-body bulk, but
we make $V$ unspecified for the moment for the sake of generality.
We represent the ground-state wavefunction of the many-body system
(the spin bath) by $|\psi_{0}\rangle$ and conveniently write (\ref{eq:Full Hamiltonian})
in the form,\begin{equation}
H=\bar{H}_{0}+\bar{V},\label{eq:Full Hamiltonian New Form}\end{equation}
where we have introduced the notation $\bar{H}_{0}=H_{0}+\langle\psi_{0}|V|\psi_{0}\rangle$
and $\bar{V}=V-\langle\psi_{0}|V|\psi_{0}\rangle$. Following the
standard condensed-matter approach, we assume that the spectrum of
$H_{0}$ consists of disjoint sectors labeled by the index $i$ in
each of which the spectrum can be either continuous or semicontinuous,
\emph{i.e.}~$|E_{i\mu}-E_{i\upsilon}|\ll|E_{i\alpha}-E_{j\beta}|$.
We denote by $\mathcal{P}_{i}=\sum_{\mu}|\psi_{i,\mu}\rangle\langle\psi_{i,\mu}|$
the projector operator into the eigenstates of $H_{0}$ with energy
$E_{i}$ and suppose that $\bar{V}$ has no matrix elements between
eigenstates in the same sector (this is always the case in the systems
of spins we study in this monograph%
\footnote{To see this is sufficient to take the rotational invariant form of
$V$ {[}Eq.~(\ref{eq:Chap5-Eq1}){]} and take the average in sector
$i$. This yields, $\alpha J\langle\psi_{i}|S_{a}\cdot\tau_{a}+S_{b}\cdot\tau_{b}|\psi_{i}\rangle$.
On the other hand, each of these averages must vanish due to the probe's
full degeneracy at $\alpha=0$, that is $\langle\psi_{i}|\tau_{a(b)}^{\alpha}|\psi_{i}\rangle=0$
{[}see Eqs.~(\ref{eq:App4-DegPerturbTh-e2})-(\ref{eq:App4-DegPerturbTh-e3})
and comments therein{]}. %
}), $\langle\psi_{i}|\bar{V}|\psi_{i}\rangle=0$. With these definitions
in mind we can write,\begin{eqnarray}
\bar{H}_{0} & = & \sum_{i}\mathcal{P}_{i}H\mathcal{P}_{i}\label{eq:Hamiltonian_with_projectors0}\\
\bar{V} & = & \sum_{i\ne j}\mathcal{P}_{i}H\mathcal{P}_{j}\label{eq:Hamiltonian_with_projectors}\end{eqnarray}
The canonical transformation $S$ will change Eq.~(\ref{eq:Full Hamiltonian New Form})
according to,\begin{equation}
e^{-i\epsilon S}He^{i\epsilon S}=\bar{H}_{0}+\epsilon\left(\bar{V}-i\left[S,\bar{H}_{0}\right]\right)+O(\epsilon^{2}),\label{eq:Hamiltonian Canonical Expanded}\end{equation}
where $\epsilon$ is formal expansion parameter that we set equal
to one at the end of the calculation. We fix the generator to be $S=\bar{S}$,
such that,\begin{equation}
\bar{V}+i\left[\bar{H}_{0},\bar{S}\right]=0.\label{eq:Choise of generator}\end{equation}
In the spirit of degenerate perturbation theory (Appendix~\ref{sec:App-Degenerate-perturbation-theory}),
we apply projectors operators to the left and right of the latter
equation. Using relation (\ref{eq:Hamiltonian_with_projectors}) we
find,\begin{eqnarray}
\mathcal{P}_{i}\left(\bar{V}+i\left[\bar{H}_{0},\bar{S}\right]\right)\mathcal{P}_{j} & = & 0\Leftrightarrow\label{eq:Aux1}\\
\Leftrightarrow\mathcal{P}_{i}(\sum_{m\ne n}\mathcal{P}_{m}H\mathcal{P}_{n})\mathcal{P}_{j}(1-\delta_{ij}) & = & i\mathcal{P}_{i}\left[\bar{S},\sum_{m}\mathcal{P}_{m}H\mathcal{P}_{m}\right]\mathcal{P}_{j}\Leftrightarrow\label{eq:Aux2}\\
\Leftrightarrow\mathcal{P}_{i}H\mathcal{P}_{j} & = & -i\left(\mathcal{P}_{i}\bar{S}\mathcal{P}_{j}H\mathcal{P}_{j}-\mathcal{P}_{i}H\mathcal{P}_{i}\bar{S}\mathcal{P}_{j}\right).\label{eq:Aux3}\end{eqnarray}
where in the last step we assumed $i\ne j$. Making the substitution
$\mathcal{P}_{j}H\mathcal{P}_{j}\rightarrow E_{j}\mathcal{P}_{j}$,
with $E_{i\mu}$ approximated by the average energy of sector $E_{i}$,
\emph{i.e.}~$E_{i\mu}\simeq\langle E_{i\mu}\rangle_{\mu}:=E_{i}$,
we finally get,\begin{equation}
i\mathcal{P}_{i}\bar{S}\mathcal{P}_{j}\simeq\frac{\mathcal{P}_{i}\bar{V}\mathcal{P}_{j}}{E_{i}-E_{j}}\left(1-\delta_{ij}\right).\label{eq:Generator_Sandwich}\end{equation}
On the other hand, from the definition of the generator $\bar{S}$
{[}Eq.~(\ref{eq:Choise of generator}){]} we easily choose $\mathcal{P}_{i}\bar{S}\mathcal{P}_{i}=0$
and thus the equation above give us all the non-zero matrix elements
of $\bar{S}$. Indeed, we return to the transformed Hamiltonian {[}Eq.~(\ref{eq:Hamiltonian Canonical Expanded}){]}
under the generator $\bar{S}$,\begin{eqnarray}
H & \rightarrow & e^{-i\bar{S}}He^{i\bar{S}}=\bar{H}_{0}+\underbrace{\bar{V}}-i\left[\underbrace{\bar{S},\bar{H}_{0}}+\bar{V}\right]-\frac{1}{2}\left[\bar{S},\underset{=-i\bar{V}}{\underbrace{\left[\bar{S},\bar{H}_{0}\right]}}\right]+O(\alpha^{3}\Delta^{-2})\label{eq:Aux4}\\
 & = & \bar{H}_{0}-\frac{i}{2}\left[\bar{S},\bar{V}\right]+O(\nu^{3}\Delta^{-2}),\label{eq:Aux5}\end{eqnarray}
where $\nu$ denotes an energy scale of $\bar{V}$ and $\Delta$ a
gap scale of the unperturbed Hamiltonian $H_{0}$. We define the effective
Hamiltonian,\emph{\begin{equation}
H_{S}:=\bar{H}_{0}-\frac{i}{2}\left[\bar{S},\bar{V}\right],\label{eq:Effective Hamiltonian Transformed}\end{equation}
}which, by the virtue of Eq.~(\ref{eq:Generator_Sandwich}), equals
the effective Hamiltonian derived earlier {[}Eq.~(\ref{eq:App4-DegPerturbTh-H(ab)}){]},
when projected onto the many-body lowest energy sector by the action
of $\mathcal{P}_{0}$. 

We are finally in position to specialize to the case of two probes,
$a$ and $b$, that locally couple to a spin bath. The projection
procedure {[}Eq.~(\ref{eq:App4-DegPerturbTh-HamProjection}){]} integrates
out the redundant degrees of freedom, and yields an effective low-dimensional
Hamiltonian describing the physics of the probes as function of $\alpha$
and all the relevant parameters of the condensed-matter bulk. The
probe-bath interaction has the following form \begin{equation}
V=J\alpha\left(\boldsymbol{\tau}_{a}\cdot\boldsymbol{S}_{A}+\boldsymbol{\tau}_{b}\cdot\boldsymbol{S}_{B}\right),\label{eq:ProbesInteracion}\end{equation}
where $\alpha J$ is the coupling strength between the probe qubits
and the lattice, and $J$ a characteristic energy scale of the lattice.
Eq.~(\ref{eq:ProbesInteracion}) describes an isotropic interaction
between the probes {[}with Pauli operators, $\vec{\boldsymbol{\tau}}_{a(b)}${]}
and the bath spins, $\vec{\boldsymbol{S}}_{A(B)}$, at specific lattice
sites $A(B)$. Denoting the probe $a(b)$ space state by $|\chi_{a(b)}^{\alpha}\rangle$
with $\alpha=\left\{ \uparrow,\downarrow\right\} $, the projection
onto the many-body system ground state reads, \begin{eqnarray}
H^{(ab)} & = & \langle\psi_{0}|\mathcal{P}_{o}H_{S}\mathcal{P}_{0}|\psi_{0}\rangle\label{eq:Aux6}\\
 & = & \sum_{\alpha,\beta,\bar{\alpha}\bar{\beta}}|\chi_{\alpha}^{\alpha},\chi_{b}^{\beta}\rangle\langle\chi_{\alpha}^{\alpha},\chi_{b}^{\beta},\psi_{0}|H_{\text{eff}}|\psi_{0},\chi_{\alpha}^{\bar{\alpha}},\chi_{b}^{\bar{\beta}}\rangle\langle\chi_{\alpha}^{\bar{\alpha}},\chi_{b}^{\bar{\beta}}|\label{eq:Aux7}\\
 & = & \sum_{\alpha,\beta,\bar{\alpha}\bar{\beta}}H_{\alpha,\beta,\bar{\alpha},\bar{\beta}}^{(ab)}|\chi_{\alpha}^{\alpha},\chi_{b}^{\beta}\rangle\langle\chi_{\alpha}^{\bar{\alpha}},\chi_{b}^{\bar{\beta}}|.\label{eq:Aux8}\end{eqnarray}
The matrix elements $H_{\alpha,\beta,\bar{\alpha},\bar{\beta}}^{(ab)}$
are given by Eq.~(\ref{eq:Effective Hamiltonian Transformed}) {[}or
equivalently by Eq.~(\ref{eq:App4-DegPerturbTh-H(ab)}){]}. In Appendix~\ref{sec:Time-correlation-functions}
we showed that $H^{(ab)}$ can be written as function of the DC adiabatic
susceptibility for the spins $A$ and $B$ of the bath, $\tilde{\chi}_{ab}^{zz}(0)$.
For $SU(2)$ systems with full rotational symmetry, the form of $H^{(ab)}$
is rather simple and reads \begin{equation}
H_{SU(2)}^{(ab)}=(J\alpha)^{2}\tilde{\chi}_{ab}^{zz}(0)\boldsymbol{\tau}_{a}\cdot\boldsymbol{\tau}_{b},\label{eq:Effective_SU2_Interaction}\end{equation}
where we have set to zero all the constants as they not change the
eigenstates.

\section{The probe operators renormalization\label{sec:RenormalizationProcedure}}

In the previous appendix we derived the effective Hamiltonian of two
probes interacting with a gapped many-body system via the Schrieffer-Wolff
canonical transformation formalism. We obtained the same result of
degenerate perturbation theory, namely Eq.~(\ref{eq:Effective_SU2_Interaction}).
In order to get the low energy of the probes, we projected the effective
Hamiltonian, $H_{S}$, onto the spin bath \ac{GS}: \begin{equation}
H_{\text{eff}}:=\langle\psi_{0}|H_{S}|\psi_{0}\rangle=\underset{>0}{\underbrace{\left(J^{2}\alpha^{2}\tilde{\chi}_{ab}^{zz}(0)+O(\alpha^{3}\Delta^{-2})\right)}}\boldsymbol{\tau}_{a}\cdot\boldsymbol{\tau}_{b},\label{eq:App5-Renormalization-1}\end{equation}
At first sight one concludes that, as long as $\tilde{\chi}_{ab}^{zz}(0)>0$,
the probes form a perfect singlet for sufficiently small $\alpha$.
Indeed, the average value of $\boldsymbol{\tau}_{a}\cdot\boldsymbol{\tau}_{b}$
in the \ac{GS} of the effective Hamiltonian, $|\Psi_{\text{eff}}\rangle$,
reads \begin{equation}
\langle\Psi_{\text{eff}}|\boldsymbol{\tau}_{a}\cdot\boldsymbol{\tau}_{b}|\Psi_{\text{eff}}\rangle=-3=\min_{\phi\in\mathbb{C}^{2}\otimes\mathbb{C}^{2}}\langle\phi|\boldsymbol{\tau}_{a}\cdot\boldsymbol{\tau}_{b}|\phi\rangle.\label{eq:App5-Renormalization-1b}\end{equation}
However, the value of $-3$ must be an approximation to the real average
of $\boldsymbol{\tau}_{a}\cdot\boldsymbol{\tau}_{b}$: \begin{equation}
\langle\Psi_{0}|\boldsymbol{\tau}_{a}\cdot\boldsymbol{\tau}_{b}|\Psi_{0}\rangle\ge-3.\label{eq:App5-Renormalization-1c}\end{equation}
The origin of the apparent incompatibility between Eqs.~(\ref{eq:App5-Renormalization-1b})
and (\ref{eq:App5-Renormalization-1c}) can be comprehended by noting
that wave functions also change according to the canonical transformation
$\hat{S}$:\begin{eqnarray}
H & \longrightarrow & e^{-i\hat{S}}He^{i\hat{S}}=H_{S}\label{eq:App5-Renormalization-1d}\\
|\Psi_{0}\rangle & \longrightarrow & e^{-i\hat{S}}|\Psi_{0}\rangle=|\Psi_{0}^{(S)}\rangle\label{eq:App5-Renormalization-1e}\end{eqnarray}
The above formulae show how operators and wave-functions change according
to $\hat{S}$. Let us inspect how the averages of a generic operator
$A$ look like in both pictures, \begin{equation}
\langle\Psi_{0}|A|\Psi_{0}\rangle=\langle\Psi_{0}^{(S)}|e^{-i\hat{S}}Ae^{i\hat{S}}|\Psi_{0}^{(S)}\rangle\label{eq:App5-Renormalization-2}\end{equation}
Given the perturbative nature of the transformation, we can make following
operator expansion \begin{equation}
e^{-i\hat{S}}Ae^{i\hat{S}}=A-\left[i\hat{S},A\right]+O\left(J^{2}\alpha^{2}\Delta^{-2}\right),\label{eq:App5-Renormalization-3}\end{equation}
and so, \begin{equation}
\langle\Psi_{0}|A|\Psi_{0}\rangle=\langle\Psi_{0}^{(S)}|A|\Psi_{0}^{(S)}\rangle-\langle\Psi_{0}^{(S)}|\left[i\hat{S},A\right]|\Psi_{0}^{(S)}\rangle+O\left(J^{2}\alpha^{2}\Delta^{-2}\right).\label{eq:App5-Renormalization-4}\end{equation}
The second term is already a $J\alpha/\Delta$ correction, so, we
would conclude that, to lowest order in the perturbation, one has
\begin{equation}
\langle\Psi_{0}|\boldsymbol{\tau}_{a}\cdot\boldsymbol{\tau}_{b}|\Psi_{0}\rangle=\langle\Psi_{0}^{(S)}|\boldsymbol{\tau}_{a}\cdot\boldsymbol{\tau}_{b}|\Psi_{0}^{(S)}\rangle+O(J\alpha/\Delta).\label{eq:App5-Renormalization-4b}\end{equation}
In fact, in this particular case, one can show that the first correction
is zero. Since the state $|\Psi_{0}^{(S)}\rangle$ is a $\mathbf{\boldsymbol{\tau}}_{a}$,
$\mathbf{\boldsymbol{\tau}}_{b}$ singlet, it is an eigenstate of
$\boldsymbol{\tau}_{a}\cdot\boldsymbol{\tau}_{b}$ with eigenvalue
$-3$. In that case, the average $\langle\Psi_{0}^{(S)}|\left[i\hat{S},A\right]|\Psi_{0}^{(S)}\rangle$
is trivially zero. Therefore, \begin{equation}
\langle\Psi_{0}|\boldsymbol{\tau}_{a}\cdot\boldsymbol{\tau}_{b}|\Psi_{0}\rangle=-3+O\left(J^{2}\alpha^{2}/\Delta^{2}\right).\label{eq:App5-Renormalization-5}\end{equation}
The value of the negativity will deviate from the value of maximum
entanglement (singlet state) with $O(J^{2}\alpha^{2}/\Delta^{2})$
corrections. This as an important consequence: when computing averages
in the canonical basis we have to properly renormalize the operators,
since, as we have just seen, \begin{equation}
\langle\Psi_{0}|\boldsymbol{\tau}_{a}\cdot\boldsymbol{\tau}_{b}|\Psi_{0}\rangle\neq\langle\Psi_{0}^{(S)}|\boldsymbol{\tau}_{a}\cdot\boldsymbol{\tau}_{b}|\Psi_{0}^{(S)}\rangle.\label{eq:App5-Renormalization-5b}\end{equation}
The correct value of the probes correlation, $\langle\Psi_{0}|\boldsymbol{\tau}_{a}\cdot\boldsymbol{\tau}_{b}|\Psi_{0}\rangle$,
must take into account the way $\boldsymbol{\tau}_{a}\cdot\boldsymbol{\tau}_{b}$
changes due to the canonical transformation:\begin{equation}
\langle\boldsymbol{\tau}_{a}\cdot\boldsymbol{\tau}_{b}\rangle=\langle\Psi_{0}^{(S)}|e^{-i\hat{S}}\boldsymbol{\tau}_{a}\cdot\boldsymbol{\tau}_{b}e^{i\hat{S}}|\Psi_{0}^{(S)}\rangle:=\langle\Psi_{0}^{(S)}|\boldsymbol{\tau}_{a}^{R}\cdot\boldsymbol{\tau}_{b}^{R}|\Psi_{0}^{(S)}\rangle,\label{eq:App5-Renormalization-5c}\end{equation}
where the superscript $R$ means that the operator is properly renormalized
by the action of $\hat{S}$. These observations make it clear that
the quantitative physics of the probes is not captured by the effective
Hamiltonian alone. 

An important conclusion of this observation is that the effective
Hamiltonian cannot be employed directly to derive the partial state
of the probes. We now compute the correct partial state of the probes
and see how it relates with $H_{\text{eff}}$, and especially with
its energy scale $J_{\text{eff}}$ {[}see Eq.~(\ref{eq:Chap4-Heff_heisenberg})
for definition{]}. The full rotational symmetry entails that the probe
density matrix can be written as function of a single invariant,\begin{equation}
\hat{\rho}_{ab}=\frac{1}{3e^{-\beta J_{ab}(\beta)}+e^{3J_{ab}(\beta)}}\exp\left(-\beta J_{ab}(\beta)\boldsymbol{\tau}_{a}\cdot\boldsymbol{\tau}_{b}\right),\label{eq: App5-eq:rho_ab}\end{equation}
where now $\boldsymbol{\tau}_{a}\cdot\boldsymbol{\tau}_{b}$ describes
real spins (not renormalized spins). This allows to parametrize the
correlations between probes as \begin{equation}
\langle\boldsymbol{\tau}_{a}\cdot\boldsymbol{\tau}_{b}\rangle=\frac{e^{-4\beta J_{ab}(\beta)}-1}{e^{-4\beta J_{ab}(\beta)}+1/3}.\label{Eq: App5-eq:tau_atau_b}\end{equation}
This tells us very little for the moment since the temperature dependence
of $J_{ab}(\beta)$ is unknown. On the other hand, by definition,\begin{equation}
\langle\boldsymbol{\tau}_{a}\cdot\boldsymbol{\tau}_{b}\rangle=\frac{\textrm{Tr}\left[e^{-\beta H}\boldsymbol{\tau}_{a}\cdot\boldsymbol{\tau}_{b}\right]}{\textrm{Tr}\left[e^{-\beta H}\right]}\label{eq:App5-Renormalization-8}\end{equation}
and, using the canonical transformation $\hat{S}$, \begin{equation}
\langle\boldsymbol{\tau}_{a}\cdot\boldsymbol{\tau}_{b}\rangle=\frac{\textrm{Tr}\left[e^{-\beta H_{S}}e^{-i\hat{S}}\boldsymbol{\tau}_{a}\cdot\boldsymbol{\tau}_{b}e^{i\hat{S}}\right]}{\textrm{Tr}\left[e^{-\beta H_{S}}\right]}\label{eq:App5-Renormalization-9}\end{equation}
The transformed Hamiltonian is $H_{S}=H_{p}+H_{b}^{'}$; the corresponding
eigenbasis is made of product states {[}Eqs.~(\ref{eq:Chap5-Eq4})-(\ref{eq:Chap5-Eq5}){]}.
Under the assumption that $k_{B}T\ll\Delta(\alpha)$ (\emph{i.e.}
that the temperature is much smaller than the gap to excited states
of the bath), we can limit the trace to the states of the form $|\psi_{0}\rangle\otimes|\chi\rangle$,
where $|\chi\rangle$is any probe state, and $|\psi_{0}\rangle$ is
the non-degenerate ground state of the spin bath. This leads to \begin{equation}
\langle\boldsymbol{\tau}_{a}\cdot\boldsymbol{\tau}_{b}\rangle=\frac{\textrm{Tr}_{p}\left[e^{-\beta H_{p}}\langle\psi_{0}|e^{-i\hat{S}}\boldsymbol{\tau}_{a}\cdot\boldsymbol{\tau}_{b}e^{i\hat{S}}|\psi_{0}\rangle\right]}{\textrm{Tr}_{p}\left[e^{-\beta H_{p}}\right]}\label{eq:tau_atau_n_canon}\end{equation}
where $\textrm{Tr}_{p}(\dots)$ is a trace over probe states only.
Since the operator $\boldsymbol{\tau}_{a}\cdot\boldsymbol{\tau}_{b}$
is diagonal in bath space, this can obviously be written as\begin{eqnarray}
\langle\boldsymbol{\tau}_{a}\cdot\boldsymbol{\tau}_{b}\rangle & = & \frac{\textrm{Tr}_{p}\left[e^{-\beta H_{p}}\sum_{m}\langle\psi_{0}|e^{-i\hat{S}}|\psi_{m}\rangle\boldsymbol{\tau}_{a}\cdot\boldsymbol{\tau}_{b}\langle\psi_{m}|e^{i\hat{S}}|\psi_{0}\rangle\right]}{\textrm{Tr}_{p}\left[e^{-\beta H_{p}}\right]}\label{eq:App5-Renormalization-10}\\
 & = & \frac{\textrm{Tr}_{p}\left[e^{-\beta H_{p}}\sum_{m}A_{m}\boldsymbol{\tau}_{a}\cdot\boldsymbol{\tau}_{b}A_{m}^{\dagger}\right]}{\textrm{Tr}_{p}\left[e^{-\beta H_{p}}\right]}\label{eq:tau_atau_b_am}\end{eqnarray}
where $A_{m}\equiv\langle\phi_{0}|e^{-i\hat{S}}|\phi_{m}\rangle$
is an operator defined in probe space. By symmetry, the operator \begin{equation}
\sum_{m}A_{m}\boldsymbol{\tau}_{a}\cdot\boldsymbol{\tau}_{b}A_{m}^{\dagger}=\langle\psi_{0}|e^{-i\hat{S}}\boldsymbol{\tau}_{a}\cdot\boldsymbol{\tau}_{b}e^{i\hat{S}}|\psi_{0}\rangle\label{eq:App5-DefinitionAm}\end{equation}
must be a scalar in probe space, and, therefore, of the form \begin{equation}
\sum_{m}A_{m}\boldsymbol{\tau}_{a}\cdot\boldsymbol{\tau}_{b}A_{m}^{\dagger}=\eta+\left(1-\Phi\right)\boldsymbol{\tau}_{a}\cdot\boldsymbol{\tau}_{b}\label{eq:eta_phi}\end{equation}
where $\eta$ and $\Phi$ are, by construction, temperature independent
renormalization constants. Since $\textrm{Tr}_{p}\left[\boldsymbol{\tau}_{a}\cdot\boldsymbol{\tau}_{b}\right]=0$,
and \begin{equation}
(\boldsymbol{\tau}_{a}\cdot\boldsymbol{\tau}_{b})^{2}=3-2\boldsymbol{\tau}_{a}\cdot\boldsymbol{\tau}_{b}\label{eq:App5-Renormalization-11}\end{equation}
we obtain\begin{eqnarray}
\textrm{Tr}_{p}\left[\sum_{m}A_{m}\boldsymbol{\tau}_{a}\cdot\boldsymbol{\tau}_{b}A_{m}^{\dagger}\right] & = & 4\eta\label{eq:eta:def}\\
\textrm{Tr}_{p}\left[\sum_{m}\boldsymbol{\tau}_{a}\cdot\boldsymbol{\tau}_{b}A_{m}\boldsymbol{\tau}_{a}\cdot\boldsymbol{\tau}_{b}A_{m}^{\dagger}\right] & = & 12(1-\Phi)\label{eq:phi_def}\end{eqnarray}
With these definitions it is clear that \begin{equation}
\langle\boldsymbol{\tau}_{a}\cdot\boldsymbol{\tau}_{b}\rangle=\eta+(1-\Phi)\langle\boldsymbol{\tau}_{a}\cdot\boldsymbol{\tau}_{b}\rangle_{\mathrm{can}}\label{eq:tau_tau_parameters}\end{equation}
where,\begin{equation}
\langle\boldsymbol{\tau}_{a}\cdot\boldsymbol{\tau}_{b}\rangle_{\mathrm{can}}:=\frac{\textrm{Tr}_{p}\left[e^{-\beta H_{p}}\boldsymbol{\tau}_{a}\cdot\boldsymbol{\tau}_{b}\right]}{\textrm{Tr}_{p}\left[e^{-\beta H_{p}}\right]}=\frac{e^{-\beta J_{can}}-1}{e^{-\beta J_{can}}+1/3}.\label{eq:tau_tau_can}\end{equation}
This looks exactly like the expression above, except that now $J_{\mathrm{can}}$,
unlike $J_{ab}$, is \emph{temperature independent.} So we achieve
a parametrization of $\langle\boldsymbol{\tau}_{a}\cdot\boldsymbol{\tau}_{b}\rangle$
in terms of $3$ temperature independent parameters $J_{\mathrm{can}},$$\Phi$
and $\eta$. This result although being simple has important consequences;
for instance, we see that symmetry implies that $J_{can}$ is in fact
the gap separating the probes singlet and the probes triplet up to
any order%
\footnote{Under the assumption of a low energy sector mapped to a probe singlet,
$|\chi^{s}\rangle$, and probe triplet, $|\chi_{m}^{t}\rangle$, via
the canonical transformation, the gap is given by:\begin{eqnarray*}
\Delta_{ab} & := & \langle\Psi_{0,m}|H|\Psi_{0,m}\rangle-\langle\Psi_{0,s}|H|\Psi_{0,s}\rangle\\
 & = & \langle\chi_{m}^{t}|\langle\psi_{0}|(H_{p}+H_{b}^{\prime})|\psi_{0}\rangle|\chi_{m}^{t}\rangle-\langle\chi^{s}|\langle\psi_{0}|(H_{p}+H_{b}^{\prime})|\psi_{0}\rangle|\chi^{s}\rangle\\
 & = & \langle\chi_{m}^{t}|H_{p}|\chi_{m}^{t}\rangle-\langle\chi_{m}^{s}|H_{p}|\chi_{m}^{s}\rangle:=J_{can}.\end{eqnarray*}
The definition in the last equality is the same found in $\langle\boldsymbol{\tau}_{a}\cdot\boldsymbol{\tau}_{b}\rangle_{\mathrm{can}}$,
namely Eq.~(\ref{eq:tau_tau_can}). This is consistent as the $SU(2)$
symmetry forces $H_{p}$ to have the form,\[
H_{p}=\text{constant}+\gamma\boldsymbol{\tau}_{a}\cdot\boldsymbol{\tau}_{b},\]
yielding a gap of $\Delta_{ab}=4\gamma$. In our notation we have
chosen to express the probe Hamiltonian in terms of the gap, directly,
$J_{can}:=4\gamma$. %
}, for neither a particular form for $\hat{S}$ was adopted, nor an
approximation was made in deriving the latter expression. We thus
can write,\begin{equation}
J_{can}=\Delta_{ab}(\alpha):=E_{triplet}(\alpha)-E_{singlet}(\alpha).\label{eq:App5-Renormalization-ExactGap}\end{equation}
Using equations~\ref{Eq: App5-eq:tau_atau_b}, \ref{eq:tau_tau_parameters}
and \ref{eq:tau_tau_can}, we can express $J_{ab}(\beta)$ explicitly
in these temperature independent parameters:\begin{equation}
J_{ab}(\beta)=\frac{1}{4\beta}\ln\left[\frac{3(\Phi-\eta)+(4-3\Phi-\eta)\exp(\beta J_{can})}{4-\Phi+\eta+(\Phi+\eta/3)\exp(\beta J_{can})}\right].\label{eq:App5-Renormalization-13}\end{equation}
In the following appendix we derive the expression for $\Phi$, and
show that it is of second order in the small parameter $J\alpha/\Delta$,
in resemblance to $J_{can}$ {[}the expression of $J_{can}$ in second
order perturbation theory can be read from Eqs.~(\ref{eq:Effective_SU2_Interaction})
and (\ref{eq:App5-Renormalization-ExactGap}){]}:\begin{equation}
\Phi=O(\frac{\alpha^{2}J^{2}}{\Delta^{2}}),\label{eq:App5-Renormalization-14}\end{equation}
 and that $\eta$ is at most of \emph{fourth order. }

\section{The canonical parameters in perturbation theory\label{sec:The-canonical-parameters}}

According to the discussion of the previous appendix, the correct
probe-probe correlation (and hence their partial state) must be computed
by expressing the renormalized spins $\boldsymbol{\tau}_{a(b)}^{R}$
{[}see Eq.~(\ref{eq:App5-Renormalization-5c}) for definition{]}
in the original basis of the spins,$\left\{ |\uparrow_{a},\uparrow_{b}\rangle,|\uparrow_{a},\downarrow_{b}\rangle,|\downarrow_{a},\uparrow_{b}\rangle,|\downarrow_{a},\downarrow_{b}\rangle\right\} $.
Here, we compute the canonical corrections, $\Phi$ and $\eta$, to
the probe canonical correlation {[}Eq.~(\ref{eq:tau_tau_can}){]}
in perturbation theory. The derivation of the effective canonical
coupling $J_{can}$ was performed in Appendix~\ref{sec:The-Schrieffer-Wolff-canonical}:
\[
J_{can}=4(J\alpha)^{2}\tilde{\chi}_{ab}^{zz}(0)+O(\frac{\alpha^{3}J^{3}}{\Delta^{3}}).\]
We assume the following conditions to hold:

\noindent %
\framebox{\begin{minipage}[t]{1\columnwidth}%
\begin{enumerate}
\item \noindent the temperature is small enough as not to generate real
excitations of the spin bath, \begin{equation}
k_{B}T\ll\Delta\equiv\Delta_{ab}(\alpha=0).\label{eq:Condition_Temp}\end{equation}

\item the probes couple weakly to the spin bath via an isotropic interaction
with strength $\alpha J$, such that\begin{equation}
(J\alpha)^{2}\ll J\Delta,\label{eq:Condition_Alpha}\end{equation}
where $J$ is a typical energy scale for the bath (\emph{e.g.} an
exchange interaction).
\end{enumerate}
\end{minipage}}

\noindent \medskip{}
These conditions allow us to write: \begin{eqnarray}
\langle\boldsymbol{\tau}_{a}\cdot\boldsymbol{\tau}_{b}\rangle_{\beta} & = & \textrm{Tr}\left[\rho_{ab}\boldsymbol{\tau}_{a}\cdot\boldsymbol{\tau}_{b}\right]\label{eq:App5-Canonical-1}\\
 & = & \textrm{Tr}\left[e^{-i\hat{S}}\rho_{ab}e^{i\hat{S}}e^{-i\hat{S}}\boldsymbol{\tau}_{a}\cdot\boldsymbol{\tau}_{b}e^{i\hat{S}}\right]\label{eq:App5-Canonical-2}\\
 & = & \textrm{Tr}\left[\frac{e^{-\beta H_{S}}}{\mathcal{Z}_{ab}}\left(\boldsymbol{\tau}_{a}\cdot\boldsymbol{\tau}_{b}-i\underset{\text{1st order}}{\underbrace{\left[\hat{S},\boldsymbol{\tau}_{a}\cdot\boldsymbol{\tau}_{b}\right]}}-\frac{1}{2}\left[\hat{S},\left[\hat{S},\boldsymbol{\tau}_{a}\cdot\boldsymbol{\tau}_{b}\right]\right]+O(J^{3}\alpha^{3}\Delta^{-2})\right)\right],\label{eq:App5-Canonical-3}\end{eqnarray}
with $\mathcal{Z}_{ab}:=\text{Tr}\left[\exp\left(-\beta H_{S}\right)\right]$.
From the latter result, we can express the $\hat{S}$ renormalized
spins as function of the original spins: \begin{equation}
\boldsymbol{\tau}_{a}\cdot\boldsymbol{\tau}_{b}\underset{\hat{S}}{\longrightarrow}\boldsymbol{\tau}_{a}^{R}\cdot\boldsymbol{\tau}_{b}^{R}=\boldsymbol{\tau}_{a}\cdot\boldsymbol{\tau}_{b}+\boldsymbol{\xi}_{ab}\label{eq:App5-Canonical-3b}\end{equation}
with,\begin{equation}
\boldsymbol{\xi}_{ab}=-\frac{1}{2}\left[\hat{S},\left[\hat{S},\boldsymbol{\tau}_{a}\cdot\boldsymbol{\tau}_{b}\right]\right]+O(J^{3}\alpha^{3}\Delta^{-2}),\label{eq:Correction to Probes Operators}\end{equation}
entailing that the \ac{GS} of the probes will never be a perfect
singlet if $\langle\boldsymbol{\xi}_{ab}\rangle_{\beta}$ is non-negligible.
The trace in Eq.~(\ref{eq:App5-Canonical-3}) can be executed in
two steps: 1. tracing out the bath by considering just the overlap
with the ground-state, which is justified by condition (\ref{eq:Condition_Temp})
and 2. performing a thermal average in the $2\otimes2$ Hilbert space
of the probes. 

\noindent The first order term does not contribute as the generator
$\hat{S}$ has null matrix elements between the ground state, \emph{i.e.\begin{equation}
\langle\psi_{0}|\left[\hat{S},\boldsymbol{\tau}_{a}\cdot\boldsymbol{\tau}_{b}\right]|\psi_{0}\rangle=\left[\langle\psi_{0}|\hat{S}|\psi_{0}\rangle,\boldsymbol{\tau}_{a}\cdot\boldsymbol{\tau}_{b}\right]=0.\label{eq:App5-Canonical-4}\end{equation}
}Then we are left with a zero-order term,\[
C_{0}=\text{Tr}\left[\frac{e^{-\beta H_{S}}}{\mathcal{Z}_{ab}}\boldsymbol{\tau}_{a}\cdot\boldsymbol{\tau}_{b}\right],\]
yielding the canonical correlation, $\langle\tau_{a}\cdot\tau_{b}\rangle_{can}$
, and with a second-order correction $C_{2}$,\begin{equation}
C_{2}=-\frac{1}{2\mathcal{Z}_{ab}}\textrm{Tr}\left\{ e^{-\beta H_{S}}\left[\hat{S},\left[\hat{S},\boldsymbol{\tau}_{a}\cdot\boldsymbol{\tau}_{b}\right]\right]\right\} .\label{eq:Correction_GeneratorForm}\end{equation}
We must have some care in order to evaluate the above thermal average.
Let us reproduce the main steps,\begin{eqnarray}
C_{2} & = & -\frac{1}{2\mathcal{Z}_{ab}}\textrm{Tr}\left[e^{-\beta H_{S}}\left(\hat{S}\hat{S}\boldsymbol{\tau}_{a}\cdot\boldsymbol{\tau}_{b}-2\hat{S}\boldsymbol{\tau}_{a}\cdot\boldsymbol{\tau}_{b}\hat{S}+\boldsymbol{\tau}_{a}\cdot\boldsymbol{\tau}_{b}\hat{S}\hat{S}\right)\right]\label{eq:App5-Canonical-5}\\
 & = & -\frac{1}{\mathcal{Z}_{ab}}\textrm{Tr}\left[e^{-\beta H_{S}}\left(\hat{S}\hat{S}\boldsymbol{\tau}_{a}\cdot\boldsymbol{\tau}_{b}-\hat{S}\boldsymbol{\tau}_{a}\cdot\boldsymbol{\tau}_{b}\hat{S}\right)\right]\label{eq:App5-Canonical-6}\\
 & \simeq & -\frac{1}{\mathcal{Z}_{ab}}\sum_{k\ge0}\textrm{Tr}_{a,b}\left[\langle\psi_{0}|e^{-\beta H_{S}}|\psi_{k}\rangle\langle\psi_{k}|\left(\hat{S}\hat{S}\boldsymbol{\tau}_{a}\cdot\boldsymbol{\tau}_{b}-\hat{S}\boldsymbol{\tau}_{a}\cdot\boldsymbol{\tau}_{b}\hat{S}\right)|\psi_{0}\rangle\right].\label{eq:App5-Canonical-7}\end{eqnarray}
The effective Hamiltonian has no matrix elements between eigenstates
belonging to different sectors (up to the order we are working at),
which simplifies the above summation as only the \ac{GS} contributes
{[}compare with Eqs.~(\ref{eq:App5-Renormalization-9})-(\ref{eq:tau_atau_n_canon}){]}:
\begin{equation}
\langle\psi_{0}|e^{-\beta H_{S}}|\psi_{k}\rangle=e^{-\beta H_{p}}\langle\psi_{0}|e^{-\beta H_{b}^{\prime}}|\psi_{k}\rangle\delta_{0k}.\label{eq:App5-Canonical-7b}\end{equation}
We further get, \begin{eqnarray}
C_{2} & = & -\frac{1}{\mathcal{Z}_{ab}}\text{Tr}_{a,b}\left\{ \langle\psi_{0}|e^{-\beta H_{S}}|\psi_{0}\rangle\langle\psi_{0}|\left(\hat{S}\hat{S}\boldsymbol{\tau}_{a}\cdot\boldsymbol{\tau}_{b}-\hat{S}\boldsymbol{\tau}_{a}\cdot\boldsymbol{\tau}_{b}\hat{S}\right)|\psi_{0}\rangle\right\} \label{eq:App5-Canonical-8}\\
 & = & -\frac{1}{\mathcal{Z}_{ab}}\text{Tr}_{a,b}\left\{ e^{-\beta H_{p}}\langle\psi_{0}|\left(\hat{S}\hat{S}\boldsymbol{\tau}_{a}\cdot\boldsymbol{\tau}_{b}-\hat{S}\boldsymbol{\tau}_{a}\cdot\boldsymbol{\tau}_{b}\hat{S}\right)|\psi_{0}\rangle\right\} \label{eq:App5-Canonical-9}\\
 & := & -\frac{1}{\mathcal{Z}_{ab}}(A_{1}-A_{2}).\label{eq:App5-Canonical-10}\end{eqnarray}
The averages of the quadratic terms $\sim\hat{S}\hat{S}$ must be
done separately as $\hat{S}$ does not commute with the probe's operators
in general,\begin{eqnarray}
A_{1} & = & \sum_{k>0}\textrm{Tr}_{a,b}\left[e^{-\beta H_{p}}\langle\psi_{0}|\hat{S}|\psi_{k}\rangle\langle\psi_{k}|\hat{S}|\psi_{0}\rangle\boldsymbol{\tau}_{a}\cdot\boldsymbol{\tau}_{b}\right]\label{eq:App5-Canonical-11}\\
 & = & \sum_{k>0}\textrm{Tr}_{a,b}\left[e^{-\beta H_{p}}\boldsymbol{S}_{ab}(0,k)\boldsymbol{S}_{ab}(k,0)\boldsymbol{\tau}_{a}\cdot\boldsymbol{\tau}_{b}\right]\label{eq:App5-Canonical-12}\\
A_{2} & = & \sum_{k>0}\textrm{Tr}_{a,b}\left[e^{-\beta H_{p}}\boldsymbol{S}_{ab}(0,k)\boldsymbol{S}_{ab}(k,0)\boldsymbol{\tau}_{a}\cdot\boldsymbol{\tau}_{b}\right]\label{eq:App5-Canonical-13}\\
 & = & \sum_{k>0}\textrm{Tr}_{a,b}\left[e^{-\beta H_{p}}\boldsymbol{S}_{ab}(0,k)\boldsymbol{S}_{ab}(k,0)\boldsymbol{\tau}_{a}\cdot\boldsymbol{\tau}_{b}\right],\label{eq:Corrections}\end{eqnarray}
where we have conveniently introduced the operator:\begin{eqnarray}
\boldsymbol{S}_{ab}(m,n): & = & \langle\psi_{m}|\hat{S}|\psi_{n}\rangle\label{eq:App5-Canonical-14}\\
 & = & \langle\psi_{m}|\mathcal{P}_{m}\hat{S}\mathcal{P}_{n}|\psi_{n}\rangle\label{eq:App5-Canonical-15}\\
 & =J\alpha & \frac{\langle\psi_{m}|\left(\boldsymbol{\tau}_{a}\cdot\boldsymbol{S}_{A}+\boldsymbol{\tau}_{b}\cdot\boldsymbol{S}_{B}\right)|\psi_{n}\rangle}{i(E_{m}-E_{n})}\label{eq:App5-Canonical-16}\\
 & := & \frac{J\alpha}{i\Delta_{mn}}\left(\boldsymbol{\tau}_{a}\cdot\langle\psi_{m}|\boldsymbol{S}_{A}|\psi_{n}\rangle+\boldsymbol{\tau}_{b}\cdot\langle\psi_{m}|\boldsymbol{S}_{B}|\psi_{n}\rangle\right).\label{eq:Correction_aux}\end{eqnarray}
The matrix elements of the operator $\mathcal{P}_{m}\hat{S}\mathcal{P}_{n}$
in second order perturbation theory were derived in Appendix~\ref{sec:The-Schrieffer-Wolff-canonical}
via the Schrieffer-Wolff formalism {[}see Eq.~(\ref{eq:Generator_Sandwich}){]}.
Inserting the last equality in the corrections $A_{1}$ and $A_{2}$,
we find, after some algebra,\begin{equation}
A_{1}-A_{2}=4\sum_{k>0}\frac{|\langle\psi_{0}|\left(\boldsymbol{S}_{A}^{z}-\boldsymbol{S}_{B}^{z}\right)|\psi_{k}\rangle|^{2}}{\Delta_{k0}^{2}}.\label{eq::App5-Canonical-17}\end{equation}
We finally get the desired result\begin{eqnarray}
\langle\boldsymbol{\tau}_{a}\cdot\boldsymbol{\tau}_{b}\rangle & = & \langle\boldsymbol{\tau}_{a}\cdot\boldsymbol{\tau}_{b}\rangle_{can}(1-\Phi)+\eta.\label{eq::App5-Canonical-18}\\
\Phi & := & 4(J\alpha)^{2}\sum_{k>0}\frac{|\langle\psi_{0}|\left(\boldsymbol{S}_{A}^{z}-\boldsymbol{S}_{B}^{z}\right)|\psi_{k}\rangle|^{2}}{\Delta_{k0}^{2}},\label{eq::App5-Canonical-19}\end{eqnarray}
with $\langle\boldsymbol{\tau}_{a}\cdot\boldsymbol{\tau}_{b}\rangle_{\mathrm{can}}$
given previously in Eqs.~(\ref{eq:tau_tau_parameters})-(\ref{eq:tau_tau_can}),
and where $\eta=0$ up to second-order perturbation theory since no
constant term has emerged from our expansion. In fact this term is
at most of fourth order, $O[(J\alpha/\Delta)^{4}]$. Let us show this
result more carefully. 

We take the definition of $\eta$ {[}Eq.~(\ref{eq:eta:def}){]} and
Eq.~(\ref{eq:App5-DefinitionAm}) to get:\begin{eqnarray*}
4\eta & = & \frac{1}{2}\textrm{Tr}_{p}\langle\psi_{0}|\left[i\hat{S},\left[i\hat{S},\boldsymbol{\tau}_{a}\cdot\boldsymbol{\tau}_{b}\right]\right]|\psi_{0}\rangle\\
 & = & -\frac{1}{2}\textrm{Tr}_{p}\left[\langle\psi_{0}|\hat{S}^{2}|\psi_{0}\rangle\boldsymbol{\tau}_{a}\cdot\boldsymbol{\tau}_{b}+\boldsymbol{\tau}_{a}\cdot\boldsymbol{\tau}_{b}\langle\psi_{0}|\hat{S}^{2}|\psi_{0}\rangle\right.\\
 &  & \left.-2\sum_{m}\langle\psi_{0}|\hat{S}|\psi_{m}\rangle\boldsymbol{\tau}_{a}\cdot\boldsymbol{\tau}_{b}\langle\psi_{m}|\hat{S}|\psi_{0}\rangle\right]\\
 & = & -\frac{1}{2}\textrm{Tr}_{p}\left[\langle\psi_{0}|\hat{S}|\psi_{m}\rangle\langle\psi_{m}|\hat{S}|\psi_{0}\rangle\boldsymbol{\tau}_{a}\cdot\boldsymbol{\tau}_{b}\right.\\
 &  & +\boldsymbol{\tau}_{a}\cdot\boldsymbol{\tau}_{b}\langle\psi_{0}|\hat{S}|\psi_{m}\rangle\langle\psi_{m}|\hat{S}|\psi_{0}\rangle\\
 & - & \left.2\sum_{m}\langle\psi_{0}|\hat{S}|\psi_{m}\rangle\boldsymbol{\tau}_{a}\cdot\boldsymbol{\tau}_{b}\langle\psi_{m}|\hat{S}|\psi_{0}\rangle\right]\end{eqnarray*}
denoting $S_{m}=\langle\phi_{0}|\hat{S}|\phi_{m}\rangle$ (a probe
operator) we recognize\begin{equation}
4\eta=-\frac{1}{2}\sum_{m}\textrm{Tr}_{p}\left[S_{m}S_{m}^{\dagger}\boldsymbol{\tau}_{a}\cdot\boldsymbol{\tau}_{b}+\boldsymbol{\tau}_{a}\cdot\boldsymbol{\tau}_{b}S_{m}S_{m}^{\dagger}-2S_{m}\boldsymbol{\tau}_{a}\cdot\boldsymbol{\tau}_{b}S_{m}^{\dagger}\right]\label{eq:App5-Canonical-20}\end{equation}
Using the cyclic invariance of the trace, this reduces to \begin{equation}
4\eta=-\sum_{m}\textrm{Tr}_{p}\left[S_{m}S_{m}^{\dagger}\boldsymbol{\tau}_{a}\cdot\boldsymbol{\tau}_{b}-S_{m}\boldsymbol{\tau}_{a}\cdot\boldsymbol{\tau}_{b}S_{m}^{\dagger}\right]\label{eq:App5-Canonical-21}\end{equation}
To proceed, we must specify the operator $S_{m}$, \begin{eqnarray}
S_{m} & = & \frac{i\alpha}{\Delta_{m0}}\langle\psi_{0}|\left[\boldsymbol{\tau}_{a}\cdot\mathbf{S}_{A}+\boldsymbol{\tau}_{b}\cdot\mathbf{S}_{B}\right]|\psi_{m}\rangle\label{eq:App5-Canonical-22}\\
 & = & i\alpha\left(\boldsymbol{\tau}_{a}\cdot\mathbf{u}_{ma}+\boldsymbol{\tau}_{b}\cdot\mathbf{u}_{mb}\right)\label{eq:App5-Canonical-23}\end{eqnarray}
with $\mathbf{u}_{ma(b)}$ (a c-number in probe space) is defined
as \begin{equation}
\mathbf{u}_{ma(b)}=\frac{1}{\Delta_{m0}}\langle\psi_{0}|\mathbf{S}_{A(B)}|\psi_{m}\rangle.\label{eq:App5-Canonical-24}\end{equation}
We arrive at\begin{eqnarray}
4\eta & = & -\alpha^{2}\sum_{m}\textrm{Tr}_{p}\left[\left(\boldsymbol{\tau}_{a}\cdot\mathbf{u}_{ma}+\boldsymbol{\tau}_{b}\cdot\mathbf{u}_{mb}\right)\left(\boldsymbol{\tau}_{a}\cdot\mathbf{u}_{ma}^{*}+\boldsymbol{\tau}_{b}\cdot\mathbf{u}_{mb}^{*}\right)\boldsymbol{\tau}_{a}\cdot\boldsymbol{\tau}_{b}\right.\nonumber \\
 & - & \left.\left(\boldsymbol{\tau}_{a}\cdot\mathbf{u}_{ma}+\boldsymbol{\tau}_{b}\cdot\mathbf{u}_{mb}\right)\boldsymbol{\tau}_{a}\cdot\boldsymbol{\tau}_{b}\left(\boldsymbol{\tau}_{a}\cdot\mathbf{u}_{ma}^{*}+\boldsymbol{\tau}_{b}\cdot\mathbf{u}_{mb}^{*}\right)\right]\label{eq:dad}\end{eqnarray}
At this point we can again use the cyclic invariance of the trace
on the $\boldsymbol{\tau}$ operators, and obtain \begin{eqnarray}
4\eta & = & -\alpha^{2}\sum_{m}\textrm{Tr}_{p}\left[(\tau_{a}^{i}\tau_{a}^{j}-\tau_{a}^{j}\tau_{a}^{i})u_{ma}^{i}(u_{ma}^{*})^{j}\right.\nonumber \\
 & + & \left.(\tau_{b}^{i}\tau_{b}^{j}-\tau_{b}^{j}\tau_{b}^{i})u_{mb}^{i}(u_{mb}^{*})^{j}\right]\boldsymbol{\tau}_{a}\cdot\boldsymbol{\tau}_{b}=0\label{eq:dad2}\end{eqnarray}
The cross terms in $\boldsymbol{\tau}_{a}$ and $\boldsymbol{\tau}_{b}$
are zero because the corresponding operators commute; in this form,
the trace over probe space kills this expression because there is
always an alone $\tau_{a}^{i}$ or $\tau_{b}^{i}$ operator in all
terms.

\chapter*{List of publications by the author}

\mychaptermark{List of publications by the author}
\begin{itemize}
\item \emph{Emergence of robust gaps in 2D antiferromagnets via additional
spin-$1/2$ probes}, AIRES FERREIRA, J. Viana Lopes, and J. M. B.
Lopes dos Santos, submitted to PRL (2009).
\item \emph{Production of bright entangled photons from optical moving boundaries},
A. Guerreiro, AIRES FERREIRA, and J. T. Mendonça. arXiv: 0906.0522,
to be submitted (2009).
\item \emph{Analytical results in long distance entanglement mediated by
gapped spin chains}, AIRES FERREIRA, and J. M. B. Lopes dos Santos,
Phys. Rev. A \textbf{77}, 034301 (2008).
\item \emph{Optomechanical entanglement between a movable mirror and a cavity
field, }D. Vitali, S. Gigan, AIRES FERREIRA, H. R. Bohm, P. Tombesi,
A. Guerreiro, V. Vedral, A. Zeilinger, and M. Aspelmeyer, Phys. Rev.
Lett. \textbf{98}, 030405 (2007).
\begin{onehalfspace}
\item \emph{Macroscopic thermal entanglement due to radiation pressure},
AIRES FERREIRA, A. Guerreiro, and V. Vedral, Phys. Rev. Lett. \textbf{96},
060407 (2006).\end{onehalfspace}
\end{itemize}
\begin{lyxcode}

\end{lyxcode}
\begin{onehalfspace}
\bibliographystyle{apsrev}
\cleardoublepage\addcontentsline{toc}{chapter}{\bibname}\bibliography{bibliography,bibliography-part2}
\end{onehalfspace}

\end{document}